\documentclass[11pt,a4paper]{article}
\usepackage[dvipdfmx]{graphicx} 
\usepackage{bmpsize} 
\usepackage{jheppub}

\usepackage[utf8]{inputenc}

\usepackage{bm}

\usepackage{subcaption}
\usepackage{textcomp}

\makeatletter

\@ifundefined{textcolor}{}
{
	\definecolor{BLACK}{gray}{0}
	\definecolor{WHITE}{gray}{1}
	\definecolor{RED}{rgb}{1,0,0}
	\definecolor{GREEN}{rgb}{0,1,0}
	\definecolor{BLUE}{rgb}{0,0,1}
	\definecolor{CYAN}{cmyk}{1,0,0,0}
	\definecolor{MAGENTA}{cmyk}{0,1,0,0}
	\definecolor{YELLOW}{cmyk}{0,0,1,0}
}
\makeatletter
\def\simgt{\mathrel{\lower2.5pt\vbox{\lineskip=0pt\baselineskip=0pt
			\hbox{$>$}\hbox{$\sim$}}}}
\def\simlt{\mathrel{\lower2.5pt\vbox{\lineskip=0pt\baselineskip=0pt
			\hbox{$<$}\hbox{$\sim$}}}}

\makeatother
\numberwithin{equation}{section}

\def\sect#1{Sec.~{\ref{#1}}}
\def\sects#1#2{Secs.~{\ref{#1}} and~{\ref{#2}}}
\def\app#1{Appendix~{\ref{#1}}}

\def\tab#1{Table~{\ref{#1}}}

\def\fig#1{Fig.~\ref{#1}}
\def\Fig#1{Fig.~\ref{#1}}
\newcommand{\eq}[2]{\be\begin{aligned}#1 \label{#2}\end{aligned}\ee}
\newcommand{\Ref}[1]{Ref.~\cite{#1}}
\newcommand{\Eq}[1]{Eq.~\eqref{#1}}

\newcommand{\Sec}[1]{Sec.~\ref{#1}}

\def\nn{\nonumber}
\def\spa#1.#2{\left\langle#1\,#2\right\rangle}
\def\spb#1.#2{\left[#1\,#2\right]}
\def\sand#1.#2.#3{%
\left\langle#1{\vphantom1}\right|{#2}\left|#3\right]}%
\def\sandmp#1.#2.#3{%
\left\langle#1{\vphantom1}\right|{#2}\left|#3\right]}%
\def\sandpm#1.#2.#3{%
\left[#1{\vphantom1}\right|{#2}\left|#3\right\rangle}%
\def\sandmm#1.#2.#3{%
\left\langle#1{\vphantom1}\right|{#2}\left|#3\right\rangle}%
\def\sandpp#1.#2.#3{%
\left[#1{\vphantom1}\right|{#2}\left|#3\right]}%
\def\sandmx#1.#2.#3{%
\left\langle#1{\vphantom1}\right|{#2}\left|#3\right]}%

\def\eps{\epsilon}
\def\Ord{{\cal O}}
\def\Tr{{\rm Tr}}
\def\tr{{\rm tr}}
\def\tree{{\rm tree}}
\def\id{\protect{{1 \kern-.28em {\rm l}}}}
\def\pol{\varepsilon}

\def\oneloop{{\rm 1\hbox{-}loop}}
\def\pol{\varepsilon}
\newcommand{\eqn}[1]{Eq.~\eqref{#1}}
\newcommand{\eqns}[2]{Eqs.~(\ref{#1}) and~(\ref{#2})}

\def\f{\tilde f}
\def\t{\tau}
\def\pp{\sigma}

\newbox\charbox
\newbox\slabox
\def\s#1{{      
        \setbox\charbox=\hbox{$#1$}
        \setbox\slabox=\hbox{$/$}
        \dimen\charbox=\ht\slabox
        \advance\dimen\charbox by -\dp\slabox
        \advance\dimen\charbox by -\ht\charbox
        \advance\dimen\charbox by \dp\charbox
        \divide\dimen\charbox by 2
        \raise-\dimen\charbox\hbox to \wd\charbox{\hss/\hss}
        \llap{$#1$} }}

\newcommand{\be}{\begin{equation}}
	\newcommand{\ee}{\end{equation}}

\newcommand{\colorA}[1]{\mathbb{#1}}

\def\log{\ln}

\usepackage{array}   
\newcolumntype{L}{>{$}l<{$}}

\allowdisplaybreaks
\title{Black Hole Binary Dynamics from the Double Copy and Effective Theory}
\author[a,b]{Zvi Bern,}
\author[c]{Clifford Cheung,}
\author[d]{Radu Roiban,}
\author[a]{Chia-Hsien Shen,}
\author[c]{Mikhail P.\ Solon,}
\author[e]{Mao Zeng}

\affiliation[a]{Mani L. Bhaumik Institute for Theoretical Physics, Department of Physics and Astronomy, UCLA, Los Angeles, USA}
\affiliation[b]{Theoretical Physics Department, CERN, 1211 Geneva 23, Switzerland}
\affiliation[c]{Walter Burke Institute for Theoretical Physics, California Institute of Technology, Pasadena, USA}
\affiliation[d]{Institute for Gravitation and the Cosmos, Pennsylvania State University, University Park, PA 16802, USA}
\affiliation[e]{Institute for Theoretical Physics, ETH Z\"urich, 8093 Z\"urich, Switzerland}

\emailAdd{bern@physics.ucla.edu}
\emailAdd{clifford.cheung@caltech.edu}
\emailAdd{radu@phys.psu.edu}
\emailAdd{chshen@physics.ucla.edu}
\emailAdd{mpsolon@gmail.com}
\emailAdd{mzeng@phys.ethz.ch}

\preprint{CERN-TH-2019-128, CALT-TH 2019-026, UCLA/TEP/2019/103}
\arxivnumber{1908.01493}

\abstract{We describe a systematic framework for computing the conservative
potential of a compact binary system using modern tools from scattering
amplitudes and effective field theory. Our approach combines methods for
integration and matching adapted from effective field theory,
generalized unitarity, and the double-copy construction, which relates
gravity integrands to simpler gauge-theory expressions. With these methods we 
derive the third post-Minkowskian correction to the conservative
two-body Hamiltonian for spinless black holes.  We describe
in some detail various checks of our integration methods and the resulting Hamiltonian.}

\begin{document}
\maketitle

\section{Introduction}
\label{sec:intro}

The extraordinary detection of gravitational waves by the LIGO and
Virgo collaborations~\cite{1602.03837, 1710.05832} has opened a new window into the
cosmos. Gravitational wave astronomy is now a critical tool for
answering longstanding questions in astronomy and cosmology, and
offers a test of gravity in violent environments never before probed
by experiment.
 
The LIGO and Virgo detectors boast an exquisite precision which will
grow in future upgrades, thus demanding commensurately accurate
theoretical predictions encoded in waveform templates utilized for
detection and extraction of source parameters.  These waveforms are
constructed from an array of complementary approaches, including the
effective one-body (EOB) formalism \cite{gr-qc/9811091, gr-qc/0001013}, numerical
relativity~\cite{gr-qc/0507014, gr-qc/0511048, gr-qc/0511103}, the self-force formalism~\cite{gr-qc/9606018, gr-qc/9610053}, and
a number of perturbative methods for the inspiral phase, including the
post-Newtonian (PN)~\cite{Droste,EIH} and post-Minkowskian
(PM)~\cite{PM1, PM2, PM3, PM4, PM5, PM6, PM7, PM8, Westpfahl2PM, Damour:2016gwp, DamourTwoLoop}
approximations, as well as the nonrelativistic general relativity
(NRGR) formalism~\cite{NRGR} based on effective field theory
(EFT). For recent reviews see Refs.~\cite{1310.1528, 1601.04914, 1805.07240, 1805.10385, 1806.05195, 1807.01699} and
references therein.  In the coming years, further improvements in
high-precision theoretical predictions from general relativity will be
essential given expected improvements in detector sensitivity.

\begin{figure}
\begin{center}

\includegraphics[width=\textwidth]{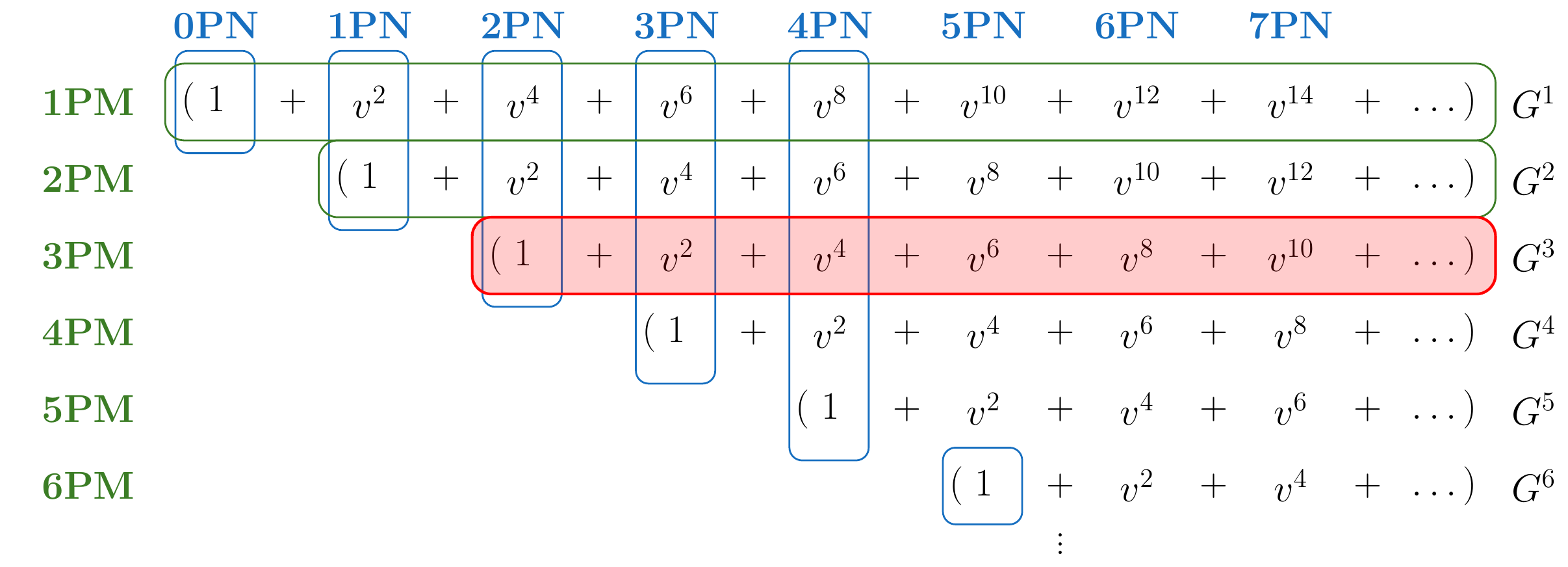}
\end{center}
\vspace*{-0.5cm}
\caption{\small A summary of known results for the two-body potential for 
spinless black holes in the PN and PM expansions, outlined in blue and 
green regions respectively. The new 3PM result summarized in \Ref{3PMPRL} 
and discussed at length in this paper is highlighted in the shaded (red) region. 
The overlap gives strong crosschecks on any calculations in either approach.}
\label{fig:PN_PM}
\end{figure}

During the early inspiral phase, the gravitational field is weak and
the constituents of binary black hole system are non-relativistic.  In the PN
approximation, we organize the interaction Hamiltonian as an expansion in
\eq{
\bm v^2 \sim \frac{Gm}{|\bm r|} \ll 1 \,,
}{} 
where $G$ is Newton's constant and $m$ is the total mass of the
binary system, while $\bm v$ and $\bm r$ are the relative velocity and
position between the black holes in units with $c=1$. The PN expansion
is a double expansion in the velocity squared and the inverse separation
in units of the Schwarzschild radius, which are of order each other
due to the virial theorem. The powers in $G$ and $\bm v$ corresponding
to each PN order are depicted in \fig{fig:PN_PM}.

In the present work we focus on conservative dynamics.  
The PN perturbative framework is well-established
with a long history dating back to the leading 1PN correction to the
Newtonian gravitational potential, which was computed by Einstein,
Infeld, and Hoffman~\cite{EIH}.  Later pioneering work derived the
full 2PN~\cite{2PNOhta}, 3PN~\cite{Jaranowski:1997ky, gr-qc/9912092, gr-qc/0004009, gr-qc/0105038},
and 4PN~\cite{JaranowskSchafer1,JaranowskSchafer2, Bernard:2015njp, Marchand:2017pir, Foffa:2012rn, Foffa:2016rgu, Foffa:2019rdf, arXiv:1703.06433, arXiv:1903.05118}
expressions for the conservative potential.  More recently, 5PN static
contributions have also been computed~\cite{1902.10571, 1902.11180}.

In contrast, the PM expansion is organized differently, 
including instead contributions to all orders in velocity at fixed order in
$G$. So when we refer to the $n$PM correction, we refer to a
contribution which, when expanded in $\bm v$, generates {\it all} PN
terms at order $G^n$. In particular, when expanded to ${\bm v}^6$
order, the 3PM result gives a previously unknown contribution to the
5PN potential.  The powers in $G$ and $\bm v$ corresponding to each PM
order are shown \fig{fig:PN_PM}. This expansion has recently received
new attention~\cite{Damour:2016gwp, 1805.00813, 1806.08347,CachazoGuevara, 1709.00590, 1709.06016, 1805.10809, 1812.06895, 1812.00956, 1812.08752, 1906.09260, 1906.10071, Plefka,
CliffIraMikhailClassical, 3PMPRL, Paolo2PM, Cristofoli:2019neg,
1905.05657, 1906.05209}, based in part on the connection of classical
physics to quantum scattering amplitudes~\cite{Iwasaki:1971vb, Iwasaki:1971iy, Gupta:1979br, gr-qc/9405057, hep-th/0405239,
RothsteinClassical, Vaidya:2014kza,OConnellObservables}.
Relativistic scattering amplitudes are naturally organized as a series
in powers of the coupling $G$, keeping all orders in the velocity, and
for this purpose we define the PM potential to be
\eq{
V(\bm p, \bm r) &= \sum_{n=1}^\infty \left(\frac{G}{|\bm r|}\right)^n
c_n (\bm p^2) \, ,
}{eq:VPR} 
where the coefficients $c_n$ are functions of $\bm p \sim \bm v$ which
contain arbitrarily high powers in the velocity.  Of course, whether
the new information in PM dynamics can be directly used to improve
gravitational wave templates for inspiraling binary systems requires detailed study, e.g. along the lines of
Ref.~\cite{Antonelli:2019ytb}. Nonetheless, at the very least, as can
be seen from \fig{fig:PN_PM} the PM approximation is complementary to
the PN approximation, providing results for a subset of terms at each
PN order.

The primary goal of this paper is to develop efficient methods for
high-precision predictions of the dynamics of gravitationally bound
compact objects. By using scattering amplitudes as the starting point,
we take advantage of the enormous progress in the past decade for
computing and understanding them in gravitational theories, with
systematically improvable precision.  This includes applying
generalized unitarity~\cite{hep-ph/9403226, hep-ph/9409265, hep-ph/9708239, hep-th/0412103, 0705.1864, hep-ph/9602280, 1103.1869, 1103.3298} and
double-copy constructions~\cite{KLT, BCJ, BCJLoop}, which have enabled
explicit (super)gravity calculations at remarkably high orders of perturbation
theory~\cite{SimplifyingBCJ, 0905.2326, SimplifyingBCJ, 1309.2498, 1409.3089, 1708.06807, 1804.09311, 1507.06118, 1701.02422}. The
double copy allows us to express gravitational scattering amplitudes
in terms of corresponding simpler gauge theory amplitudes, while
generalized unitarity gives a means for building loop amplitudes from
simpler tree amplitudes. As we shall see, these can also be combined
with spinor-helicity methods~\cite{Berends:1981rb, Berends:1981uq, Xu:1986xb} which then yield amazingly
compact expressions for unitarity cuts that contain all information
required to build the classical potential at 3PM.

The central idea in relating scattering amplitudes to the orbital
dynamics of compact binaries is that both processes are governed by
the same underlying theory.  By construction, the effective two-body
potential in \eqn{eq:VPR} reproduces the same physics as the full
gravitational theory for kinematics defined by massive bodies
interacting via a classical long-range force. We can therefore extract
the effective potential from scattering amplitudes, which are
convenient to calculate using modern field theory tools.  This was
demonstrated long ago~\cite{Iwasaki:1971vb, Iwasaki:1971iy, Gupta:1979br, gr-qc/9405057, hep-th/0405239}, and recently revived
using EOB~\cite{DamourTwoLoop} and
EFT~\cite{RothsteinClassical,Vaidya:2014kza} methods, including those
that incorporate corrections to all orders in velocity. The EFT
approach has led to new results for the PM potential at higher
orders~\cite{CliffIraMikhailClassical, 3PMPRL}.  As we shall see, EFT
methods are not only useful for systematically mapping scattering
amplitudes to classical potentials but also for efficiently dealing
with the integrals encountered in the full theory~\cite{CliffIraMikhailClassical}.

The combination of these key ingredients from the modern amplitudes
program and effective field theory led to a remarkably compact
expression for the classical 3PM conservative two-body
potential~\cite{3PMPRL}. This result is state of the art. As shown
in \fig{fig:PN_PM}, it provides new information not obtained
previously by PN or effective one-body methods, and a strong
independent crosscheck of known terms in the PN
expansion. Furthermore, these results have already been examined by
LIGO theorists~\cite{Antonelli:2019ytb} and compared against lower
order PM calculations, numerical relativity, and various effective
one-body models.

The present paper is a companion to the Letter~\cite{3PMPRL}
summarizing our results for the 3PM conservative Hamiltonian. Our aim
is to fill in the various technical details, and the analysis will be
divided into several parts. First, we introduce the basic tools for
extracting classical potentials from quantum scattering
amplitudes. Quantum gravitational amplitudes encode in principle the
physics of both bound quasi-elliptic orbits and unbound quasi-hyperbolic orbits in
a relativistic manner that is well suited for the PM expansion. They
are however quite complicated, and important simplifications arise by
truncating away quantum contributions as early as possible in the
calculation. In \sect{QuantumVsClassicalSection}, we discuss the
kinematics, hierarchies of scales, and power counting that allow us to
identify a precise demarcation between classical and quantum
contributions to the scattering amplitude at the integrand level. The
latter distinction is crucial for the scalability of the method to
higher-loop orders since integration of full quantum integrands prior
to classical expansion is not viable with current technology.

Second, we use the double copy and generalized unitarity to obtain the
relativistic integrands relevant for one- and two-loop classical
scattering of massive gravitationally interacting scalars.  As
discussed in \sect{IntegrandsSection}, the starting point of our
construction are remarkably compact four- and five-point gauge-theory
tree-level scattering amplitudes. Using double-copy methods, these are
converted to appropriate tree-level gravitational scattering
amplitudes, which are then combined into generalized unitarity cuts in
order to build loop integrands.  By identifying terms that cannot
contribute to the classical potential, we are able to vastly reduce
the complexity of these expressions. Details of the construction using
various helicity and double-copy methods as well as explicit results
at one and two loops are given
in \sect{OneLoopSection}, \sect{OneLoopDCutsSection},
and \sect{TwoLoopSection}.

Third, to obtain the parts of the scattering amplitudes needed for
building the classical potential, we integrate the relativistic
integrands via an assortment of old and newly developed tools. We
consider both nonrelativistic and relativistic methods of integration,
which are discussed in \sect{sec:NRint} and \sect{sec:Rint},
respectively. The former approach is an adaptation of the method of
regions~\cite{Beneke:1997zp, Smirnov:2004ym} and mimics the mechanics
of NRGR~\cite{NRGR} in that integration occurs via a reduction to
three-dimensional bubble integrals. While this method obscures
relativistic covariance, it is very efficient and by design scalable
to high loop order.  The latter approach includes the methods of
differential equations and Mellin-Barnes integration which produce
exact results to all orders in velocity for certain diagram
topologies. In \sect{sec:scattering} we give the integrated answers
for the contributing diagrams, discuss the resummation of the
results from nonrelativistic integration, and present the final 
amplitude containing all contributions to the 3PM potential in \Eq{eq:PM_amp}

Fourth, in \sect{sec:EFT} we use effective field theory to extract the
classical conservative potential from the resulting scattering
amplitude.  This procedure systematically implements the subtraction
of infrared divergent iterated contributions in the amplitude, leaving
behind the desired new contribution to the potential.  
The 3PM coefficients for the classical potential \eqref{eq:VPR} are given in \Eq{eq:HamiltonianCoefficients}.
A convenient
byproduct of the matching is that we can choose a frame in which the
potential is in a much more compact form compared to previous
expressions.

Lastly, in \sect{sec:checks} we validate our result through various checks
against the existing literature. In the probe limit, our result
reduces to the known potential from the Schwarzschild solution. As
shown in \fig{fig:PN_PM}, our 3PM Hamiltonian overlaps with the known
4PN result~\cite{JaranowskSchafer2}, and we confirm their physical
equivalence by providing the canonical transformation that maps the
result of Ref.~\cite{JaranowskSchafer2} to ours. We also compare
results for scattering amplitudes and scattering angles
computed from classical potentials. In \sect{DiscussionSection}, we discuss various features and
subtleties.  This includes the appearance of a mass singularity in the
3PM two-body potential, that four-dimensional constructions of the
integrands are sufficient through 3PM order despite using dimensional
regularization, and the lack of contributions from radiation modes to
the conservative potential through 3PM order.

We conclude in \sect{sec:conclusion}, and provide several
appendices. In \app{appendix:notation}, we collect notation used in
the paper.  In \app{TreeAmplitudesAppendix} we provide the
gauge-theory amplitudes that are necessary for constructing the
gravitational amplitudes. In \app{appendix:series}, we collect the
series that appear in our nonrelativistic integration and their
resummation. In \app{HxH} we extract the classical limit of a two-loop
Feynman integral by starting from the fully integrated result in
Ref.~\cite{Bianchi:2016yiq}, and then take the limit in the final
expression.  This evaluation matches the results obtained with our
methods, confirming the presence of the mass singularity.  It also
displays an imaginary part, connected to the presence of on-shell
radiation, which does not contribute to the 3PM conservative
potential.

\section{Classical Versus Quantum}
\label{QuantumVsClassicalSection}

The goal of this section is to introduce the basic ideas for
efficiently identifying the parts of quantum scattering amplitudes
that contribute to the classical potential. We discuss the kinematics,
scale hierarchies, power counting, and truncation of graph structures
that allow us to drop quantum contributions at the integrand
level. This leads to enormous simplifications that are crucial for the
scalability to high loop orders.

\subsection{External Matter Kinematics}
\label{sec:matter_kinematics}
Gravitationally interacting spinless compact bodies with masses $m_1$ and $m_2$ can be described 
by a system of two real scalar fields $\phi_1$ and $\phi_2$
minimally coupled to gravity: 
\begin{align}
S_{\textrm{GR}} = \int d^Dx \sqrt{-g} \left[ -\frac{1}{16\pi G} R+\frac{1}{2} \sum_{i=1,2}
\left( D^{\mu}\phi_i D_{\mu}\phi_i - m_i^2 \phi^2_i \right)
\right] \, ,
\label{eq:gr_action}
\end{align}
where the first term is the usual Einstein-Hilbert action.  Here we
consider the point-particle approximation, although finite size
corrections can be systematically included using
higher-dimension operators~\cite{Damour:1998jk, NRGR}.\footnote{These effects are important, especially for
  neutron-star  mergers~\cite{0711.2420, 0906.0096, 0906.1366, 1110.3764, 1503.03240, 1606.08895}.} Moreover, we exclude local interactions
between matter fields, which violate the classical assumption that the
inter-particle separation is larger than their de~Broglie wavelength.

The main focus of our analysis is the elastic-scattering amplitude of
$\phi_1$ and $\phi_2$ in the center of mass frame, where the incoming
states have four-momenta $(E_1, \bm p)$ and $(E_2, -\bm p)$ while the
outgoing states have four-momenta $(E_1, \bm p')$ and $(E_2, -\bm
p')$. The energies are defined in the usual way, e.g. $E_1 =
\sqrt{\bm p^2 + m_1^2}$, and the conservation of energy for each
matter field implies $\bm p^2 = \bm p^{\prime 2}$.  We define the
four-momentum transfer in the scattering process as
\begin{equation}
q = (0, \bm q)  = ( 0, \bm p - \bm p' ) \,. 
\label{eq:defq}
\end{equation}

Classical physics applies whenever the minimal inter-particle separation is larger than the de~Broglie wavelength, $\lambda$, of each particle. For a scattering process we may take the impact parameter $|\bm b|$ as a measure of the minimal separation, while for a bound state we 
may take it to be the periastron or the average radius for quasi-circular orbits. Thus, in the classical regime we have
\eq{
	|\bm b| \gg \lambda = \frac{1}{|\bm p|}  \,,
}{}
in natural, $\hbar = 1$, units. An immediate consequence is that, 
for any such two-body classical system, the angular momentum is large
\eq{
	J \sim  |\bm p\times \bm b| \gg 1\, .
}{}
Since the impact parameter is of order of the inverse momentum
transfer in a scattering process, $|\bm b| \sim 1/|\bm q|$, the
classical limit implies the kinematic hierarchy\footnote{This
  hierarchy implies that our results should not be expected to be
  valid for massless particles; indeed as we discuss in some detail in
  \sect{DiscussionSection}, the classical and massless limits do not
  commute. This results in the massless limit of the 3PM classical
  potential not being smooth. }
\eq{
	m_1, m_2 , |\bm p| \sim J\,  |\bm q| \, \gg |\bm q|\,.
}{eq:classical_limit}
Classical and quantum contributions to scattering processes enter at different orders in an expansion in large $J$, or equivalently, in small $|\bm q|$.
For example, from the form of the effective potential in
Eq.~\eqref{eq:VPR}, the classical term in scattering amplitudes at
${\cal O}(G)$, ${\cal O}(G^2)$ and ${\cal O}(G^3)$ correspond
respectively to the coefficient of $1/\bm q^2$, $1/| \bm q|$ and $\ln
{\bm q}^2$.  

It is worth noting that at first sight \eqn{eq:classical_limit}
appears to be in contradiction with usual classical intuition that
during the process of a closed orbit, the momenta of the two bodies
are deflected by an amount comparable to their original momenta.
However, such long-time classical processes, which are solutions of
the classical equations of motion, are comprised of a large number of elementary two-particle interactions mediated
by graviton exchanges. Each such interaction transfers a momentum $|\bm q|$
far less than the center-of-mass momentum $|\bm p|$ of the bodies, while the
complete classical solution transfers a momentum commensurate with
$|\bm p|$. In the case of scattering at linear order in $G$, this is concretely described by the exponentiation of tree-level graviton exchange in the eikonal approximation~\cite{eikonal}.

While we are ultimately interested in relativistic classical dynamics, expanding in the nonrelativistic limit is an important tool, and is defined by an expansion in small relative velocity $\bm v$, or equivalently by the hierarchy 
\eq{
	|\bm p| \ll m_1 ,m_2 \, .
}{eq:NR_limit}
This limit is essential for comparing our PM Hamiltonian with known PN results, and for the method of nonrelativistic integration described in \sect{sec:NRint}.

To summarize, the small parameters that define the classical and nonrelativistic regimes are
\eq{
	{\cal O} (1/J) &\sim {\cal O}(\bm q) \qquad (\textrm{classical expansion})\,,\\
	{\cal O}(\bm v) &\sim {\cal O} (\bm p) \qquad (\textrm{nonrelativistic expansion}) \, .
}{}
Throughout the paper we will use the above power counting to expand in
the appropriate variables where convenient.  The PM expansion, giving
analytic results at fixed order in $G$ (or $1/J$) and to all orders in
velocity, will be defined as the resummation of the small velocity
expansion.\footnote{In principle, one may worry about exponentially
  small terms in velocity not captured by the PN expansion,
  e.g. $\exp(-1/|{\bm v}|^2)$.  However, such terms do not arise in any of the fully relativistic
  expressions that we have computed.
See Ref.~\cite{LeTiec:2011dp} for additional insight on the validity of perturbation theory.}  

\subsection{Graviton Kinematics}
\label{sec:graviton_kinematics}
While the classical part of an integrated amplitude can be extracted
by taking the small $\bm q$ limit, we would like to truncate away
quantum contributions already at the integrand level in order to
reduce the complexity of integration. This is especially important for
the scalability of the method to high orders in the PM expansion. We
therefore require power counting rules that implement the classical
limit for loop momenta.

Consider an internal graviton line with four-momentum $\ell = (\omega, \bm \ell)$. Following the method of regions~\cite{Beneke:1997zp, Smirnov:2004ym}, we consider the possible scalings of its momentum components:
\eq{
{\rm hard}: \quad (\omega, \bm \ell) &\sim (m,m) \,,\\
{\rm soft}: \quad (\omega, \bm \ell) &\sim  (|\bm q|,|\bm q|) \sim J^{-1}\,(m |\bm v|,m |\bm v|) \,, \\
{\rm potential}: \quad  (\omega, \bm \ell) &\sim (|\bm q| |\bm v|, |\bm q|) \sim J^{-1}\,(m |\bm v|^2,m |\bm v|) \,, \\
{\rm radiation} : \quad  (\omega, \bm \ell) &\sim (|\bm q| |\bm v|, |\bm q| |\bm v|) \sim J^{-1}\,(m |\bm v|^2, m |\bm v|^2)\,,
}{eq:modes}
where we take as reference scale $m = m_1 + m_2$, and we use
\Eq{eq:classical_limit} to arrive at the second set of scalings in the
above equation. Note that we consider the nonrelativistic limit $|{\bm
  v}| \ll 1$ to define these modes, and this is sufficient for
determining the potential to arbitrary order in the velocity
expansion. The full PM result, containing all orders in velocity, is
then obtained through resummation (see \sect{sec:scattering}), and for
some graph topologies we verify the result using relativistic
integration methods (see \sect{sec:Rint}).

The modes in \Eq{eq:modes} identify the dominant contribution from
each region, which is computed by expanding a loop momentum about the
given scaling then integrating over the {\it full} phase space using
dimensional regularization.  The method of regions is a powerful tool
that has close connections with effective field theory, and there is a
large body of literature dedicated to its formulation and various
applications. For further details we refer the reader to
Ref.~\cite{Smirnov:2004ym}. Here we simply use it as a means for
expanding the integrand in the potential region to extract the
classical potential.

Potential modes have several key properties that are characteristic of
a classical force mediator. First, as we already mentioned, the
overall momentum scaling is parametrically determined by the total
momentum transfer $\bm q$ of the classical scattering process. In
particular, it follows from the scaling $|\bm \ell| \sim |\bm q| \sim
|\bm b|^{-1}$ that they mediate long range interactions of the order
of impact parameter.  Second, following from the scaling $\omega \sim
|\bm q| |\bm v|$, with $| \bm v | \ll 1$, the graviton exchange only
mediates a relatively small amount of energy compared to spatial
momentum. So the interaction is approximately instantaneous,
consistent with the description of the usual classical potential.  In
other words, because the potential modes are off shell, $\omega \ll
|\bm \ell|$, we can integrate them out to define an effective
potential~\cite{NRGR}. Finally, with exchanges of potential modes,
internal matter lines in loops are close to being on shell, conforming
with the physical intuition that classical particles cannot fluctuate
off their mass shell. In particular, as we will see in
\Sec{sec:NRint}, restricting gravitons to be in the potential region
enforces that for each loop there is one matter line on shell. This is
also consistent with the underlying mechanics of various other methods
for solving classical binary dynamics, such as the use of equations of
motion and worldline actions.

Gravitons with hard momenta lead to quantum-mechanical contributions
because their energy component is too large, causing the matter fields
to be far off shell. Moreover, the interaction length of a hard mode
is of order the de~Broglie wavelength of matter, and therefore
corresponds to a short-distance contact interaction and not a
long-range force.  On the contrary, the other three regions have
wavelength around or greater than the impact parameter or orbital
radius, $|\bm b|\sim |\bm q|^{-1}$, and therefore may contribute to
the classical potential.

The soft mode can be used to extract classical
contributions~\cite{RothsteinClassical,StermanEikonal,BjerrumClassical}, and plays a
central role in the eikonal
approximation~\cite{eikonal,BjerrumClassical,StermanEikonal}.  However, the
contribution to the classical nonrelativistic potential is actually
dominated by the potential mode within the soft region.  After the
overlap is subtracted, the residual is expected to be quantum
mechanical because the energy transfer is too large to keep
the matter fields on shell.  In the following analysis, we focus on
extracting the classical conservative dynamics from the potential
region.

Gravitons in the radiation region correspond to emitted radiation,
which are of course critical in the context of gravitational-wave
physics.
At 3PM order, which is the focus of this paper, such modes cannot
contribute to the conservative potential. We will therefore only be
interested in effects induced by potential-mode gravitons.  As is
well-known, this distinction between potential and radiation modes,
i.e.~near zone and far zone dynamics, becomes subtle at
sufficiently high order due to radiation reaction effects. See
\sect{sec:radiation} for details.

The purpose of the method of regions is to identify the dominant
contribution in loop integration, given the external kinematics. To
verify results obtained by restricting to the potential region, we use
fully relativistic integration methods when possible. For such
methods, the integration is over the full domain, i.e. effectively
including all regions, and the classical contribution can be retrieved
by taking the classical limit given in \Eq{eq:classical_limit}.
Indeed, for all available cases we find that relativistic integration
confirms that the potential region captures all contributions upon
resummation to all orders in velocity. See more details in
\sect{sec:Rint}.  A nontrivial two-loop example is also given in
\app{HxH}, based on the fully integrated results of
Ref.~\cite{Bianchi:2016yiq}

To summarize, we can use the loop momentum scaling for the potential
region given in \Eq{eq:modes} to consistently expand in the classical
and nonrelativistic limits at the integrand level. For example, the
leading order contribution that leads to the 1PM potential has a
single graviton exchange with momentum $q$ given in \Eq{eq:defq},
which obviously satisfies the scalings in \Eq{eq:modes}.  For the
conservative potential at 3PM order, it is sufficient to take all
gravitons to be potential modes.


\subsection{Truncation to Potential Region}

\label{sec:classical_limit}
\begin{figure}[tb]
	\begin{center}
		\includegraphics[scale=.4]{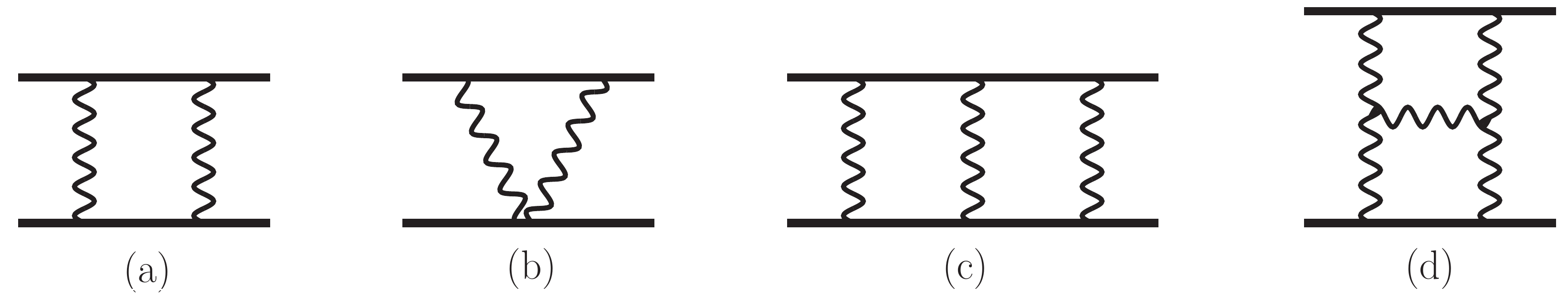}
	\end{center}
	\vskip -.5 cm
	\caption{\small Examples of one- and two-loop diagram that
          contribute to the classical potential.  Wiggly lines
          represent gravitons and straight lines scalars.  Here the
          diagrams are not Feynman diagrams, but demonstrating the
          singularity structure from propagators in the graphs.  }
	\label{DiagramKeptFigure}
\end{figure}

\begin{figure}[tb]
	\begin{center}
		\includegraphics[scale=.4]{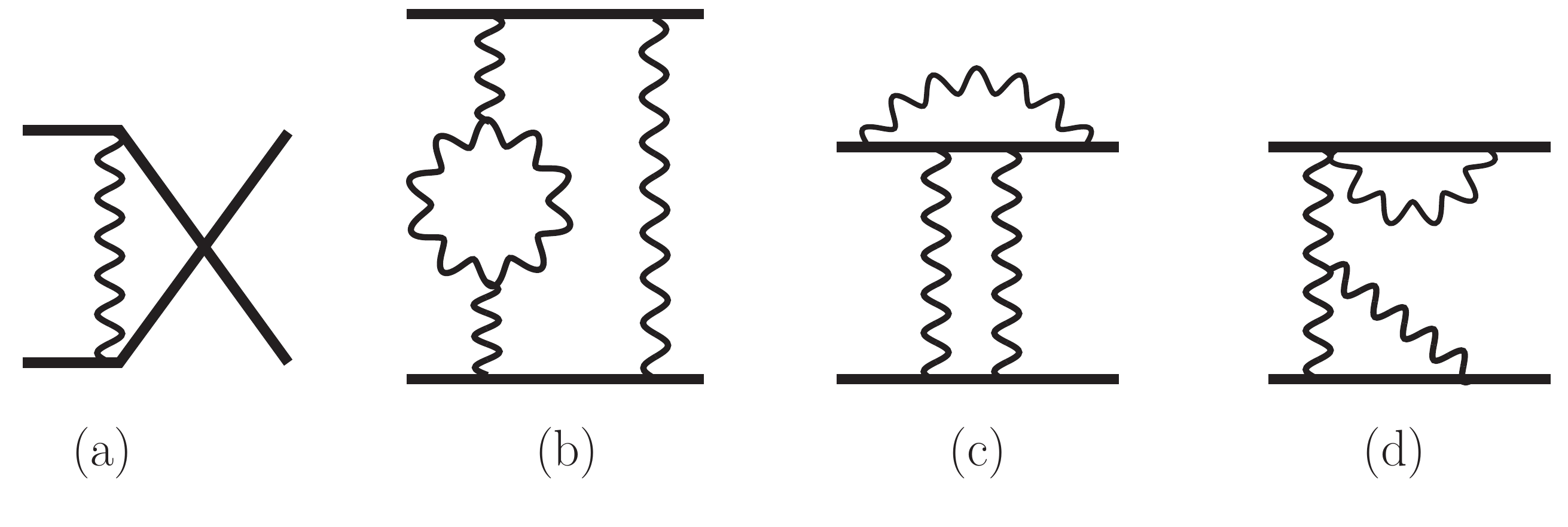}
	\end{center}
	\vskip -.5 cm
	\caption{\small Examples of one- and two-loop diagrams that do
          not contribute to the classical potential. Wiggly lines
          represent gravitons and straight lines scalars.  The meaning
          of diagrams here is the same as in \Fig{DiagramKeptFigure}.
        }
	\label{DiagramsDroppedFigure}
\end{figure}

A full quantum mechanical calculation of the scattering amplitude
would require a proper accounting of all contributing diagrams.
However, when taking the classical limit only a subset
of diagrams survive and determine the contributions to the classical potential.
Examples of one- and two-loop diagrams that may contain classical
contributions are shown in \fig{DiagramKeptFigure}. Others, such as
those in \fig{DiagramsDroppedFigure}, can be immediately discarded. By
applying classical truncation at every step---from the construction of
the integrand to integration---we can achieve massive simplifications,
which are especially crucial for more challenging higher-loop
calculations.  Here we briefly outline the specific truncations we
use.  We will elaborate on these points substantially throughout the
rest of the paper.

\begin{figure}[tb]
	\begin{center}
		\includegraphics[scale=.5]{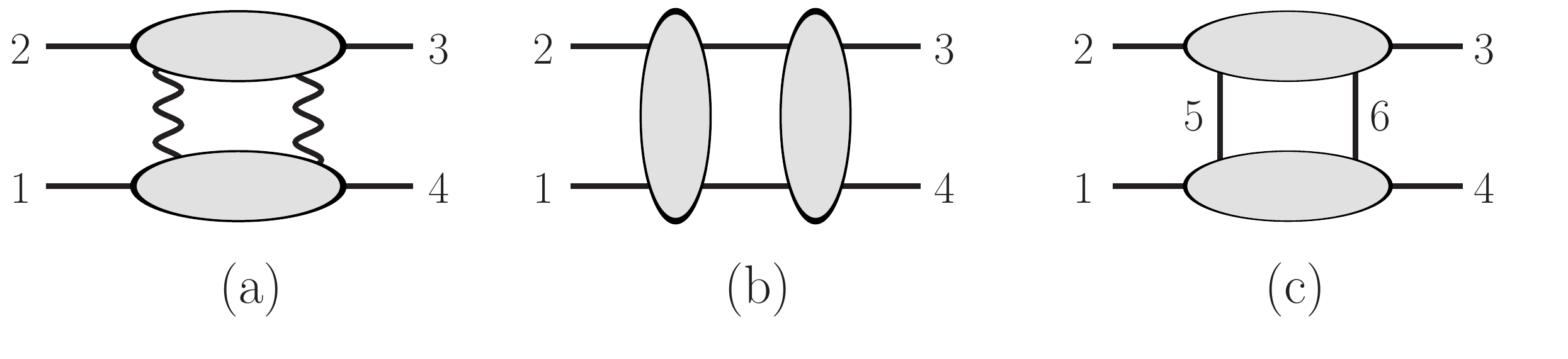}
	\end{center}
	\vskip -.5cm
	\caption{\small Generalized two-particle cuts for a elastic
          scattering of two distinct scalars $\phi_1$ and $\phi_2$
          with masses $m_1$ and $m_2$.  The pairs of legs (1,4) and
          (2,3) correspond to $\phi_1$ and $\phi_2$ respectively. The
          blobs represent tree amplitudes, which can have several
          diagrams in a given blob, and exposed lines are all on
          shell.  Cut (a) separates the two matter fields by cutting
          graviton lines and contributes to the classical
          potential. The top and bottom internal lines in cut (b) are
          scalar $\phi_1$ and $\phi_2$, respectively.  
          For cut (c) the internal lines are either both $\phi_1$ or 
          $\phi_2$.  Neither of the cuts (b)
          and (c) contains any new classical potential contributions.  }
	\label{SampleOneLoop2CutsFigure}
\end{figure}

Following arguments that led to the absence of matter contact terms in
the classical Lagrangian, an obvious contribution to discard is any
diagram or part of a diagram in which matter fields come together at a
local contact interaction, as illustrated at one loop in
\fig{DiagramsDroppedFigure}(a).  Because the classical Lagrangian
\eqref{eq:gr_action} does not contain such terms (since classical
physics requires that the particles are always sufficiently
separated), such contact interactions can appear only through some
quantum processes, which are of no interest to us.
From the perspective of generalized unitarity, this amounts to
building the integrand only from those cuts that split the amplitude
such that the two matter lines are on opposite sides of cut gravitons.
At one loop, for example, this requirement amounts to keeping those
terms that arise from the generalized unitarity cut (a) in
\fig{SampleOneLoop2CutsFigure}, but not including any new
contributions from cuts (b) or (c).  In
\fig{SampleOneLoop2CutsFigure}, there are two pairs of distinct
scalars, (1,4) and (2,3), with masses $m_1$ and $m_2$.

Another contribution that can be discarded arises from any diagram or 
contribution to a diagram that contains a closed loop of momentum $\ell$ that
never flows through a matter line, but only through graviton
propagators, as illustrated in \fig{DiagramsDroppedFigure}(b).
The only singularities in the closed loop come from graviton poles, when $|\ell_0| \sim |\bm \ell|$, which is outside 
the potential region that contributes to the classical potential.
Conversely, this implies that classical contributions only arise from
diagrams in which all closed loops include at least one matter line.
Note however that diagrams that pass this criterion may still be
quantum mechanical, such as diagrams (c) and (d) in
\fig{DiagramsDroppedFigure}.

A third discarded contribution contains graviton lines which start and end on the 
same matter line.
This implies, for example, that diagrams (c) and (d) in
\fig{DiagramsDroppedFigure} can be discarded. One may intuitively
understand this by noticing that these diagrams represent quantum
mechanically-induced gravitational form factors for the matter
fields. Alternatively, as we will discuss in \sect{sec:NRint}, from
the mechanics of integration one finds that the three-momentum
component of such graviton exchanges is not parametrically set by the
momentum transfer $\bm q$.
From the perspective of generalized unitarity at one loop discarding
such diagrams amounts to ignoring cuts that pass twice through the
same matter line, such as the one illustrated in
\fig{SampleOneLoop2CutsFigure}(c).  In this case, the two cut internal 
lines are the
same scalar, as identified by its mass and the Lagrangian
\eqref{eq:gr_action}.

\begin{figure}[tb]
	\begin{center}
		\includegraphics[scale=.5]{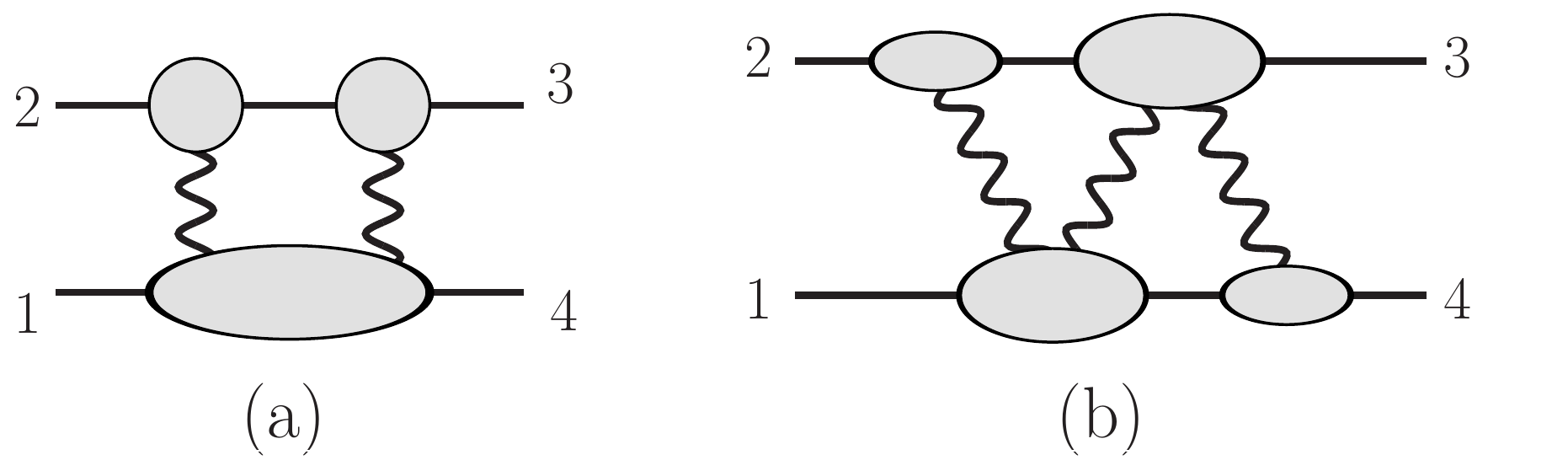}
	\end{center}
	\vskip -.5 cm
	\caption{\small Some generalized unitarity cuts for extracting the
		conservative two-body potential at (a) 2PM order and (b) at 3PM order. The blobs represent
		tree amplitudes and exposed lines are all on shell. Straight lines
		represent massive scalars and the wiggly lines are either gluons or
		gravitons, depending on whether we are considering a gauge theory or
		a gravity cut. 
        }
	\label{SampleCutsFigure}
\end{figure}

As we have seen in \sect{sec:graviton_kinematics} and will further
discuss in \sect{sec:NRint}, contributions to the classical potential
come only from the region where internal matter lines are close to on
shell or, mathematically, from regions where each energy integral is
localized to a matter
pole~\cite{RothsteinClassical,BjerrumClassical,CliffIraMikhailClassical}. This
implies that we can effectively cut one matter line per loop. As
discussed in \sect{sec:graviton_kinematics}, this is consistent with
exchanged gravitons having momenta in the potential region.
At one loop the net effect is that instead of using the two-particle
cut in \fig{SampleOneLoop2CutsFigure}(a), all diagrams contributing to
the classical potential are determined by generalized unitarity cut in
\fig{SampleCutsFigure}(a) and its relabelings, where an additional
matter line is cut, compared to \fig{SampleOneLoop2CutsFigure}(a).  As
usual, all exposed lines in the figure are placed on shell.  While
this provides only a modest simplification at one loop, it is much
more powerful at higher loops.

Let us elaborate on how to apply the classical truncation via small
$|\bm q|$ expansion for loop-level integrands.
Recall that, as summarized in the previous sections, classical power
counting implies the hierarchy of scales $m_1, m_2 , |\bm p| \gg |\bm
q| \sim |\bm \ell|$, where ${\bm \ell}$ stands for the spatial part of
a generic loop momentum. We therefore scale $q\rightarrow \alpha q$
and $\ell\rightarrow \alpha \ell$ by $\alpha \sim J^{-1} \ll 1$,
including the measure and propagators\footnote{For any matter
  propagators with mixed $\alpha$ counting, we only count the leading
  scaling and do not expand the denominators.}.
For the amplitude at ${\cal O}(G^n)$, all terms with $\alpha^{k>n-3}$
do not have classical contributions, cf. \eqref{eq:VPR}. In general,
this simple scaling argument implies that terms with many powers of
loop momenta in the numerators will have too high a power counting in
$\alpha$ and can therefore be ignored.  Such terms are responsible for
the UV properties of general relativity, so it is intuitively clear
that they must be quantum mechanical.
For example, a consequence of this power counting is that the relevant
numerators for diagrams (a) and (b) in \fig{DiagramKeptFigure} scale
at most as $\alpha$ and $\alpha^0$, respectively. At two loops, this
power counting implies that terms with more than four powers of loop
momentum in any diagram numerator cannot contribute to the classical
potential, and for some diagrams the bound is even tighter (see
\sect{sec:PMexpand} for a related discussion).
Although we keep these terms when building the integrands to make
numerical cross checks easier, we drop them prior to integration to
take advantage of enormous simplifications.

An important feature of this expansion is that the leading term of a
diagram is not always the classical one. Rather, there exist diagrams
which exhibit terms larger than the classical ones in the small
$\alpha$ expansion, i.e. at ${\cal O}(G^n)$ they scale as $\alpha^{k <
  n-3}$.  For example, with a numerator that is independent of $q$ or
$\ell$, the one-loop diagram in \fig{DiagramKeptFigure}(a) scales as
$\alpha^{-2}$, while that in \fig{DiagramKeptFigure}(b) scales as
$\alpha^{-1}$ and contributes to the classical potential.
The former is an example of terms with such enhanced $\alpha$ scaling, which we refer to as
``superclassical''. As we will see in \sects{sec:NRint}{sec:EFT}, they
have a natural interpretation as iterations of terms
determined at lower orders that do not contribute directly to the
classical potential at ${\cal O}(G^n)$. They are infrared divergent and cancel in the
matching between full theory and effective theory amplitudes so that
the remainder is a classical contribution to the
potential.

To summarize, the guidelines for efficiently applying generalized
unitarity for obtaining integrands from which we can extract the classical potential are

\begin{enumerate}
	\item Generalized unitarity cuts must separate the two matter lines to
	opposite sides of a cut.
	
	\item Every independent loop must have one cut matter line.
	
	\item Contributions where both ends of a graviton propagator attach
	to the same matter line are dropped.
	
	\item Terms with too high a scaling in $q$ or $\ell$ are dropped. At
	two loops this implies that any term in a diagram numerator with
	more than four powers of loop momentum yields only quantum mechanical contributions; some diagrams
	require fewer loop-momentum factors.
\end{enumerate}

We have confirmed these rules to be valid through two loops and will
use them to organize our 3PM calculation. They determine which
generalized unitarity cuts need to be included to obtain the classical
potential, and which can be ignored. They also enable enormous
simplifications at the integrand level by truncating away quantum
contributions.
\medskip

\section{Building Integrands from Tree Amplitudes}
\label{IntegrandsSection}

\subsection{General Considerations}

The first step towards obtaining the classical potential is to obtain
an appropriate loop integrand containing all desired classical
contributions.  To build the integrand we use the generalized
unitarity method~\cite{hep-ph/9403226, hep-ph/9409265, hep-ph/9708239, hep-th/0412103, 0705.1864, BRY, BDDPR}.  This method
meshes well with double-copy relations~\cite{KLT,BCJ,BCJLoop} that
allow us to express gravity tree amplitudes and loop integrands in
terms of the corresponding gauge-theory quantities.
It has proven to be especially successful for carrying out high-loop
computations in gravity theories, including the determination of the
ultraviolet properties of extended supergravity theories at four and
five loops~\cite{0905.2326, SimplifyingBCJ, 1309.2498, 1409.3089, 1708.06807, 1804.09311,SimplifyingBCJ} and of Einstein gravity at
two loops~\cite{1507.06118, 1701.02422}.
Further details of this methods may be found in various reviews~\cite{hep-ph/9602280, 1103.1869, 1103.3298}.

The basic input into our construction are the gauge-theory tree-level
scattering amplitudes collected in \app{TreeAmplitudesAppendix}.  They
are then converted into gravity tree amplitudes through the
Kawai-Lewellen-Tye (KLT) and Bern-Carrasco-Johansson (BCJ) forms of
the double copy and are subsequently used to construct generalized
unitarity cuts.
The KLT form of the double copy is convenient when using
four-dimensional helicity states which give remarkably compact
expressions for tree  amplitudes.  
The BCJ double copy is the
natural choice when organizing the calculation in terms of diagrams,
as we do in our $D$-dimensional constructions. 

While a purely four-dimensional approach would not lead to the correct
integrand in the full quantum theory when using dimensional
regularization for ultraviolet and infrared singularities, for the
purpose of extracting the classical potential we shall find that four-dimensional helicity methods are sufficient.  While we do encounter
infrared divergences, we will see that in contrast to the full quantum
theory, a simple continuation of the four-dimensional expressions to
$D$ dimensions by extending the loop integration measure as well as
all momentum invariants
regularizes infrared singularities and does not result in any lost
contributions.

While many strategies can be applied at one-loop (2PM) order, at two loops (3PM) and 
beyond calculations become more challenging, and require methods that
scale well with increasing complexity.  Our methods were designed with this in mind.
For 3PM calculations both the four- and $D$-dimensional approach work
well; we, however, anticipate that, given the remarkable simplicity of
tree-level helicity amplitudes, as the perturbative order increases
the four-dimensional helicity approach will be the method of
choice. While we do not have a general proof that four-dimensional
helicity states are sufficient to all orders, based on our
investigations here, it seems plausible that it will continue beyond 3PM.

\subsection{Building Integrands using Generalized Unitarity}

The generalized unitarity method gives us a means for constructing 
integrands of loop-level amplitudes in terms of sums of products of tree amplitudes.
Integrands are rational functions with simple poles for 
momentum configurations where internal lines in Feynman graphs go on shell.
A propagator corresponding to such a particle is cut, i.e.\ it is replaced with 
a delta function enforcing the corresponding on-shell constraint. The residues of these poles---or generalized cuts---are 
given by sums of products of integrands of amplitudes whose external lines are 
the totality of the initial external lines and the cut lines.
An important set of generalized cuts is that in which all factors are tree-level amplitudes; 
then, the residue is:
\begin{equation}
C \equiv \sum_{\rm states} A^\tree_{(1)} A^\tree_{(2)} A^\tree_{(3)} \cdots  
A^\tree_{(m)} \, .
\label{GenericCut}
\end{equation}
We will normalize the generalized cut to be the product over
the tree amplitudes that compose the cut.
The sum runs over all intermediate physical states that
can contribute, i.e.\ for which all the tree-level amplitude factors are nonvanishing.
The generalized unitarity method assembles generalized cuts into complete amplitudes,
by constructing a unique function whose generalized cuts reproduce those of the amplitude's integrand.
An example of generalized cut at one loop is shown in
\fig{SampleCutsFigure}(a). In this figure the exposed lines are all
on-shell delta functions and the blobs represent on-shell tree
amplitudes.  The expression for this generalized cut is given by the
sum over the intermediate states of the product these two three-point
amplitudes and one four-point amplitude.
The full amplitude satisfies a spanning set of generalized cuts, which
determine all integrands of fixed loop order and multiplicity.

The parts of amplitudes that contribute to the
classical potential are determined by a rather restricted set of generalized unitarity cuts, as explained in \sect{sec:classical_limit}.
While the advantage may not obvious at one loop, this restriction greatly simplifies the generalized unitarity cuts at two loops.

\subsection{Gravity Tree Amplitudes from the Double Copy}

The generalized unitarity method is especially powerful in gravity
theories because it meshes well with double-copy constructions,
allowing us to express gravity amplitudes in terms of much
simpler gauge-theory ones.  The KLT relations~\cite{KLT}, which were
originally derived in string theory, give us a simple means for
obtaining gravity tree amplitudes in terms of color-ordered
gauge-theory partial amplitudes.  Through five points, which is all we need
in this paper, these relations are\footnote{Note that $A,M$ here are defined as the amplitudes that include all factors of $i$ from Feynman diagrams, in contrast with $M=i\mathcal{M}$ used later.}
\begin{align}
M_3^\tree (1,2,3) & =  i  A_3^\tree(1,2,3) \,
   A_3^\tree(1,2,3)\,, \cr
M_4^\tree (1,2,3,4) & = - i s_{12} A_4^\tree(1,2,3,4) \,
   A_4^\tree(1,2,4,3)\,, \cr
M_5^\tree(1,2,3,4,5)
& = i s_{12} s_{34}  A_5^\tree(1,2,3,4,5) \, A_5^\tree(2,1,4,3,5) \cr
& \hskip 2 cm
+ i s_{13}s_{24} A_5^\tree(1,3,2,4,5) \, A_5^\tree(3,1,4,2,5) \,,
\label{KLT}
\end{align}
where the $M_n^\tree$ are tree-level gravity amplitudes, the
$A_n^\tree$ are gauge-theory partial amplitudes stripped
of color factors and coupling constant, and $s_{ij} = (p_i+p_j)^2$.
We have suppressed a factor of the coupling $(\kappa/2)^{m-2}$ at $m$
points, where the coupling is given in terms of Newton's constant via
$\kappa^2 = 32\pi G$.  These relations hold in any space-time
dimension.  An extension valid for any number of
external legs may be found in Ref.~\cite{MultiLegKLT}.

The KLT relations are usually formulated in terms of massless
amplitudes, but for the tree amplitudes used in this paper, with a
single massive scalar pair, dimensional reduction shows they
hold in this case as well.  To show this we can start from, say, six
dimensional massless gauge or gravity amplitudes and dimensionally
reduce them to a four-dimensional ones with a massive scalar pair.  To
do so we can start with pure Yang-Mills amplitudes in, say $D=6$ and
then choose the polarization vectors of legs we wish to be massive
scalars, say $1$ and $2$, to be
\begin{equation}
  \varepsilon_1 = \varepsilon_2 = (0,0,0,0,0,1) \,,
  \label{eq:reduction_1}
\end{equation}
and their momenta to be
\begin{equation}
p_1^A = (p_1^\mu,  m, 0)\,,  
\hskip 3 cm 
p_2^A = (p_2^\mu, -m, 0) \,,
  \label{eq:reduction_2}
\end{equation}
where the $p_i^A$ are six dimensional momenta and $p_i^\mu$
four-dimensional ones.  From the four-dimensional perspective, legs 1
and 2 are massive scalars with mass $m$ in both the gauge and gravity
theories.
The polarization vectors and momenta of the remaining legs live in
the four-dimensional subspace specified by the first four entries.
Dimensional regularization fits naturally into this 
framework by analytically continuing the four-dimensional subspace to $D=4-2\epsilon$ dimensions, as usual.  The
conclusion is that we can directly apply the KLT relations \eqref{KLT} to
tree amplitudes with $n$ external gravitons two external massive scalar.

Our calculation of the part of the two-loop amplitude that contributes to the 3PM
potential will require only up to five-point tree amplitudes.  When
evaluated using the KLT relations, the gravity tree amplitudes
automatically inherit the remarkable simplicity of the gauge-theory
helicity amplitudes presented in \app{TreeAmplitudesAppendix}.

A subtlety of dimensional regularization which requires careful
analysis, especially beyond one-loop level and in non-supersymmetric
theories, relates to finite terms of the type $\eps/\eps$, where the
$1/\eps$ originates from an infrared singularity \footnote{Ultraviolet
  $1/\eps$ singularities do not need to be considered because these
  are of quantum origin and should not contribute to the classical
  potential.}  while the $\epsilon$ in the numerator comes from the
component of loop momenta or polarization states outside four
dimensions.
While it seems unlikely that this could affect the classical potential, it is nevertheless 
important to confirm this.  

To cross-check our calculation and to confirm the absence of any
dimensional regularization subtleties, we also made use of BCJ form of
$D$-dimensional gauge-theory tree amplitudes.
While the BCJ approach generalizes to loop level integrands,
here we only use it for tree-level amplitudes that enter into
unitarity cuts contributing to the classical potential.  

Consider an $m$-point tree-level gauge-theory
scattering amplitude with all particles in the adjoint representation.
We can write any such amplitude as a sum over diagrams with only cubic
vertices:
\begin{equation}
\colorA{A}^{\tree }_{m} = g^{m-2} \sum_{j}
\frac{c_{j} n_{j}}{D_j} \,,
\label{CubicRepresentation}
\end{equation}
where $g$ is gauge-theory coupling constant.\footnote{The gauge-theory action is 
	$S=-\frac{1}{4}\int d^D x F^a_{\mu\nu}F^{a\mu\nu}$, where 
	$F^a_{\mu\nu}=\partial_{\mu}A^a_{\nu}-\partial_{\nu}A^a_{\mu} - g f^{abc}A^b_{\mu}A^c_{\nu}$.
	}
The denominator $1/D_j$
is given by the product of Feynman propagators of graph $j$.
The sum runs over the $(2m-5)!!$ distinct,
$m$-point graphs with only cubic vertices.
Such graphs are sufficient because the contribution of any diagram with
quartic vertices can be assigned to a graph with only
cubic vertices by multiplying and dividing by appropriate
propagators.  The nontrivial kinematic information is contained in the
numerators $n_{j}$, which generically depend on momenta, polarizations, and spinors.
The color factor $c_j$ is obtained by dressing every vertex in graph
$j$ with the relevant gauge-group structure constant,
\begin{equation}
\tilde{f}^{abc}=i\sqrt{2}f^{abc} = \mathrm{Tr}([T^{a},T^{b}] T^{c}) \,,
\label{Fabcadjoint}
\end{equation}
where the gauge-group generators are normalized via
$\mathrm{Tr}(T^{a}T^{b})=\delta^{a b}$, 
which is different than the textbook ones~\cite{PeskinSchroeder},
so as to be compatible with Ref.~\cite{BCJ}.

\begin{figure*}[tb]
\begin{center}
\includegraphics[scale=.55]{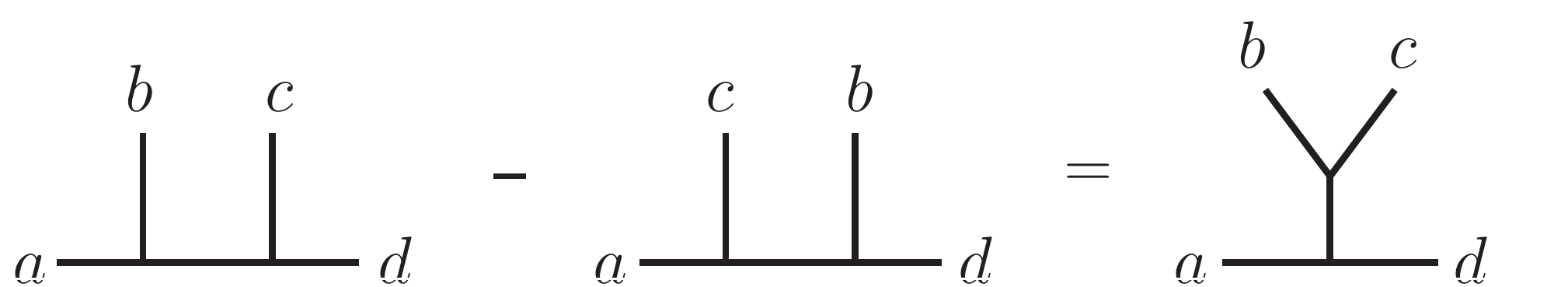}
\end{center}
\vskip -.1 cm
\caption[a]{\small The Jacobi relation for a group of three
  graphs. The graphs can represent color factors or numerator factors
  and can be thought of as embedded in a larger graph.
\label{GeneralJacobiFigure}
}
\end{figure*}

In a BCJ representation, kinematic numerators obey the same generic algebraic relations as the color
factors~\cite{BCJ, BCJLoop, SimplifyingBCJ}.  For theories with
only fields in the adjoint representation there are two key 
properties that the kinematic numerators satisfy.  
The first is {\em antisymmetry} under graph vertex flips:
\begin{equation}
 c_{\overline{\imath}} = - c_{i}\ \Rightarrow\ n_{\overline{\imath}}=- n_{i}\, ,
\label{BCJFlipSymmetry}
\end{equation}
where the graph $\overline{\imath}$ has same graph connectivity as
graph $i$, except an odd number of vertices have been cyclically
reversed.  
The second property is that we require that graph numerators obey dual (kinematic) Jacobi identities whenever the
color factors obey the group theoretic Jacobi identities:
\begin{equation}
c_{i}= c_{j}-c_{k} \ \Rightarrow\ n_{i} =  n_{j} - n_{k}  \,.
\label{BCJDuality}
\end{equation}
Here $i$, $j$, and $k$ refer to three graphs which are identical except for one internal edge,
as illustrated in \fig{GeneralJacobiFigure}.  
The relative signs in the color Jacobi identity are dictated by the
particular way that we drew the four-point subgraphs. For any choice,
the signs of the color and kinematic Jacobi identities are
identical.  In general, beyond four point amplitudes, Feynman diagrams
do not immediately give tree amplitudes with kinematic numerators
obeying the duality.  There are various systematic ways to reorganize
them and manifest the duality~\cite{1010.3933, 1104.5224, 1608.00006, 1702.08158, 1703.01269}. Low-point tree
amplitudes are sufficiently simple to use an ansatz to enforce desired
properties, including the duality, locality and gauge invariance.

Once gauge-theory tree amplitudes or loop integrands have been arranged
into a form where the duality is manifest~\cite{BCJ,BCJLoop},
corresponding gravity tree amplitudes or loop integrands are given simply by 
replacing color factors of a 
the diagrams of a second gauge theory with the kinematic numerators of 
the first gauge-theory:
\begin{equation}
c_{i}\ \rightarrow\ {n}_{i}\,.
\label{ColorSubstitution}
\end{equation}
This immediately gives the double-copy form of a gravity tree amplitude,
\begin{equation}
{M}^{\tree}_{m} = i 
\sum_{j}\frac{\tilde{n}_{j}n_{j}}{D_j} \, ,
\label{DoubleCopy}
\end{equation}
where $\tilde{n}_j$ and $n_j$ are the kinematic numerator factors of
the two gauge theories. As usual we have not included factors of the
gravitational coupling.  In general, the two gauge theories can be
distinct; since however we are interested in 
Einstein gravity coupled minimally to massive scalars, the two gauge
theories will be identical and correspond to a Yang-Mills gauge field
coupled to scalar fields in the fundamental
representation \cite{Johansson:2014zca, Johansson:2015oia, Johansson:2019dnu}. We will therefore have $\tilde{n}_{j} = n_{j}$.  Since,
however, we will be interested only in the double copy of tree
amplitudes with two scalars, we may also take them to be in
the adjoint representation.

Following the same dimensional reduction procedure as for the KLT
relations, we immediately see that the double copy
relations \eqns{ColorSubstitution}{DoubleCopy} hold for the case of tree
amplitudes with a single massive scalar pair and the rest being gravitons.
Remarkably, these ideas extend to full integrands at loop
level~\cite{BCJLoop}, though in this paper we will not make use of
this property.  We will apply the duality and double copy only for tree-level
amplitudes used as input to generalized unitarity cuts.

Let us finish this section with a remark on obtaining pure Einstein
gravity as a double copy.  Typically the double copy of two vector
fields contains more than just the graviton. For example, a gluon in
four dimensions has two helicities $\pm 1$, so the square has four
states: the $\pm 2$ correspond to graviton, and the zero-helicity
states which are identified as the dilaton and axion.  The projection
to pure Einstein gravity, which eliminates the dilaton and axion, also
meshes well with our generalized unitarity construction. We can simply
apply the projection to graviton on each of the massless cut legs,
which are external lines for tree-level blobs in a generalized
unitarity cut.  In any of the tree-level amplitudes, the projection on
the external legs ensures that the dilaton or axion do not appear
internally either, because they have to be produced in pairs when
interacting with the gravitons.\footnote{Note that a single dilaton
  can be generated by massive scalars.  However, this dilaton has to
  appear as an external state in a tree-level amplitude with only two
  massive external scalars.}

In practice, we use two approaches for the graviton projection
depending on the details of the unitarity construction.  When using
the four-dimensional helicity method, we simply correlate the
helicities of the two copies of gauge theory amplitudes.\footnote{This
  may be also understood as orbifolding by the $U(1)$ global symmetry
  whose charges are given by the difference of helicities of fields in
  the two gauge theories entering the double copy.}  For
$D$-dimensional constructions, we often use the efficient projector in
\Eq{SimplifiedGravitonProjector}, as opposed to the standard one in
\Eq{eq:full_projector}. Such simplification is made possible by
choosing a special form of tree-level amplitudes in gauge theory. See
the discussion in \sect{OneLoopDCutsSection}.

\section{Integrands at One Loop Using Four-Dimensional Helicity}
\label{OneLoopSection}

\begin{figure}[tb]
        \begin{center}
                \includegraphics[scale=.5]{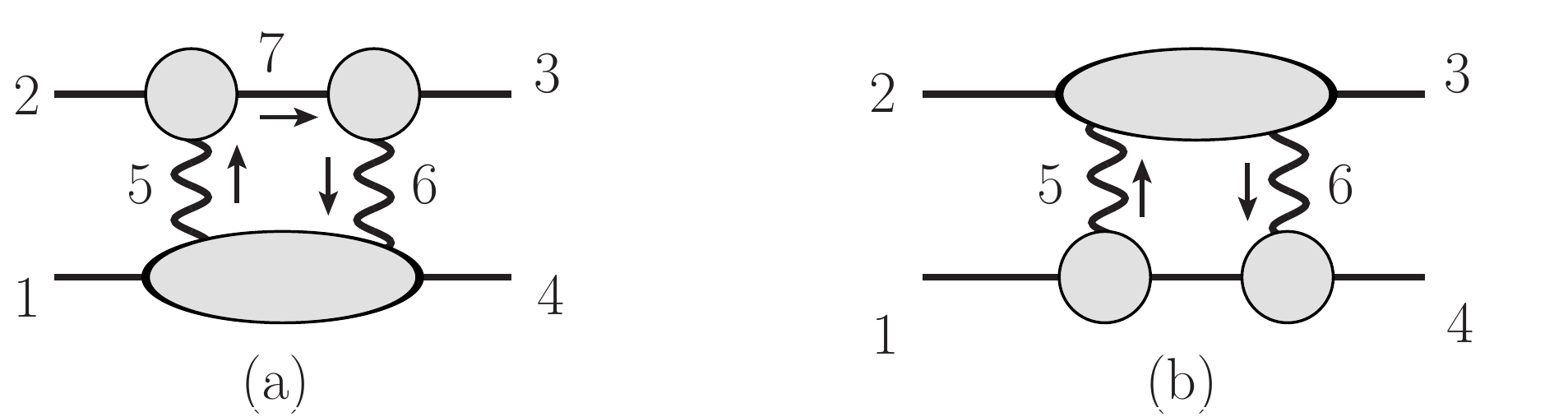}
        \end{center}
        \vskip -.5 cm
        \caption{\small The two generalized unitarity cuts for extracting the
                conservative two-body potential at 2PM order. The blobs represent
                tree amplitudes and exposed lines are all on shell. Straight lines
                represent massive scalars and the wiggly lines are either gluons or
                gravitons, depending on whether we are considering a gauge theory or
                a gravity cut.}
        \label{OneLoopCutsFigure}
\end{figure}

To illustrate some essential features of our procedure for
constructing amplitudes' integrands in the post-Minkowskian expansion,
we first describe the 2PM order, discussed earlier in
Refs.~\cite{Westpfahl2PM, CachazoGuevara, DamourTwoLoop, BjerrumClassical, CliffIraMikhailClassical, Paolo2PM, Cristofoli:2019neg, 1905.05657, 1906.05209}.
As discussed in \sect{QuantumVsClassicalSection}, the part of loop integrands that contributes to the
classical potential is determined by a restricted set of generalized unitarity cuts: they 
separate the two matter lines and also they place on shell one matter line 
per loop~\cite{RothsteinClassical, BjerrumClassical, CliffIraMikhailClassical}. 
Thus, at one loop, corresponding to the 2PM case,  we need only compute the 
 two generalized cuts illustrated in \fig{OneLoopCutsFigure}. 

At one-loop the cuts can be evaluated in various ways.
In particular, the calculation is not simplified substantially by making  use of on-shell conditions
on the matter lines.  Similarly, its complexity is not affected much by judicious choices for the graviton physical
state projector. 
Beyond one loop, however, care is necessary to ensure that
the calculation scales well~\cite{CliffIraMikhailClassical, 3PMPRL}.  
This includes adjusting the tree-level amplitude factors so that the physical-state projectors trivialize and also making sure 
that on-shell conditions are used to their maximal extent. 
Therefore, to illustrate all subtleties of the amplitudes' construction, in our one-loop calculation below we 
will make use of the cut conditions on the matter lines,  even though it causes the slight additional complication 
that different cuts double count certain terms.  This must be taken into account when the cuts are combined into a single
integrand.  Such double counts are natural beyond one loop, so the one-loop case offers a simple 
illustration of this issue and its resolution.

\subsection{Warm-up: Gauge-Theory Integrands}

As a warm-up, we consider gauge theory and construct the one-loop integrand for the 
classical potential between two massive scalars due to gluon exchange. 
As we shall see, once we have the gauge-theory integrand,
obtaining the gravity one is quite straightforward.

We first evaluate the generalized cut in \fig{OneLoopCutsFigure}(a), for the
case of a one-loop color-ordered YM amplitude. In color-ordered
amplitudes the external legs follow a cyclic ordering and are stripped
of color factors.  This  generalized cut is,
\begin{align}
C_{\rm YM}^{\rm (a)} & = \sum_{h_5, h_6 = \pm}
A^\tree_3(3^s,6^{h_6}, -7^s)  \, A^\tree_3(7^s , -5^{h_5}, 2^s) \, 
A^\tree_4(1^s,5^{-h_5}, -6^{-h_6},4^s) \,,
\label{OneloopCutA}
\end{align}
where $\pm$ superscripts denote the gluon helicity and $s$ stands for scalar, 
and the minus sign on the label indicates that the sign of the momentum 
should be reversed when treated as an outgoing momentum of the amplitude.
The state sum runs over the gluon helicities.  Since each propagating 
scalar corresponds to a single state there is no sum associated with it.  In four
dimensions each gluon can either be of positive or negative helicity, giving a
total of four helicity configurations in the state sum.  
 We label them as
follows:
\begin{equation}
\{h_5, h_6\} :  \; \; \{++\}, \; \{--\}, \; \{-+\}, \; \{+-\} \,,
\label{TwoCutHelicities}
\end{equation}
where the helicities of the gluon legs are labeled with an outgoing momentum convention.
In $D$ dimensions
each gluon has $D-2$ states.  At one-loop we will evaluate the cuts in
four dimensions and compare it to the result obtained via $D$-dimensional
methods. We show that the two methods give same four-dimensional classical potential.

The four-point tree amplitudes appearing in the cut \eqref{OneloopCutA} are given in \eqn{SGGS}.
One may evaluate the product of three-point tree amplitudes appearing in cuts by starting from 
Feynman three-point vertices and evaluating them on helicity states.  Alternatively,
we may extract their product from the four-point amplitude \eqref{SGGS}, by removing the scalar 
propagator and placing its momentum on shell. Below we use the second approach.

Starting with the two-scalar two-gluon amplitude in \eqn{SGGS} with the gluons in the
$\{++\}$ helicity configuration and removing the scalar propagator $i/\t_{36}$ we have
\begin{equation}
A^\tree_3 (3^s, 6^{+}, -5^{+}, 2^s)  \rightarrow  A^\tree_3 (3^s, 6^{+}, -7^s) \, A^\tree_3 (7^s, -5^{+}, 2^s)  = \frac{m_2^2 \spb{-5}.{6}} {\spa{-5}.{6}
 } \,,
\end{equation}
where we use the helicity notation of Ref.~\cite{hep-th/0509223, hep-ph/9601359}.
For simplicity, we only keep the momentum label in spinor expressions, e.g.\ $\spb{-5}.{6} \equiv \spb{(-\ell_5)}.{\ell_6}$.
Using the on shell condition for that propagator, $p_3\cdot \ell_6= 0$, the cut for
the $\{++\}$ helicity configuration is
\begin{align}
C_{\rm YM}^{\{++\}} & = 
A^\tree_3 (3^s, 6^{+}, -7^s) \, A^\tree_3 (7^s, -5^{+}, 2^s) \, 
A^\tree_4 (1^s, 5^{-}, -6^{-}, 4^s)  \nonumber \\
& = i \frac{m_2^2 \spb{-5}.{6}} {\spa{-5}.{6} }
\, \frac{m_1^2 \spa{-6}.{5}} {\spb{-6}.{5} 
 \, \t_{15}} \nonumber \\
& = i\frac{m_1^2 m_2^2} {\t_{15}} \, ,
\label{YMPP}
\end{align}
where 
\begin{equation}
\t_{ij} = 2 p_i \cdot \ell_j \,,
\label{tDef}
\end{equation}
and we again used the two-scalar two-gluon amplitude in \eqn{SGGS}  for the second four-point amplitude in the cut.  

Similarly, for the $\{--\}$ helicity configuration we obtain the same 
expression, 
\begin{align}
C_{\rm YM}^{\{--\}} & = 
A^\tree_3 (3^s, 6^{-}, -7^s) \, 
A^\tree_3 (7^s, -5^{-}, 2^s) \,
A^\tree_4 (1^s, 5^{+}, -6^{+}, 4^s)  \nonumber \\
& = i \frac{m_1^2 m_2^2} { \t_{15}} \,.
\label{YMMM}
\end{align}

Next consider the $\{-+\}$ helicity configuration. Evaluating it along
the same lines of taking the residue of the $\tau_{36}$ pole of a
two-particle cut and evaluating the result at $\tau_{36}=0$, we find
\begin{align}
C_{\rm YM}^{\{-+\}} & =
A^\tree_3(3^s,6^{+}, -7^s) \,
A^\tree_3(7^s, -5^{-}, 2) \,
A^\tree(1^s,5^{+}, -6^{-},4^s) \nonumber \\
& = i \frac{ \sand{-5}. {2}. {6}^2 }{(\ell_5 - \ell_6)^2 } \times
\frac{ \sand{-6}. {4}. {5}^2}{(\ell_5 - \ell_6)^2 \t_{15}} \nonumber \\
& =  \frac{i}{t^2} \frac{\tr_+^2[5462]} { \t_{15}}\,,
\label{YMMP}
\end{align}
where $t \equiv (p_2 + p_3)^2 = (\ell_5 - \ell_6)^2$ and
\begin{equation}
\tr_\pm [abcd] \equiv 
\frac{1}{2} \tr[abcd] \pm \frac{1}{2} \tr_5[abcd] \equiv 
\frac{1}{2} \tr[(1 \pm \gamma_5) \s a \s b \s c \s d] \,.
\end{equation}

The $\{+-\}$ helicity configuration is identical except we have to switch
angle and square brackets, which has the effect of interchanging a $\tr_-$ and a
$\tr_+$.  This gives
\begin{align}
\hskip 1 cm 
C_{\rm YM}^{\{+-\}} & = 
A^\tree_3(3^s,6^{-}, -7^s) \,
A^\tree_3(7^s, -5^{+}, 2) \, A^\tree_4(1,5^{-}, -6^{+},4) \nonumber \\
& =  \frac{i}{t^2} \frac{\tr_-^2[5 46  2]} { \t_{15}}\,.
\label{YMPM}
\end{align}
The parity odd terms cancel, as expected, after summing the $\{-+\}$ and
$\{+-\}$ helicities configurations in \eqns{YMMP}{YMPM}.  

Summing over the four helicity contributions in Eqs.~\eqref{YMPP},
\eqref{YMMM}, \eqref{YMMP} and \eqref{YMPM}  to the cut in \fig{OneLoopCutsFigure}(a)
and simplifying the resulting expression, we find
\begin{align}
C_{\rm YM}^{\rm (a)}  & = 
C_{\rm YM}^{\{++\}} + C_{\rm YM}^{\{--\}} + C_{\rm YM}^{\{-+\}}  + C_{\rm YM}^{\{+-\}} \nn\\
& =  \frac{i}{\t_{15}} \Bigl( \frac{1}{t^2} (\tr_+^2[5 4 6 2] +  \tr_-^2[5 46  2] ) + 2 m_1^2 m_2^2 \Bigr)\,.
\label{CutAFinal}
\end{align}
The cut in \fig{OneLoopCutsFigure}(b) is given by a simple relabeling, $(1,4)\leftrightarrow (2,3)$, $5\rightarrow -5$ and $6\rightarrow -6$,
\begin{align}
C_{\rm YM}^{\rm (b)} 
& =  \frac{i}{\t_{36}} \Bigl( \frac{1}{t^2} (\tr_+^2[5 1 6 3] +  \tr_-^2[5 1 6 3] ) + 2 m_1^2 m_2^2 \Bigr)\nn\\
& =  \frac{i}{\t_{36}} \Bigl( \frac{1}{t^2} (\tr_+^2[5 4 6 2] +  \tr_-^2[5 46  2] ) + 2 m_1^2 m_2^2 \Bigr)\,,
\label{CutBFinal}
\end{align}
where in the second equal sign we  used that $p_2 = -p_3 +\ell_5 - \ell_6$ and
$p_4 = -p_1 -\ell_5 + \ell_6$ and that legs $5$ and $6$ are on shell.

\begin{figure}[tb]
        \begin{center}
                \includegraphics[scale=.5]{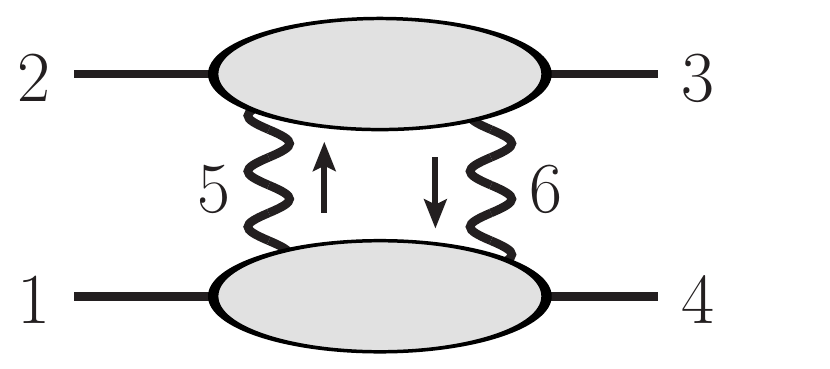}
        \end{center}
        \vskip -.5 cm
        \caption{\small The two particle cut at one loop containing classical potential. The blobs represent
                tree amplitudes and exposed lines are all on shell.}
        \label{OneLoop2CutFigure}
\end{figure}

\begin{figure}[tb]
	\begin{center}
		\includegraphics[scale=.6,trim={0 0.5cm 0.5cm 0cm},clip]{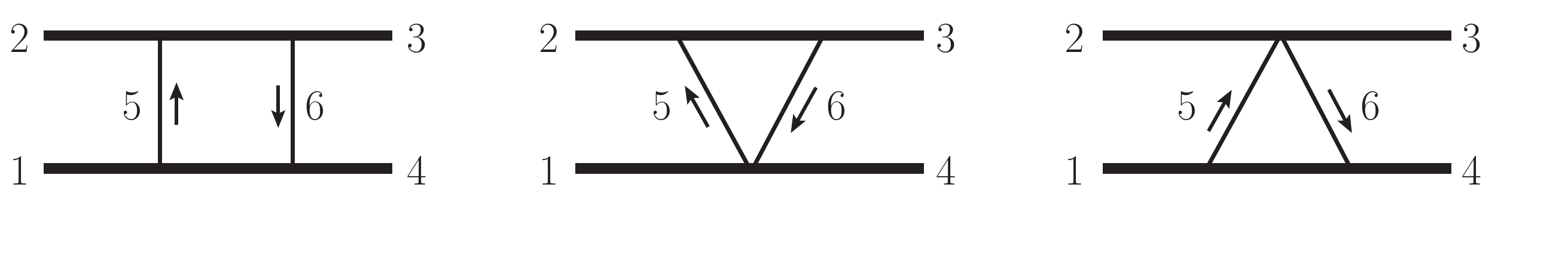}
	\end{center}
	\vskip -.5 cm
	\caption{\small The one-loop box and triangle integrals. The top and bottom
          thick lines are massive scalars $1$ and $2$, respectively,
          and the thin lines are massless.
            }
	\label{fig:oneloopbox}
\end{figure}

The last step in the construction of the integrand requires that we combine
the two cuts in \fig{OneLoopCutsFigure}, with values given in \eqns{CutAFinal}{CutBFinal}, into a single function.
This needs to be done such that any double counting coming from the same terms appearing
in both cuts is eliminated.  At two loops we will do this by constructing an ansatz
whose generalized cuts are constrained to match all required cuts.  
At one loop  such an approach is overly laborious.  The overlap can be more easily seen by inspection.
Because the two cuts have identical numerator factors, it is straightforward to construct a function which
reduces to \eqns{CutAFinal}{CutBFinal} upon imposing cut conditions on one of the propagators $i/\t_{15}$ and $i/\t_{36}$, respectively.
It is 
\begin{align}
C_{\rm YM} & = 
 - \frac{1}{\t_{15} \t_{36}} \Bigl( \frac{1}{t^2} (\tr_+^2[5 4 6 2] +  \tr_-^2[5 46  2] ) + 2 m_1^2 m_2^2 \Bigr)\, .
\label{OneLoop4DNoBubble}
\end{align}
It is worth stressing that this function needs not \textit{a priori} be (and in fact it is not) the two-gluon cut of the one-loop four-scalar amplitude displayed
in \fig{OneLoop2CutFigure} because it may not contain {\it all}
terms from which $1/\t_{15}$ and $1/\t_{36}$ are absent, corresponding 
to a graviton
bubble connected to the matter lines.\footnote{It however turns out that
Eq.~\eqref{OneLoop4DNoBubble} {\it is} the complete two-graviton cut
due to accidental features related to the simplicity of the one-loop
amplitude.}  It however does capture, by construction, all the box and
triangle contributions to the one-loop amplitude, which are the parts
needed for the extraction of the classical limit of the amplitude. See \Fig{fig:oneloopbox} for the diagrams that 
correspond to the box and triangle integrals.

Restoring the loop integration and the graviton propagators while
relaxing the on shell condition for their momenta gives the part of
the one-loop amplitude which contributes in the classical limit of the
gauge-theory four-scalar amplitude:
\begin{align}
A^{\oneloop}_4 & = \int \frac{d^D \ell_5}{(2 \pi)^D} \,
 \frac{N_{\rm YM}}{\ell_5^2 ((\ell_5 - p_2)^2 - m_2^2) (\ell_5 - p_2 -p_3)^2 ((\ell_5 + p_1)^2 - m_1^2)}\, .
\label{OneLoopFinalGR}
\end{align}
This last step does not capture terms in which either one or both
graviton propagators are canceled; while such terms do appear in the
complete quantum amplitude, which we discuss in
Sec.~\ref{MassSingularitySubsection}, they do not contribute to its
classical limit because they correspond to contact rather than
long-distance interaction of the matter fields.
The gauge-theory numerator is 
\begin{equation}
N_{\rm YM} =  \frac{1}{t^2} (\tr_+^2[5 4 6 2] +  \tr_-^2[5 46  2]) + 2 m_1^2 m_2^2 \,.
\label{NumeratorTraceYM}
\end{equation}
We also promoted the integration measure to $D=4-2\epsilon$
dimensions, dimensionally regularizing the infrared divergences of the
amplitude.

We can simplify somewhat Eq.~\eqref{NumeratorTraceYM} by evaluating the 
$\gamma$-traces.  Using standard $\gamma$-matrix identities, they evaluate to
\begin{align}
\tr_\pm[5 46  2] &  = {\cal E} \pm  {\cal O}\,,
\label{TrPlusMinus}
\end{align}
where the parity even part is
\begin{align}
 {\cal E} & = \frac{1}{2} \tr[5 4 6 2]\nn \\
 & =  2 \ell_5\cdot p_4 \ell_6 \cdot p_2 - 2\ell_5\cdot \ell_6 p_4 \cdot p_2
    + 2\ell_5\cdot p_2 \ell_6 \cdot p_4  \nonumber \\
&  = \frac{1}{2}  \Bigl( -st + t m_1^2 + t m_2^2 + t \, \t_{15} + t \t_{36} + 2 \t_{15}\t_{36} \Bigr) \, ,
\label{Even}
\end{align}
which we simplified using the graviton cut conditions. 
The parity odd part is 
\begin{equation}
 {\cal O} =  2 i \varepsilon_{\mu\nu\rho\sigma} \ell_5^\mu  p_4^\nu \ell_6^\rho  p_2^\sigma \,.
\end{equation}
This term integrates to zero because there are insufficient independent external momenta to saturate all the indices 
of the Levi-Civita symbol.
While the parity-odd term cancels between the two traces in Eq.~\eqref{NumeratorTraceYM}, 
\begin{eqnarray}
\tr_-^2[5 4 6 2] +  \tr_+^2[5 46  2] = 2( {\cal E}^2+  {\cal O}^2 )\,,
\label{YMTraces}
\end{eqnarray}
its square does not;
it gives a Gram determinant, 
\begin{align}
 {\cal O}^2 &=   G_4  \equiv 4 \det
\left( \begin{matrix}
  \ell_5 \cdot \ell_5 &  \ell_5 \cdot p_4 &  \ell_5 \cdot \ell_6 &   \ell_5 \cdot p_2 \\
   p_4\cdot\ell_5     &  p_4 \cdot p_4    &  p_4 \cdot \ell_6 &   p_4 \cdot p_2 \\
   \ell_6\cdot\ell_5     &  \ell_6\cdot p_4    &  \ell_6 \cdot \ell_6 &   \ell_6 \cdot p_2 \\
   p_2\cdot\ell_5     &  p_2\cdot p_4    &  p_2 \cdot \ell_6 &   p_2 \cdot p_2 
\end{matrix} \right)  .
\label{GramDetExample}
\end{align}
After being evaluated and simplified it becomes
 \begin{align}
 \nn\\
  {\cal O}^2  = {\cal E}^2 - (t m_1^2 + t\, \t_{15} + \t_{15}^2)(t m_2^2 + t\, \t_{36} + \t_{36}^2)\,,
\label{Odd2}
\end{align}
with ${\cal E}^2$ being the square of eq.~\eqref{Even}.

We can rewrite the numerator \eqref{NumeratorTraceYM} in terms of Mandelstam variables via \eqn{YMTraces}
\begin{equation}
N_{\rm YM} =  \frac{2}{t^2}\left({\cal E}^2 +{\cal O}^2 \right)+  2m_1^2 m_2^2  \, ,
\end{equation}
where ${\cal E}^2$ and ${\cal O}^2$ can be found from \eqns{Even}{Odd2}.
After inserting this into \eqn{OneLoopFinalGR} and keeping only 
the box and triangle integrals, as shown in \Fig{fig:oneloopbox}, we find that the one-loop four scalar amplitude is
\begin{align}
A^{\oneloop}_4 & = \int \frac{d^D \ell_5}{(2 \pi)^D} \, \frac{1}{\ell_5^2 (\ell_5-p_1-p_2)^2}
\biggl[ \frac{ (s - m_1^2 - m_2^2)^2}{((\ell_5 + p_1)^2 - m_1^2)\, ((\ell_5-p_2)^2 - m_2^2)}
 \label{OneLoopFinalYM2}\\
& \null \hskip 2. cm 
 + \frac{2m_1^2 t - 2m_2^2\t_{15} + \t_{15} t - 2 s t} {((\ell_5-p_2)^2 - m_2^2)\, t}
+ \frac{2 m_2^2 t-2 m_1^2 \t_{36} + \t_{36} t - 2 s t}{ ((\ell_5+p_1)^2 -m_1^2)\, t} \biggr]
 + \hbox{quantum}\, . \nn
\end{align}
The integrals are to be evaluated in $D=4 - 2\eps$ dimensions.

Given that the integrand of \eqn{OneLoopFinalYM2} is determined in
four dimensions, yet when integrated it is infrared divergent, one
might worry that $\Ord(\eps)$ terms which are dropped from the integrand may
actually contribute when multiplied by an infrared divergence. In the
full quantum theory a four-dimensional integrand evaluation would miss
certain rational pieces, as it happens in QCD~\cite{BernMorgan}.
However, as we shall see in \sect{sec:evanescence}, for extracting the classical
potential it is sufficient to evaluate the integrand in four dimensions
not only at one loop, but also at two loops.

\subsection{Gravity Integrands}

Using the double copy we now convert the above gauge-theory results for the integrand
into a corresponding gravity integrand.  Given the simplicity of the gauge-theory helicity amplitudes,
it is natural to apply the KLT relations \eqref{KLT} for tree amplitudes in the cuts.
This strategy is used in Ref.~\cite{BjerrumQuantum} to study PN and quantum corrections to 
the Newtonian potential.

Using the KLT relations, the generalized unitarity cut in \fig{OneLoopCutsFigure}(a) for the
gravity amplitude is
\begin{align}
C_{\rm GR}^{\rm (a)} & = \sum_{h_5, h_6 = \pm}
M^\tree_3(3^s,6^{h_6}, -7^s)   \, 
M^\tree_3(7^s , -5^{h_5}, 2^s) \, 
M^\tree_4(1^s,5^{-h_5}, -6^{-h_6},4^s) \nn \\
& = \sum_{h_5, h_6 = \pm} it [A^\tree_3(3^s,6^{h_6}, -7^s) \,
A^\tree_3(7^s , -5^{h_5}, 2^s) \, 
A^\tree_4(1^s,5^{-h_5}, -6^{-h_6},4^s)]  \nonumber\\
& \hskip 2 cm \null \times 
 [A^\tree_3(3^s,6^{h_6}, -7^s)   \,
  A^\tree_3(7^s , -5^{h_5}, 2^s) \,
   A^\tree_4(4^s,5^{-h_5}, -6^{-h_6},1^s)] \,,
\end{align}
where we use the BCJ amplitude relation~\cite{BCJ}
\begin{equation}
	s_{12} A^\tree(1,2,4,3)= s_{23} A^\tree(4,2,3,1)\,,
	\label{eq:fund_BCJ}
\end{equation}
for all four-point tree amplitudes in this paper.  So, in the end, the
gravity cut is expressed directly in terms of the four helicity
components of the gauge-theory cut.  We project out the dilaton and
axion ubiquitous in the double-copy relations and enforce that only
gravitons propagate across cuts simply by correlating the helicities
in the two gauge-theory copies.  As in the gauge-theory calculation,
some care is needed to ensure that the difference between $D$ and four
dimensions does not affect the classical limit which we discuss in
\sect{sec:evanescence}.

Following the gauge-theory analysis, we step, one by one, through the four helicity
configurations \eqref{TwoCutHelicities} of the cut gravitons.  For the $\{++\}$ configuration we have
\begin{align}
C_{\rm GR}^{\rm \{++\}} & = M^\tree_3(3^s,6^{+}, -7^s)   \,
M^\tree_3(7^s , -5^{+}, 2^s) \, 
M^\tree(1,5^{-}, -6^{-},4) \nonumber \\
& = i t \, [A^\tree_3(3^s,6^{+}, -7^s)   \,
A^\tree_3(7^s , -5^{+}, 2^s)  \,
        A^\tree_4(1,5^{-}, -6^{-}, 4^s)]  \nonumber\\
& \hskip .76 cm \null \times
 [A^\tree_3(3^s,6^{+}, -7^s)   \, 
  A^\tree_3(7^s , -5^{+}, 2^s)  \,
          A^\tree_4(4,5^{-}, -6^{-}, 1^s)] \,.
\end{align}
We simply read off the answer from the YM result in \eqn{YMPP}
after appropriate relabeling.  It is
\begin{align}
C_{\rm GR}^{\{++\}} & = -i t \frac{m_1^2 m_2^2}{\t_{15}} \,
                               \frac{m_1^2 m_2^2}{\t_{45}} \,.
\end{align}
We partial fraction the product of propagators, to make it compatible with a diagrammatic interpretation,
\begin{align}
\frac{1}{\t_{15} \t_{45}} & = 
-\frac{1}{t} \Bigl(\frac{1}{\t_{15}} + \frac{1}{\t_{45}} \Bigr) \,,
\label{PartialFaction}
\end{align}
where $t  = (\ell_5 - \ell_6)^2 = (p_1 + p_4)^2$.
In carrying out this partial fractioning we use the fact that $\ell_5$ and $\ell_6$ are on-shell momenta 
and, as in the gauge-theory evaluation of cuts, we have an all-outgoing convention for external momenta.  
The momentum flow of
$\ell_5$ and $\ell_6$ is indicated in \fig{OneLoopCutsFigure}.
This gives the remarkably simple result for the generalized cut with the $\{++\}$ helicity configuration,
\begin{equation}
C_{\rm GR}^{\{++\}}  = i m_1^4 m_2^4
   \Bigl(\frac{1}{ \t_{15}} + \frac{1}{\t_{45}} \Bigr)  \,.
\label{GRPP}
\end{equation}
As for gauge theory, the result for the $(--)$ helicity configuration is identical to that of the $\{++\}$ configuration:
\begin{equation}
C_{\rm GR}^{\{--\}} =  i m_1^4 m_2^4
   \Bigl(\frac{1}{ \t_{15}} + \frac{1}{\t_{45}} \Bigr)  \,.
\label{GRMM}
\end{equation}

Next, we evaluate the cut with the $\{-+\}$ helicity configuration for the cut gravitons. We obtain
\begin{align}
C_{\rm GR}^{\{-+\}} =& 
M^\tree_3(3^s,6^{+}, -7^s) \,
M^\tree_3(7^s , -5^{-}, 2^s) \,
M^\tree_4(1^s,5^{+}, -6^{-},4^s) \nonumber \\
=& i t \, [A^\tree_3(3^s,6^{+}, -7^s) \,
A^\tree_3(7^s , -5^{-}, 2^s) \,
              A^\tree_4(1^s,5^{+}, -6^{-},4^s)]  \nonumber\\
&\quad \times
        [A^\tree_3(3^s,6^{+}, -7^s) \,
         A^\tree_3(7^s , -5^{-}, 2^s) \,
         A^\tree_4(4^s,5^{+}, -6^{-},1^s)] \,.
\end{align}
Reading off the YM result in \eqn{YMMP} and appropriately relabeling, we find
\begin{align}
C_{\rm GR}^{\rm (-+)} & =
 -\frac{i}{t^3} \frac{\tr_-^2[5 46  2]} {\t_{15}} \,
 \frac{\tr_-^2[5 46  2]} {\t_{45}} \,.
\end{align}
In addition, partial fractioning the product of propagators using \eqn{PartialFaction} gives
\begin{align}
C_{\rm GR}^{\rm (-+)}  = &
\frac{i}{t^4} \tr_-^4[5 46  2] \,
   \Bigl(\frac{1}{  \t_{15}} + \frac{1}{\t_{45}} \Bigr) \,.
   \label{GRMP}
\end{align}
Following similar steps for the final $(+-)$ helicity configuration (or simply conjugating the result for the $\{-+\}$ configuration), we have
\begin{align}
C_{\rm GR}^{\rm (+-)}  = &
\frac{i}{t^4} \tr_+^4[5 46  2] 
\, \Bigl(\frac{1}{  \t_{15}} + \frac{1}{\t_{45}} \Bigr)
   \,.
   \label{GRPM}
\end{align}

Combining Eqs.~\eqref{GRPP}, \eqref{GRMM}, \eqref{GRMP} and \eqref{GRPM} yields the cut in \fig{OneLoopCutsFigure}(a):
\begin{align}
C_\text{GR}^{\rm (a)} & = C_{\rm GR}^{\rm (++)} + C_{\rm GR}^{\rm (--)} + C_{\rm GR}^{\rm (-+)} 
      + C_{\rm GR}^{\rm (+-)} \nn \\
& =  i \Bigl(\frac{1}{t^4} \, \bigl(\tr_-[5 46  2]^4 + \tr_+[5 46  2]^4 \bigr) + 2 m_1^4 m_2^4 \Bigr)
  \Bigl(\frac{1}{  \t_{15}} + \frac{1}{\t_{45}} \Bigr) \,.
\label{GRCutTotalA}
\end{align}
Again the parity odd terms cancel in the integrand between the contributions of the various helicity configurations. 
We obtain the second required cut, shown in \fig{OneLoopCutsFigure}(b), by a simple relabeling and applying on-shell conditions, as in the gauge-theory case:
\begin{align}
C_\text{GR}^{\rm (b)} & = i\Bigl(\frac{1}{t^4} \, \bigl(  \tr_-[5 46  2]^4 +  \tr_+[5 46  2]^4 \bigr) 
   + 2 m_1^4 m_2^4 \Bigr)  \Bigl(\frac{1}{\t_{36}} + \frac{1}{\t_{26}} \Bigr) \,.
\label{GRCutTotalB}
\end{align}

Relaxing the on-shell condition for the cut momenta, restoring the cut propagators and combining
Eqs.~\eqref{GRCutTotalA} and \eqref{GRCutTotalB} into a single expression whose two cuts 
in \fig{OneLoopCutsFigure} match \eqns{GRCutTotalA}{GRCutTotalB} leads to
\begin{align}
M^\oneloop(1,2,3,4) & =  \frac{1}{2}  \int \frac{d^D \ell_5}{(2 \pi)^D} \, 
N_{\rm GR} \,
\biggl[\frac{1}{\ell_5^2 ((\ell_5 - p_2)^2 - m_2^2) (\ell_5 - p_2 - p_3)^2 ( (\ell_5 +p_1)^2 -m^2)} \nn\\
& \hskip 4 cm 
           + \{2 \leftrightarrow 3\}  + \{1 \leftrightarrow 4\} +\{2 \leftrightarrow 3, 1 \leftrightarrow 4\}\biggr] 
 + \hbox{quantum}\,,
\label{OneLoopGRAmplitude}
\end{align}
where $\ell_6 = \ell_5 -p_2 - p_3$ and we included an overall 1/2 for
the two-graviton phase-space symmetry factor. The numerator $N_{\rm GR}$ is 
\begin{align}
N_{\rm GR} & = 
\frac{1}{t^4} \, (\tr_-^4[5 46  2] +  \tr_+^4[5 46  2]) + 2 m_1^4 m_2^4 \nn \\
& = \frac{2}{t^4} \, ( ({\cal E}^2 + {\cal O}^2)^2  + 4 {\cal E}^2 {\cal O}^2) + 2 m_1^4 m_2^4 \,,
\label{NumeratorGR}
\end{align}
where ${\cal E}^2$ and ${\cal O}^2$ can be found from \eqns{Even}{Odd2}.
The numerator is symmetric under the relabelings indicated in \eqn{OneLoopGRAmplitude}.
We see that each integral appears twice, and we can simplify
the amplitude \eqref{OneLoopGRAmplitude} to 
\begin{align}
M^\oneloop(1,2,3,4) & =  \int \frac{d^D \ell_5}{(2 \pi)^D} \, N_{\rm GR}
\biggl[\frac{1}{\ell_5^2 ((\ell_5 - p_2)^2 - m_2^2) (\ell_5 - p_2 - p_3)^2 ((\ell_5 + p_1)^2 - m_1^2)} \nn\\
& \hskip 3 cm  + \{1 \leftrightarrow 4\} \Bigr] + \hbox{quantum}\,.
\label{FinalGrav}
\end{align}
This immediately can be reduced to the standard scalar box, triangle
and bubble integrals.  The bubble integrals are all quantum and can be
dropped without affecting the classical potential.  The box integral
is obtained by setting $\t_{15}, \t_{36} \rightarrow 0$ since these factors cancel
matter propagators that would give us triangle or bubble integrals.
This gives us the box-integral contribution
\begin{equation}
-t^2 (M_4^\tree)^2
\int \frac{d^D \ell_5}{(2 \pi)^D} \, 
\frac{1}
 {\ell_5^2 ((\ell_5 - p_2)^2 - m_2^2) (\ell_5 - p_2 - p_3)^2}
\biggl(\frac{1} {(\ell_5 + p_1)^2 - m_1^2} + 
\frac{1}{(\ell_5 + p_4)^2 - m_1^2} \biggr) \,, 
\label{D4BoxContribution}
\end{equation}
where the four-dimensional tree amplitude is
\begin{equation}
M_4^\tree = -i \, \frac{( s - m_1^2 - m_2^2 )^2
       - 2 m_1^2  m_2^2}{t} \,.
\label{TreeFourD}
\end{equation}
The appearance of the square of the tree amplitude in front of the box
integral is not an accident and is intimately connected to the fact
that, from the effective field theory perspective, the box contribution is an infrared-divergent iteration of
tree-level exchange.  This will be subtracted in matching with
effective field theory and will not have an independent contribution
to the classical potential.

As a simple check, it is not difficult to add back the dilaton and axion (antisymmetric
tensor) contributions.  To get these we simply sum over the
contributions where, for at least one cut leg, the helicities
of two gluons corresponding to a particle in the double copy theory
are anti-correlated. Repeating the above steps for this 
case then gives the dilaton and axion contributions to the
numerator,
\begin{equation}
N_{\rm \phi a} =
\frac{1}{t^4} \, \bigl(2 \tr_+^2[5 4 6  2] \,\tr_-^2[5 4 6  2] \bigr)
    + 2 m_1^4 m_2^4  + \frac{4}{t^2} \bigl(\tr_+^2[5 4 6  2] + \tr_-^2[5 46  2]\bigr) m_1^2 m_2^2  \,.
\label{NumeratorDA}
\end{equation}
Combining the graviton contribution to the numerator \eqref{NumeratorGR}, with that of the dilaton and axion \eqref{NumeratorDA} 
gives exactly  the simple double-copy form for the numerator,
\begin{align}
N_{\rm GR} +  N_{\rm \phi a}
& = \Bigl(\frac{\tr_+^2[5 46  2] + \tr_-^2[5 4 6  2]}{t^2}  +  2 m_1^2 m_2^2 \Bigr)^2 , 
\end{align}
which is the square of the gauge-theory numerator in \eqn{NumeratorTraceYM}, as expected.

Although we calculated the
integrand in four dimensions, the result in \eqn{FinalGrav} is
perfectly valid for extracting the classical potential, despite of the
fact that we will use dimensional regularization to regularize
infrared divergences. The details can be found in \sect{sec:evanescence}.

\section{Integrands at One Loop Using $D$-Dimensional Methods}
\label{OneLoopDCutsSection}

When constructing scattering amplitudes we encounter
infrared-divergent integrals.  To evaluate them we use dimensional
regularization, where we continue the space-time dimension from four to $D= 4 - 2 \epsilon$.  Infrared singularities appear as poles in $\epsilon$,
which can interfere with ${\cal O}(\epsilon)$ terms to produce finite
contributions.  Indeed, in the full quantum theory such terms do occur
and need to be kept, as described in e.g. Ref.~\cite{BernMorgan}. We therefore need to confirm
that ${\cal O} (\eps)$ terms in the integrand do not alter our results
for the classical potential in some way.\footnote{Such interference
terms also appear in certain integral identities used
in \sect{sec:diffEq}, where they are crucial for obtaining the correct
classical limit when evaluating the integrals via the differential
equations method.}

When applying dimensional regularization, the simplest scheme is the
so-called conventional dimensional regularization (CDR) scheme~\cite{CDR},
where all states and loop momenta are analytically continued from four
to $D$ dimensions. The four-dimensional helicity (FDH) scheme~\cite{Bern:1991aq, hep-ph/0202271} is
an alternative that meshes well with helicity methods.  In this scheme,
we distinguish between the dimension of loop momenta and the space
where physical states live.  We treat any factor of $D$
arising in the integrand from contracting Lorentz indices, $D_s
= \eta_\mu^{\ \mu} = 4$, differently from loop momenta which we take
to live in $D=4-2\eps$.  We find it useful to distinguish between
these two sources of dimension dependence to help track how dependence
on the dimensional regularization prescriptions drop out in the classical 
potential, as discussed
in \sect{sec:evanescence}.  While a technical point, this is of some
importance, especially beyond two loops where four-dimensional
helicity methods, which implicitly choose both $D=4$ and $D_s = 4$, are expected to be more
efficient than $D$-dimensional ones.

In this section, as a warm-up for the more complicated two-loop case, 
we investigate the dependence on regularization scheme by constructing a $D$-dimensional version of the one-loop integrand which
we compare to the one obtained in Sec.~\ref{OneLoopSection} using four-dimensional methods. 
Such a $D$-dimensional one-loop integrand has also been recently constructed using similar
methods in Ref.~\cite{Paolo2PM}.  The one-loop case is especially simple
because, as we shall see, the $D$- and four-dimensional integrands are
identical up to differences in the $D_s$ state-counting
parameters. As we discuss in \sect{TwoLoopSection}, at two loops the
difference is nontrivial because of the appearance of Gram
determinants of momentum invariants that vanish in four dimensions but
not in $D >4$ dimensions.  Nevertheless, we will show that they do not
affect the classical potential.

\subsection{Warm-up: Gauge-Theory Integrands}

Our $D$-dimensional construction makes use of BCJ duality, reviewed
in \sect{IntegrandsSection}, so it is natural to start with
color-dressed amplitudes, instead of color-ordered ones.  As in the
four-dimensional discussion, once we obtain a $D$-dimensional
gauge-theory integrand, the double-copy construction gives
the gravitational one with minimal additional effort.

An essential component of the evaluation of generalized unitarity cuts is the sum over physical 
states. At two loops they can become quite involved so it is useful to devise methods for simplifying them. 
The sum over the physical states of a gluon in general dimension
is less straightforward than summing over positive and negative helicities in four dimensions.
It is given by the so-called physical state projector,
\begin{equation}
P^{\mu\nu}(p,q) =  \sum_{\rm pols.} \pol^\mu(-p) \pol^\nu(p)
 = \eta^{\mu\nu} - { q^\mu p^\nu + p^\mu q^\nu \over q\cdot p } \,,
\label{YMProjector}
\end{equation} 
where the sum runs over the physical gluon polarizations, $p$ is the gluon momentum and $q$ is an arbitrary null reference momentum.  
We can use the projector to replace a pair of polarization vectors with kinematic invariants.
Note that there is a spurious propagator,  $1/q\cdot p$,  which must cancel out once all on-shell 
conditions on the cut are applied.

While including this projector in one-loop unitarity cuts poses no
technical challenges, by two loops using it becomes cumbersome,
especially for gravity.  However, as we now explain, we can adjust
the form of the tree amplitudes to automatically set to zero all terms
that depend on the reference momentum.  The essential idea is to
choose a form of the tree-level amplitudes used in cuts such that, when a polarization vector
is replaced by the corresponding momentum (i.e. under a gauge
transformation), the amplitude vanishes {\it without} using the
transversality of the remaining polarization vectors~\cite{Paolo2PM}.
This significantly cleans up $D$-dimensional unitarity cuts beyond one 
loop because it allows us to use a simpler physical state projector.

To illustrate this idea, we start with the three-point tree amplitude with one gluon and 
two massive scalars in the adjoint representation.  It is given by the Feynman
three-point vertex,
\begin{equation}
\colorA{A}_3^\tree(1^s,2,3^s) = \frac{i}{\sqrt{2}}\, \f^{a_1 a_2 a_3} (p_3 - p_1)\cdot\pol_2 \,.
\label{YMThreeAmplitude}
\end{equation}
This amplitude automatically satisfies the on-shell Ward identity
\begin{equation}
\colorA{A}_3^\tree(1^s,2^s,3) \bigr|_{\pol_2 \rightarrow p_2} = 0 \,,
\label{3PtWard}
\end{equation}
because $p_2 \cdot (p_3 - p_1) = - (p_3 + p_1) \cdot (p_3 - p_1) = 0$.

\begin{figure}
\begin{center}
\includegraphics[scale=.53]{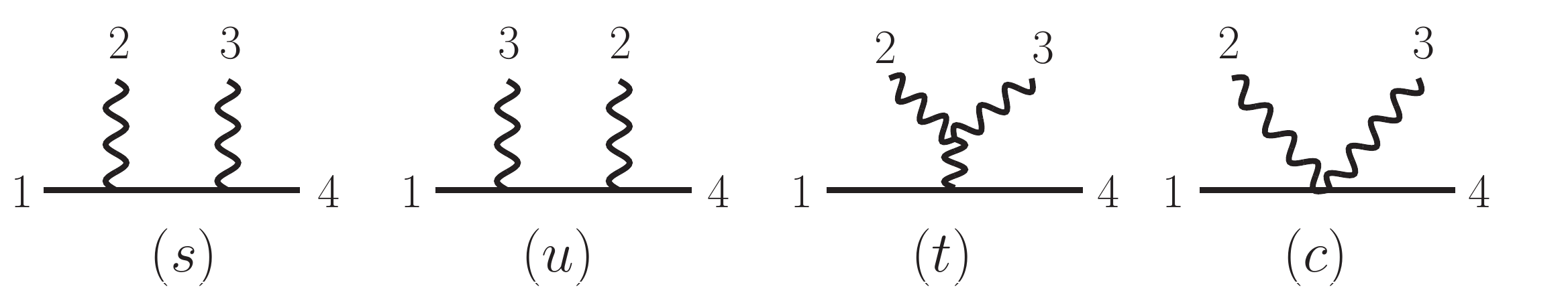}
\end{center}
\vskip -.5cm
\caption{\small Feynman diagrams for the four-point amplitudes. Absorbing
the four-point contact term $(c)$ into diagrams $(s)$, $(t)$, and $(u)$ gives a BCJ representation of
the amplitude.
}
\label{FourPtFeynmanFigure}
\end{figure}

At four points it is straightforward to obtain gauge-theory tree amplitudes that manifest 
BCJ duality, because {\it all} representations of the amplitude in term of diagrams with 
only cubic vertices have this property~\cite{BCJ}.  
We, however, also want that, simultaneously, the amplitude satisfies the on-shell Ward identities
\begin{equation}
\colorA{A}^\tree_4(1^s,2,3,4^s) \bigr|_{\pol_2 \rightarrow p_2} = 0 \,, 
\hskip 1.5 cm 
\colorA{A}^\tree_4(1^s,2,3,4^s) \bigr|_{\pol_3 \rightarrow p_3} = 0\,,
\label{4PtWard}
\end{equation}
{\em without} using the transversality of the asymptotic state of the remaining gluon,
\begin{equation}
\pol_3\cdot p_3 = 0
\,, \hskip 1.5 cm 
\pol_2\cdot p_2 = 0 \,,
\label{OnShellFourPt}
\end{equation}
respectively.
To find such a representation we can start, for example, with Feynman
diagrams in Feynman gauge, illustrated in \fig{FourPtFeynmanFigure};
for four-point trees there is no need for more sophisticated approaches.
We then rearrange the diagrams in two ways.  First we 
absorb the four-vertex into the diagrams with only three vertices by 
 matching the color factors and multiplying and dividing by the appropriate propagator.
Then we add terms that vanish on shell while demanding that
the on-shell Ward identity hold automatically on each gluon leg without using 
physical state conditions on the other external gluon.   The result is an amplitude
of the form
\begin{equation}
\colorA{A}^\tree_4(1^s,2,3,4^s) = \frac{n_s \, c_s}{s-m^2}+ \frac{n_t \, c_t}{t}
+ \frac{n_u \, c_u}{u-m^2}\,,
\label{YMFourAmplitude}
\end{equation}
where the Mandelstam invariants are 
\begin{equation}
s = (p_1 + p_2)^2\,, \hskip 1 cm 
t = (p_2 + p_3)^2\,, \hskip 1 cm 
u = (p_1 + p_3)^2\,,
\end{equation}
and $m^2 = k^2_1 = k^2_4$ is the mass of the scalar.
The color factors are easily read off from the diagrams in
\fig{FourPtFeynmanFigure},
\begin{equation}
c_s=  \f^{a_1 a_2 b} \f^{b a_3 a_4}\,, \hskip 1 cm 
c_t=  \f^{a_2 a_3 b} \f^{b a_4 a_1}\,, \hskip 1 cm 
c_u=  \f^{a_1 a_3 b} \f^{b a_2 a_4}\,,
\label{color4pt}
\end{equation}
where the color group structure constants are defined in \eqn{Fabcadjoint}. The color factors satisfy the Jacobi identity
\begin{equation}
c_t = c_s - c_u \,.
\label{ColorJacobiDDim}
\end{equation}
The $s$-channel kinematic numerator is
\begin{align}
n_s = n(1^s,2,3,4^s) =
\frac{i}{2}\Bigl\{ (s-m^2) \,  \pol_2 \cdot \pol_3
- 2 (p_1 \cdot \pol_2)\, (p_{12} \cdot \pol_3)
- 2 (p_{34} \cdot \pol_2) \, (p_{4} \cdot \pol_3)
\Bigr\}, 
\label{ImprovedNumerators}
\end{align}
with $p_{ij} =p_i +p_j$. The other two numerators follow from the above: the $u$-channel numerator is obtained by
 swapping labels $2$ and $3$ in $n_s$,
\begin{align}
	n_u &= n(1^s,3,2,4^s),
\end{align}
and the $t$-channel numerator follows from the kinematic Jacobi relation
\begin{equation}
n_t = n_s -n_u\,.
\label{KinematicJacobiDDim}
\end{equation}
The duality between color and kinematics is manifest in Eq.~\eqref{ColorJacobiDDim} and Eq~\eqref{KinematicJacobiDDim}.

Compared to the result of a Feynman diagram calculation, we have added to the numerators terms proportional to $p_2\cdot \pol_2$ and  $p_3\cdot \pol_3$ 
that vanish for asymptotic physical states. 
While this choice has no physical effect, it greatly simplifies the
unitarity construction, especially beyond one loop. Observe that the
second term in the projector in Eq.~\eqref{YMProjector} replaces one
of the polarization vectors with its momentum, which resembles the left-hand
side of the on shell Ward identity. The only difference is that the rest of legs
may not satisfy physical state conditions, $p_i \cdot \pol_i =0$,
after they have been sewed with other tree amplitudes in the unitarity
cut.  However, if we can choose the amplitude to satisfy the on-shell Ward
identity without demanding physical state conditions, then the Ward
identity can be applied term by term in the state sum.  So, with the numerators
chosen as above, we can simplify the projector to~\cite{3PMPRL,Paolo2PM}
\begin{equation}
\sum_\lambda \pol^\mu(-p) \pol^\nu(p) \rightarrow \eta^{\mu\nu}\,,
\label{ImprovedYMPropagator}
\end{equation}
which not only reduces the number of terms, but also eliminates the
appearance of the reference momentum and the corresponding spurious
propagator.  While the propagator
numerator \eqref{ImprovedYMPropagator} is analogous to the one in
Feynman gauge, there is no need for Fadeev-Popov ghosts or
physical-state projectors even when constructing full quantum amplitudes
via generalized unitarity.
Although the one-loop case in gauge theory is simple enough for the
advantage of this reorganization to not be obvious, we will see that,
at two loops, the same technique enormously simplifies the $D$
dimensional cut calculations, especially in gravity.

Using the adjusted four-point amplitude, we now construct the one-loop
color-dressed gauge-theory cuts. We will then extract their
color-ordered components and compare them with the color-ordered
four-dimensional cuts. The color-dressed cut in
\fig{OneLoopCutsFigure}(a) is given by
\begin{align}
 {\cal  C}_{\rm YM}  
 = & \sum_{\rm pols.}
 \colorA{A}_3^\tree(3^s,6,-7^s)\, \colorA{A}_3^\tree(7^s,-5,2^s)\,
 \colorA{A}_4^\tree(1^s,5,-6,4^s)  \nn \\
 = & \sum_{\rm pols.}
   \biggl(2 \f^{a_3 a_6 a_7}\f^{a_7 a_5 a_2}\, (p_2 \cdot \pol_{-5})\,
   (p_3 \cdot \pol_{6})  \biggr) \times 
   \biggl(\frac{c'_t \, n'_t}{t}  
   + \frac{c'_s \, n'_s}{\t_{15}} 
   + \frac{c'_u \, n'_u}{\t_{45}} \biggr )\,,
\label{YMFullColorCut}
\end{align}
where the color factors $c'_{s,t,u}$ are obtained from
Eq.~\eqref{color4pt} by swapping the labels $(2,3)$ into $(5,-6)$
while the kinematic numerator factors $n_{s,t,u}'$ are obtained from
$n_{s,t,u}$ by the same relabeling.  Because the on-shell Ward
identities are all automatically satisfied, the state sum simplifies
and we only need to replace the product of two polarization vectors by
the improved state sum in \eqn{ImprovedYMPropagator}.
It is not difficult to simplify the resulting expressions using the
cut conditions on the gluon lines, $\tau_{55}=\tau_{66}=0$, and on the
matter line, $\tau_{36}=0$.

To compare with the four-dimensional color-ordered cut calculation
discussed in \sect{OneLoopSection} we need to extract the
color-ordered components of \eqref{YMFullColorCut}. For example, the
coefficient of $\Tr[T^{a_1}T^{a_2}T^{a_3}T^{a_4}]$ receives
contributions from the numerators $n_s'$ and $n_t'$; it is
\begin{align}
C_{{\rm YM}}^{{\rm (a)}}  = \sum_{\rm pols.}
  \Bigl( 2 (p_2 \cdot \pol_{-5})\,
 (p_3 \cdot \pol_{6})  \Bigr) 
 \left( \frac{n_s'}{\tau_{15}} + \frac{n_t'}{t} \right)  .
\end{align}
Simplifying it using the on shell conditions for the cut lines and the improved gluon physical 
state projector \eqref{ImprovedYMPropagator}, gives 
\begin{align}
C_{\rm YM}^{\rm (a)} & =
 \frac{i}{t\,\t_{15}}\left(t(s - m_1^2 - m_2^2)^2 + \t_{15} (2 t\, m_1^2 - 2 \t_{15}\, m_2^2  + \t_{15} t - 2 s t) \right) \, .
\label{OneLoopDNoBubble1}
\end{align}
By relabeling we also obtain the cut in \fig{OneLoopCutsFigure}(b)
\begin{align}
C_{\rm YM}^{\rm (b)} & =
\frac{i}{t\,\t_{36}}\left(t(s - m_1^2 - m_2^2)^2 + \t_{36} (2 t\, m_2^2 - 2 \t_{36}\, m_1^2  + \t_{36} t - 2 s t) \right) \,.
\label{OneLoopDNoBubble2}
\end{align}
Combining the two cuts reproduces the amplitude obtained from four-dimensional cuts in \eqn{OneLoopFinalYM2}. 

The match between one-loop integrands obtained from four- and $D$-dimensional methods is not accidental.  Discrepancies can arise from
two sources: (1) Gram determinants involving five independent vectors,
which vanish in four dimensions but not $D$ dimensions and (2) factors
of dimension arising from the trace of the metric, $\eta_\mu{}^\mu =
D_s$. The former cannot appear, because the one-loop problem contains
only four independent momenta. The latter can occur only in bubble
diagrams which do not contribute to the classical limit.

The color-dressed cut in \eqn{YMFullColorCut} immediately gives us the
cut with the same topology of the gravity amplitude by replacing each
color factor by its corresponding numerator factor, and taking into
account the physical state projector that results from the sum over
graviton polarizations, which we discuss next.

\subsection{Gravity Integrands via BCJ Double Copy}

Consider now the gravity generalized cuts in $D = 4 -2\eps$
dimensions. We begin here with the less involved 2PM case as a warm-up for 3PM.
Given that the gauge-theory numerators in Eq.~\eqref{ImprovedNumerators}
respect BCJ duality, it is a simple matter to recycle the gauge
theory generalized cuts to gravity ones.  It is nevertheless important to examine them because, 
similarly with gauge theory,  gravity also exhibits infrared singularities, so it is possible that 
interference terms between ${\cal O}(\eps)$ numerator pieces and $1/\eps$ infrared singularities may 
yield finite terms in addition to those found through a $D=4$ cut calculation.
The 2PM scattering amplitude in $D$ dimensions was recently presented in \cite{Paolo2PM}.  Here we discuss 
a similar analysis of the integrand, focusing on the question of whether any dimensional regularization subtleties 
can affect the four-dimensional cut construction.

Naively, it would seem that we need the graviton physical-state 
projector in order to prevent the dilaton and antisymmetric tensor
from contributing in double-copy constructions.  The $D$-dimensional 
graviton physical state projector is
\begin{align}
P^{\mu\nu \rho \sigma}(p,q)= \sum_{\rm pols.} \pol^{\mu\nu}(-p) \pol^{\rho\sigma}(p) =  &\, \frac{1}{2} \bigl( P^{\mu\rho} P^{\nu\sigma} + P^{\mu\sigma} P^{\nu\rho} \bigr)
- \frac{1}{D_s-2} P^{\mu\nu} P^{\rho\sigma} \,,
\label{eq:full_projector}
\end{align}
where $P^{\mu\nu}$ is the gluon physical state projector in Eq.~\eqref{YMProjector} with momentum $p$ 
and a null reference momentum $q$.
The sum runs over the $D_s(D_s -3)/2$ physical states of the graviton.
As usual we denote any factor $D$ associated with state count as $D_s$, so that
we may distinguish it from the dimension of the loop integration.

We can simplify the cuts considerably, by using in the double copy construction gluon amplitudes that automatically project out all 
longitudinal polarizations. In this way the graviton generalized cuts inherit the simplicity of the gluon ones.
Indeed, the net effect of using such representations of tree amplitudes is that, in all unitarity cuts, we can replace the 
physical state projectors with a much simpler one,  equivalent to the projector in de Donder gauge,
\begin{equation}
\sum_{\rm pols.} \pol_{\mu\nu}(-p) \pol_{\rho\sigma}(p)
= \frac{1}{2} \eta_{\mu \rho} \eta_{\nu \sigma} 
+ \frac{1}{2} \eta_{\nu \rho} 
\eta_{\mu \sigma} - \frac{1}{ D_s-2}
\eta_{\mu \nu} \eta_{\rho \sigma}\,.
\end{equation}
We can simplify it even further by observing that the antisymmetric tensor
does not couple directly to scalar fields. This implies
that, for external scalars, the antisymmetric tensor can appear only
in closed loops that do not contain a matter line.
Since this violates the rule that every loop needs at least one cut matter line, 
the antisymmetric tensor cannot, in fact, contribute to the classical potential.
We therefore do not need to explicitly symmetrize the projector
indices, leaving us with the remarkably simple projector
\begin{equation}
\sum_{\rm pols.} \pol_{\mu\nu}(-p) \pol_{\rho\sigma}(p)
=  \eta_{\mu \rho} \eta_{\nu \sigma} 
- \frac{1}{ D_s-2} \eta_{\mu \nu} \eta_{\rho \sigma}\,.
\label{SimplifiedGravitonProjector}
\end{equation}
The first term is precisely the one of a naive double copy.  The
second term is a correction needed to subtract out the dilaton or,
alternatively, the trace of the metric fluctuation.  Because this
projector preserves much of the double copy structure, it is much
simpler to use and makes more transparent the transference of
gauge-theory properties to gravity.  Remarkably, with this
reformulation, the propagator for sewing amplitudes is even
simpler than the standard de Donder gauge one.

To obtain the gravity cut, we apply the double-copy
procedure \eqref{ColorSubstitution} to \eqn{YMFullColorCut}, i.e.  we
replace the color factors by another copy of the kinematic numerator
factors, and find
\begin{align}
{\cal C}_{\rm GR}^{\rm (a)} &= \sum_{\rm pols.}
\biggl( -4(p_2 \cdot \pol_{-5})^2\,
(p_3 \cdot \pol_{6})^2  \biggr) \times 
i\,\biggl(\frac{n'_t \, n'_t}{t}  
+ \frac{n'_s \, n'_s}{\t_{15}} 
+ \frac{n'_u \, n'_u}{\t_{45}} \biggr )\, .
\end{align}
The sum over polarizations generates the simplified physical state projector 
given in \eqn{SimplifiedGravitonProjector}.

We have explicitly verified that, when sewing tree amplitudes using the
graviton physical-state projectors with $D_s = 4$, we obtain {\em
precisely} the result found by summing over four-dimensional
helicity states in cuts, given in~\eqn{GRCutTotalA}.  The reason for
this match is the same as in gauge theory: at one loop there is
no kinematic object that vanishes in four dimensions but not in $D$
dimensions.  The only source of dimensional dependence is then
the state-counting parameter $D_s$ in the graviton
propagator~\eqref{SimplifiedGravitonProjector}.  

The dependence on $D_s$ in the physical state projector does however imply that 
there is a difference between the $D$-dimensional integrand 
and the four-dimensional one. For the cut in \fig{OneLoopCutsFigure}(a) 
this difference is
\begin{align}
{\cal  C}^{\rm (a)}_{\rm GR} - {\cal  C}^{\rm (a)}_{\rm GR}|_{D_s=4} =&
\frac{4 m_1^2 m_2^2(D_s -4)}{(D_s-2)} \Bigg[
\frac{t}{\t_{45}}(2s-2m_1^2-2m_2^2-\t_{15})-\frac{m_2^2 \t_{15} \t_{45}}{2m^2_1 t}
\nn\\
& \hskip 1 cm \null 
\left(
(s-m_1^2-m_2^2)^2 
+ m_2^2 (m_1^2 \,t-\t_{15}\t_{45})\frac{D_s}{t(D_s-2)} \right)
\left(\frac{1}{\t_{15}}+\frac{1}{\t_{45}}\right)
\Bigg] \,.
\label{OneloopDCutA}
\end{align}
Using $\t_{15} + \t_{45} = -t$ on the cut, we can organize the above equation into standard cut integrals.
As before, the cut in \fig{OneLoopCutsFigure}(b) may be obtained by a simple relabeling.
In sec.~\ref{DiscussionSection} we will show that this difference between the results of the $D$-dimensional 
and four-dimensional cut construction has no effect on the classical potential.

\section{Integrands at Two Loops}
\label{TwoLoopSection}     

\begin{figure}[tb]
\begin{center}
\includegraphics[scale=.45]{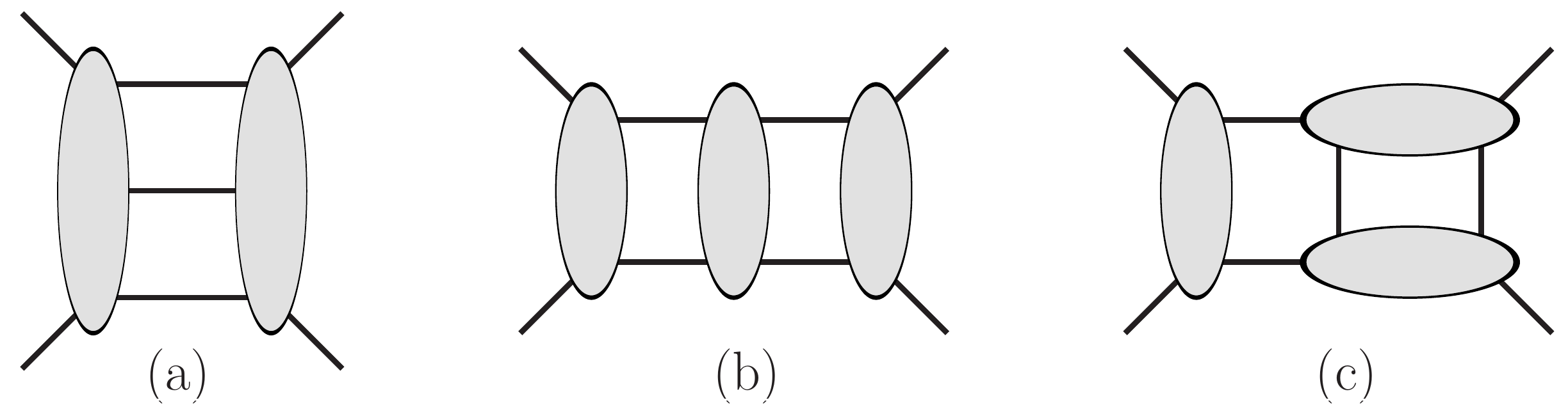}
\end{center}
\vskip -.5cm
\caption{\small A spanning cuts that determine the full integrand other than quantum bubble
  on external leg contributions.  The complete set of cuts is given
  including all distinct labels and routing of different particles through the
  cuts. The lines represent on-shell particles of any type.
}
\label{TwoLoopSpanningCutsFigure}
\end{figure}

In this section  we describe the construction of the two-loop gravity
integrand using both four-dimensional helicity and $D$-dimensional methods 
and show explicitly that the four-dimensional construction is sufficient to capture
all contributions to the classical potential in four space-time dimensions. 
As a warm-up, we first analyze the simpler case of color-ordered gauge-theory 
four-scalar amplitude, pointing out various features of the construction that carry 
over to the gravity case.

As at one loop, we use the generalized unitarity
method~\cite{hep-ph/9403226, hep-ph/9409265, hep-ph/9708239, hep-th/0412103, 0705.1864}, briefly summarized in
\sect{IntegrandsSection}, to construct the two-loop integrand.  The complete
quantum integrand can be obtained using a spanning set of cuts,
which amounts to the set of cuts from which every term in the loop
integrand can be determined.  For the two loop massless case such a set is
shown in \fig{TwoLoopSpanningCutsFigure}, where all distinct
labeling and routings of different particles need to be
included. In the massive case there are additional contributions not captured by these cuts, related 
to bubbles on external legs, but these are purely quantum effects
which we ignore.

\begin{figure}[tb]
\begin{center}
\includegraphics[scale=.5]{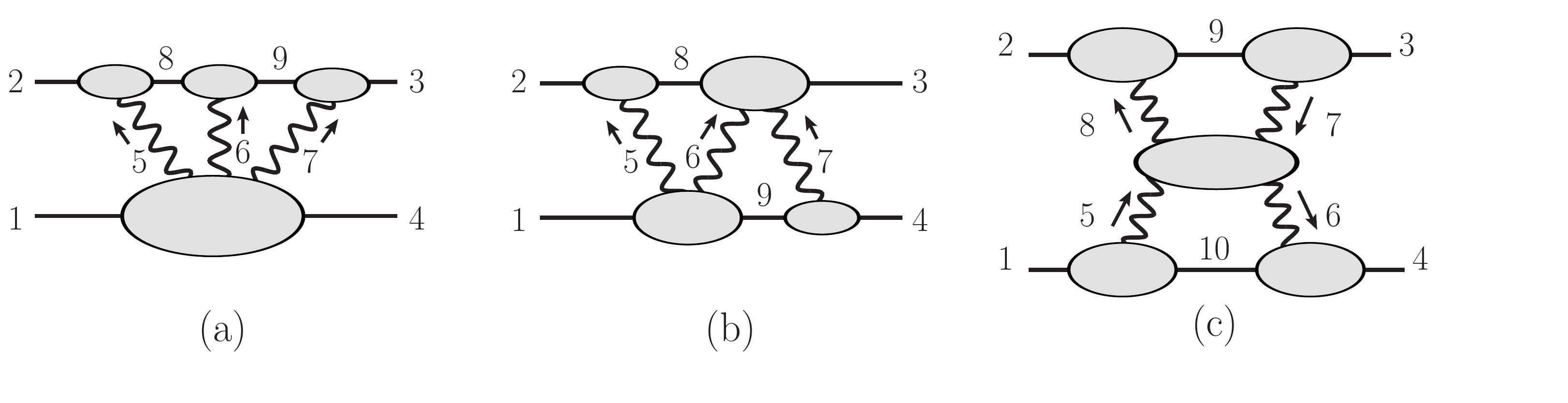}
\end{center}
\vskip -.5cm
\caption{\small The independent generalized cuts needed at two loops for the classical potential.
The remaining contributing cuts are given by simple relabeling of external legs.  Here the straight
lines represent on-shell scalars and the wiggly lines correspond to on-shell gravitons or gluons.}
\label{TwoLoopCutsFigure}
\end{figure}

\begin{figure}[tb]
\begin{center}
\includegraphics[scale=.5]{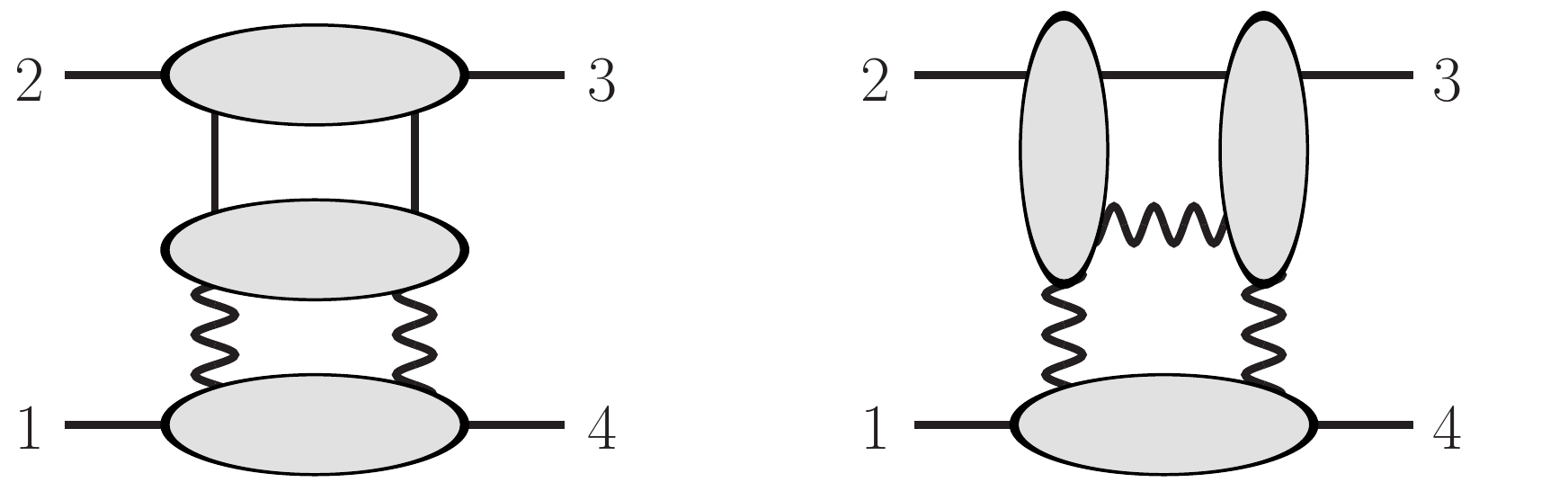}
\end{center}
\vskip -.5cm
\caption{\small Examples of cuts that are discarded because they have no
  further classical or iteration contributions to the potential that
  are not already found in the cuts of \fig{TwoLoopCutsFigure}.}
\label{TwoLoopDiscardedCutsFigure}
\end{figure}

The set of cuts needed to determine the classical potential is, in fact,
a subset of the spanning set.  As explained in \sect{QuantumVsClassicalSection}, the only
unitarity cuts that can contain pieces of the classical potential
separate the two massive lines on opposite sides of the cut and have
one cut matter line in each loop.  In addition there are no
contributions from diagrams containing a graviton propagator starting and ending 
on the same scalar line.  
After dropping from the spanning set all unitarity cuts that do not contain any 
new contributions that satisfy these criteria, we are left with the cuts 
in \fig{TwoLoopCutsFigure}, together with independent ones obtained from relabeling 
the external lines.  
The first two cuts,  (a) and (b), are just the three-particle cut shown in \fig{TwoLoopSpanningCutsFigure}(a), 
where the three cut lines are gravitons, but with the additional requirement that one matter line per loop should
also be cut.  Similarly, the cut in
\fig{TwoLoopCutsFigure}(c) is just the iterated two-particle cut in \fig{TwoLoopSpanningCutsFigure}(b), 
with all cut lines being gravitons, but with the additional condition imposed that one matter line per
loop is cut.  Any other cuts, such as the ones shown in
\fig{TwoLoopDiscardedCutsFigure}, will contain pieces either already determined by
the cuts in \fig{TwoLoopCutsFigure}, or diagrams that do not contribute to the conservative potential.  
In particular, there are no new classical potential pieces in any iterated two particle cut of the form 
in \fig{TwoLoopSpanningCutsFigure}(c).

\subsection{Warm-up: Gauge-Theory Generalized Cuts in Four Dimensions}

Before turning to gravity, we first evaluate the cuts in
\fig{TwoLoopCutsFigure} for a scalar-coupled gauge theory.  Once the
gauge-theory cuts have been determined, the gravity ones are obtained
easily through double copy.

As in the four-dimensional one loop construction, we take the amplitude to be color ordered
so we preserve the cyclic ordering of legs when writing out the tree amplitudes that compose the cut. 
The first two cuts in \fig{TwoLoopCutsFigure} are given by
\begin{align}
C_{\rm YM}^{\rm (a)} & = \!\sum_{h_5, h_6, h_7 = \pm} \!
A_3^\tree(3^s, -7^{-h_7}, -9^s) \, A_3^\tree(9^s, -6^{-h_6}, -8^s) \, 
\nonumber \\
&\hspace{3.5cm} \times A_3^\tree(8^s, -5^{-h_5}, 2^s)  A^\tree_5(1^s, 5^{h_5}, 6^{h_6}, 7^{h_7}, 4^s)  \,,
\label{WCutGT} \\
C_{\rm YM}^{\rm (b)} & =\! \sum_{h_5, h_6, h_7 = \pm} \!
A_4^\tree(3^s, -7^{-h_7}, -6^{-h_6},-8^s) \, A_3^\tree(8^s, -5^{-h_5}, 2^s) \nonumber \\
&\hspace{3.5cm} \times A^\tree_4(1^s, 5^{h_5}, 6^{h_6}, 9^{s})   \, A_3^\tree(-9^s, 7^{h_7}, 4^s) \,.
\label{NCutGT}
\end{align}
The helicity sums run over $2^3=8$ possible configurations:
\begin{equation}
\{ h_5, h_6, h_7 \} : \,
\{+++ \}, 
\{--- \},
\{+--\}, 
 \{-++\}, 
\{+-+\}, 
\{-+-\}, 
\{++-\}, 
\{--+\} \,.
\label{3CutHelicities}
\end{equation}
Similarly, the cut in \fig{TwoLoopCutsFigure}(c) is, 
\begin{align}
C_{\rm YM}^{\rm (c)} = \sum_{h_5, h_6, h_7, h_8 = \pm} \!
A^\tree_3(3^s, 7^{h_7}, -9^s) \, A^\tree_3(9^s,  -8^{-h_8}, 2^s) \,
 A^\tree_4(-5^{-h_5}, 8^{h_8},  -7^{-h_7},  6^{h_6}) \nn \\
\times A^\tree_3(1^s,5^{h_5}, -10^s) A^\tree_3(10^s, -6^{-h_6},4^s) \,,
\label{HCutGT}
\end{align}
where legs $ 5, 6, 7, 8$ are cut gluon lines and $9$ and
$10$ are cut matter lines.  The sum over helicity configurations runs 
over
\begin{equation}
\{ h_5, h_6, h_7, h_8 \} : \,
\{++++ \},  \; \{----\}, \; \{+--+\}, \; \{++--\}, \; \{--++\}, \; \{-++-\} \,.
\label{HCutHelicitiesGT}
\end{equation}
The helicity labels correspond to the momentum flow indicated in \fig{TwoLoopCutsFigure}(c).  
All other configurations give vanishing contributions, because the four-gluon amplitude represented by the central blob vanishes
except for configurations with two legs of positive helicity and two of negative helicity for all outgoing momenta.  

As we encountered at one loop, each cut contains both terms that do not 
appear in any other cut  and terms that do. For example, the cut in \fig{TwoLoopCutsFigure}(a)
contains a term in which the lines 1, 4, 5, 6, 7 meet at a single five-point vertex which does not appear
in any other cut. It also contains a term corresponding to a two-loop planar double-box graph topology, 
which also appears in the cut in \fig{TwoLoopCutsFigure}(b).
While we can simplify the cuts following a similar strategy as at one loop,
the process of organizing the result in terms of local diagrams and removing such double-counted terms
is more complex at two loops.
Instead, we will use the simple strategy where the cuts in Eqs.~\eqref{WCutGT}, \eqref{NCutGT} and~\eqref{HCutGT} are 
directly matched onto the cuts of an ansatz for the integrand with the desired local properties.  
In this way we simultaneously remove any overcount of terms in cuts and organize the result in terms of Feynman integrals.

To illustrate some important features and to explain the terms that can be
dropped when evaluating the cuts in four dimensions, we will summarize the calculation of the cut in \fig{TwoLoopCutsFigure}(c).
In particular, we explain in \sect{sec:evanescence} ambiguities in the integrand when working in four dimensions  and why they 
do not affect the classical potential.

Following similar algebraic steps as at one loop and using standard spinor
manipulations we evaluate the terms in the cut in \fig{TwoLoopCutsFigure}(c) given by the different 
helicity configurations in \eqn{HCutGT}.  We find
\begin{align}
&C_{\rm YM}^{\{++++\}} = 
 - i \frac{m_1^2 m_2^2 \,t } {(- \t_{58}) } \,, 
&C_{\rm YM}^{\{+--+\}} =
   -i\frac{\tr^2_-[7 2 8 6 1 5] }
   {t^3  (- \t_{58}) } \, , \quad
&C_{\rm YM}^{\{+-+-\}} =
-i\frac{ \tr^2_-[8 2 7 6 1 5] }
{t^3 (- \t_{58})}\,,  \nn\\ 
&C_{\rm YM}^{\{----\}}  =
- i \frac{m_1^2 m_2^2 \, t } { (- \t_{58})} \,,
%
&C_{\rm YM}^{\{-++-\}} = 
  -i\frac{\tr^2_+[7 2 8 6 1 5]}
  {t^3  (- \t_{58}) } \, ,\quad
&C_{\rm YM}^{\{-+-+\}}  =
-i\frac{ \tr^2_+[8 2 7 6 1 5] }
{t^3 (- \t_{58})} \, .
\label{YMHCutResult}
\end{align}

While not necessary for the construction of the color-ordered two-loop
scalar amplitude with the ordering ${\rm
  Tr}[T^{a_1}T^{a_2}T^{a_3}T^{a_4}]$, to construct the gravity cut
corresponding to \fig{TwoLoopCutsFigure}(c) using the KLT relations we
also need the twisted color-ordered cut, obtained by exchanging $\{-5
\leftrightarrow 6 \}$ in the four-point tree-amplitude factor,
\begin{align}
C_{\rm YM}^{\prime \rm (c)} = \sum_{h_5, h_6, h_7, h_8 = \pm} \!
A_3^\tree(2^s,  9^s,  -8^{-h_8}) \, A_3^\tree(7^{h_7},-9^s, 3^s) \, 
 A^\tree_4(6^{h_6}, 8^{h_8},  -7^{-h_7}, -5^{h_5}) \nn \\
\times A^\tree_3(1^s,5^{h_5}, -10^s) A^\tree_3(10^s, -6^{-h_6},4^s) \, .
\label{YMHCutPrimeResult}
\end{align}
The sum runs over the helicity configurations in \Eq{HCutHelicitiesGT}. Using the BCJ amplitudes relation in \eqn{eq:fund_BCJ},
\begin{equation}
A^\tree_4(6^{h_6}, 8^{h_8},  -7^{-h_7}, -5^{h_5}  ) = -\frac{\t_{58}}{\t_{68}} 
    A^\tree_4(-5^{h_5}, 8^{h_8}, -7^{-h_7}, 6^{h_6} ) \,,
\end{equation} 
it is not difficult to find that the twisted cut and its helicity-labeled components are simply
\begin{equation}
C_{\rm YM}^{\prime \{h_5, h_6, h_7, h_8\}}  =  \frac{(- \t_{58})} {\t_{68}} C_{\rm YM}^{\{h_5, h_6, h_7, h_8\}} \, ,
\hskip 1cm
C_{\rm YM}^{\prime \rm (c)}  =  \frac{(- \t_{58})} {\t_{68}} C_{\rm YM}^{\rm (c)} \, .
\label{FinalTwisted}
\end{equation}
The rational factor effectively replaces the $-1/\t_{58}$ propagator with the  $1/\t_{68}$ one, corresponding to the relabeled
four-point amplitude in the middle blob in \fig{TwoLoopCutsFigure}(c).

Next, we evaluate the two remaining traces in \Eq{YMHCutResult}. We split them into parity-even and
parity-odd parts, as at one-loop, resulting in
\begin{align}
\tr_\pm^2[8 2 7 6 1 5]  & =  {\cal E}_1^2 + {\cal O}_1^2 \pm 2 {\cal O}_1 {\cal E}_1\, ,
\label{TraceTwoLoop1}
\\
\tr_\pm^2[7 2 8 6 1 5]  & =  {\cal E}_2^2 + {\cal O}_2^2 \pm  2 {\cal O}_2 {\cal E}_2\, .
\label{TraceTwoLoop2}
\end{align}
It is however clear that, in the complete cut, the parity-odd part cancels out, so we only need to compute ${\cal O}_i^2$.
Evaluating the trace using standard gamma-matrix identities we find  
\begin{align}
{\cal E}_1^2 & = 
\frac{1}{4}  t^2 \bigl(\t_{1 2} \t_{5 8} - \t_{1 8}\t_{2 5} \bigr)^2 \,,
\qquad
{\cal O}_1^2 = {\cal E}_1^2 - \t_{5 8}^2 m_1^2  m_2^2 \, t^2  \,.
\label{EvenOddOne}
\\
{\cal E}_2^2 & = 
\frac{1}{4} t^2 \bigl( \t_{24} \t_{5 7} + \t_{1 7} \t_{57} \bigr)^2 \,,
\qquad
{\cal O}_2^2 = {\cal E}_2^2 - \t_{5 7}^2 m_1^2 m_2^2 \, t^2 \,,
\label{EvenOddTwo}
\end{align}
where $t = (p_2 + p_3)^2$ and $\t_{i j} = 2 p_i \cdot \ell_j$.

Adding up the contributions of the six helicity configurations, we find that the complete 
four-dimensional cut in \fig{TwoLoopCutsFigure}(c) is given by
\begin{align}
\label{full4dHcutGT}
C_{\rm YM}^{\rm (c)}
& = -2  i \frac{1} {(-\t_{58})}
   \biggl[m_1^2 m_2^2 t + \frac{1}{t^3} \bigl(
 {\cal E}_1^2 + {\cal O}_1^2  +  {\cal E}_2^2 + {\cal O}_2^2 \bigr) \biggr] \\
& = -2  i \frac{1} {(-\t_{58})}
   \biggl[m_1^2 m_2^2 t + \frac{1}{2 t} \bigl(
     (\t_{1 2} \t_{5 8} - \t_{1 8}\t_{2 5})^2 - 2 \t_{5 8}^2 m_1^2  m_2^2 + 
     (\t_{24} \t_{5 7} + \t_{1 7} \t_{57})^2 - 2 \t_{5 7}^2 m_1^2 m_2^2 \bigr) \biggr] \, .
     \nn
\end{align}
The twisted cut is simply obtained through \eqn{FinalTwisted}.

If naively continued outside of four dimensions, the expression
\eqref{full4dHcutGT} contains a spurious singularity.  To expose this
issue we construct the maximal cut obtained by imposing the additional
cut condition
\begin{equation}
\t_{5 8} = 0 \,.
\end{equation}
Together with the initial six cut conditions, this corresponds to
cutting the maximum number of propagators for a four-point two-loop
amplitude; consequently, the result should be local, as can be checked
by solving the seven cut conditions in four dimensions. However, with
this additional cut condition, we get the maximal cut interpreted as a
$D$-dimensional expression
\begin{equation}
C_{\rm YM}^{\rm max cut}  = -  \frac{1}{ t} \,\t_{1 8}^2 \, \t_{2 5}^2 \, .
\end{equation}
It contains a $1/t$ pole.  Maximal cuts should be local, since they are nothing more than 
sums of products of three vertices. This strongly suggests that the correct $D$-dimensional expression 
is missing a term proportional to a Gram determinant that vanishes in four
dimensions but not in $D$ dimensions.

This is indeed the case. To remove the spurious $1/t$ singularity we add a term proportional to the Gram determinant
 \begin{equation}
G_5  = 16 \det
\left( \begin{matrix}
  p_1  \cdot p_1 &  p_1  \cdot p_2 &  p_1 \cdot p_3 &   p_1\cdot \ell_5 &   p_1\cdot \ell_8 \\
  p_2  \cdot p_1 &  p_2  \cdot p_2 &  p_2 \cdot p_3 &   p_2\cdot \ell_5 &   p_2\cdot \ell_8 \\
  p_3  \cdot p_1 &  p_3  \cdot p_2 &  p_3 \cdot p_3 &   p_3\cdot \ell_5 &   p_3\cdot \ell_8 \\
  \ell_5  \cdot p_1 &  \ell_5  \cdot p_2 &  \ell_5 \cdot p_3 &   \ell_5\cdot \ell_5 &   \ell_5\cdot \ell_8 \\
  \ell_8  \cdot p_1 &  \ell_8  \cdot p_2 &  \ell_8 \cdot p_3 &   \ell_8 \cdot \ell_5 &   \ell_8\cdot \ell_8 
\end{matrix} \right),
\label{GramDete}
\end{equation}
which vanishes in four dimensions.  By adjusting the coefficient of $G_5$ we can 
remove the spurious singularity.  This results in a cut that is now valid in $D$-dimensions,
\begin{align}
\label{improvedcutcGT}
C_{\rm YM} = -2  i \frac{1} {(-\t_{58}) }
   \biggl[ m_1^2 m_2^2 t + \frac{1}{t^3} \Bigl({\cal E}_1^2 + {\cal O}_1^2  +  {\cal E}_2^2 + {\cal O}_2^2  - t\,  G_5 \Bigr)\biggr] \,,
\end{align}
where the state-counting parameter is $D_s =4$. 
An alternative way to arrive at the same result is to match an ansatz written in terms of local 
diagrams (which by definition have no spurious singularities) onto $C_\text{YM}^{(c)}$.
The twisted cut can be obtained from  \eqn{improvedcutcGT} via \eqn{FinalTwisted}.

\subsection{Gravity Generalized Cuts in Four Dimensions}

The KLT relations~\eqref{KLT} provide a simple path for assembling the gauge-theory 
cut components constructed above into analytic expression for the three gravity cuts in \fig{TwoLoopCutsFigure}.
Those corresponding to \fig{TwoLoopCutsFigure}(a,b) are
\begin{align}
C_{\rm GR}^{\rm (a)} & = \sum_{h_5, h_6, h_7 = \pm} \!
M_3^\tree(2^s, 8^s, -5^{-h_5}) \, M_3^\tree(-6^{-h_6},-8^s, 9^s) \nn\\
& \hskip 3 cm \null \times
  M_3^\tree(7^{h_7},-9^s, 3^s) \, M_5^\tree(1^s, 5^{h_5}, 6^{h_6}, 7^{h_7}, 4^s) \,, 
\label{WCutGR} \\
C_{\rm GR}^{\rm (b)} & = \sum_{h_5, h_6, h_7 = \pm} \!
M_3^\tree(2^s, 8^s, -5^{-h_5}) \, M_4^\tree(-7^{-h_7}, -6^{-h_6},-8^s, 3^s) \nn \\
& \hskip 3 cm \null \times
 M^\tree_4(1^s, 5^{h_5}, 6^{h_6}, 9^s) \, M_3^\tree(-9^s, 7^{h_7}, 4^s)  \,,
\label{NCutGR}
\end{align}
where the helicity configurations run over the eight possibilities in \eqn{3CutHelicities}, except that here 
the $\pm$ refer to the spin states of gravitons instead of those of gluons.  The gravity tree amplitudes
can be simply evaluated using the KLT relations~\eqref{KLT}.
Similarly, the remaining cut in~\fig{TwoLoopCutsFigure}(c) is 
\begin{align}
C_{\rm GR}^{\rm (c)} & = \sum_{h_5, h_6, h_7, h_8 = \pm} \!
M^\tree_3(2^s,  9^s,  -8^{-h_8}) \, M^\tree_3(7^{h_7},-9^s, 3^s) \, 
 M^\tree_4(-5^{-h_5}, 6^{h_6}, -7^{-h_7}, 8^{h_8} ) \nn \\
& \null \hskip 4 cm 
\times M^\tree_3(1^s,5^{h_5}, -10^s) M^\tree_3(10^s, -6^{-h_6},4^s) \nn \\
& = -i t
\sum_{h_5, h_6, h_7, h_8 = \pm} 
\biggl[ A^\tree_3(2^s,  9^s,  -8^{-h_8}) \, A^\tree_3(7^{h_7},-9^s, 3^s) \, 
 A^\tree_4(-5^{-h_5}, 6^{h_6}, -7^{-h_7}, 8^{h_8} ) \nn \\
& \null \hskip 4 cm 
\times A^\tree_3(1^s,5^{h_5}, -10^s) \, A^\tree_3(10^s, -6^{-h_6},4^s) \biggr] \nn\\
& \hskip 2.75 cm \null \times
\biggl[ A^\tree_3(2^s,  9^s,  -8^{-h_8}) \, A^\tree_3(7^{h_7},-9^s, 3^s) \, 
 A^\tree_4(6^{h_6}, 8^{h_8},  -7^{-h_7}, -5^{h_5} ) \nn \\
& \null \hskip 4 cm 
\times A^\tree_3(1^s,5^{h_5}, -10^s) \, A^\tree_3(10^s, -6^{-h_6},4^s) \biggr]
\,,
\label{HCutGR}
\end{align}
where legs $ 5, 6, 7, 8$ are cut graviton lines and $9$ and $10$ are cut matter lines.  
As for the gluon case, some helicity configurations have vanishing contributions because of the properties 
of the four-graviton amplitude factor, 
so the  sum over helicities  in \eqn{HCutGR} effectively runs over the configurations in \eqn{HCutHelicitiesGT}.
As at one loop, we project out the dilaton and axion simply by requiring that the helicities of gluons in the two 
gauge-theory factors  in Eqs.~\eqref{WCutGR}, \eqref{NCutGR} and \eqref{HCutGR}   be identical.

Upon using the gauge-theory tree amplitudes given in \app{TreeAmplitudesAppendix} we find compact expressions
for the three types of cuts in \fig{TwoLoopCutsFigure}, which can subsequently be assembled into a single integrand, as we discuss below.
It is instructive to inspect in some detail the cut in \fig{TwoLoopCutsFigure}(c),  in order to understand potential issues with 
carrying out the calculation in four dimensions compared to $D$ dimensions.  In general, we prefer the four-dimensional 
expressions because their complexity at higher loops increases slower than that of their $D$-dimensional counterparts.

The cut components, labeled by the helicities of the cut graviton lines in \fig{TwoLoopCutsFigure}(c) 
are obtained through the KLT relations in terms of the direct and twisted YM cut components 
in \eqns{YMHCutResult}{YMHCutPrimeResult}
\begin{align}
C_{\rm GR}^{\{h_1 h_2 h_3 h_4\}} 
&  = -i \, t \, C_{\rm YM}^{\{h_1 h_2 h_3 h_4\}} C_{\rm YM}^{\prime \{h_1 h_2 h_3 h_4\}} \, .
\end{align}
As explained above, by identifying the helicity labels in the two gauge-theory amplitudes we project out the dilaton and axion 
contributions to cuts. 
Using the relation in \eqn{FinalTwisted}, we can read off the results in different helicity configurations as
\begin{align}
&C_{\rm GR}^{\{++++\}}   = 
 i \frac{m_1^4 m_2^4 t^3 } { (-\t_{58})  \t_{68}} \,,
&C_{\rm GR}^{\{+--+\}}   =
 i\frac{\tr^4_-[7 2 8 6 1 5] }
{t^5  (-\t_{58}) \t_{68}} \,, \quad\;
&C_{\rm GR}^{\{+-+-\}}   =
i\frac{ \tr^4_-[8 2 7 6 1 5] }
{t^5 (-\t_{58}) \t_{68}} \,, \nn \\ 
&C_{\rm GR}^{\{----\}}   =
 i \frac{m_1^4 m_2^4 t^3 } { (-\t_{58})  \t_{68}} \,,
&C_{\rm GR}^{\{-++-\}}   = 
i\frac{\tr^4_+[7 2 8 6 1 5]}
{t^5  (-\t_{58})  \t_{68}} \,, \quad\;
&C_{\rm GR}^{\{-+-+\}}   =
   i\frac{ \tr^4_+[8 2 7 6 1 5] }
   {t^5 (-\t_{58}) \t_{68}} \,.
\label{GRHCutResult}
\end{align}
Summing over all helicity configurations of gravitons crossing the iterated two-particle cuts, we find
\begin{align}
C_{\rm GR}^{\rm (c)} & = 
 -i \biggl[ 2 t^2 m_1^4 m_2^4 + \frac{1}{t^6} \Bigl(
  \tr^4_-[7 2 8 6 1 5] +
  \tr^4_-[8 2 7 6 1 5] +
  \tr^4_+[7 2 8 6 1 5] +
  \tr^4_+[8 2 7 6 1 5] \Bigr)
\biggr] \nonumber \\
&\quad \times \biggl[\frac{1}{(-\t_{58})} + \frac{1}{\t_{68}}
          \biggl] \,,
\label{GRcutC}          
\end{align}
where we used the partial fractioning identity (recall that momenta are flowing as indicated in~\fig{TwoLoopCutsFigure}(c))
\begin{align}
\frac{1}{ (-\t_{58}) \t_{68}}
 & = - \frac{1}{t} \biggl[\frac{1}{(-\t_{58})} + \frac{1}{\t_{68}}
          \biggl]\, .
\end{align}
By inspecting \eqn{GRcutC} it is easy to see that all parity-odd terms cancel out.
We can evaluate the traces using \eqns{TraceTwoLoop1}{TraceTwoLoop2}
\begin{align}
\label{GCfullC}
C_{\rm GR}^{\rm (c)} & =
  -2i \biggl[  t^2 m_1^4 m_2^4 + \frac{1}{t^6} \Bigl(
  {\cal E}_1^4 + {\cal O}_1^4 + 6 {\cal O}_1^2 {\cal E}_1^2 +
  {\cal E}_2^4 + {\cal O}_2^4 + 6 {\cal O}_2^2 {\cal E}_2^2  \Bigr)
\biggr]
 \biggl[\frac{1}{(-\t_{58})} + \frac{1}{\t_{68}}
          \biggl] \,,
\end{align}
where the even-squared and odd-squared terms are given in
Eqs.~~\eqref{EvenOddOne} and  \eqref{EvenOddTwo}.

As for the YM cut with the same topology, if we naively extend \eqn{GCfullC} to $D$ dimensions we encounter 
a spurious singularity in the maximal cut, which we can remove by adding in \eqn{GramDete} a term proportional to the Gram determinant, 
that vanishes in four dimensions.  Doing so gives,
\begin{align}
C_{\rm GR}^{\rm (c)} & =
  -2i \biggl[  t^2 m_1^4 m_2^4 + \frac{1}{t^6} \Bigl(
  {\cal E}_1^4 + {\cal O}_1^4 + 6 {\cal O}_1^2 {\cal E}_1^2 +
  {\cal E}_2^4 + {\cal O}_2^4 + 6 {\cal O}_2^2 {\cal E}_2^2
  \nonumber \\
  & \qquad \qquad -2 t G_5  ( {\cal E}_1^2 + {\cal O}_1^2  +  {\cal E}_2^2 + {\cal O}_2^2) 
   + t^2 G_5^2\Bigr) \biggr]
 \biggl[\frac{1}{(-\t_{58})} + \frac{1}{\t_{68}}
          \biggl] \,.
          \label{GCfullCnaiveDextension}
\end{align}
As we will see, further terms that vanish in four dimensions (and therefore proportional to $G_5$) are necessary to 
obtain the complete $D$-dimensional cut. 
Nevertheless, all terms proportional to $G_5$ do not contribute to the classical  potential, as we discuss in \sect{sec:evanescence}.

For comparison, the cut in \fig{TwoLoopCutsFigure}(c) for the scattering of scalars coupled to a graviton, dilaton and antisymmetric tensor
(or axion) is given by the simple double copy:
\begin{align}
C_{\rm GR}^{\rm (c)} +  C_{\rm \phi a}^{\rm (c)} & =
  -4i \biggl[  t m_1^2 m_2^2 + \frac{1}{t^3} \Bigl(
  {\cal E}_1^2 + {\cal O}_1^2  + {\cal E}_2^2 + {\cal O}_2^2  - t G_5  \Bigr) \biggr]^2
 \biggl[\frac{1}{(-\t_{58})} + \frac{1}{\t_{68}} \biggl]\, .
\end{align}
This expression holds for kinematics in general dimension, but with $D_s = 4$.


\subsection{$D$-Dimensional Generalized Cuts}

The evaluation of the $D$-dimensional version of the generalized cuts in \fig{TwoLoopCutsFigure} 
is similar to the evaluation of the one-loop $D$-dimensional cuts in \sect{OneLoopSection}, except that the relevant gauge-theory 
trees are somewhat more complicated.  Besides the two-scalar three-gluon tree-level amplitude needed for the cut in
\fig{TwoLoopCutsFigure}(a), we also need the four-gluon amplitude in a
BCJ form that also satisfies the on-shell Ward identity \eqn{GeneralizedWardIdentity} without use of transversality of 
the external states.  
These tree amplitudes are given in \app{TreeAmplitudesAppendix}.  
By using a good basis choice and BCJ duality, the amplitudes are determined by specifying a single
numerator.  Because BCJ duality is manifest, the desired gravity cuts follow immediately from the gauge-theory ones through
the double-copy substitution in \Eq{ColorSubstitution}.

As at one-loop, we exploit the compact physical-state projector in \eqn{SimplifiedGravitonProjector}, which is a consequence of
the on-shell Ward identities holding without imposing the transversality of gluon and graviton asymptotic states and of the decoupling of antisymmetric tensor.
This trick simplifies the two-loop calculation enormously.
We will not include the details since the calculations are straightforward extensions of the one-loop ones. 
We will however comment on the result for the cut in \fig{TwoLoopCutsFigure}(c) and compare it with the four-dimensional
expression in \Eq{GCfullC}.

Taking $D_s = \eta_\mu^{\ \mu} = 4$, the cut in \fig{TwoLoopCutsFigure}(c) evaluates to the rather compact expression
\begin{align}
C_{\rm GR}^{\rm (c)}  =&
  -2i \biggl[  t^2 m_1^4 m_2^4 + 
  \frac{1}{t^6} \Bigl(
  {\cal E}_1^4 + {\cal O}_1^4 + 6 {\cal O}_1^2 {\cal E}_1^2 +
  {\cal E}_2^4 + {\cal O}_2^4 \nn \\
  & \qquad \qquad + 6 {\cal O}_2^2 {\cal E}_2^2
   -2 t G_5  ( {\cal E}_1^2 + {\cal O}_1^2  +  {\cal E}_2^2 + {\cal O}_2^2)
   + t^2 G_5^2 \Bigr) \nn \\
  & \qquad \qquad
   - \frac{1}{t}  ( 2 m_1^2 m_2^2 + (\t_{1 2} - \t_{1 8} + \t_{2 5} - \t_{5 8})^2 ) G_5
 \biggr]  \biggl(\frac{1}{(-\t_{58})} + \frac{1}{\t_{68}} \biggl) \,,
\end{align}
where $G_5$ is the Gram determinant in \eqn{GramDete}.  We note that,
apart from the $G_5$ terms we found by requiring absence of a spurious
pole in the maximal cut of the naive $D$-dimensional continuation of
the four-dimensional result in \Eq{GCfullCnaiveDextension}, the
complete $D$-dimensional calculation yields additional terms linear in
$G_5$.
The other cuts are also straightforward to evaluate, although we have not obtained compact $D$-dimensional expressions for them.

\subsection{Merging the Cuts into an Integrand}
\label{MergingSubsection}

\begin{figure}
\begin{center}
\includegraphics[scale=.5]{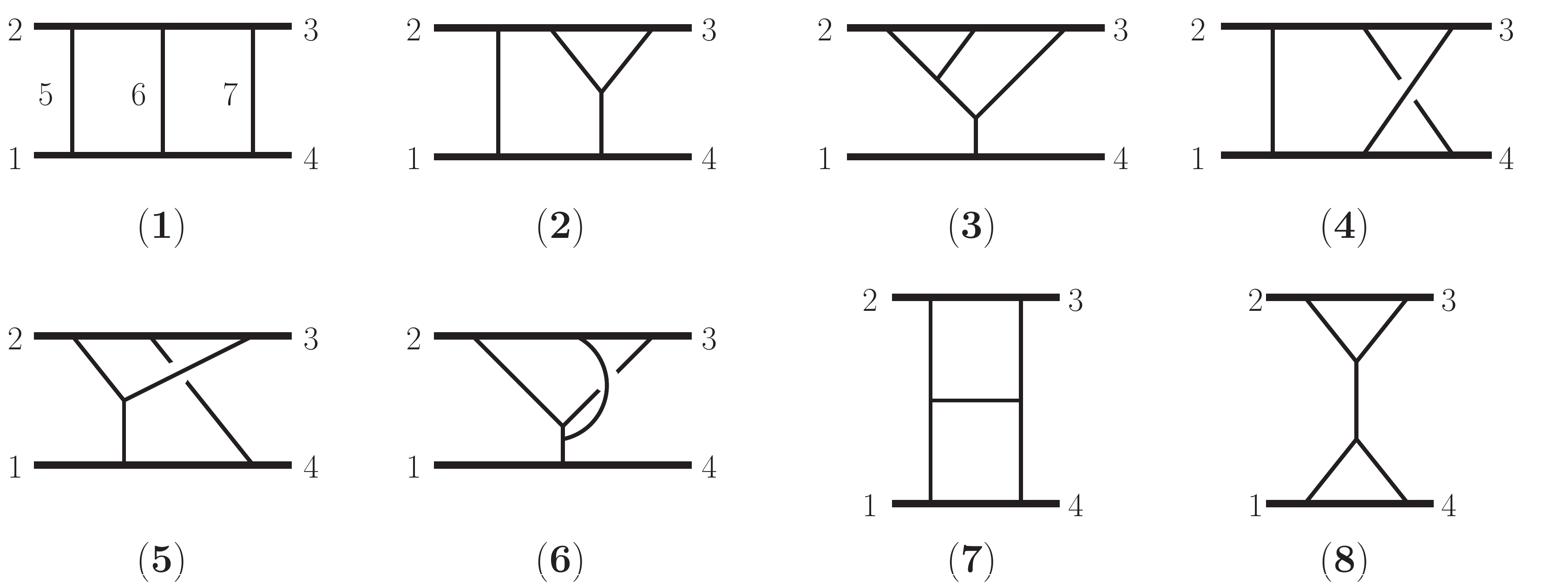}
\end{center}
\vskip -.5cm
\caption{\small The eight independent diagrams with only cubic vertices
  showing the propagator structure of integrals which contribute to
  the classical potential. The thick lines correspond to massive
  particles.  All other diagrams are given by relabeling the external legs. In the main text, we will denote each diagram using bold label, e.g. diagram $\bm 1$ for the first diagram.
  }
\label{MasterDiagsFigure}
\end{figure}

Once the cuts have been evaluated, the next step is to reorganize them into an integrand.  
We do this for both their four and $D$-dimensional versions. 
While obtaining the integrand from the four-dimensional cuts turns out to be sufficient, it is nevertheless important 
to verify that no subtleties arise from truncating the integrand to four dimensions, given the fact that the
integrated amplitude is infrared divergent.

In general, different unitarity cuts contain information about the same terms in the integrand.  
For example, contributions to diagram $\bm 1$ in \fig{MasterDiagsFigure} are found in both cuts (a) 
and (b) of \fig{TwoLoopCutsFigure} while contributions to diagram $\bm 7$ in \fig{MasterDiagsFigure} 
appear both in cuts (b) and (c) of \fig{TwoLoopCutsFigure}.  
Such double counting can be removed either during the construction of
the integrand ~\cite{BDDPR,TwoLoopSplitting} or after
integration~\cite{Bern:2002tk}. We will choose the former approach, as
it meshes well with our integration methods described in the next
sections.

To determine the numerators for the diagrams in
\fig{MasterDiagsFigure} we construct an ansatz for each of them. While
contact diagrams with four- or higher-point vertices can be included
in the ansatz, we choose to minimize the number of independent graphs
by instead using diagrams that have only cubic vertices as shown in
\fig{MasterDiagsFigure}, and capturing contact terms by multiplying
and dividing by appropriate propagators.
The highest number of loop-momentum factors in each term of the ansatz
follows from the two-derivative nature of gravity couplings, counting
two such factors for each three-point vertex. As we will see, it is
not necessary to saturate the maximal power counting for all diagrams.
After including relabelings of external lines, the eight diagram
topologies shown in \fig{MasterDiagsFigure} are the minimal set needed
for matching the three unitarity cuts in~\fig{TwoLoopCutsFigure}.
We impose the natural requirement that the symmetries of a diagram
are also symmetries of its numerator factor.  The parameters of the
ansatze are then fixed by computing their generalized cuts (including
contributions from the relabelings) and match to those
in~\fig{TwoLoopCutsFigure} computed earlier via multiplying tree
amplitudes.  Once this is done, the merged integrand simultaneously
satisfies all other generalized unitarity cuts related to
\fig{TwoLoopCutsFigure} by exchanging the external legs, because the
relabeling is built-in explicitly.

An observation that leads to simplifications when imposing diagram
symmetries is that on-shell conditions may be imposed on a given leg
in a given diagram if that leg is on shell in {\em all} cuts that
contribute to that diagram.
For example, the on-shell conditions for the three gravitons can be
applied to the numerator of diagram $\bm 1$ because these gravitons
are on shell in both cuts (a) and (b) in \fig{TwoLoopCutsFigure}.  The
four gravitons that connect to matter lines in diagram $\bm 7$ must
however be treated off shell because they are off shell in cut (b) in
\fig{TwoLoopCutsFigure}.
On the other hand, we can use the on shell condition for the two
matter lines in diagrams $\bm 7$ and $\bm 8$ without affecting the
classical potential because only cuts (b) and (c) in
\fig{TwoLoopCutsFigure} source them and they both place the matter
lines on shell.

In the construction of the ansatz for each diagram's numerator, we only require a 
sufficiently high power of loop momenta
such that a solution exists and matches either to four-dimensional cuts
or to $D$-dimensional cuts.  
For example, diagram $\bm 1$ in \fig{MasterDiagsFigure} does not require
any powers of loop momenta in the numerators, while the numerator of
diagram $\bm 6$ requires the maximum of 12 powers of loop momenta to
match the cuts, prior to dropping quantum terms containing more than 4
powers of loop momenta in the numerator.  

In general, the numerators contain free parameters because of the possibility of
assigning the same contact term to different diagrams without changing
the value of integrand.
We choose the remaining undetermined parameters to simplify the matching with the effective field 
theory amplitudes in \sect{sec:EFT}. For example, we chose diagram (1) to have a numerator proportional to the third power 
of the numerator in tree-level four-scalar amplitude, i.e.
\begin{equation}
N^{(1)}_{\rm GR}(1,2,3,4) =  \left((s - m_1^2 - m_2^2)^2 -\frac{4}{D_s-2} m_1^2 m_2^2 \right)^3 \,.
\end{equation}
By imposing diagram's symmetry on all numerators, simple relabeling gives us the value of the numerators 
with swapped legs.  For example,  the numerator corresponding to the case where legs 1 and 4 are swapped 
is simply,
\begin{equation}
N^{(1)}_{\rm GR}(1,3,2,4) =  \left((u - m_1^2 - m_2^2)^2 -\frac{4}{D_s-2} m_1^2 m_2^2 \right)^3 \,.
\end{equation}
The numerators of the other diagrams are more involved. The free parameters also provide a nontrivial check 
since they should cancel out in the final integrated amplitude. We have performed extensive checks to confirm this property. 

\begin{table}[t]
	\centering 
	\def\arraystretch{1.2}
	\begin{tabular*}{0.7 \textwidth}{c  @{\extracolsep{\fill}} l}
		\hline\hline
		Graph & External relabeling \\
		\hline
		1 & (1 2 3 4),\, (1 3 2 4)\\
		2 & (1 2 3 4),\, (1 3 2 4),\, (4 3 2 1),\, (4 2 3 1),\, $+\{1,4\} \leftrightarrow \{2,3\}$\\
		3 & (1 2 3 4),\, (1 3 2 4),\, $+\{1,4\} \leftrightarrow \{2,3\}$\\
		4 & (1 2 3 4),\, (1 3 2 4),\, (4 3 2 1),\, (4 2 3 1)\\
		5 & (1 2 3 4),\, (4 2 3 1),\, $+\{1,4\} \leftrightarrow \{2,3\}$\\
		6 & (1 2 3 4),\, $+\{1,4\} \leftrightarrow \{2,3\}$\\
		7 & (1 2 3 4),\, (1 3 2 4) \\
		8 & (1 2 3 4) \\
		\hline \hline
	\end{tabular*}
	\caption{\small List of 27 cubic diagrams from relabeling of the eight graphs in \fig{MasterDiagsFigure}. The right column gives all independent labels by exchanging the original external legs $(1 2 3 4)$ in each graph in \fig{MasterDiagsFigure}. Here $\{1,4\} \leftrightarrow \{2,3\}$ means swapping $1,4$ with $2,3$ in all the previous labels. For example, this yields (2 1 4 3) and (2 4 1 3) for graph 5.  }
	\label{table:assembly}
\end{table}

Once a valid set of numerators with the properties described above has been 
determined, we can treat the diagrams as if they were ordinary off-shell Feynman diagrams.  
In particular, the  diagram numerators are functions of the external labels and can 
be taken as respecting diagram symmetry. While on-shell conditions have been 
used in their construction, we can view their use as merely affecting terms with no 
contributions to the classical potential.

We write the part of the two-loop four-scalar amplitude necessary for
the extraction of the complete 3PM conservative potential as a sum
over 27 diagrams,
\begin{equation}
  M^{\rm 2\hbox{-}loop} = (4E_1 E_2)\,i\mathcal{M}_3 =  i (8 \pi G)^3 \int \frac{d ^D \ell_5}{(2 \pi)^D} \frac{d ^D\ell_6}{(2 \pi)^D}
           \sum_{i=1}^{27} {\cal I}^{(i)} \, ,
\end{equation}
where $\mathcal{M}_3$ is the 3PM amplitude including a nonrelativistic normalization $1/4E_1 E_2$. We will use this normalization for the amplitudes $\mathcal{M}_n$ presented in \sect{sec:scattering} and \sect{sec:EFT}.
The sum runs over 27 distinct diagrams arising from the independent relabelings of the external legs of the eight diagrams in~\fig{MasterDiagsFigure}, which are taken to correspond to $i=1,\dots,8$ and are found in the ancillary Mathematica text file {\tt Integrand\_TwoLoop.m}~\cite{AttachedFile}.
The relabelings are listed in Table~\ref{table:assembly}.
The integrands are expressed 
in terms of numerators and denominators, in the usual way:
\begin{equation}
{\cal I}^{{i}} = \frac{N^{(i)}}{D^{(i)}} \, .
\end{equation}
The denominators are simply the Feynman propagators of the corresponding diagram.  
For example, for diagram (1) of 
\fig{MasterDiagsFigure},
\begin{equation}
D^{(1)} =  \ell_5^2 \ell_6^2  (\ell_5 +\ell_6 -p_2 - p_3)^2  (\ell_5 -p_2)^2  (\ell_5+ \ell_6 - p_2) ^2  (\ell_5 +p_1)^2  (\ell_5 +  \ell_6 + p_1)^2 \,,
\end{equation}
where we have suppressed the standard Feynman $i\eps$ prescription,
which should be included for each propagator.  The numerators in this
file are valid in $D$ dimensions, but with the state-counting
parameter $D_s = 4$ for simplicity.

We have also constructed  several alternative integrands starting both from cuts constructed in four and $D$ 
dimensions, as described above. As already noted, expressions obtained with these two starting points differ
by terms proportional
to the Gram determinant $G_5$ in \Eq{GramDete}, which vanishes in four dimensions. In addition, when we use helicity methods, the state-counting parameter is automatically taken to be $D_s = 4$. We have explicitly checked
that integrands extracted from four dimensions and $D$ dimensions and
using $D_s=4$ or $D_s = 4 - 2 \eps$ lead to identical final results
for the potential.

The independence of the potential on the details of the dimensional
regularization prescription is not accidental.  In
\sect{sec:evanescence}, we explain why any two-loop integral
containing a Gram determinant that vanishes in four dimensions cannot contribute to the
classical potential and why the results are independent of the choice
of prescription for the $D_s$ parameter.  The net effect is that, 
analogous to the way as it works at one loop, the difference between
constructing the integrand in four and in $D$ dimensions does not affect the classical potential.  This, of course, agrees with naive expectations.

\section{Nonrelativistic Integration}
\label{sec:NRint}

In the previous sections we constructed fully relativistic expressions for the integrands relevant to classical potential scattering.  It would then seem natural to 
integrate these objects using fully relativistic methods.
While such a strategy is straightforward and efficient
at one loop, see e.g.  \sect{MassSingularitySubsection}, it becomes
considerably more difficult at two-loop order.  Indeed from the
point of view of scalability to high loop orders, the situation is even more dire.  To match
the current state of the art in the PN expansion requires computational power
at 5PN order, corresponding to full integration at five
loops~\cite{1902.10571, 1902.11180}. This is a tall order, already extending far beyond the capabilities of any known fully relativistic 
loop integration method.

Clearly, a simpler method of integration must exist. After all,
conventional methods using equations of motion or NRGR have already
achieved results up to 5PN, and these calculations certainly did not
necessitate the evaluation of difficult relativistic integrals. 
 The
simple fact that these existing formalisms are by construction
nonrelativistic and classical suggests an alternative path to
integration.

In this section, we describe such a method, which we dub
``nonrelativistic integration''.  This approach is by design as similar as
possible to the method of integration which appears in NRGR~\cite{NRGR}.  Roughly
speaking, it involves performing all energy integrations first, and
then integrating spatial $(D-1)$-momenta.

We begin with a brief summary of the complete procedure.  We then go
through the details carefully by working through several explicit
examples at one and two loops.

\subsection{General Procedure}
\label{sec:general}

Our integration procedure is constructed to extend mechanically to arbitrary loop orders.  In this section we outline the core strategy of this method, which has been tested explicitly at one and two loops.  We believe that this basic approach should apply beyond this order, though as past experience suggests, subtleties may arise due to new topologies and certain physical phenomena which only appear at higher orders, e.g. the tail effect~\cite{TailEffect1, TailEffect2, TailEffect3, TailEffect4, TailEffect5, TailEffect6}. See \sect{sec:radiation} for further comments.

Consider a general multi-loop integral characterizing the scattering of matter fields through graviton exchange:
\eq{
I
 = \bigg[ \prod_{i=1}^{n_L} \int \frac{d^{D-1} \bm \ell_i}{(2\pi)^{D-1}}  \bigg] \bigg[ \prod_{i=1}^{n_L} \int \frac{d\omega_i }{2\pi}  \bigg] \,{\cal I} = \bigg[ \prod_{i=1}^{n_L} \int \frac{d^{D-1} \bm \ell_i}{(2\pi)^{D-1}}  \bigg]  \, \widetilde {\cal I},
}{eq:integrands}
where $i$ runs over $n_L$ loop momenta $\ell_i = (\omega_i, \bm \ell_i)$ and we have split the integration over the energy and $(D-1)$-momentum components. For convenience, we will typically choose the $\ell_i$ to be a subset of the momenta flowing through internal graviton lines.  We follow the standard dimensional regularization prescription for the spatial directions only.

We pause to define notation which will be employed for the rest of the
paper.  From here on, the script quantity ${\cal I}$ will denote a
full relativistic integrand, e.g. as computed earlier
in \sect{TwoLoopSection}.  The integrand ${\cal I}$ is a function of
$n_L$ loop energies and $n_L$ loop $(D-1)$-momenta, and will appear
below with subscripts labeling the particular diagram. Meanwhile, the
tilded script quantity $ \widetilde {\cal I}
= \prod_{i=1}^{n_L} \int \frac{d\omega_i }{2\pi} \,{\cal I} $ will
denote a spatial integrand, which is defined to be the energy integral
of a corresponding full relativistic integrand. The spatial integrand
is a function of $n_L$ loop $(D-1)$-momenta.  Throughout, the spatial
integration will appear with subscripts labeling the corresponding
diagram.

The process of integration has three key steps which we now discuss
broadly. The detailed mechanics will be illustrated with one-loop
examples in \sect{sec:NRint_oneloop} and two-loop examples
in \sect{sec:NRint_twoloop}.

\subsubsection*{Step 1: Determine the Effective Numerator}
The integrand takes the general form 
\eq{
{\cal I} &= \bigg[\prod_{i=1}^{n_M}\frac{1}{\varepsilon_i^2 - \bm k_i^2 -m_i^2}\bigg]\bigg[ \prod_{j=1}^{n_G} \frac{1}{\omega_j^2 - \bm \ell_j^2} \bigg] \, {\cal N} \,,
}{}
where $i$ runs over $n_M$ internal scalar field lines whose energy, $(D-1)$-momentum, and mass are $\varepsilon_i$, $\bm k_i$, and $m_i$, and $j$ runs over $n_G$ internal graviton lines whose energy and $(D-1)$-momentum are $\omega_j$ and $\bm \ell_j$.   Of course, these energies and $(D-1)$-momenta are not independent.  All kinematic parameters depend implicitly on the external masses $m_1$ and $m_2$, external $(D-1)$-momenta $\bm p$ and $\bm p'$, as well as $n_L$ independent loop energies and $(D-1)$-momenta.  

We can factorize the scalar propagators into matter and antimatter poles,
\eq{
\frac{1}{\varepsilon_i^2 - \bm k_i^2 -m_i^2} &= \frac{1}{\varepsilon_i -  \sqrt{\bm k_i^2 + m_i^2}} \, \frac{1}{\varepsilon_i +  \sqrt{\bm k_i^2 + m_i^2}} \,.
}{}
The matter poles occur at positive energy values and correspond to singularities which can genuinely appear in the low-energy scattering of interest. On the other hand, the antimatter poles occur at negative energy values.   These states are never on shell in the nonrelativistic regime and correspond to high-energy modes.  Note that even though the antimatter poles never become singular for low energy kinematics, they still have an effect since these factors should still be expanded in the nonrelativistic limit, yielding classical velocity corrections.

Next, we combine all contributions from the original numerator, the force-carrier (graviton) propagators, and the antimatter poles into an ``effective numerator'',
\eq{
\widetilde{\cal N}= \bigg[ \prod_{i=1}^{n_M}  \frac{1}{\varepsilon_i +  \sqrt{\bm k_i^2 + m_i^2}} \bigg]\bigg[ \prod_{j=1}^{n_G} \frac{1}{\omega_j^2 - \bm \ell_j^2} \bigg]\, {\cal N}\,,
}{}
so the integrand takes the form
\eq{
{\cal I} &= \bigg[\prod_{i=1}^{n_M} \frac{1}{\varepsilon_i -  \sqrt{\bm k_i^2 + m_i^2}} \bigg]\,  \widetilde{\cal N}\,.
}{eq:integrand_eff}
The above representation manifests the structure of all matter poles,
while treating everything else in the integrand as a regular,
nonsingular function of the matter poles.  Of course $\widetilde{\cal
N}$ has poles in $\varepsilon_i$, but none of these singularities
contribute to the conservative potential. Consequently we are
justified in series expanding $\widetilde{\cal N}$ in loop energies
and $(D-1)$-momenta where needed.

\subsubsection*{Step 2: Energy Loop Integration}
To evaluate the energy integrals we have devised two independent and
equivalent methods.  The first approach we dub ``energy-integral
reduction''.  This procedure is analogous to the reduction of
relativistic tensor integrals down to a basis of scalar integrals,
except in this context it is applied to energy integration. It exploits the
fact that all energy integrals with no matter poles are scaleless
integrals in the classical limit, and can thus be effectively
set to zero:
\eq{
\int d\omega \, \omega^n  \rightarrow 0 \,, \qquad n \geq 0 \, ,
}{eq:scaleless}
for some loop energy $\omega$. Said another way, these integrals do not have support in the potential region.

For example, consider a diagram that yields a scaleless loop integral
which has no matter particle propagators when we pinch some
combination of matter propagators.  Then any term in the numerator
which is proportional to this combination is effectively zero, and we
can freely apply constraints of the form
\eq{
\prod_i \left[  \varepsilon_i -  \sqrt{\bm k_i^2 + m_i^2}  \right] \to 0 
 }{eq:energy_reduction_constraints}
on the effective numerator $\widetilde {\cal N}$.  Recall that each
$\varepsilon_i $ denotes the energy flowing through a matter line, and
is a linear function of the energies $\omega_i$ flowing through
internal gravitons.  Here $i$ runs over a particular subset of matter
propagators which varies depending on the precise structure of the
diagram.  In particular, this subset is defined such that if all
matter propagators labeled by $i$ are canceled, e.g.~due to
compensating factors in the numerator, then the resulting integral
either vanishes or does not contribute classically. A typical
higher-loop diagram exhibits several constraints of the
type \eqref{eq:energy_reduction_constraints} We will elaborate on the
mechanics of this at length when we consider specific examples.

Importantly, the above replacement {\it does not} imply that we are
permitted to set to zero a single one of the linear factors
in \Eq{eq:energy_reduction_constraints}.  Instead, the statement is
that when the full product in \Eq{eq:energy_reduction_constraints}
appears in a numerator, it can be zeroed out.

By repeated application of replacements
like \Eq{eq:energy_reduction_constraints} and by canceling matter
poles in the denominator, we can systematically reduce the degree of
the numerator in all loop energy variables until we are left with a
sum over integrals with no loop energy dependence in the numerator. We
identify these integrals as master energy integrals and they take the
form of \Eq{eq:integrand_eff} with $\widetilde{\cal N}$ set to unity,
i.e. they are comprised entirely of simple matter energy poles.

We then directly evaluate the master energy integrals, crucially
keeping track of the contour prescriptions dictated by $i\epsilon$
factors.  If the integrals are convergent we can simply do them
explicitly.  However, many of the integrals naively diverge due to
contributions from poles at infinity.  As we will see, by averaging
over routings of the loop energies, we obtain finite answers which
automatically incorporate various computable symmetry factors.  A
similar approach is discussed in
Refs. \cite{StermanEikonal,BjerrumClassical} and developed in
Refs.~\cite{CliffIraMikhailClassical,3PMPRL}.  However, in the
previous relativistic approaches, symmetrization occurs via summing
over reroutings of the full relativistic momenta.  In our case, we
symmetrize purely over the energy components.

All of our examples in \sects{sec:NRint_oneloop}{sec:NRint_twoloop}
will employ this method of energy-integral reduction.  However, there
is a second simpler approach which we dub the ``residue method''.  In this
case, we evaluate the energy integral as a weighted sum over residues
on the matter poles.  The residues are reweighted by various symmetry
factors that can be derived systematically and are actually equal to
the corresponding master energy integrals encountered in the method of
energy-integral reduction.  We present the details of the residue
method in \sect{sec:residue}.

\subsubsection*{Step 3: Momentum Loop Integration}

Our last step is to perform the $(D-1)$-momentum integration. The
spatial integrand $ \widetilde{\cal I}$ is a complicated nonanalytic
function of the loop $(D-1)$-momenta.  Its functional form involves
square roots coming from the evaluation of the energy integrals.
Direct evaluation of these integrals is nontrivial.  Our approach is
instead to expand $\widetilde{\cal I}$ in the nonrelativistic limit,
$|\bm p| \ll m_1,m_2$ (cf. see \eqn{eq:NR_limit}), up to some
order. As we will see, each term in the series is a simple rational
function of the loop $(D-1)$-momenta, and in fact, the form of these
objects is identical to those which appear in NRGR.

Upon expansion, every integral---in some cases after straightforward integration by parts (IBP) reduction~\cite{Chetyrkin:1981qh, Laporta:2001dd}---can be written in the form
\eq{
\widetilde{\cal I} = \sum_{\alpha} \sum_{\beta} \sum_\gamma
\frac{f^{(\alpha\beta\gamma)}(\bm \ell)}{ [\bm \ell^2]^\alpha\,
[(\bm \ell +\bm w)^2]^{\beta} \, [2\bm z \bm \ell  +\bm \ell^2]^{\gamma}} \,,
}{eq:integrable}
where $\bm \ell$ is one of the loop $(D-1)$-momenta, $\bm w$ and
$\bm z$ denote vectors built from other loop $(D-1)$-momenta or the
external $(D-1)$-momenta, and $\bm z \bm \ell$ is a shorthand for $\bm z \cdot \bm \ell$. 
 Here $\alpha$ and $\beta$ can take
positive fractional values but $\gamma$ is one or zero.  The function
$f^{(\alpha\beta\gamma)}$ is a polynomial in $\bm \ell$.  The
$\bm \ell^2$ and $(\bm \ell + \bm w)^2$ poles are generated by
internal graviton propagators while the $2 \bm z \bm \ell
+ \bm \ell^2$ poles arise from internal matter propagators.

Remarkably, sequential integration of each loop $(D-1)$-momentum is sufficient to evaluate all two-loop integrals. To begin, we check whether 
there exists a loop $(D-1)$-momentum $\bm \ell$ such that $\widetilde{\cal I}$ takes the form in \Eq{eq:integrable} with $\gamma=0$. This occurs 
whenever a diagram contains a triangle subdiagram.  In this case we trivially evaluate the integral using the following analytic formula for an arbitrary tensor numerator~\cite{Smirnov:2004ym}, 
\eq{
\int \frac{d^{D-1}\bm \ell}{(2\pi)^{D-1}} \, \frac{\bm \ell^{\mu_1} \bm \ell^{\mu_2} \cdots \bm\ell^{\mu_n}}{ [\bm \ell^2]^\alpha[(\bm \ell +\bm w)^2]^{\beta}} =
\frac{(-1)^{n} (4\pi)^{\frac{1-D}{2}}}{ [\bm w^2]^{\alpha+\beta -\frac{D-1}{2}}}
\sum_{m=0}^{\lfloor{n/2}\rfloor} A(\alpha,\beta; n,m) \left[\frac{\bm w^2}{2} \right]^m \left\{[\delta]^m [\bm w]^{n-2m} \right\}^{\mu_1 \mu_2 \cdots \mu_n},
}{eq:Smirnov}
where the quantity in curly brackets denotes a fully symmetric tensor built from $m$ powers of the spatial metric, i.e.~the $(D-1)$-dimensional Kronecker delta function, and $n-2m$ powers of $\bm w$.  This tensor is normalized so that each distinct term has unit coefficient.  Here we have also defined
\eq{
A(\alpha,\beta; n,m) =  \frac{\Gamma(\alpha+\beta  -m - \frac{D-1}{2}) \Gamma(n-m - \alpha+\frac{D-1}{2})\Gamma(m-\beta +\frac{D-1}{2})}{\Gamma(\alpha)\Gamma(\beta) \Gamma(n-\alpha-\beta +D-1)} \,.
}{}
Note that starting at two loops, it is necessary to use dimensional regularization in order to properly deal with ultraviolet divergences encountered in spatial integrals.\footnote{As usual in the method of regions, the contribution from a particular region (here the potential region) may be ultraviolet or infrared divergent even though the full relativistic integrand is finite. } We will discuss this aspect in detail in \sect{sec:evanescence}. 

We identify and evaluate all triangle subdiagrams sequentially until the only remaining loop integrals take the form of \Eq{eq:integrable} with $\gamma=1$.  These contributions correspond to box subdiagrams, are infrared divergent, and scale with additional factors of $|\bm q|^{-1}$ relative to classical contributions. As mentioned in \sect{QuantumVsClassicalSection}, we refer to such terms as ``superclassical''. In principle, one can evaluate these integrals. We will see, however, that this is unnecessary because they are infrared artifacts that exactly cancel between the full theory and EFT contributions to the matching.

Before working through explicit examples, we first offer a note of caution on interpreting and comparing results from nonrelativistic integration discussed here and relativistic integration discussed in \sect{sec:Rint}. Nonrelativistic integration runs over the momentum configurations of potential gravitons, so no internal gravitons are ever on shell.  Typically,  relativistic methods run over the soft region, which also includes on-shell graviton contributions.  In the classical limit, these methods will produce the same final answer, however, only after summing all contributions.  Separate graph contributions which are not individually gauge invariant will in general produce different answers depending on which integration method is applied.  For this reason, when we make comparisons later between nonrelativistic and relativistic integration, we will always compute certain sums of graph topologies. 

\subsection{One-Loop Examples}
\label{sec:NRint_oneloop}

In this section we consider the explicit example of classical scattering at one loop in order to illustrate the methods we use at two loops.  As per our previous discussion in \sect{sec:classical_limit}, the classical contributions come from triangle and box type diagrams, which we consider in turn below.  While the calculation presented here is less familiar than other approaches to one-loop integration, the methodology has the crucial advantage that it scales well to higher-loop orders.

\subsubsection{Triangle Diagram}

\newcommand{\tri}{{\rm T}}

\begin{figure}[t]
\begin{center}
\includegraphics[scale=.6]{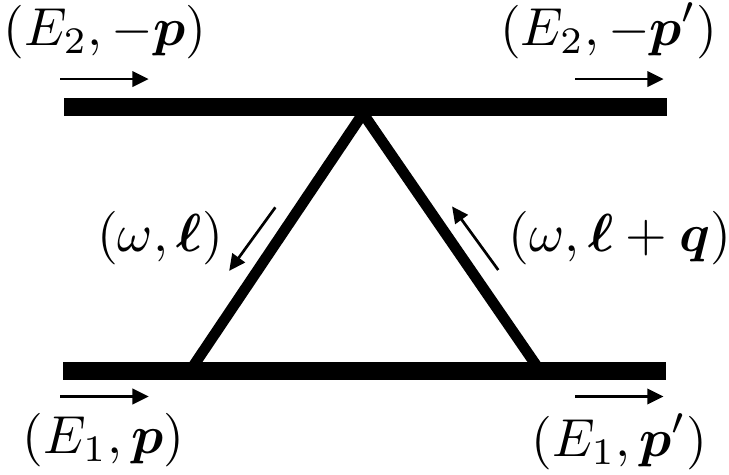}
\end{center}
\vspace*{-0.5cm}
\caption{\small The triangle diagram. The thick and thin lines represent massive and massless particles, respectively.
 }\label{fig:NRint_triangle}
\end{figure}

As a warm up, we study a general triangle diagram, defined here to be a one-loop integral comprised of an arbitrary numerator together with one $\phi_1$ propagator and two graviton propagators arranged as in \fig{fig:NRint_triangle}.
In the notation of \Eq{eq:integrands}, the corresponding integrand takes the form
\eq{
{\cal I}_\tri &=\frac{1}{(E_1+\omega)^2 - (\bm p+\bm \ell)^2 -m_1^2} \, \frac{1}{\omega^2 - \bm \ell^2}\, \frac{1}{\omega^2 - (\bm \ell+ \bm q)^2} \, {\cal N}_\tri \,.
}{}

First we determine the effective numerator. The scalar propagators can be factored into matter and antimatter components,
\eq{
\frac{1}{(E_1+\omega)^2 - (\bm p+\bm \ell)^2 -m_1^2}= \frac{1}{(\omega -\omega_{P_1})(\omega-\omega_{A_1})}
\,, \qquad \omega_{P_1},\omega_{A_1} =  -E_1 \pm \sqrt{E_1^2 +2\bm p \bm \ell+\bm \ell^2} \,.
}{eq:split_prop1}
Thus, the spatial integrand is given in terms of the effective numerator by 
\eq{
\widetilde {\cal I}_{\tri} = \int \frac{d\omega}{2\pi}\, \frac{\widetilde{\cal N}_\tri(\omega)}{\omega-\omega_{P_1}} \,, \qquad
\widetilde{\cal N}_\tri(\omega) = \frac{1}{\omega -\omega_{A_1}} \, \frac{1}{\omega^2 - \bm \ell^2}\, \frac{1}{\omega^2 - (\bm \ell+ \bm q)^2} \, {\cal N}_\tri(\omega) \,,
}{eq:triangle_eN}
where for emphasis we have made explicit all dependence on $\omega$.

Second we use energy-integral reduction to evaluate the energy integral.   
From \Eq{eq:triangle_eN}, we see that any numerator term which cancels the $\omega- \omega_{P_1}$ pole leads to a scaleless integral. 
Thus, the general constraint in \Eq{eq:energy_reduction_constraints} for this example is simply
\eq{
\omega - \omega_{P_1} \rightarrow 0 \,.
}{eq:triangle_zero}
Applying this replacement is of course equivalent to sending
\eq{
 \widetilde{\cal N}_\tri(\omega) \rightarrow  \widetilde{\cal N}_\tri(\omega_{P_1}) \,.
 }{}
The spatial integrand then takes the form
\eq{
\widetilde {\cal I}_\tri = \widetilde{\cal N}_\tri(\omega_{P_1})\int \frac{d\omega}{ 2\pi}\frac{1}{\omega-\omega_{P_1}} \,.
}{}
The energy integral has no more energy dependence in the numerator, and we identify this as the triangle master energy integral.

The triangle master energy integral can be computed in various ways
which will be instructive when we eventually encounter more
complicated objects.  The first option is to compute the integral by
direct evaluation on a symmetric interval:
\eq{
 \int  \frac{d\omega}{\omega -\omega_{P_1}+i\epsilon} \equiv \lim_{R\rightarrow \infty}   \int\limits_{-R}^R \frac{d\omega}{\omega -\omega_{P_1}+i\epsilon} = \frac{1}{2}\times (-2\pi i) \,,
}{}
where on the left hand side we have reintroduced the appropriate
$i\epsilon$ to regularize the integral coming from the original propagators, following the standard Feynman $i\epsilon$ prescription in \Eq{eq:split_prop1}.  A
heuristic for the sign in front of the $i\epsilon$ is that it
designates whether positive $\omega$ increases or decreases the energy
of the matter particle corresponding to this pole.  We have expressed
our answer in terms of a 1/2 symmetry factor multiplying the quantity
one would naively extract from computing the residue on the matter
pole including signs.

One approach is to compute the master energy integral directly via
residues.  At one loop there is a single integration variable and the
analysis is straightforward~\cite{CliffIraMikhailClassical}. This is
not the case for multivariable integration. Pushing the contour of
integration into the upper half complex plane, we pick up the residue
at $\omega= \omega_{P_1}-i\epsilon$ as well as the contribution from
the upper half arc at infinity.  Pushing the contour down, there are
no residues to pick up but we include the contribution from the
lower-half arc at infinity. Averaging over the two equivalent
prescriptions, the half-arc contributions cancel and we are again left
with the 1/2 symmetry factor relative to the residue on the matter
pole.

The following approach will be our standard method of choice since it
scales nicely to higher loop.  First we assign to each graviton a
unique energy component $\omega_i$ with $i=1$ to $n_G$. Then we
introduce delta functions to rewrite the expression as an integral
over a subset of the $n_G$ loop energy variables. Finally, we average
over permutations of the loop energies that preserve isometries of the
delta functions. If the resulting integrand is such that it vanishes
in the $i\epsilon \to 0$ limit, then we have a convergent integral
that can be directly performed. Applying this to the triangle gives
\eq{
\int   \frac{d\omega}{\omega-\omega_{P_1} + i\epsilon} &\equiv \int   d\omega_1 \, d\omega_2 \, \delta(\omega_1 + \omega_2)\frac{1}{2!} \left[ \frac{1}{\omega_1-\omega_{P_1} + i\epsilon}  +\frac{1}{\omega_2-\omega_{P_1} + i\epsilon}  \right]  \\
&=\frac{1}{2} \int  \left(   \frac{d\omega }{\omega-\omega_{P_1} + i\epsilon} +   \frac{d\omega }{-\omega-\omega_{P_1} +i\epsilon} \right) = \frac{1}{2} \times (-2\pi i) \,,
}{eq:triangle_3}
where in the final equality we have evaluated the integral over the real domain, which is well-defined and yields the 1/2 symmetry factor. 

The purpose of the symmetrization procedure is to make the integrand
manifestly convergent.  Without symmetrization, the original integrand
scales as $1/\omega$ at large $\omega$ and thus receives boundary
contributions at infinity.  By symmetrizing over two equivalent energy
routings, we obtain a new integrand that falls of manifestly as
$1/\omega^2$ and has no contribution at infinity.  The same feature
will emerge in all of our other examples which include triangle
subdiagrams.  In the soft eikonal expansion, it is known that this
type of averaging over graviton permutations effectively eliminates
the singular ``principal value'' contribution that arises when
$i\epsilon$ is set to zero~\cite{StermanEikonal,BjerrumClassical}.

Finally, we evaluate the spatial integral. After carrying out the energy integral we obtain
\eq{
\widetilde{\cal I}_\tri(\bm p, \bm q,\bm \ell) = -\frac{i}{2}\, \frac{1}{\omega_{P_1} -\omega_{A_1}} \, \frac{1}{\omega_{P_1}^2 - \bm \ell^2}\, \frac{1}{\omega_{P_1}^2 - (\bm \ell+ \bm q)^2} \, {\cal N}_\tri(\omega_{P_1}) \, .
}{}
Note that  $\widetilde{\cal I}_\tri$ is a quite complicated function of $\bm p$, $\bm q$, and $\bm \ell$ since $\omega_{P_1}$ and $\omega_{A_1}$ contain square root functions. As a result, direct integration would be difficult. We therefore first expand the spatial integrand in the nonrelativistic limit, $|\bm p| \ll m_1 , m_2$.  In this limit the various denominators become
\eq{
\frac{1}{\omega_{P_1} -\omega_{A_1}} &= \frac{1}{2m_1} + \cdots\,, \\
 \frac{1}{\omega_{P_1}^2 - \bm \ell^2} &= -\frac{1}{\bm \ell^2}+ \cdots\,, \\
  \frac{1}{\omega_{P_1}^2 - (\bm \ell+ \bm q)^2} &= -\frac{1}{(\bm \ell+\bm q)^2}+ \cdots  \, .
}{}
Thus we see that, in the nonrelativistic expansion of $\widetilde{\cal
I}_\tri$, the only singularities arise from powers of $\bm \ell^2$ and
$(\bm \ell + \bm q)^2$ in the denominator.  Consequently
$\widetilde{\cal I}_\tri$ takes exactly the form of \Eq{eq:integrable}
with $\gamma=0$ and can thus be evaluated straightforwardly
using \Eq{eq:Smirnov}.

As a concrete illustration, consider the scalar triangle integral for
which ${\cal N}_\tri =1$.  From relativistic considerations it is
obvious that the scalar triangle integral is only a function of $\bm
q^2$ and $m_1^2$ since these are the only invariants that can be
formed.  Hence the scalar triangle integral must be a series expansion
in $|\bm q| /m_1$, and the classical contribution is entirely given
by the first term
\eq{
\widetilde{\cal I}_\tri (\bm p, \bm q,\bm \ell)  = -\frac{i}{4m_1\bm \ell^2 (\bm \ell+ \bm q)^2} \, .
}{eq:tildeI_triangle}
By direct integration with \Eq{eq:Smirnov}, we obtain the 
known result for the classical part of the one-loop triangle diagram
\eq{
I_\tri  &= -\frac{i}{32 m_1 |\bm q|} \,.
}{onelooptriangle}
Note that this result is not the full relativistic scalar triangle
integral, but rather the leading classical contribution arising from
exchanges of potential gravitons. This is derived as well using
Mellin-Barnes integration in \sect{sec:MB}.

\subsubsection{Box Diagram \label{1loopbox}}

\newcommand{\boxdiag}{{\rm B}}

\begin{figure}[t]
\begin{center}
\includegraphics[scale=.8]{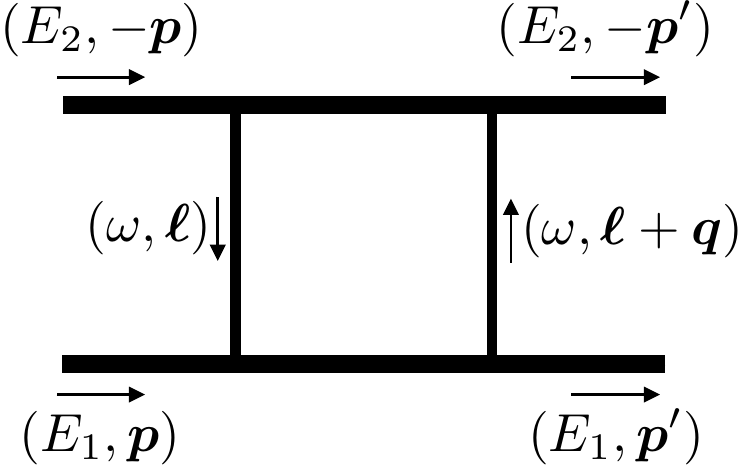}
\end{center}
\vspace*{-0.5cm}
\caption{\small The one-loop box diagram. The thick and thin lines represent massive and massless particles, respectively.
}\label{fig:NRint_box}
\end{figure}

The next simplest example is the box diagram, defined as an arbitrary numerator together with one $\phi_1$ propagator, one $\phi_2$ propagator, and two graviton propagators as shown in \fig{fig:NRint_box}.
Since the numerator is arbitrary, this example has as a subcase the triangle diagram.  The box integrand is
\eq{
{\cal I}_\boxdiag &=\frac{1}{(E_1+\omega)^2 - (\bm p+\bm \ell)^2 -m_1^2} \, \frac{1}{(E_2-\omega)^2 - (\bm p+\bm \ell)^2 -m_2^2}\, \frac{1}{\omega^2 - \bm \ell^2}\, \frac{1}{\omega^2 - (\bm \ell+ \bm q)^2} \, {\cal N}_\boxdiag \,.
}{eq:boxstart}
As before, we split the scalar propagators into matter and antimatter components, obtaining \Eq{eq:split_prop1} together with the analogous formula for the second massive propagator
\eq{
\frac{1}{(E_2-\omega)^2 - (\bm p+\bm \ell)^2 -m_2^2}&= \frac{1}{(\omega -\omega_{P_2})(\omega-\omega_{A_2})}\, , \qquad \omega_{P_2},\omega_{A_2} =  E_2 \mp \sqrt{E_2^2 +2\bm p \bm \ell+\bm \ell^2} \, .
}{}
The spatial integrand is then given in terms of the effective numerator by
\eq{
\widetilde{\cal I}_\boxdiag =   \int \frac{d\omega}{2\pi}\, \frac{\widetilde{\cal N}_\boxdiag(\omega)}{(\omega-\omega_{P_1})(\omega-\omega_{P_2})}\,, 
}{eq:box_eff_num}
where 
\eq{
\widetilde{\cal N}_\boxdiag(\omega) &= \frac{1}{\omega -\omega_{A_1}}\frac{1}{\omega -\omega_{A_2}} \, \frac{1}{\omega^2 - \bm \ell^2}\, \frac{1}{\omega^2 - (\bm \ell+ \bm q)^2} \, {\cal N}_\boxdiag(\omega) \,.
}{}

Next, we apply energy-integral reduction to recast the integrand in
terms of a set of master energy integrals.  Since simultaneously
eliminating both poles in $\widetilde{\cal I}_\boxdiag$ produces a
scaleless energy integral, the general constraint
in \Eq{eq:energy_reduction_constraints} for this example is given by
\eq{
(\omega - \omega_{P_1})(\omega - \omega_{P_2 }) \rightarrow 0 \, .
}{eq:box_constraints}
The constraint in \Eq{eq:box_constraints} defines the zero locus for a
certain quadratic polynomial in $\omega$.  Note how this contrasts
with the simpler linear constraint we saw earlier
in \Eq{eq:triangle_zero}.  Previously, to evaluate the effective
numerator on the linear constraint it sufficed to trivially plug in
for the single solution for $\omega$.  However, imposing the quadratic
constraint in \Eq{eq:box_constraints} is more involved.  In
particular, \Eq{eq:box_constraints} does {\it not} indicate that we
can set either $\omega = \omega_{P_1}$ or $\omega = \omega_{P_2}$ in
the numerator.  Instead, the claim is that any time the full quadratic
function in \Eq{eq:box_constraints} appears in the effective
numerator, we can drop it.  This is required because only if both
matter poles, $\omega - \omega_{P_1}$ and $\omega - \omega_{P_2}$, are
canceled do we have a scaleless integral that can be dropped, cf. \eqn{eq:scaleless}.

To apply energy reduction on the effective numerator we simply compute the remainder of $\widetilde {\cal N}_{\rm B}$ under modular division by the quadratic polynomial in \Eq{eq:box_constraints}.  If $\widetilde {\cal N}_{\rm B}$ were a polynomial, e.g.~obtained by expanding \Eq{eq:box_constraints} as a series expansion in $\omega$ up to some high order, then it is obvious how to perform this modular division, i.e., simply by repeated application of  \Eq{eq:box_constraints}.  However, if we want to retain \Eq{eq:box_eff_num} to all orders, then the effective numerator is not simply a polynomial in $\omega$ but has complicated denominator poles.  Nevertheless, we can still perform modular division by the following procedure. First we shift the integration variable $\omega$ such that the constraint in \Eq{eq:box_constraints} is of the form $\omega^2 - \zeta =0$, where $\zeta$ is a constant. Then we express $\widetilde {\cal N}_{\rm B}$ in terms of even and odd functions in $\omega$, whose denominators are even in $\omega$. Then all appearances of $\omega^2$ can simply be set to $\zeta$. In particular, the even functions become constants, while the odd functions become linear in $\omega$. In this way, 
an arbitrary numerator is reduced to the form
\eq{
\widetilde{\cal N}_\boxdiag(\omega) \rightarrow f_0 + f_1 \omega \,.
}{}

Next, we consider further expanding this linear function for
$\widetilde{\cal N}_\boxdiag(\omega)$ about the point $\omega_{P_1}$,
so there is a term proportional to $\omega - \omega_{P_1}$ and a
remainder term.  The former will cancel one of the matter poles,
yielding an expression of the form of the triangle diagram tackled in
the previous section. Meanwhile, the latter is a new
ingredient since it does not cancel any of the matter poles. This
contribution yields the box master energy integral. As before, we
evaluate it by averaging over the two permutations of the graviton
labels,
\eq{
\int   \frac{d\omega}{(\omega-\omega_{P_1} + i\epsilon)(\omega-\omega_{P_2} - i\epsilon)}  &= \frac{1}{2} \int\left(  \frac{ d\omega_1 d\omega_2 \delta(\omega_1 + \omega_2) }{(\omega_1-\omega_{P_1} + i\epsilon)(\omega_1-\omega_{P_2} - i\epsilon)} +\{ \omega_1 \leftrightarrow -\omega_2 \} \right) \\
&= \frac{1}{\omega_{P_1}-\omega_{P_2}} \times (- 2\pi i) \,, 
}{eq:energy_box}
where again we use the standard Feynman $i\epsilon$ prescription. This agrees with the heuristic mentioned earlier, which assigns opposite signs to each $i\epsilon$ factor since for the box, increasing $\omega$ will increase the energy of one matter line and decrease that of the other. Here we have symmetrized over routings of the loop energies to parallel our earlier analysis, but since the original integrand falls off as $1/\omega^2$ it is also straightforward to evaluate it via contour integration without symmetrization.

In the above discussion we made the choice of expanding $\widetilde{\cal N}_\boxdiag(\omega)$ about $\omega_{P_1}$, and we obtained contributions from the master triangle and master box energy integrals.  Both contributions must be included to obtain the correct answer.  Alternatively, we could have chosen to expand about the point $\omega_{P_2}$ instead, in which case the various integrals would slightly change but yield the same final answer.

The important feature of \Eq{eq:energy_box} is that it introduces a new kind of singularity coming from the $(\omega_{P_2} - \omega_{P_1})$ denominator factor. For example, expanding in the classical limit using the scaling $(\bm \ell^2 + 2 \bm p \bm \ell) \sim {\cal O}(\bm q)$, this pole becomes
\eq{
 \frac{1}{\omega_{P_1}-\omega_{P_2}} =  \frac{2E_1E_2}{E_1+E_2}  \, \frac{1}{2\bm p \bm \ell+ \bm \ell^2} + \cdots \,, 
}{} 
which is singular when the internal $\phi_1$ and $\phi_2$ particles are on shell. The ellipsis here denotes terms higher order in $(\bm \ell^2 + 2 \bm p \bm \ell)$, which may contribute to the final classical result.

As an example consider the scalar box integral for which ${\cal N}_\boxdiag =1$.  Applying the procedure described above, we obtain the spatial integrand 
\eq{
\widetilde{\cal I}_\boxdiag  &= {i \over 2E \bm \ell^2 ( \bm \ell + \bm q)^2 (\bm \ell^2 + 2 \bm p \bm \ell)} + {i (E_1^2 - 3E_2^2) \left(\bm \ell^2 ( \bm \ell + \bm q)^2 - (\bm \ell^2 +  ( \bm \ell + \bm q)^2)(\bm \ell^2 + 2 \bm p \bm \ell) \right) \over 16 E_1^2 E_2^2 E  \bm \ell^4 ( \bm \ell + \bm q)^4} \\
&- {i (E_1^4 - 3E_2^4) \left[(\bm \ell^4 +\bm  \ell^2( \bm \ell + \bm q)^2 +( \bm \ell + \bm q)^4)(\bm \ell^2 + 2 \bm p \bm \ell) -3  \bm\ell^2( \bm \ell + \bm q)^2 (\bm \ell^2 +  ( \bm \ell + \bm q)^2)  \right]  (\bm \ell^2 + 2 \bm p \bm \ell)^2 \over 64 E_1^4 E_2^4 E   \bm \ell^6 ( \bm \ell + \bm q)^6} \\
& + \cdots \,,
}{eq:box1}
where the asymmetry in the labels 1 and 2 is due to the choice of expanding about $\omega_{P_1}$. The final answer below is symmetric in the labels 1 and 2, as it should be.
To obtain \Eq{eq:box1} we performed two expansions. First, we expanded in the classical limit of large angular momentum $J$, or equivalently small momentum $\bm q$, using the scalings $\bm \ell \sim ( \bm \ell + \bm q) \sim (\bm \ell^2 + 2 \bm p \bm \ell) \sim {\cal O}(\bm q)$. Second, we expanded in the nonrelativistic limit, but keeping existing energy factors intact. By direct integration of \Eq{eq:box1} using \Eq{eq:Smirnov}, we obtain
\eq{
I_\boxdiag  &= \int \frac{d^{D-1} \bm \ell }{(2\pi)^{D-1}} \, {i \over 2E \bm \ell^2 ( \bm \ell + \bm q)^2 (\bm \ell^2 + 2 \bm p \bm \ell)} \, ,
}{}
where the classical terms vanish order by order in the nonrelativistic expansion. The terms in \Eq{eq:box1} that are antisymmetric in the labels 1 and 2 have vanished. The above remaining integral has the form of \Eq{eq:integrable} with $\gamma=1$. It is infrared divergent and superclassical, having an additional factor of $|\bm q|^{-1}$ relative to the classical scaling. As noted previously, we choose not to evaluate these quantities explicitly since they are IR artifacts that are guaranteed to subtract exactly with matching contributions from the EFT amplitude.

\subsubsection{Crossed-Box Diagram}

\newcommand{\xboxdiag}{\overline{\rm B}}

\begin{figure}
\begin{center}
\includegraphics[scale=.8]{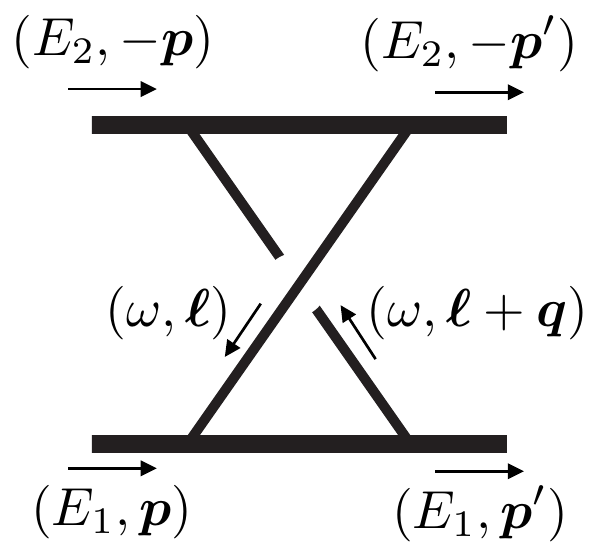}
\end{center}
\vspace*{-0.5cm}
\caption{\small The crossed-box diagram. }\label{fig:NRint_crossed}
\end{figure}

Next, consider the crossed box diagram, defined as an arbitrary
numerator together with one $\phi_1$ propagator, one $\phi_2$ propagator, and
two graviton propagators as shown in \Fig{fig:NRint_crossed}.  The
integrand of the crossed-box diagram is
\eq{
{\cal I}_{\xboxdiag} &=\frac{1}{(E_1+\omega)^2 - (\bm p+\bm \ell)^2 -m_1^2} \, \frac{1}{(E_2+\omega)^2 - (\bm p-\bm \ell-\bm q)^2 -m_2^2}\, \frac{1}{\omega^2 - \bm \ell^2}\, \frac{1}{\omega^2 - (\bm \ell+ \bm q)^2} \, {\cal N}_{\xboxdiag} \,.
}{}
The analysis here is almost identical to the box integral.  We again apply
energy-integral reduction, exploiting the fact that when the energy
denominators for both the $\phi_1$ and $\phi_2$ propagators are
canceled, then the resulting scaleless energy integral is quantum and can be
discarded.  As for the box integral, we will obtain contributions proportional
to the master triangle energy integral.  The only difference from the
box is that upon energy-integral reduction, we obtain another new
crossed box master energy integral,
\eq{
\int   \frac{d\omega}{(\omega-\omega_{P_1} + i\epsilon)(\omega-\omega_{P_2} + i\epsilon)}  &= \frac{1}{2} \int  \left(  \frac{d\omega}{(\omega-\omega_{P_1} + i\epsilon)(\omega-\omega_{P_2} + i\epsilon)} +\{ \omega \leftrightarrow -\omega\} \right) =0 \,.
}{eq:energy_xbox}
Note the same sign in front of each $i\epsilon$ factor.  This occurs
because for the crossed box, increasing $\omega$ will increase the
energy flow through both matter propagators.  Crucially, this slight
difference means that the crossed-box master energy integral is zero.

This vanishing can be understood by deforming the integration contour.
The integrand in \Eq{eq:energy_xbox} obviously vanishes at infinity,
so by closing the contour of integration into the upper half plane,
we obtain zero.  This type of cancellation is the calling card for
contributions which vanish due to causality in nonrelativistic field
theory. There is a simple diagrammatic diagnostic to determine which
energy integrals are zero.  Since every graviton of interest is a
potential mode, it connects two points of equal time.  We can
then divide all topologies according to whether they are connected or
disconnected after cutting all matter lines.  For connected diagrams,
like the triangle, all vertices connected to a graviton are
simultaneous interactions.  Meanwhile, disconnected diagrams like the
box and crossed box can be thought of describing a series of two
simultaneous events, corresponding to each graviton exchange.  Any two
events must occur in some time order.  If we then draw the worldlines
of the scalar fields for the box and crossed box, we then see that the
latter requires backwards in time propagating matter particles,
i.e. antimatter, to be consistent.  Consequently these diagrams vanish
by causality.

We now briefly discuss the scalar example, setting $ {\cal
N}_{\xboxdiag} = 1$. Applying energy-integral reduction gives a
contribution proportional to the crossed-box master integral and a
contribution proportional to the triangle master integral. The former
vanishes since the crossed-box master integral is zero, while the
latter vanishes by direct integration of the spatial integrand, order
by order in the velocity expansion. Hence, the scalar crossed-box
diagram has no classical contribution.

The vanishing of classical contributions from the sum of the box and
crossed-box diagrams can be shown to all orders using other
integration methods, such as direct integration of the soft region or
the use of differential equations. Our focus here is to illustrate the
application of our nonrelativistic integration procedure.

\subsection{Two-Loop Examples}
\label{sec:NRint_twoloop}

We are now prepared to consider some examples at two loops.  As
advertised, our methodology will be identical to the one-loop cases, only here
applied sequentially to each loop.

\subsubsection{Double-Triangle Diagram \label{sec:TT}}

\newcommand{\TT}{{\rm TT}}

\begin{figure}
\begin{center}
\hspace{2cm}\includegraphics[scale=.8]{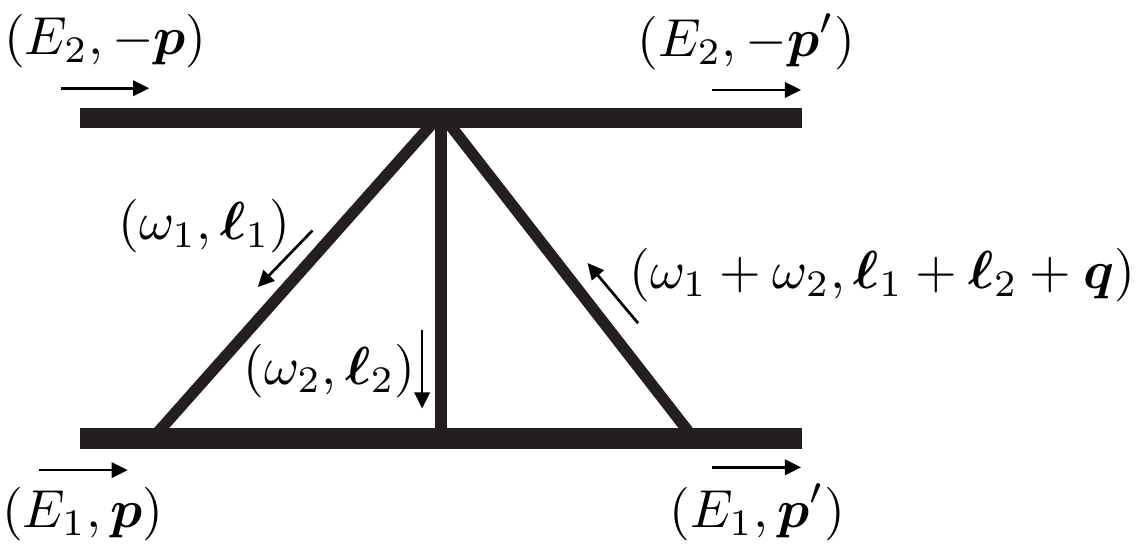}
\end{center}
\vspace*{-0.5cm}
\caption{\small The double-triangle diagram. }\label{fig:NRint_W}
\end{figure}

To begin, consider the general double-triangle diagram, which is a
two-loop integral with an arbitrary numerator together with two
$\phi_1$ propagators and three graviton propagators arranged as
in \Fig{fig:NRint_W}.
The integrand corresponding to this diagram is
\eq{
{\cal I}_{\TT} =&\frac{1}{(E_1+\omega_1)^2 - (\bm p+\bm \ell_1)^2 -m_1^2}\, \frac{1}{(E_1+\omega_1+\omega_2)^2 - (\bm p+\bm \ell_1+\bm \ell_2)^2 -m_1^2} \\
& \times \frac{1}{\omega_1^2 - \bm \ell_1^2}\, \frac{1}{\omega_2^2 - \bm \ell_2^2}\, \frac{1}{(\omega_1+\omega_2)^2 - (\bm \ell_1 + \bm \ell_2+ \bm q)^2} \, {\cal N}_\TT \,.
}{}
Both $\phi_1$  propagators can be factored into matter and antimatter components,
\eq{
\frac{1}{(E_1+\omega)^2 - (\bm p+\bm \ell)^2 -m_1^2}&= \frac{1}{(\omega_1 -\omega_{P_1})(\omega_1-\omega_{A_1})} \,, \\
 \frac{1}{(E_1+\omega_1+\omega_2)^2 - (\bm p+\bm \ell_1+\bm \ell_2)^2 -m_1^2}&= \frac{1}{(\omega_1 +\omega_2 -\omega_{P_1'})(\omega_1 +\omega_2 -\omega_{A_1'})} \,.
}{}
The spatial integrand is then given in terms of the effective numerator by
\eq{
\widetilde {\cal I}_{\TT} &= \int \frac{d\omega_1}{2\pi}\frac{d\omega_2}{2\pi}\, \frac{\widetilde{\cal N}_\TT(\omega_1 ,\omega_2)}{(\omega_1-\omega_{P_1})(\omega_1+\omega_2-\omega_{P_1'})} \,, 
}{}
where
\eq{
\widetilde{\cal N}_\TT(\omega_1 ,\omega_2) &=
\frac{1}{\omega_1-\omega_{A_1}} \, \frac{1}{\omega_1+\omega_2-\omega_{A_1'}}\, \frac{1}{\omega_1^2 - \bm \ell_1^2}\, \frac{1}{\omega_2^2 - \bm \ell_2^2}\, \frac{1}{(\omega_1+\omega_2)^2 - (\bm \ell_1 + \bm \ell_2+ \bm q)^2} \, {\cal N}_\TT(\omega_1,\omega_2) \,.
}{}

Next, we use the fact that canceling either energy denominator will produce a scaleless integral in either $\omega_1$ or $\omega_2$.  Consequently, we can freely send 
\eq{
\omega_1 - \omega_{P_1} \rightarrow 0 \,,  \qquad \omega_1 +\omega_2 - \omega_{P_1'} \rightarrow 0\,,
}{}
so we can effectively evaluate the numerator 
\eq{
 \widetilde{\cal N}_\TT(\omega_1,\omega_2) \rightarrow  \widetilde{\cal N}_\TT (\omega_{P_1},\omega_{P_1'} - \omega_{P_1}) \,.
 }{}
Upon this replacement, the spatial integrand becomes
\eq{
\widetilde {\cal I}_\TT &=  \widetilde{\cal N}_\TT(\omega_{P_1},\omega_{P_1'} - \omega_{P_1}) \int \frac{d\omega_1}{2\pi}\frac{d\omega_2}{2\pi}\, \frac{1}{(\omega_1-\omega_{P_1})(\omega_1+\omega_2-\omega_{P_1'})} \,.
} 
{}
Here the last factor is the double-triangle master energy integral.  We evaluate it by averaging over permutations of labelings of the three exchanged gravitons,
\eq{
\int  \frac{d\omega_1 \, d\omega_2}{(\omega_1-\omega_{P_1}+i\epsilon)(\omega_1+\omega_2-\omega_{P_1'}+i\epsilon)} & \equiv \frac{1}{3!} \left( \int  \frac{d\omega_1 \, d\omega_2}{(\omega_1-\omega_{P_1}+i\epsilon)(\omega_1+\omega_2-\omega_{P_1'}+i\epsilon)}  + \{ \textrm{perm.}\} \right)\\
&= \frac{1}{6} \times (-2\pi i)^2 \,,
}{eq:double_master}
where we include all permutations of $\omega_1$, $\omega_2$, and $\omega_3$, and we set $\omega_3=-\omega_1-\omega_2$ by energy conservation in the end. Again we use the standard Feynman $i\epsilon$ prescription, which agrees with the same heuristic as before: we include the same sign in front of each $i\epsilon$ factor since increasing $\omega_1$ increases the energy flowing through each matter line. By explicit calculation we have obtained a 1/6 symmetry factor relative to the naive expression one would obtain by sequential application of the residue theorem on each pole.  

As before, the original integrand scales as $1/\omega_2$ at infinity, but the symmetrization improves the asymptotic behavior to $1/\omega_2^2$ so there is no boundary term to consider and integration is mechanical. 

Having evaluated the energy integral, we obtain the spatial integrand
\eq{
\widetilde {\cal I}_\TT(\bm p,\bm q, \bm \ell_1,\bm \ell_2) &=
\frac{1}{\omega_{P_1}-\omega_{A_1}} \, \frac{1}{\omega_{P_1'}-\omega_{A_1'}}\, \frac{1}{\omega_{P_1}^2 - \bm \ell_1^2}\, \frac{1}{(\omega_{P_1'}-\omega_{P_1})^2 - \bm \ell_2^2}\, \frac{1}{\omega_{P_1'}^2 - (\bm \ell_1 + \bm \ell_2+ \bm q)^2} \\
&\qquad \times {\cal N}_\TT(\omega_{P_1},\omega_{P_1'} - \omega_{P_1}) \times \left( - \frac16 \right).
}{}
As before, consider for illustration the simple case of the scalar double-triangle integral, for which ${\cal N}_\TT =1$.
Expanding the spatial integrand in the nonrelativistic limit, ${\bm p}\ll m_{1,2}$, 
implemented here as $m_1\rightarrow \infty$,  we obtain
\eq{
\widetilde {\cal I}_\TT(\bm p,\bm q, \bm \ell_1,\bm \ell_2) 
 =  {1 \over 24 m_1^2 {\bm \ell_1}^2 {\bm \ell_2}^2 {( \bm \ell_1 + \bm \ell_2 + \bm q )}^2  }  \, .
}{eq:W_spatial_integrand}
As in the case of the one-loop triangle, the scalar double-triangle
integral is only a function of $\bm q^2$ and $m_1^2$, since these are
the only rotational invariants that can be constructed from the
momenta flowing into the integral.  For this reason, we know that the
integral is a series expansion in $\bm q^2/m_1^2$, and we have kept
only the classical term in \Eq{eq:W_spatial_integrand}.

We immediately see that the dependence on both $\bm \ell_1$ and $\bm \ell_2$ is of the form of \Eq{eq:integrable}, which follows trivially since the double triangle is made entirely of triangle subdiagrams.  Applying \Eq{eq:Smirnov} sequentially to each integral, we obtain our final result for the classical contribution from the scalar double-triangle integral to all orders in velocity,
\eq{
I_\TT  &=  - \frac{1}{768 \pi^2 m_1^2} \log \bm q^2 \, .
}{eq:integrated_DT}
This is also derived using Mellin-Barnes integration in \sect{sec:MB}.

In obtaining \Eq{eq:integrated_DT} we used the general formula in \Eq{eq:Smirnov} applied to this example,
\eq{
\int \frac{d^{D-1}\bm \ell}{(2\pi)^{D-1}} \, \frac{1}{ |\bm \ell | (\bm \ell +\bm q)^2} = \frac{1}{4\pi^2 \epsilon} - \frac{1}{4\pi^2} \log {{\bm q}^2 \over \mu^2} \,,
}{}
where we employ dimensional regularization in $D=4-2\epsilon$ and the
$\overline{\hbox{MS}}$ scheme. The $\log \bm q^2$ term is the
classical contribution at two-loop order since its Fourier transform
to position space is $\left[ G^3 \log \bm q^2 \right]_{\rm FT} =
-{1 \over 2\pi} \left( {G \over |{\bm r}|} \right)^3$.  Crucially all
terms that are constant in ${\bm q}$, such as the $1/\epsilon$ and
$\log \mu^2$ terms, are contact terms whose Fourier transforms to
position space yield $\delta(|{\bm r}|)$. Hence they do not contribute
to the long-distance classical potential. These ultraviolet-sensitive
contributions are quantum mechanical, and can be absorbed into an
appropriate counterterm.

\subsubsection{Double-Triangle Prime Diagram}
\label{sec:N}

\newcommand{\TTp}{{\rm TT'}}

\begin{figure}
\begin{center}
\hspace{2cm}\includegraphics[scale=.8]{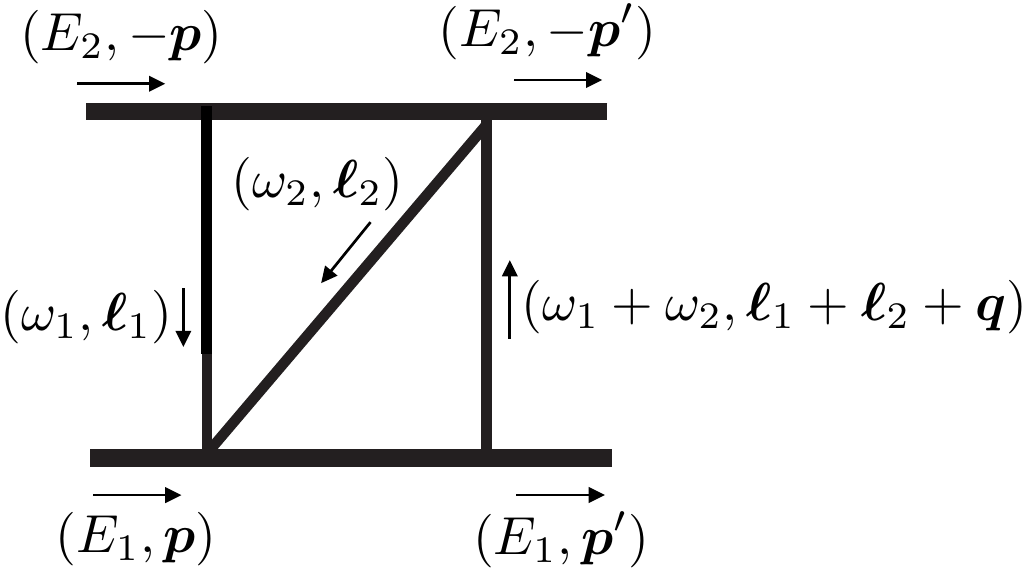}
\end{center}
\vspace*{-0.5cm}
\caption{\small The double-triangle prime diagram.
 }\label{fig:NRint_N}
\end{figure}

Consider next the double-triangle-prime diagram, which is a two-loop
integral with an arbitrary numerator together with one $\phi_1$ propagator, one
$\phi_2$ propagator, and three graviton propagators arranged as
in \Fig{fig:NRint_N}.
The integrand corresponding to this diagram is
\eq{
{\cal I}_\TTp =\null &
\frac{1}{(E_1+\omega_1 + \omega_2)^2 - (\bm p+\bm \ell_1 + \bm \ell_2)^2 -m_1^2}\, \frac{1}{(E_2- \omega_1)^2 - (\bm p+\bm \ell_1)^2 -m_2^2} \\
& \times \frac{1}{\omega_1^2 - \bm \ell_1^2}\, \frac{1}{\omega_2^2 - \bm \ell_2^2}\, \frac{1}{(\omega_1+\omega_2)^2 - (\bm \ell_1 + \bm \ell_2+ \bm q)^2} \, {\cal N}_\TTp \,.
}{}
Both $\phi_1$ and $\phi_2$  propagators can be factored into matter and antimatter components,
\eq{
 \frac{1}{(E_1+\omega_1+\omega_2)^2 - (\bm p+\bm \ell_1+\bm \ell_2)^2 -m_1^2}&= \frac{1}{(\omega_1 +\omega_2 -\omega_{P_1})(\omega_1 +\omega_2 -\omega_{A_1})} \,, \\
\frac{1}{(E_2 - \omega_1)^2 - (\bm p+\bm \ell_1)^2 -m_2^2}&= \frac{1}{(\omega_1 -\omega_{P_2})(\omega_1-\omega_{A_2})} \, .
}{}
The spatial integrand in terms of the effective numerator is then
\eq{
\widetilde {\cal I}_\TTp &= \int \frac{d\omega_1}{2\pi}\frac{d\omega_2}{2\pi}\, \frac{\widetilde{\cal N}_\TTp(\omega_1 ,\omega_2)}{(\omega_1-\omega_{P_2})(\omega_1+\omega_2-\omega_{P_1})} \,, 
}{}
where
\eq{
\widetilde{\cal N}_\TTp(\omega_1 ,\omega_2) &=
\frac{1}{\omega_1-\omega_{A_2}} \, \frac{1}{\omega_1+\omega_2-\omega_{A_1}}\, \frac{1}{\omega_1^2 - \bm \ell_1^2}\, \frac{1}{\omega_2^2 - \bm \ell_2^2}\, \frac{1}{(\omega_1+\omega_2)^2 - (\bm \ell_1 + \bm \ell_2+ \bm q)^2} \, {\cal N}_\TTp(\omega_1,\omega_2).
}{}

Next, we use the fact that canceling either energy denominator will produce a scaleless integral in either $\omega_1$ or $\omega_2$.  Consequently, we can freely send 
\eq{
\omega_1 - \omega_{P_2} \rightarrow 0 \qquad {\rm and} \qquad \omega_1 +\omega_2 - \omega_{P_1} \rightarrow 0 \,,
}{}
so we can effectively evaluate the numerator 
\eq{
 \widetilde{\cal N}_\TTp(\omega_1,\omega_2) \rightarrow  \widetilde{\cal N}_\TTp(\omega_{P_2},\omega_{P_1} - \omega_{P_2}) \,.
 }{}
Upon this replacement, the spatial integrand becomes
\eq{
\widetilde {\cal I}_\TTp &=  \widetilde{\cal N}_\TTp(\omega_{P_2},\omega_{P_1} - \omega_{P_2}) \int \frac{d\omega_1}{2\pi}\frac{d\omega_2}{2\pi}\, \frac{1}{(\omega_1-\omega_{P_2})(\omega_1+\omega_2-\omega_{P_1})} \,.
} 
{}
Here the last factor is the double-triangle prime master energy integral.  As before, we evaluate it by averaging over permutations of labelings of the three exchanged gravitons,
\begin{align}
\int  \frac{d\omega_1 \, d\omega_2}{(\omega_1-\omega_{P_1}+i\epsilon)(\omega_1+\omega_2-\omega_{P_1'}-i\epsilon)} & \equiv
 \frac{1}{3!} \left( \int  \frac{d\omega_1 \, d\omega_2}{(\omega_1-\omega_{P_1}+i\epsilon)(\omega_1+\omega_2-\omega_{P_1'}-i\epsilon)}  + \{ \textrm{perm.}\} \right)\nn \\
&= - \frac{1}{3} \times (-2\pi i)^2 \,. 
\label{eq:flipped_double_master}
\end{align}
Note the sign difference in the $i\epsilon$ compared to the
double-triangle master integral. By explicit calculation we have
obtained a $-1/3$ symmetry factor relative to the naive expression one
would obtain by sequential application of the residue theorem on each
pole.

Having evaluated the energy integral, we obtain the spatial integrand,
\eq{
\widetilde {\cal I}_\TTp(\bm p,\bm q, \bm \ell_1,\bm \ell_2) &= \frac13 \widetilde{\cal N}_\TTp(\omega_{P_2},\omega_{P_1} - \omega_{P_2}) \, .
}{}
As before, we consider the simple case of a scalar integral for which
${\cal N}_\TTp =1$. We proceed by expanding the integrand in the
classical limit and dropping the quantum contributions. Then to put
the integrand in the form of \Eq{eq:integrable}, we may expand in the
nonrelativistic limit implemented as $m_{1,2} \to \infty$ as
before. Alternatively, we choose instead to expand in
$E_{1,2} \to \infty$ in order to keep the functions of $\sqrt{{\bm
p}^2 + m_i^2}$ intact. Unlike the double-triangle integral, this
integral depends on the invariant $p_1 \cdot p_2 = E_1 E_2 + \bm
p^2$. Expanding the spatial integrand this way, we obtain
\eq{
\widetilde {\cal I}_\TTp &= {1 \over 12 E_1 E_2 {\bm \ell_1^2} (\bm \ell_2-\bm \ell_1)^2  (\bm \ell_2 + \bm q)^2 } \bigg[ 1 + { (\bm \ell_1^2 + 2 \bm p \bm \ell_1)^2 \over 4 E_2^2 {\bm \ell_1^2} } + { (\bm \ell_1^2 + 2 \bm p \bm \ell_1)^2 \over 4 E_2^2 (\bm \ell_2-\bm \ell_1)^2 }    \\[10pt]
&\quad+ { (\bm \ell_1^2 + 2 \bm p \bm \ell_1)  (\bm \ell_2^2 + 2 \bm p \bm \ell_2) \over 2 E_1 E_2  (\bm \ell_2-\bm \ell_1)^2}  +  { (\bm \ell_2^2 + 2 \bm p \bm \ell_2)^2 \over 4 E_1^2 (\bm \ell_2-\bm \ell_1)^2  } + { (\bm \ell_2^2 + 2 \bm p \bm \ell_2)^2 \over 4 E_1^2 (\bm \ell_2+ \bm q)^2 }  + \cdots \bigg] \,, 
}{eq:N_spatial_integrand}
where we have shifted the integration variable $\bm \ell_2 \to \bm \ell_2 - \bm \ell_1$, and the ellipsis denotes higher-order terms in the nonrelativistic expansion. Applying \Eq{eq:Smirnov} sequentially to each integral, we obtain
\eq{
I_\TTp  &=  - \frac{ \log \bm q^2 }{384 \pi^2 E_1 E_2 } \bigg[ 1 + {\left(E_1^3+ E_2^3 \right) \over 3 E_1^2 E_2^2 E} {\bm p}^2  + \frac{\left(E_1^5+E_2^5\right)}{5 E_1^4 E_2^4 E} {\bm p}^4 + \frac{\left(E_1^7+E_2^7\right)}{7 E_1^6 E_2^6 E} {\bm p}^6 + \cdots \bigg] \,,
}{eq:N_series}
where $E = E_1 + E_2$ and the ellipsis denotes higher-order terms in the nonrelativistic expansion. We have included in \Eq{eq:N_series} results for higher-order terms not explicitly shown in \Eq{eq:N_spatial_integrand}, and have checked to sufficiently high orders that the obvious pattern persists, allowing us to resum (see \app{appendix:series}) the result as
\eq{
I_\TTp  & = - \frac{ \log \bm q^2 }{192 \pi^2 m_1 m_2} {{\rm arcsinh} \sqrt{ {\sigma-1 \over 2} } \over  \sqrt{\sigma^2 - 1} } \, .
}{eq:Nscalar}
For later convenience we have defined the quantity
\eq{
\sigma = {p_1 \cdot p_2 \over m_1 m_2}  = { E_1 E_2 + \bm p^2  \over m_1 m_2} \,,
}{eq:sigma_def}
where $p_1$ and $p_2$ are the incoming four-momenta associated with $\phi_1$ and $\phi_2$. In the nonrelativistic limit, $\sigma$ approaches unity. 

\begin{figure}
\begin{center}
\includegraphics[scale=.8]{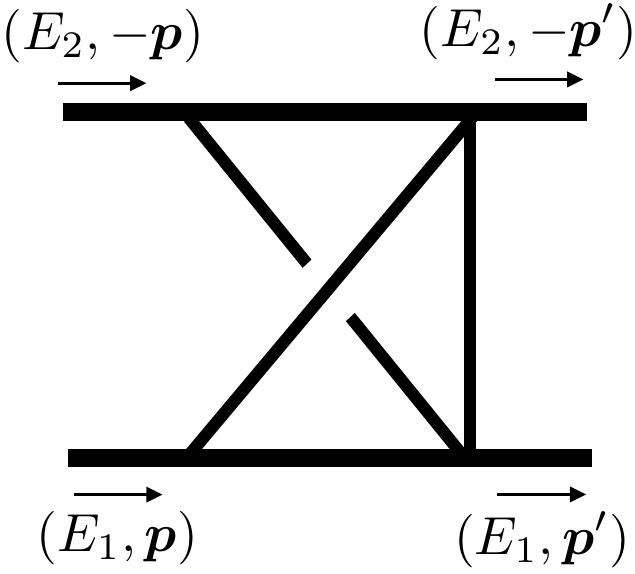}
\end{center}
\vspace*{-0.5cm}
\caption{\small The nonplanar version of the double-triangle prime diagram. }\label{fig:NRint_xN}
\end{figure}

The classical contribution from the scalar double-triangle prime
integral is given to all orders in velocity in \Eq{eq:Nscalar}. As we
cautioned at the end of \sect{sec:general}, results for individual
diagrams may be prescription dependent. Such is the case here, and
only the sum of the double-triangle prime diagram with its nonplanar
version can be unambiguously determined and compared to results
obtained from other methods. Of course, our aim in this section is to
illustrate the mechanics of our nonrelativistic integration method for
general applicability, whether it be for a single diagram or for a sum
of diagrams. Nonetheless, we briefly remark on the nonplanar
diagram shown in \fig{fig:NRint_xN}. The calculation is similar to
that of the planar case except that the symmetry factor is -1/6
instead of -1/3. The result is
\eq{
I_{\overline{\TTp}}  & =  - \frac{ \log \bm q^2 }{384 \pi^2 m_1 m_2} {{\rm arcsinh} \sqrt{ {\sigma-1 \over 2} } \over  \sqrt{\sigma^2 - 1} } \, .
}{eq:xNscalar}
The sum $I_\TTp  + I_{\overline{\TTp}}$ agrees with the result from relativistic integration given in the first line of \Eq{eq:NmasterIntegrals}. 
 
\subsubsection{H and Crossed-H Diagrams}
\label{sec:NRint_H}

\begin{figure}
\begin{center}
\includegraphics[scale=.8]{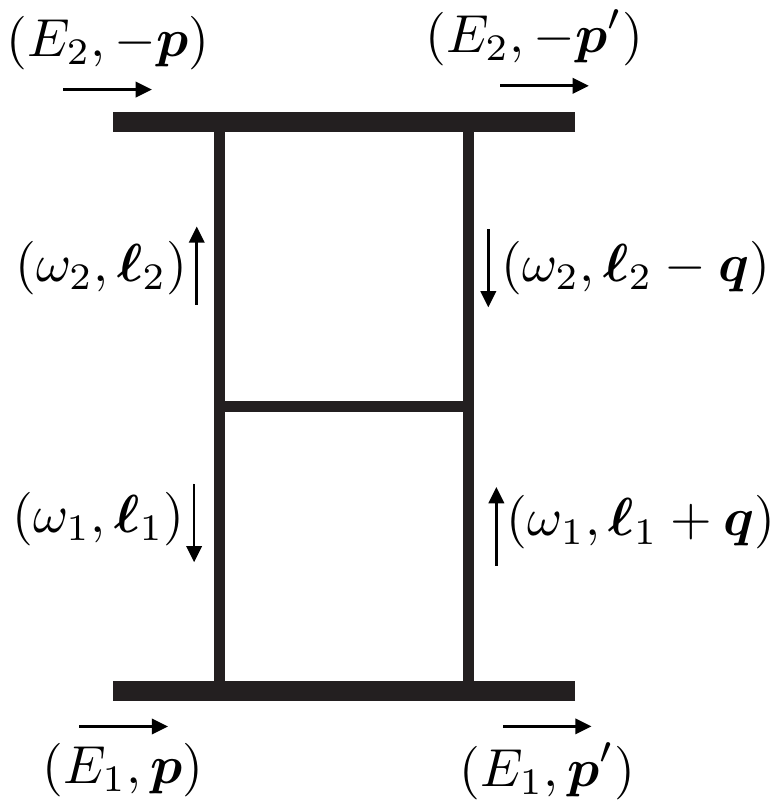}
\hspace{1cm}\includegraphics[scale=.8]{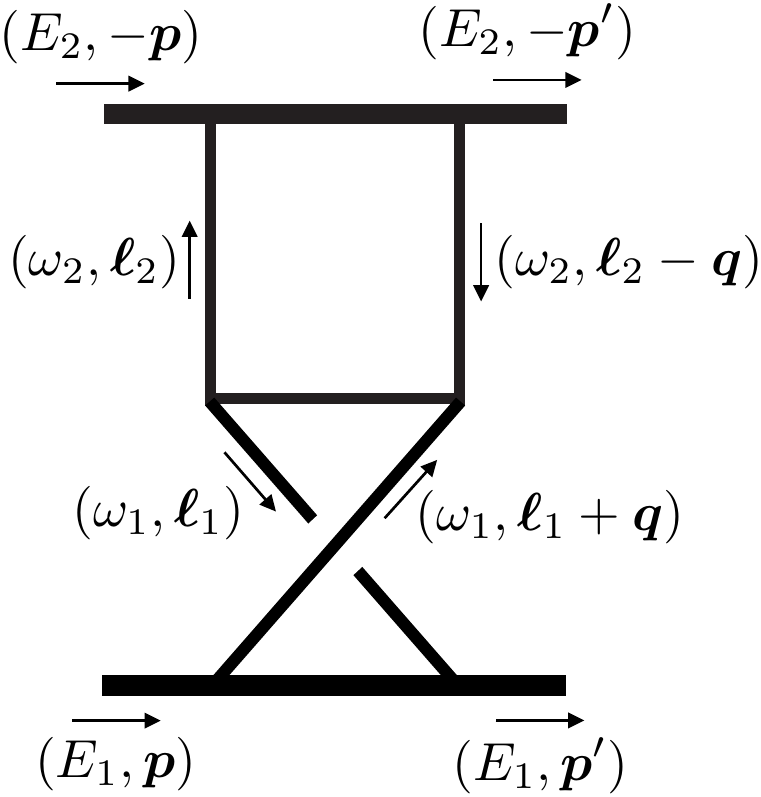}
\end{center}
\vspace*{-0.5cm}
\caption{\small The H and crossed-H diagrams. }\label{fig:NRint_H}
\end{figure}

The next example we consider is the sum of the general H and crossed-H
diagrams. We will denote the latter by ${\overline{\rm H}}$.
These are two-loop integrals with arbitrary numerators and
one $\phi_1$ propagator, one $\phi_2$ propagator, and five graviton
propagators as shown in Fig.~\ref{fig:NRint_H}.  The corresponding
integrands are
\eq{
{\cal I}_{{\rm H}}  =&\frac{1}{(E_1+\omega_1 )^2 - (\bm p+\bm \ell_1)^2 -m_1^2}\, \frac{1}{(E_2+ \omega_2)^2 - (\bm p-\bm \ell_2)^2 -m_2^2} \\
& \times \frac{1}{\omega_1^2 - \bm \ell_1^2}\, \frac{1}{\omega_1^2 - (\bm \ell_1 + \bm q)^2} \, \frac{1}{\omega_2^2 - \bm \ell_2^2}\, \frac{1}{\omega_2^2 - (\bm \ell_2 - \bm q)^2} \, \frac{1}{(\omega_1+\omega_2)^2 - (\bm \ell_1 + \bm \ell_2)^2} \, {\cal N}_{\rm H} \,, \\[3pt]
{\cal I}_{\overline{\rm H}}  =&\frac{1}{(E_1- \omega_1 )^2 - (\bm p' - \bm \ell_1)^2 -m_1^2}\, \frac{1}{(E_2+ \omega_2)^2 - (\bm p-\bm \ell_2)^2 -m_2^2} \\
& \times \frac{1}{\omega_1^2 - \bm \ell_1^2}\, \frac{1}{\omega_1^2 - (\bm \ell_1 + \bm q)^2} \, \frac{1}{\omega_2^2 - \bm \ell_2^2}\, \frac{1}{\omega_2^2 - (\bm \ell_2 - \bm q)^2} \, \frac{1}{(\omega_1+\omega_2)^2 - (\bm \ell_1 + \bm \ell_2)^2} \, {\cal N}_{\overline{\rm H}} \,.
}{}
As before the $\phi_1$ and $\phi_2$  propagators can be factored into matter and antimatter components,
\eq{
 \frac{1}{(E_1+\omega_1 )^2 - (\bm p+\bm \ell_1)^2 -m_1^2}&= \frac{1}{(\omega_1 -\omega_{P_1})(\omega_1  -\omega_{A_1})} \,, \\
  \frac{1}{(E_1-\omega_1 )^2 - (\bm p'-\bm \ell_1)^2 -m_1^2}&= \frac{1}{(\omega_1 -\omega_{\overline{P}_1})(\omega_1  -\omega_{\overline{A}_1})} \,, \\
\frac{1}{(E_2+ \omega_2)^2 - (\bm p-\bm \ell_2)^2 -m_2^2} &= \frac{1}{(\omega_2 -\omega_{P_2})(\omega_2-\omega_{A_2})} \, .
}{}
The spatial integrands in terms of the effective numerator are then
\eq{
\widetilde {\cal I}_{\rm H} &= \int \frac{d\omega_1}{2\pi}\frac{d\omega_2}{2\pi}\, \frac{\widetilde{\cal N}_{\rm H}(\omega_1 ,\omega_2)}{(\omega_1-\omega_{P_1})(\omega_2-\omega_{P_2})} \,, \\
\widetilde {\cal I}_{\overline{\rm H}} &= \int \frac{d\omega_1}{2\pi}\frac{d\omega_2}{2\pi}\, \frac{\widetilde{\cal N}_{\overline{\rm H}}(\omega_1 ,\omega_2)}{(\omega_1-\omega_{\overline{P}_1})(\omega_2-\omega_{P_2})} \,,
}{}
where 
\eq{
\widetilde{\cal N}_{\rm H}(\omega_1 ,\omega_2) &=
\frac{1}{\omega_1-\omega_{A_1}} \, \frac{1}{\omega_2-\omega_{A_2}}\, \frac{1}{\omega_1^2 - \bm \ell_1^2}\, \frac{1}{\omega_1^2 - (\bm \ell_1 + \bm q)^2} \, \frac{1}{\omega_2^2 - \bm \ell_2^2}\, \frac{1}{\omega_2^2 - (\bm \ell_2 - \bm q)^2} \\
&\qquad \times  \frac{1}{(\omega_1+\omega_2)^2 - (\bm \ell_1 + \bm \ell_2)^2} \, {\cal N}_{\rm H}(\omega_1 ,\omega_2)\,, \\
\widetilde{\cal N}_{\overline{\rm H}}(\omega_1 ,\omega_2) &=
\frac{1}{\omega_1-\omega_{\overline{A}_1}} \, \frac{1}{\omega_2-\omega_{A_2}}\, \frac{1}{\omega_1^2 - \bm \ell_1^2}\, \frac{1}{\omega_1^2 - (\bm \ell_1 + \bm q)^2} \, \frac{1}{\omega_2^2 - \bm \ell_2^2}\, \frac{1}{\omega_2^2 - (\bm \ell_2 - \bm q)^2} \\
&\qquad \times  \frac{1}{(\omega_1+\omega_2)^2 - (\bm \ell_1 + \bm \ell_2)^2} \, {\cal N}_{\overline{\rm H}}(\omega_1 ,\omega_2)\, .
}{}

Next, we use the fact that in each diagram canceling either energy denominator will produce a scaleless integral in either $\omega_1$ or $\omega_2$.  Consequently, we can freely send  
\eq{
{\rm H}&: \quad \omega_1 - \omega_{P_1} \rightarrow 0 \qquad {\rm and} \qquad \omega_2 - \omega_{P_2} \rightarrow 0 \,, \\
\overline{{\rm H}}&: \quad \omega_1 - \omega_{\overline{P}_1} \rightarrow 0 \qquad {\rm and} \qquad \omega_2 - \omega_{P_2} \rightarrow 0 \,, 
}{}
so we can effectively evaluate the numerators
\eq{
 \widetilde{\cal N}_{\rm H}(\omega_1,\omega_2) \rightarrow  \widetilde{\cal N}_{\rm H}(\omega_{P_1},\omega_{P_2})\,, \qquad   \widetilde{\cal N}_{\overline{\rm H}}(\omega_1,\omega_2) \rightarrow  \widetilde{\cal N}_{\overline{\rm H}}(\omega_{\overline{P}_1},\omega_{P_2}) \,.
 }{}
Upon this replacement, the spatial integrands become
\eq{
\widetilde {\cal I}_{\rm H} &=  \widetilde{\cal N}_{\rm H}(\omega_{P_1}, \omega_{P_2}) \int \frac{d\omega_1}{2\pi}\frac{d\omega_2}{2\pi}\, \frac{1}{(\omega_1-\omega_{P_1})(\omega_2-\omega_{P_2})}\,, \\
\widetilde {\cal I}_{\overline{\rm H}} &=  \widetilde{\cal N}_{\overline{\rm H}}(\omega_{\overline{P}_1}, \omega_{P_2}) \int \frac{d\omega_1}{2\pi}\frac{d\omega_2}{2\pi}\, \frac{1}{(\omega_1-\omega_{\overline{P}_1})(\omega_2-\omega_{P_2})}\,.
}{}

Before proceeding with the energy master integrals, we note a few properties that are helpful in combining the H and $\overline{\rm H}$ integrands. First, the relativistic numerators and the antimatter poles for each diagram are related as
\eq{
{\cal N}_{\rm H} = {\cal N}_{\overline{\rm H}} + {\cal O}(\bm q)\,, \qquad \frac{1}{\omega_1-\omega_{A_1}}= \frac{1}{2E_1} + {\cal O}(\bm q)  \,, \qquad   - \frac{1}{\omega_1-\omega_{\overline{A}_1}}  = \frac{1}{2E_1} + {\cal O}(\bm q) \, .
}{}
These are due essentially to the fact that H and $\overline{\rm H}$ are related by relabeling $p_1 \leftrightarrow p_4$ and that $p_1 = p_4 + q$, where the momentum transfer $q$ is defined in \Eq{eq:defq}. Now since the graviton propagators and the matter propagator for $\phi_2$ are the same for both diagrams, the effective numerators satisfy
\eq{
\widetilde{\cal N}_{\rm H} &= - \widetilde{\cal N}_{\overline{\rm H}} + {\cal O}(\bm q) \, .
}{eq:numer_H_xH}
Second, the triple-graviton vertex comes with a scaling of $\bm q^2$, and there are two of them in both H and $\overline{\rm H}$. We thus have the scalings ${\cal N}_{\rm H} \sim \bm q^4$ and ${\cal N}_{\overline{\rm H}} \sim \bm q^4$. This is consistent with a classical contribution given the $\bm q$ scalings of the integration measure and propagators, and implies that the ${\cal O}(\bm q)$ difference in \Eq{eq:numer_H_xH} is quantum and can be dropped. We thus combine the integrands as 
\eq{
\widetilde {\cal I}_{\rm H} + \widetilde {\cal I}_{\overline{\rm H}}&=  \widetilde{\cal N}_{\rm H}(\omega_{P_1}, \omega_{P_2}) \int \frac{d\omega_1}{2\pi}\frac{d\omega_2}{2\pi}\, \left[ \frac{1}{(\omega_1-\omega_{P_1})(\omega_2-\omega_{P_2})} -  \frac{1}{(\omega_1-\omega_{\overline{P}_1})(\omega_2-\omega_{P_2})}\right]  . \\
}{eq:Hcombine}

The last factor in \Eq{eq:Hcombine} is the master energy integral. The integrals over $\omega_1$ and $\omega_2$ factorize. The integral over $\omega_1$ is convergent, and direct integration gives
\eq{
\int d \omega_1  \left[ \frac{1}{(\omega_1-\omega_{P_1})} -  \frac{1}{(\omega_1-\omega_{\overline{P}_1})}\right] = - 2 \pi i \,.
}{}
The integral over $\omega_2$ is done similar to the triangle master energy integral in \Eq{eq:triangle_3},
\eq{
\int \frac{d \omega_2 }{\omega_2-\omega_{P_2}} \equiv \frac12 \times (  - 2 \pi i) \, .
}{}
We thus obtain a 1/2 symmetry factor relative to the naive expression
one would obtain by sequential application of the residue theorem on
each pole.

Having evaluated the energy integral, we obtain the spatial integrand,
\eq{
\widetilde {\cal I}_{\rm H} + \widetilde {\cal I}_{\overline{\rm H}}&= -\frac12 \widetilde{\cal N}_{\rm H}(\omega_{P_1},\omega_{P_2}) \,.
}{}
Consider for illustration the scalar case ${\cal N}_{\rm H} =\bm q^4$, with $\bm q$ scaling chosen to give a classical result. Upon expanding the spatial integrand in the classical limit and then in large $E_{1,2}$ we obtain
\eq{
\widetilde {\cal I}_{\rm H} + \widetilde {\cal I}_{\overline{\rm H}} &= {\bm q^4 \over 8 E_1 E_2 {\bm \ell_1^2} (\bm \ell_1 + \bm q)^2 (\bm \ell_1+\bm \ell_2)^2 {\bm \ell_2^2} (\bm \ell_2 - \bm q)^2 } \bigg[ 1  + { (\bm \ell_1^2 + 2 \bm p \bm \ell_1)^2 \over 4 E_1^2 {\bm \ell_1^2} } + { (\bm \ell_1^2 + 2 \bm p \bm \ell_1)^2 \over 4 E_1^2 (\bm \ell_1 + \bm q)^2 } + { (\bm \ell_1^2 + 2 \bm p \bm \ell_1)^2 \over 4 E_1^2 (\bm \ell_1+\bm \ell_2)^2 } \\[5pt]
&\quad + { (\bm \ell_2^2 - 2 \bm p \bm \ell_2)^2 \over 4 E_2^2 {\bm \ell_2^2} } + { (\bm \ell_2^2 - 2 \bm p \bm \ell_2)^2 \over 4 E_2^2 (\bm \ell_2 - \bm q)^2 }  + { (\bm \ell_2^2 - 2 \bm p \bm \ell_2)^2 \over 4 E_2^2 (\bm \ell_1+\bm \ell_2)^2 } + { (\bm \ell_1^2 + 2 \bm p \bm \ell_1)  (\bm \ell_2^2 - 2 \bm p \bm \ell_2) \over 2 E_1 E_2  (\bm \ell_1+\bm \ell_2)^2}  + \cdots  \bigg] \,,
}{eq:H_spatial_integrand}
where the ellipsis denotes higher-order terms in the expansion. This result is not of the form given in \Eq{eq:integrable} since each one of the loop variables $\bm \ell_{1}$ and $\bm \ell_{2}$ appear in three types of graviton factors in the denominator. We proceed by first applying standard techniques for tensor reduction to express \Eq{eq:H_spatial_integrand} in terms of scalar integrands. Then we use the IBP identity
\eq{
(D-1) - a_1 -a_2 -2a_3 +a_1({\bm 1}^+ {\bm 4}^- -  {\bm 1}^+ {\bm 3}^-) +a_2({\bm 2}^+ {\bm 5}^- -  {\bm 2}^+ {\bm 3}^-) = 0\,,
}{eq:IBP_for_H}
where we have used standard notation (see e.g. Ref.~\cite{Smirnov:2004ym}) for raising and lowering operators, i.e., ${\bm 1}^+ {\bm 4}^- F[a_1,a_2,a_3,a_4,a_5] = F[a_1+1,a_2,a_3,a_4-1,a_5] $ with the definition
\eq{
F[a_1,a_2,a_3,a_4,a_5] = {1 \over {\bm \ell_1^{2a_1}} (\bm \ell_1 + \bm q)^{2a_2} (\bm \ell_1+\bm \ell_2)^{2a_3} {\bm \ell_2^{2a_4}} (\bm \ell_2 - \bm q)^{2a_5} } \, .
}{}
Repeated application of this identity (and those related by relabelings of the loop variables) puts the integrand in the form of \Eq{eq:integrable}. Finally, applying \Eq{eq:Smirnov} sequentially, we obtain
\eq{
I_{\rm H} + I_{\overline{\rm H}}  &=   \frac{ \log \bm q^2 }{128 \pi^2 E_1 E_2  } \bigg[ 1 + {\left(E_1^3+ E_2^3 \right) \over 3 E_1^2 E_2^2 (E_1+E_2) } {\bm p}^2  + \frac{\left(E_1^5+E_2^5\right)}{5 E_1^4 E_2^4 (E_1+E_2)} {\bm p}^4 + \frac{\left(E_1^7+E_2^7\right)}{7 E_1^6 E_2^6 (E_1+E_2)} {\bm p}^6 + \cdots \bigg] \, .
}{eq:H_series}
As in the previous section, we can resum this as
\eq{
I_{\rm H} + I_{\overline{\rm H}}    & = \frac{ \log \bm q^2 }{64 \pi^2 m_1 m_2 } {{\rm arcsinh} \sqrt{ {\sigma-1 \over 2} } \over  \sqrt{\sigma^2 - 1} } \, ,
}{eq:Hresummed}
where $\sigma$ is defined as in \Eq{eq:sigma_def}.  This is our final result for the classical contribution from the sum of scalar H and scalar $\overline{\rm H}$ diagrams to all orders in velocity. Individually these diagrams give results that depend on the integration prescription, but the sum can be meaningfully compared, e.g. to the result from relativistic integration in \sect{sec:Rint}. Indeed, the result in \Eq{eq:Hresummed} agrees with that in \Eq{eq:resultI1}. 

\subsubsection{Box-Triangle Diagram}
\label{sec:BTdiagram}

\newcommand{\BT}{{\rm BT}}

\begin{figure}
\begin{center}
\hspace{2cm}\includegraphics[scale=.8]{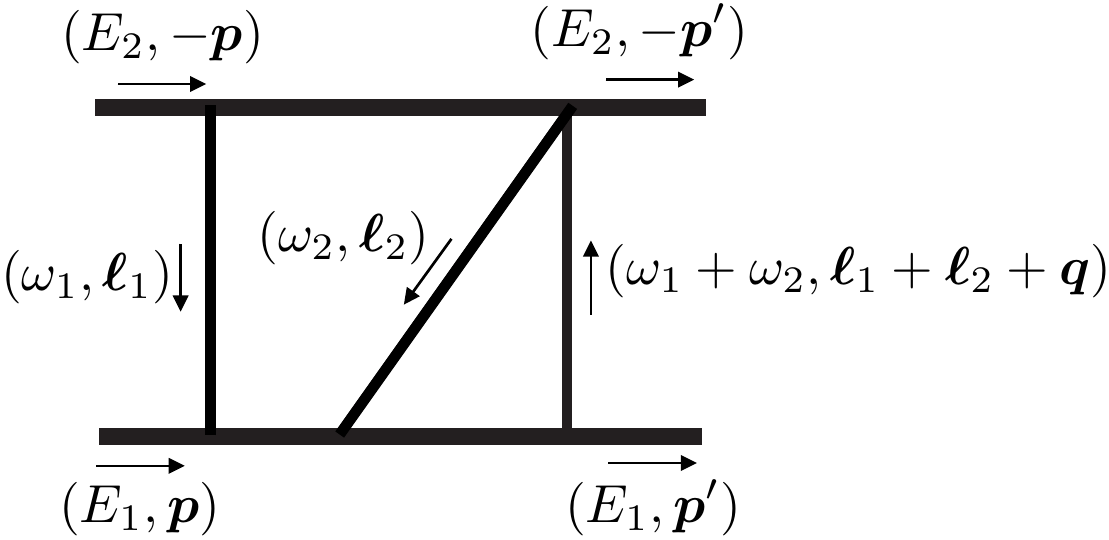}
\end{center}
\vspace*{-0.5cm}
\caption{\small The box-triangle diagram. }\label{fig:NRint_BT}
\end{figure}

Consider next the general box triangle diagram, which is a two-loop
integral with an arbitrary numerator and two $\phi_1$ propagators, one
$\phi_2$ propagator, and three graviton propagators as arranged
in \fig{fig:NRint_BT}.
 The integrand corresponding to this diagram is
\eq{
{\cal I}_{\BT} =&\frac{1}{(E_1+\omega_1 )^2 - (\bm p+\bm \ell_1 )^2 -m_1^2} \, \frac{1}{(E_1+\omega_1 + \omega_2)^2 - (\bm p+\bm \ell_1 + \bm \ell_2)^2 -m_1^2}\\
& \times \frac{1}{(E_2- \omega_1)^2 - (\bm p+\bm \ell_1)^2 -m_2^2} \, \frac{1}{\omega_1^2 - \bm \ell_1^2}\, \frac{1}{\omega_2^2 - \bm \ell_2^2}\, \frac{1}{(\omega_1+\omega_2)^2 - (\bm \ell_1 + \bm \ell_2+ \bm q)^2} \, {\cal N}_{\BT} \,.
}{}
We factor all matter propagators into matter and antimatter components,
\eq{
  \frac{1}{(E_1+\omega_1)^2 - (\bm p+\bm \ell_1)^2 -m_1^2}&= \frac{1}{(\omega_1 -\omega_{P_1})(\omega_1  -\omega_{A_1})} \,, \\
\frac{1}{(E_2 - \omega_1)^2 - (\bm p+\bm \ell_1)^2 -m_2^2}&= \frac{1}{(\omega_1 -\omega_{P_2})(\omega_1-\omega_{A_2})} \,, \\
 \frac{1}{(E_1+\omega_1+\omega_2)^2 - (\bm p+\bm \ell_1+\bm \ell_2)^2 -m_1^2}&= \frac{1}{(\omega_1 +\omega_2 -\omega_{P'_1})(\omega_1 +\omega_2 -\omega_{A'_1})} \, .
}{}
The spatial integrand is then given in terms of the effective numerator by
\eq{
\widetilde {\cal I}_{\BT} &= \int \frac{d\omega_1}{2\pi}\frac{d\omega_2}{2\pi}\, \frac{\widetilde{\cal N}_{\BT}(\omega_1 ,\omega_2)}{(\omega_1-\omega_{P_1})(\omega_1-\omega_{P_2})(\omega_1+\omega_2-\omega_{P_1'})} \,,
}{}
where
\eq{
\widetilde{\cal N}_{\BT}(\omega_1 ,\omega_2) &=
\frac{1}{\omega_1-\omega_{A_1}} \frac{1}{\omega_1-\omega_{A_2}} \, \frac{1}{\omega_1+\omega_2-\omega_{A_1'}}\, \frac{1}{\omega_1^2 - \bm \ell_1^2}\, \frac{1}{\omega_2^2 - \bm \ell_2^2}\, \\
& \hskip 4 cm \null \times 
 \frac{1}{(\omega_1+\omega_2)^2 - (\bm \ell_1 + \bm \ell_2+ \bm q)^2} \, {\cal N}_{\BT}(\omega_1,\omega_2) \,.
}{}

Next, we implement energy-integral reduction. From the diagram's topology, we see that the constraints are a combination of those of the one-loop box and triangle integrals:
\eq{
(\omega_1 - \omega_{P_1} )(\omega_1 - \omega_{P_2}) \rightarrow 0 \qquad {\rm and} \qquad \omega_1 +\omega_2 - \omega_{P_1'} \rightarrow 0 \, .
}{eq:boxtriangle_constraints}
Hence we can effectively evaluate the numerator 
\eq{
 \widetilde{\cal N}_{\BT}(\omega_1,\omega_2) \rightarrow  \widetilde{\cal N}_{\BT}(\omega_{1},\omega_{P_1'} - \omega_{1}) \, ,
 }{}
 and then treat the remaining dependence on $\omega_1$ as we would for the one-loop box. That is, we reduce it to a linear function  using the first constraint in \Eq{eq:boxtriangle_constraints}, and then expand about either $\omega_{P_1}$ or $\omega_{P_2}$. This would cancel one of the factors in $(\omega_1 - \omega_{P_1} )(\omega_1 - \omega_{P_2}) $ and lead to either the master for the double-triangle diagram in \Eq{eq:double_master} or the master for the double-triangle prime diagram in \Eq{eq:flipped_double_master}. We of course have terms with all three matter poles uncanceled, and thus the box triangle master energy integral:
\eq{
&\int  \frac{d\omega_1 \, d\omega_2}{(\omega_1-\omega_{P_1}+i\epsilon)(\omega_1-\omega_{P_2}-i\epsilon)(\omega_1+\omega_2-\omega_{P_1'}+i\epsilon)} \\
&\quad \equiv \frac{1}{3!} \left( \int  \frac{d\omega_1 \, d\omega_2}{(\omega_1-\omega_{P_1}+i\epsilon)(\omega_1-\omega_{P_2}-i\epsilon)(\omega_1+\omega_2-\omega_{P_1'}+i\epsilon)}  + \{ \textrm{perm.}\} \right)\\
&\quad =  \frac{1}{2(\omega_{P_1} - \omega_{P_2})} \times (-2\pi i)^2 \, .
}{eq:boxtriangle_master}

Consider for illustration the case ${\cal N}_{\BT} =1$. Upon implementing the above procedure, we obtain the spatial integrand,
\eq{
\widetilde {\cal I}_{\BT} &= -\frac{1}{8 E_1 (E_2+E_1)  \bm \ell_1 ^2 (  \bm \ell_2 - \bm \ell_1)^2 (\bm \ell_2 + \bm q)^2 ( \bm \ell_1^2 + 2 \bm p \bm \ell_1 )} \bigg[ 1 +  \frac{( \bm \ell_2^2 + 2 \bm p \bm \ell_2 ) ( \bm \ell_1^2 + 2 \bm p \bm \ell_1 )}{3 E_2 E_1 (  \bm \ell_2 - \bm \ell_1)^2} \\[5pt]
&\quad -\frac{( \bm \ell_2^2 + 2 \bm p \bm \ell_2 ) ( \bm \ell_1^2 + 2 \bm p \bm \ell_1 )}{6 E_1^2 (  \bm \ell_2 - \bm \ell_1)^2}+\frac{( \bm \ell_2^2 + 2 \bm p \bm \ell_2 )^2}{4 E_1^2 (\bm \ell_2 + \bm q)^2}+\frac{( \bm \ell_2^2 + 2 \bm p \bm \ell_2 )^2}{4 E_1^2 (  \bm \ell_2 - \bm \ell_1)^2}-\frac{( \bm \ell_2^2 + 2 \bm p \bm \ell_2 )}{2 E_1^2}+\frac{( \bm \ell_1^2 + 2 \bm p \bm \ell_1 )^2}{6 E_2^2 (  \bm \ell_2 - \bm \ell_1)^2}\\[5pt]
&\quad +\frac{( \bm \ell_1^2 + 2 \bm p \bm \ell_1 )^2}{12 E_1^2 (  \bm \ell_2 - \bm \ell_1)^2}+\frac{( \bm \ell_1^2 + 2 \bm p \bm \ell_1 )^2}{6 E_2^2 \bm \ell_1 ^2}+\frac{( \bm \ell_1^2 + 2 \bm p \bm \ell_1 )^2}{12 E_1^2 \bm \ell_1 ^2}-\frac{( \bm \ell_1^2 + 2 \bm p \bm \ell_1 )}{6 E_2^2}-\frac{( \bm \ell_1^2 + 2 \bm p \bm \ell_1 )}{12 E_1^2} + \cdots \bigg]\,,
}{eq:box_triangle_spatial_integrand}
where we have shifted the integration variable $\bm \ell_2 \to \bm \ell_2 - \bm \ell_1$, and the ellipsis 
denotes higher-order terms in the nonrelativistic expansion. 
Applying \Eq{eq:Smirnov} sequentially to each integral, we obtain,
\eq{
I_{\BT}  &= -\frac{1}{64 E_1 E} \int \frac{d^{D-1} \bm \ell_1 }{(2\pi)^{D-1}}  \frac{1}{  \bm \ell_1 ^2 |\bm \ell_1 - \bm q|  ( \bm \ell_1^2 + 2 \bm p \bm \ell_1 )} \left[   1 + \frac{\bm p^2}{2E_1^2}    + \frac{3\bm p^4}{8E_1^4}    + \frac{5\bm p^6}{16E_1^6}  +\frac{35\bm p^8}{128 E_1^8} + \cdots  \right] \\[5pt]
&\quad + { \log {\bm q}^2 \over 2048 \pi^2 E_1^3 E } \bigg[ 1 + {5 \bm p^2 \over 4 E_1^2} +{ 11 \bm p^4 \over 8 E_1^4} + {93 \bm p^6 \over 64E_1^6}   + \cdots \bigg] \, ,
}{eq:box_triangle_NRresult}
where we have included results for higher-order terms not shown explicitly in \Eq{eq:box_triangle_spatial_integrand}. Resummation of the series (see \app{appendix:series} for details) yields
\eq{
I_{\BT}  &= -\frac{1}{64 m_1 E} \int \frac{d^{D-1} \bm \ell_1 }{(2\pi)^{D-1}}  \frac{1}{  \bm \ell_1 ^2  |\bm \ell_1 - \bm q| ( \bm \ell_1^2 + 2 \bm p \bm \ell_1 )}  + { \log {\bm q}^2 (E_1 - m_1) \over 1024 \pi^2 m_1^2 E  \bm p^2 } \, ,
}{eq:box_triangle_resum}
where the second term is the classical contribution. The remaining
integral has the form of \Eq{eq:integrable} with $\gamma=1$. It is
superclassical, having an additional factor of $|\bm q|^{-1}$ relative
to the classical scaling, and represents the iteration of the
lower-order tree-level and one-loop triangle diagrams. It is infrared
divergent and will cancel with the same infrared artifact appearing in
the effective-theory contribution to the matching.

\subsubsection{Double-Box Diagram}

\newcommand{\BB}{{\rm BB}}

\begin{figure}
\begin{center}
\hspace{2cm}\includegraphics[scale=.8]{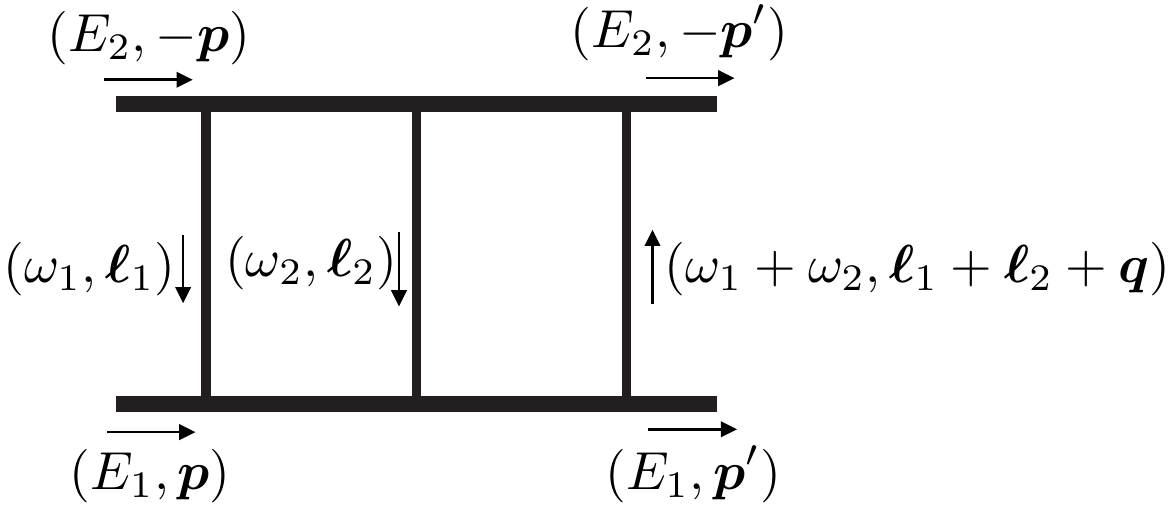}
\end{center}
\vspace*{-0.5cm}
\caption{\small The double-box diagram. }\label{fig:NRint_BB}
\end{figure}

Consider the general double-box diagram which is a two-loop
integral with an arbitrary numerator and two $\phi_1$ propagators,
two $\phi_2$ propagators, and three graviton propagators arranged as
in \fig{fig:NRint_BB}.
The integrand corresponding to this diagram is
\eq{
{\cal I}_{\BB} =&\frac{1}{(E_1+\omega_1 )^2 - (\bm p+\bm \ell_1 )^2 -m_1^2} \, \frac{1}{(E_2- \omega_1)^2 - (\bm p+\bm \ell_1)^2 -m_2^2} \\
& \times 
 \frac{1}{(E_1+\omega_1 + \omega_2)^2 - (\bm p+\bm \ell_1 + \bm \ell_2)^2 -m_1^2} \,
 \frac{1}{(E_2-\omega_1 - \omega_2)^2 - (\bm p+\bm \ell_1 + \bm \ell_2)^2 -m_2^2}  \\
& \times 
\frac{1}{\omega_1^2 - \bm \ell_1^2} \, 
\frac{1}{\omega_2^2 - \bm \ell_2^2}\, \frac{1}{(\omega_1+\omega_2)^2 - (\bm \ell_1 + \bm \ell_2+ \bm q)^2} \, {\cal N}_{\BB} \, .
}{}

As before, we factor all propagators into matter and antimatter
components. The spatial integrand is then given in terms of the
effective numerator by
\eq{
\widetilde {\cal I}_{\BB} &= \int \frac{d\omega_1}{2\pi}\frac{d\omega_2}{2\pi}\, \frac{\widetilde{\cal N}_{\BB}(\omega_1 ,\omega_2)}{(\omega_1-\omega_{P_1})(\omega_1-\omega_{P_2})(\omega_1+\omega_2-\omega_{P_1'})(\omega_1+\omega_2-\omega_{P_2'})} \,,
}{}
where
\eq{
\widetilde{\cal N}_{\BB}(\omega_1 ,\omega_2) &=
\frac{1}{\omega_1-\omega_{A_1}} \frac{1}{\omega_1-\omega_{A_2}} \, \frac{1}{\omega_1+\omega_2-\omega_{A_1'}} \,  \frac{1}{\omega_1+\omega_2-\omega_{A_2'}} \, \frac{1}{\omega_1^2 - \bm \ell_1^2}\, \frac{1}{\omega_2^2 - \bm \ell_2^2}\\
&\qquad \times \frac{1}{(\omega_1+\omega_2)^2 - (\bm \ell_1 + \bm \ell_2+ \bm q)^2} \, {\cal N}_{\BB}(\omega_1,\omega_2) \,.
}{}
The energy-reduction constraints take the form
\eq{
(\omega_1-\omega_{P_1})(\omega_1-\omega_{P_2})&\rightarrow 0 \,, \\
(\omega_1+\omega_2-\omega_{P_1'})(\omega_1+\omega_2-\omega_{P_2'}) &\rightarrow 0 \, .
}{eq:double_box_constraints}
These constraints reduce the effective numerator to the form
\eq{
{\cal N}_{\BB}(\omega_1,\omega_2) = f_0 + f_1 \omega_1 + f_2 \omega_2 + f_3 \omega_1 \omega_2 \, .
}{}
As before, the remaining dependence on  $\omega_1$ and $\omega_2$ can be expanded about a chosen energy pole, leading to the previously encountered energy master integrals for the box-triangle, double-triangle, and double-triangle prime diagrams. The term independent of $\omega_1$ and $\omega_2$ require the double-box master integral given by
\eq{
\int  \frac{d\omega_1 \, d\omega_2}{(\omega_1-\omega_{P_1}+i\epsilon)(\omega_1-\omega_{P_2}-i\epsilon)(\omega_1+\omega_2-\omega_{P_1'}+i\epsilon) (\omega_1+\omega_2-\omega_{P_2'}-i\epsilon)} \\
=  \frac{1}{(\omega_{P_1} - \omega_{P_2}) (\omega_{P_1'} - \omega_{P_2'})} \times (-2\pi i)^2 \, .
}{}

As before, consider for illustration the case ${\cal N}_{\BB} =1$. Upon expanding in the classical and nonrelativistic limits, we obtain the spatial integrand
\eq{
\widetilde {\cal I}_{\BB} &= \frac{1}{4 E^2  \bm \ell_1 ^2 (  \bm \ell_2 - \bm \ell_1)^2 (\bm \ell_2 + \bm q)^2 ( \bm \ell_1^2 + 2 \bm p \bm \ell_1 ) ( \bm \ell_2^2 + 2 \bm p \bm \ell_2 )} + \cdots \,,
}{eq:BB_spatial_integrand}
where we have shifted the integration variable $\bm \ell_2 \to \bm \ell_2 - \bm \ell_1$, and the ellipsis denotes higher-order terms in the nonrelativistic expansion. 
Applying \Eq{eq:Smirnov} sequentially to each integral, we obtain,
\eq{
I_{\BB}  &= \frac{1}{4 E^2} \int \frac{d^{D-1} \bm \ell_1 }{(2\pi)^{D-1}} \frac{d^{D-1} \bm \ell_2 }{(2\pi)^{D-1}}  \frac{1}{  \bm \ell_1 ^2 (  \bm \ell_2 - \bm \ell_1)^2 (\bm \ell_2 + \bm q)^2 ( \bm \ell_1^2 + 2 \bm p \bm \ell_1 ) ( \bm \ell_2^2 + 2 \bm p \bm \ell_2 )}   \, .
}{eq:BB_result}
The remaining integral here is infrared divergent and superclassical,
having an additional factor of $|\bm q|^{-2}$ relative to the
classical scaling. It represents the iteration of the lower-order
tree-level diagram, and we will leave the integral unevaluated since
it will simply cancel with the same infrared artifact appearing in the
effective theory contribution to the matching. As in the case of the
one-loop box diagram, the classical contributions vanish order by
order in the nonrelativistic expansion.

\subsubsection{Crossed-Double-Box Diagram}

\newcommand{\BBbar}{{\rm B}\overline{\rm B}}

\begin{figure}
\begin{center}
\hspace{2cm}\includegraphics[scale=.8]{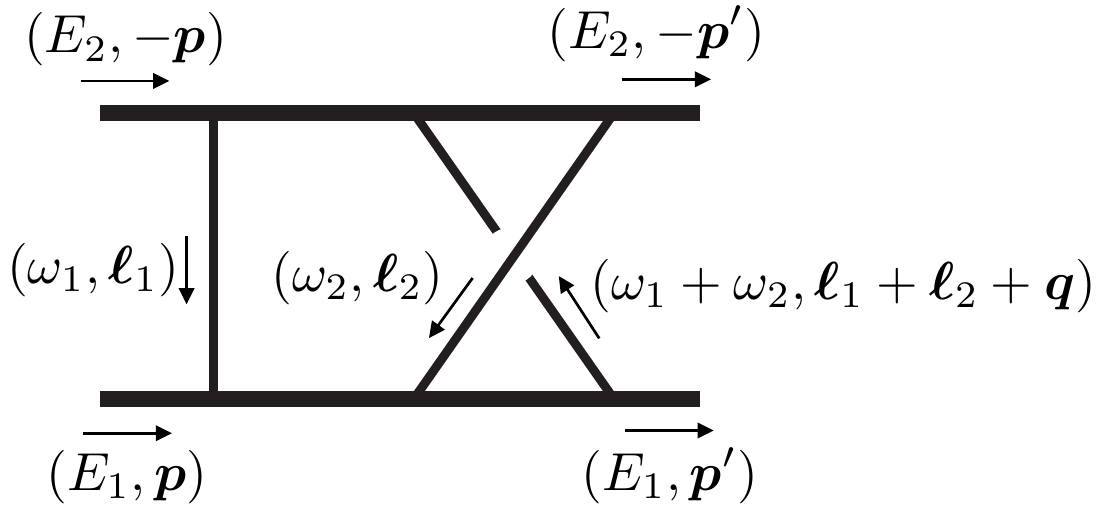}
\end{center}
\vspace*{-0.5cm}
\caption{\small The nonplanar double-box diagram. }\label{fig:NRint_BBbar}
\end{figure}

Finally we consider the general nonplanar double-box diagram, which is
a two-loop nonplanar integral with an arbitrary numerator and two
$\phi_1$ propagators, two $\phi_2$ propagators, and three graviton
propagators arranged as shown in \fig{fig:NRint_BBbar}.
 The integrand corresponding to this diagram is
\eq{
{\cal I}_{\BBbar} =&\frac{1}{(E_1+\omega_1 )^2 - (\bm p+\bm \ell_1 )^2 -m_1^2} \, \frac{1}{(E_2- \omega_1)^2 - (\bm p+\bm \ell_1)^2 -m_2^2} \\
& \times 
\frac{1}{(E_1+\omega_1 + \omega_2)^2 - (\bm p+\bm \ell_1 + \bm \ell_2)^2 -m_1^2}
\frac{1}{(E_2+ \omega_2)^2 - (\bm p - \bm \ell_2 - \bm q)^2 -m_2^2} \, \frac{1}{\omega_1^2 - \bm \ell_1^2}\, \\
& \times
\frac{1}{\omega_2^2 - \bm \ell_2^2}\, \frac{1}{(\omega_1+\omega_2)^2 - (\bm \ell_1 + \bm \ell_2+ \bm q)^2} \,
 {\cal N}_{\BBbar} \,.
}{}
As before, we factor all propagators into matter and antimatter components.
The spatial integrand is then 
\eq{
\widetilde {\cal I}_{\BBbar} &= \int \frac{d\omega_1}{2\pi}\frac{d\omega_2}{2\pi}\, \frac{\widetilde{\cal N}_{\BBbar}(\omega_1 ,\omega_2)}{(\omega_1-\omega_{P_1})(\omega_1-\omega_{P_2})(\omega_1+\omega_2-\omega_{P_1'})(\omega_2-\omega_{P_2'})} \,, 
}{}
where the effective numerator is
\eq{
\widetilde{\cal N}_{\BBbar}(\omega_1 ,\omega_2) &=
\frac{1}{\omega_1-\omega_{A_1}} \frac{1}{\omega_1-\omega_{A_2}} \, \frac{1}{\omega_1+\omega_2-\omega_{A_1'}} \,  \frac{1}{\omega_2-\omega_{A_2'}} \, \frac{1}{\omega_1^2 - \bm \ell_1^2}\, \frac{1}{\omega_2^2 - \bm \ell_2^2}\\
&\quad \times \frac{1}{(\omega_1+\omega_2)^2 - (\bm \ell_1 + \bm \ell_2+ \bm q)^2} \, {\cal N}_{\BBbar}(\omega_1,\omega_2) \,.
}{}
The energy-reduction constraints take the form
\eq{
(\omega_1-\omega_{P_1})(\omega_1-\omega_{P_2})(\omega_1+\omega_2-\omega_{P_1'})&\rightarrow 0 \,, \\
(\omega_1-\omega_{P_1})(\omega_1-\omega_{P_2})(\omega_2-\omega_{P_2'}) &\rightarrow 0 \,, \\
(\omega_1+\omega_2-\omega_{P_1'})(\omega_2-\omega_{P_2'}) &\rightarrow 0 \, .
}{eq:crossed_double_box_constraints}
We can solve for $\omega_1^3$ from the first constraint, $\omega_1^2$ from the second constraint, and $\omega_2^2$ from the third constraint. This allows us to reduce the effective numerator to the form
\eq{
{\cal N}_{\BBbar}(\omega_1,\omega_2) = f_0 + f_1 \omega_1 + f_2 \omega_2 + f_3 \omega_1 \omega_2 \, .
}{eq:crossed_double_box_form}
As before, the remaining dependence on $\omega_1$ and $\omega_2$ can be expanded about a chosen energy pole. In particular the last two terms in \Eq{eq:crossed_double_box_form} lead to the previously encountered energy master integrals for the box-triangle, double-triangle, and double-triangle prime diagrams. The first and second terms lead to vanishing masters given by
\eq{
\int  \frac{d\omega_1 \, d\omega_2}{(\omega_1-\omega_{P_1}+i\epsilon)(\omega_1-\omega_{P_2}-i\epsilon)(\omega_1+\omega_2-\omega_{P_1'}+i\epsilon) (\omega_2-\omega_{P_2'} + i\epsilon)} &= 0  \,, \\
\int  \frac{d\omega_1 \, d\omega_2}{(\omega_1-\omega_{P_1}+i\epsilon)(\omega_1+\omega_2-\omega_{P_1'}+i\epsilon) (\omega_2-\omega_{P_2'}+i\epsilon)} &= 0
\, .
}{eq:crossed_masters}
These were evaluated by averaging over six permutations of the three exchanged gravitons.

As in the case of the one-loop scalar crossed box, the two-loop
crossed-double-box with ${\cal N}_{\BBbar} =1$ vanishes order by
order in the velocity expansion.

\subsection{Quantum Contributions}

\begin{figure}
\begin{center}
\includegraphics[scale=0.8]{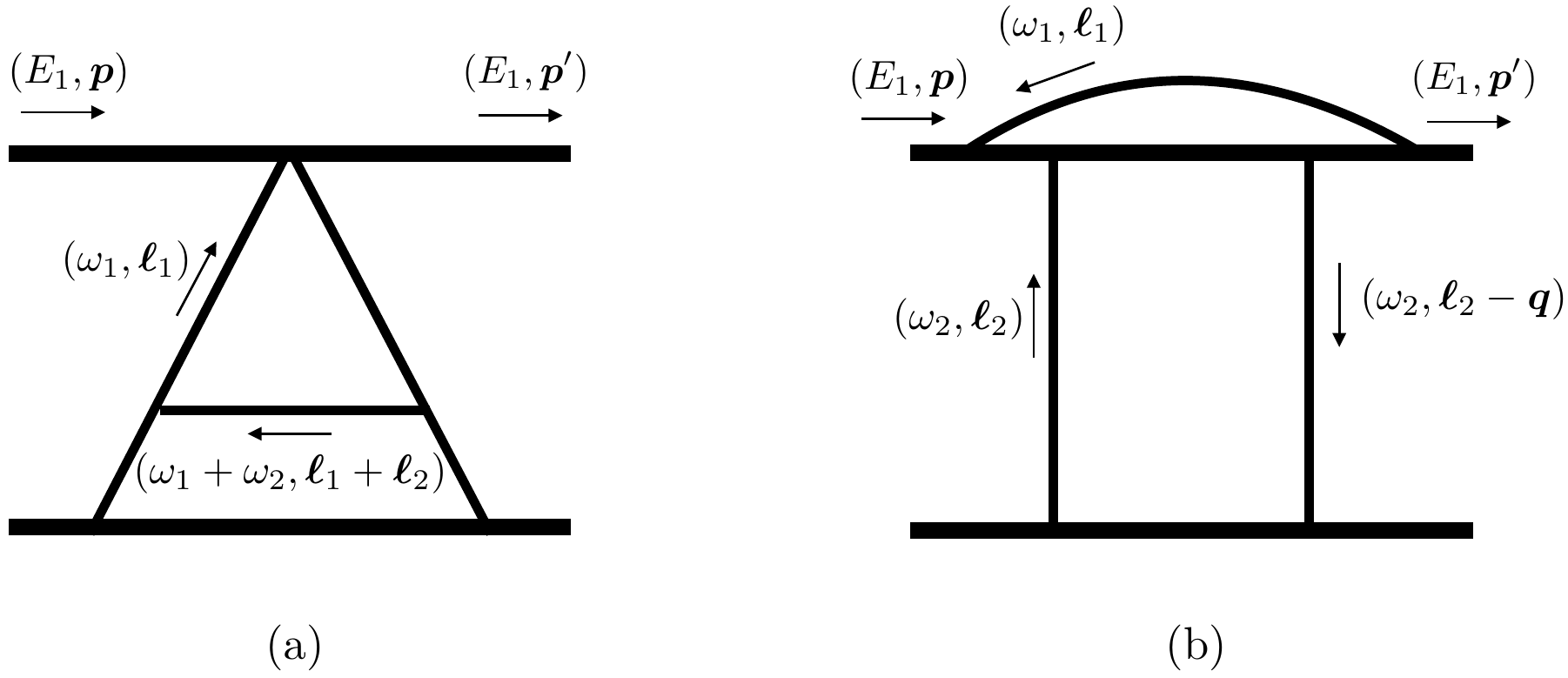}
\end{center}
\vspace*{-0.5cm}
\caption{\small Examples of diagrams that do not have classical potential contributions.}
\label{fig:AM}
\end{figure}

In the previous subsections we have described various diagrams which
contribute to classical scattering.  Conversely, we have dropped many
contributions which, as discussed in \Sec{sec:classical_limit}, would
otherwise contribute to the scattering amplitude but are
quantum mechanical.  Restricting to the classical
contribution to the conservative potential, we then restrict
consideration to diagrams in which
\begin{itemize}
\item[(i)] every loop includes at least one matter propagator,
\item[(ii)] there are no gravitons starting and ending at the same matter line.
\end{itemize}
These would exclude, for example, the diagrams shown in \Fig{fig:AM}, where diagrams (a) and (b) respectively violate the first and second criteria.

We discuss here the above criteria in terms of the mechanics of the
nonrelativistic method of integration described in this
section. Consider an internal graviton line with momentum
$(\omega, \bm \ell)$. The component $\omega$ is set through energy
integration to a matter pole, e.g. $\omega_{P_1}$, which at leading
order goes as $\omega_{P_1} \sim {\bm p \cdot \bm \ell / m_1}$. At the
same time, as seen from \Eq{eq:integrable} the component $\bm \ell$ is
set by a graviton pole in $(D-1)$-momentum variables to be of order
the momentum transfer, $| \bm \ell | \sim |\bm q|$. These, of course,
are consistent with the potential region, where a graviton loop
momentum goes as $(\omega, \bm \ell) \sim (|\bm v| |\bm q| , |\bm
q|)$. However, the energy or $(D-1)$-momentum components cannot be of
the same order as the scales $m$ or $|\bm p|$ while being in the
potential region. There is also the radiation region but, as discussed
in \sect{sec:radiation}, this only affects emission, at least at the
order at which we are computing.

From these remarks the criteria above are straightforward to understand. The
first criterion is necessary because the energy flowing through a loop
that does not include a matter propagator cannot be set to the matter
pole, and is therefore not in the potential region. For
example, consider the dependence on the graviton loop momentum
$(\omega_1 , \bm \ell_1)$ in diagram (a) in \Fig{fig:AM}. The loop
energy $\omega_1$ has no pole in $\omega_{P_1}$ because there is no
matter propagator for $\phi_1$. Note that the middle horizontal
graviton depends on the sum $\omega_1 + \omega_2$ but this is
subleading to the $(D-1)$-momentum component since it is a potential
mode, and does not correspond to a pole for $\omega_1$. This diagram
fails the first criterion. On the other hand, this diagram has a pole
in $\bm \ell_1 - \bm q$ and passes the second criterion.

The second criterion is necessary because the momentum flowing through
a graviton starting and ending on the same matter line
cannot have a scale set to the momentum transfer. As a
concrete example, consider the dependence on the graviton loop
momentum $(\omega_1 , \bm \ell_1)$ in diagram (b) in \Fig{fig:AM}. The
loop energy $\omega_1$ can be set to a matter pole since there are
matter propagators for $\phi_1$. However, there is no pole in
$\bm \ell_1 - \bm q$. Note that the matter propagators for $\phi_1$
depend on the spatial momenta $\bm p$, $\bm \ell_2$, and $\bm q$, but
there is no pole in $\bm \ell_1 - \bm q$ or $\bm \ell_1 - \bm \ell_2$.

\subsection{Residue Method}
\label{sec:residue}
In this section we discuss an alternative simpler method for
performing the energy integration.  In the previous examples our
approach was to apply energy-integral reduction to express everything
in terms of master energy integrals which were then
evaluated directly.  Here we will see that we can instead evaluate the
energy integrals using residues. This has been verified explicitly at
two loops.

Recall that the spatial integrand $\widetilde {\cal I}$ is equal to the energy integral of the full integrand,
\eq{
\widetilde {\cal I}= \bigg[ \prod_{i=1}^{n_L} \int \frac{d\omega_i }{2\pi}  \bigg]\; {\cal I}\,.
}{}
In general, ${\cal I}$ will have singularities from graviton poles as well as matter and antimatter poles.  Of course, only the matter poles reside in the kinematic region which contributes classically.  In general, we can take the $n_L$ loop energies and localize them onto $n_L$ of the $n_M$ matter poles. We label each of these solutions $\omega_{i\alpha}$ by an index $\alpha$.  Our claim is that the energy integral is equal to
\eq{
\widetilde {\cal I}= \sum_\alpha \sigma_\alpha \!\! \underset{\;\;\; \omega_i 
= \omega_{i\alpha}}{\rm Res} [{\cal I}] \,,
}{eq:residue}
where $\sigma_\alpha$ is a numerical symmetry factor associated with
each residue.  To determine $\sigma_\alpha$ we interpret the residue
as cutting certain matter propagators, while all uncut matter
propagators should be thought of as pinched.  The resulting diagram
then corresponds to one of the master energy integrals we have
considered already, and $\sigma_\alpha$ is precisely the associated
symmetry factor.

The residue method can be derived from the method of energy integral reduction. Consider expanding the integrand ${\cal I}$, written
as \Eq{eq:integrand_eff}, in the potential region. Then
the effective numerator $\widetilde {\cal N}$ is a power series in
$\omega_i$. The power series can be organized into terms that pinch all energy poles, and terms with singularities in $\omega_i$.
The first type integrates to zero via \Eq{eq:scaleless}. This is trivially equivalent
to taking the residue because the corresponding energy integrand is regular in the finite region. The second type are nothing but the energy master integrals, e.g. \Eq{eq:energy_box}
and \Eq{eq:double_master}, and evaluate to residues with appropriate symmetry factors. Hence, the energy integration for each term
in the expansion of $\widetilde {\cal N}$ is equivalent to taking the residue and including symmetry
factors, and therefore the answer should reproduce \Eq{eq:residue} after expanding in the potential region.

We briefly discuss how this works in some of our earlier examples
in \sect{sec:NRint_oneloop} and~\ref{sec:NRint_twoloop}. First, we
consider one loop, where $\omega_i$ runs over a single loop energy.
The triangle diagram has one matter pole so $\alpha$ runs over one
solution.  As we have shown the associated symmetry factor is
$\sigma_\alpha = \{ 1/2\}$.  Meanwhile, for both the box and
crossed-box diagrams there are two matter poles so $\alpha$ runs over
two solutions with symmetry factors $\sigma_\alpha = \{1/2,1/2\}$.

Moving on to two loops, we see that the double-triangle diagram has two
matter poles.  Since $\omega_i$ runs over two-loop energies, $\alpha$
runs over a single solution for which $\sigma_\alpha = \{ 1/6 \}$.
The box-triangle diagram is more complicated because it has three
matter poles.  A priori, there are three ways to localize two-loop
energies onto these matter poles. However, one of these configurations
does not exist because there is no choice of loop energy that can
localize both matter poles residing in a box subdiagram.
Consequently, for the box-triangle diagram, $\alpha$ runs over two
solutions with symmetry factors $\sigma_\alpha = \{ 1/6, 1/3\}$, which
is consistent with the energy master integral
in \Eq{eq:boxtriangle_master} since $1/3 + 1/6 = 1/2$.

\section{Relativistic Integration}
\label{sec:Rint}

Ideally, we would evaluate all integrals via fully relativistic
methods.  As we will see, this is relatively straightforward for
certain diagrams---in particular, those which are free from infrared
divergent matter singularities. One notable example is the H and
crossed-H diagrams in \sect{sec:NRint_H}, whose resummation yielded
the arcsinh function. In the present section, we describe several
relativistic methods, such as differential equations and Mellin-Barnes
integration, for analytic computation of the H diagram and other
integrals in the same class, including the diagram in
\sect{sec:N}. These methods provides strong confirmation that we have
indeed correctly resummed the amplitude to all orders in velocity.

\subsection{Method of Differential Equations \label{sec:diffEq}}
Differential equations are a powerful method for evaluating Feynman integrals~\cite{Kotikov:1990kg, hep-ph/9306240, hep-th/9711188,GehrmannRemiddi}. We will adapt the method to evaluate the classical contribution of the sum of the H and $\overline {\rm H}$ integrals, using differential equations for integrals localized on the poles of the two internal matter lines. 

\begin{figure}
\begin{center}
\includegraphics[scale=.5]{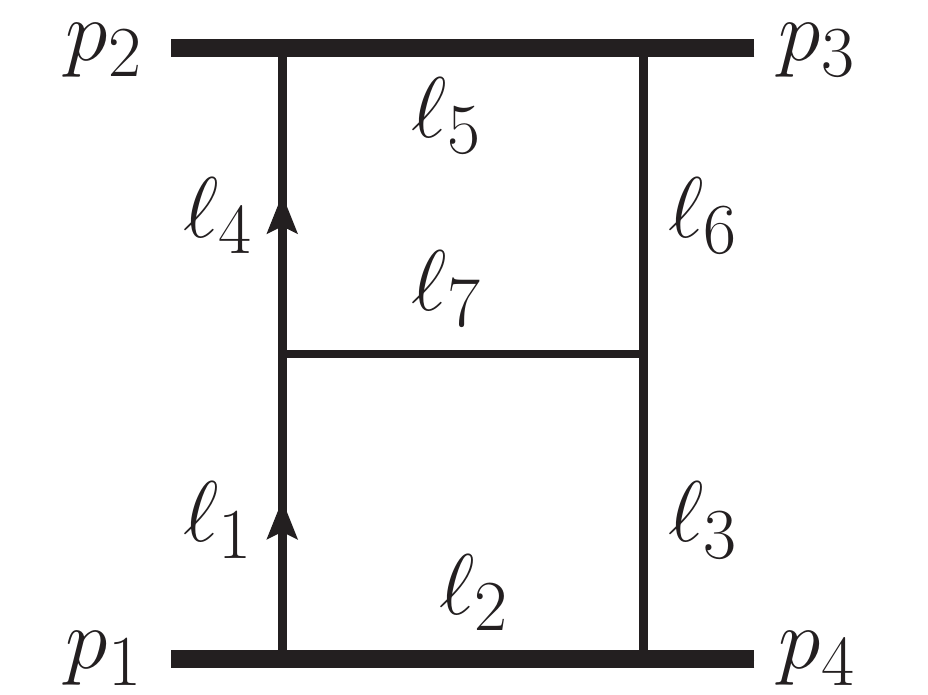}
\end{center}
\vspace*{-0.3cm}
\caption{\small The H graph with momentum assignments.  The 7 propagators carrying momentum $\ell_i$ correspond to 
the propagators in \eqn{allHints}.  The external momenta are all outgoing and the 
direction of the independent loop momenta are indicated by the arrows.
\label{Hfig}}
\end{figure}

The class of integrals we will discuss, denoted by ${\rm H}[a_1\dots a_9]$, is defined as
\begin{equation}
{\rm H}[a_1, \dots a_7, a_8, a_9] = \int \frac{d^D \ell_1 d^D \ell_4}{(2\pi)^{2D} }
\frac{((\ell_4+p_1)^2-m_1^2)^{-a_8}((\ell_1-p_2)^2-m_2^2)^{-a_9} }
{ (\ell_1^2)^{a_1}(\ell_2^2-m_1^2)^{a_2}(\ell_3^2)^{a_3}(\ell_4^2)^{a_4}(\ell_5^2-m_2^2)^{a_5}(\ell_6^2)^{a_6}(\ell_7^2)^{a_7}} \,,
\label{allHints}
\end{equation}
with the momentum labeling in Fig.~\ref{Hfig} and $\ell_{2, 3, 5, 6, 7}$ expressed in terms of $\ell_1$ and $\ell_4$ by momentum conservation.

\begin{figure}
\begin{center}
\includegraphics[scale=.5]{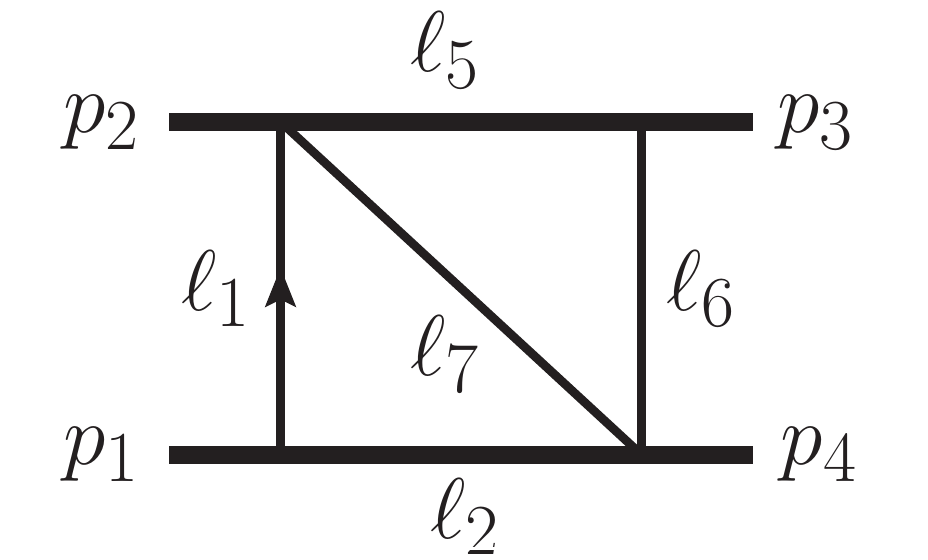}
\end{center}
\vspace*{-0.3cm}
\caption{\small The diagram ${\rm H}[1,1,0,0,1,1,1,0,0]$, with momentum assignments, 
obtained by canceling propagators 3 and 4 of the H diagram in \fig{Hfig}.
\label{Nfig}}
\end{figure}

\subsubsection{Integration-By-Parts Reduction \label{sssec:IBP}}

To obtain differential equations, we first perform
integration-by-parts (IBP) reduction~\cite{Chetyrkin:1981qh, Laporta:2001dd} in $D =
4-2\epsilon$ dimensions. There are by now many automated IBP reduction
programs. We use {\tt Kira} \cite{Maierhoefer:2017hyi} which
implements fast modern algorithms. All integrals of the H diagram and
a tower of contact diagrams with collapsed propagators are reduced to
$51$ master integrals.

The most important master integrals are
associated with the H diagram,
${\rm H}[1,1,1,1,1,1,1,0,0]$, and with ${\rm H}[1,1,0,0,1,1,1,0,0]$.
It turns out that diagrams ${\rm H}[0,1,1,1,1,1,1,0,0]$, ${\rm H}[1,1,0,1,1,1,1,0,0]$, ${\rm H}[1,1,1,0,1,1,1,0,0]$ or ${\rm H}[1,1,1,1,1,0,1,0,0]$,  obtained from H by 
canceling one of the propagators 1, 3, 4, or 6 respectively, have no master integrals. 
That is, all integrals ${\rm H}[a_1,1,1,1,1,1,1,a_8,a_9]$, ${\rm H}[1,1,a_3,1,1,1,1,a_8,a_9]$, ${\rm H}[1,1,1,a_4,1,1,1,a_8,a_9]$ or ${\rm H}[1,1,1,1,1,a_6, 1,a_8,a_9]$ 
with $a_{1, 3, 4, 6, 8, 9}\le 0$, can be reduced (via IBP) to master integrals of the  diagram ${\rm H}[1,1,0,0,1,1,1,0,0]$.

The master integrals associated with the H graph in \fig{Hfig}, i.e.\ the integrals that can be used to express any integral of the type ${\rm H}[1,1,1,1,1,1,1,a_8, a_9]$ with $a_{8,9}\le 0$ modulo contact integrals with fewer propagators, are (rescaled by appropriate powers of $t = -{\bm q}^2$ to have classical power counting) 
\begin{align}
&I_1 = t^2 \, {\rm H}[1,1,1, 1,1,1, 1, 0,0]\,, \hskip 1.9 cm I_2 = t^{3/2} \, {\rm H}[1,1,1, 1,1,1, 1, 0,-1]\,, \nonumber \\
&I_3 = t^{3/2} \, {\rm H}[1,1,1, 1,1,1, 1, -1,0]\,, \hskip 1.3 cm I_4 = t \, {\rm H}[1,1,1, 1,1,1, 1, -2,0] \, . \label{eq:mastersH}
\end{align}

The master integrals associated with the graph ${\rm H}[1,1,0,0,1,1,1,0,0]$ in \fig{Nfig}, i.e.\ the integrals that can be used to express any integral of the type 
${\rm H}[1,1,a_3, a_4,1,1,1,a_8 ,a_9]$ with $a_{3, 4, 8,9}\le 0$ modulo contact integrals with fewer propagators, are
\begin{align}
&I_5 = {\rm H}[1,1,0, 0,1,1, 1, 0,0]\,, \hskip 2.5 cm I_6 = t^{-1/2} \, {\rm H}[1,1,0, 0,1,1, 1, 0,-1]\,, \nonumber \\
&I_7 = t^{-1/2} \, {\rm H}[1,1,0, 0,1,1, 1, -1,0]\,, \hskip 1.3 cm  I_8 = t^{-1} \, {\rm H}[1,1,-1, 0,1,1, 1, 0,0]\,, \nonumber \\
&I_9 = t^{-2} \, {\rm H}[1,1,-2, 0,1,1, 1, 0,0] \,. 
\label{eq:mastersN}
\end{align}
We omit the expressions for the remaining master integrals. In fact, they are associated with diagram topologies that cannot give rise to a classical $1/ r^3$ potential; among them are topologies with collapsed matter propagators and topologies with four-scalar contact vertices. This also matches the construction of the integrand in \sect{MergingSubsection}, where the two internal matter lines in graphs of H topology are always cut and therefore never collapsed.

\subsubsection{Differential Equations \label{sssec:diffeq}}
The derivative of a master integral with respect to any kinematic variable, such as $(\partial /\partial \sigma)$, can be rewritten in the form
\begin{equation}
\Omega_{ij} \, k_i ^\mu \frac{\partial}{\partial k_j^\mu} 
\label{operator}
\end{equation}
which can then act on the propagators and numerators of an integral. Here $\Omega_{ij}$ is a linear combination of the $GL(3)$ generators.
The action of the operator \eqref{operator} generally introduces propagators raised to higher powers, but the IBP reduction re-expresses the result as a linear combination of the original master integrals. Therefore we obtain differential equations~\cite{GehrmannRemiddi}
\begin{equation}
\frac{\partial}{\partial \sigma} \Bigg|_{t,\, m_1^2, \, m_2^2} I_i = M_{ij} \, I_j , \label{eq:de-general}
\end{equation}
with some matrix $M_{ij}$ whose entries are rational functions of the kinematic variables $\sigma =(p_1 \cdot p_2) / (m_1 m_2), t = -{\bm q}^2, m_1^2, m_2^2$.

We define the ``conjugate'' master integrals from a $s\leftrightarrow u$ crossing,
\begin{equation}
\bar I_i = I_i |_{p_2 \leftrightarrow p_3}, \quad i=1,2,\dots,9
\label{IbarDef}
\end{equation}
As shown in Section \ref{sec:NRint}, in the small-$t$ limit, the conservative contribution from the sum,
\begin{equation}
  \hat I_i = \operatorname{Re}(I_i + \bar I_i)\,,
\end{equation}
is effectively a modified version of $I_i$ with the two matter propagators replaced by delta functions, see Sec.~\ref{sec:MB}.
By dropping master integrals with collapsed matter propagators, and taking the small-$t$ limits of the $M_{ij}$ entries in Eq.~\eqref{eq:de-general}, we obtain the final differential equations that are valid for integrals localized on the matter poles, similar to the energy integration procedure,
\begin{equation}
\frac{\partial}{\partial \sigma} \Bigg|_{t,\, m_1^2, \, m_2^2} \hat I_i = {\widetilde M}_{ij} \, \hat I_j ,
\label{eq:de-localized}
\end{equation}
where the values of $i$ and $j$ only run from $1$ to $9$, corresponding to the 9 master integrals in Eqs.~\eqref{eq:mastersH} and \eqref{eq:mastersN}, 
and the matrix $ \widetilde M$ on the right-hand side is defined as
\begin{equation}
\widetilde M_{ij} = M_{ij} |_{t= 0} \, .
\end{equation}
The limit on the right-hand side is well-defined because the $M_{ij}$ entries are all non-singular at $t=0$, thanks to the appropriate normalizations by powers of $t$ in Eqs.~\eqref{eq:mastersH} and \eqref{eq:mastersN}.

To solve the differential equations Eq.~\eqref{eq:de-localized}, the boundary condition is that all the integrals are non-singular in the threshold limit, 
$s \to (m_1 + m_2)^2$. This is sufficient to fix the solutions up to an overall numerical prefactor. Since the differential equations are homogeneous, the overall prefactor can be fixed by evaluating one integral (in this case, the sum of the scalar H and ${\overline {\rm H}}$ integrals) in the static limit $|\bm p|=0$, which is the first term in Eq.~\eqref{eq:H_series}.

The result for $\hat I_1$,  the sum of scalar H and ${\overline {\rm H}}$ integrals localized on matter poles and rescaled by $t^2$ as in Eq.~\eqref{eq:mastersH}, is equal to
\begin{equation}
\hat I_1= \frac{ \log \bm q^2 }{64 \pi^2 m_1 m_2} \frac{1}  {\sqrt{\sigma^2 - 1} }  \operatorname{arcsinh} \sqrt{\frac{\sigma-1}{2}}, \label{eq:resultI1}
\end{equation}
where $\sigma$ is defined as in \Eq{eq:sigma_def}. We have omitted $\mathcal O(\epsilon)$ corrections, as is the case for all results presented in this subsection. The result is in agreement with \Eq{eq:Hresummed}, but is rigorously established by the differential equations to all orders in the velocity expansion.
Using hyperbolic trigonometry identities, Eq.~\eqref{eq:resultI1} can be re-written as
\begin{equation}
\hat I_1  = \frac{ \log \bm q^2 }{128 \pi^2} \, \frac{ {\rm arcsinh} (|{\bm p}| / m_1) + {\rm arcsinh} (|{\bm p}| / m_2) } {|{\bm p}| (E_1 + E_2)},
\end{equation}
where ${\bm p}$, $E_1$ and $E_2$ are defined in the center-of-mass frame as
\begin{equation}
-p_1 = (E_1, {\bm p}), \qquad -p_2 = (E_2, - {\bm p})
\, .
\end{equation}
All other master integrals for the H diagram, combined with the conjugate integrals, have no contribution in the classical limit,
\begin{equation}
\hat I_{i} = 0, \quad i=2,3,4,5 \, .
\end{equation}
The results for the nonzero master integrals of the diagram ${\rm H}[1,1,0,0,1,1,1,0,0]$, combined with their conjugate integrals, are as follows:
\begin{align}
\hat I_5 &= - \log {\bm q}^2 \, \frac{1}{128 \pi^2 m_1 m_2} \frac{1}{\sqrt{\sigma^2 - 1}}  \operatorname{arcsinh} \sqrt{\frac{\sigma-1}{2}}, \nonumber \\
\hat I_8 &= \log {\bm q}^2 \, \frac{\sigma}{256 \pi^2 m_1 m_2} \left(
  \frac{1}{\sigma^2-1} - \frac{2\sigma}{(\sigma^2-1)^{3/2}} \operatorname{arcsinh} \sqrt{\frac{\sigma-1}{2}}
\right), \nonumber \\
\hat I_{9} &= \log {\bm q}^2\, \frac{(1+2 \sigma^2)} {1024 \pi^2 m_1 m_2 (\sigma^2-1)^{5/2}}
  \left( 3 \sigma \sqrt{\sigma^2-1} -2 (1+2\sigma^2) \right) \operatorname{arcsinh} \sqrt{\frac{\sigma-1}{2}}
\, . \label{eq:NmasterIntegrals}
\end{align}
The result for $\hat I_5$ agrees with the sum of \Eq{eq:Nscalar} and \Eq{eq:xNscalar}. The other two master integrals of the diagram ${\rm H}[1,1,0,0,1,1,1,0,0]$ 
vanish in the classical limit when localized on the matter poles,
\begin{equation}
\hat I_{j} = 0 , \quad j=6,7 \, .
\end{equation}

We emphasize that the differential equation method discussed here establishes in a rigorous way that the resummation of the velocity expansion in \Eq{eq:H_series} gives the complete result for the diagram H localized on the matter poles, which is the only part of this diagram that contributes to the conservative potential.

\subsubsection{Sample Differential Equations for the Diagram ${\rm H}[1,1,0,0,1,1,1,0,0]$}
As a simple illustration, we give sample differential equations for the integrals $\hat I_{5,8,9}$ associated 
with the double triangle prime diagram, ${\rm H}[1,1,0,0,1,1,1,0,0]$ in Fig.~\ref{Nfig}, in the equal-mass case $m_1 = m_2 =1$. 
In the limit $t \to 0, \, D \to 4$, these integrals are decoupled from the other master integrals in the differential equations,
\begin{equation}
\frac{\partial}{\partial \sigma} \Bigg|_{t,\, m_1^2, \, m_2^2} \hat I_i = \widetilde M_{ij} \, \hat I_j , 
\end{equation}
where the indices $i$ and $j$ take the values $5, 8, 9$. Setting $t=0, \, D=4, \, m_1=m_2=1$, the entries of the matrix $\widetilde M$ are
\begin{align}
\widetilde M_{5,5} &= \frac{2 \sigma ^2+1}{(\sigma -1) \sigma  (\sigma +1)},  \hskip .9 cm 
\widetilde M_{5,8} = -\frac{3 \sigma }{(\sigma -1) (\sigma +1)},  \hskip .9 cm       
\widetilde M_{5,9} = \frac{4}{3 \sigma }, \nonumber \\
\widetilde M_{8,5} &= \frac{1}{3 \sigma }, \hskip 3.3 cm 
\widetilde M_{8,8} = 0, \hskip 3.7 cm 
\widetilde M_{8,9} = -\frac{4}{3 \sigma }, \nonumber \\
\widetilde M_{9,5} &= -\frac{1}{3 \sigma }, \hskip 3.0 cm
\widetilde M_{9,8} = \frac{3 \sigma }{(\sigma -1) (\sigma +1)}, \hskip 1.2 cm
\widetilde M_{9,9} = -\frac{11 \sigma ^2+4}{3 (\sigma -1) \sigma  (\sigma +1)} \, .
\end{align}
It can be readily checked that Eq.~\eqref{eq:NmasterIntegrals} solves the differential equations given above.

The differential equations in $D=4$ are sufficient to determine the
master integrals of the diagram ${\rm H}[1,1,0,0,1,1,1,0,0]$ localized on matter poles in the leading small-$t$ limit. In general however
we need to keep dependence on $D$, for several reasons. 
First, the differential equations for the H diagram have
spurious singularities as $D \to 4$, but the actual \emph{solutions}
are non-singular. Second, the integrals in the class ${\rm H}[1,1,0,1,1,1,1,0,0]$ will be reduced (via IBP)
to integrals in the class ${\rm H}[1,1,0,0,1,1,1,0,0]$, but with coefficients that are singular in $D \to 4$.
Therefore, to determine the former as $D \to 4$, exact-$D$ solutions are needed 
for the integrals of the latter. 
The coefficients and the actual values of the integrals ${\rm H}[1,1,0,0,1,1,1,0,0]$ will conspire to cancel 
the $D \to 4$ singularities of the integrals ${\rm H}[1,1,0,1,1,1,1,0,0]$. In the end, the integral ${\rm H}[1,1,0,1,1,1,1,0,0]$
turns out to be equal to the integral ${\rm H}[1,1,0,0,1,1,1,0,0]$ up to a minus sign in the $D \to 4$
limit.
As we will see in \sect{sec:MB} by finding certain integral representations for the leading small-$t$ terms of the 
${\rm H}[1,1,0,0,1,1,1,0,0]$ and ${\rm H}[1,1,0,1,1,1,1,0,0]$ integrals, this relation holds even in the absence of localization on matter poles.


\subsubsection{Integration of Graphs ${\bf 7}$ and ${\bar {\bf 7}}$.   \label{sec:graph7}}

The complete graph $\bf 7$ in \fig{MasterDiagsFigure} contains up to rank-8 tensor integrals; while some of these tensors correspond, effectively, to inverse propagators (and thus may 
be interpreted as lower-rank tensors on a graph with quartic vertices) the rank is rarely reduced by more than two units.  A direct application of the 
IBP techniques to such high-rank tensor integrals is very time-consuming. Our case however is simpler because we are interested only in the classical contribution of this diagram,  i.e.\ only in the terms which could potentially yield $\ln{\bm q}^2$ upon integration.  
As discussed in \sect{QuantumVsClassicalSection}, simple scaling shows that any term of rank higher than 4 is subleading in the classical limit and 
can therefore be dropped.
Of the 318 terms in the original integrand of graph ${\bf 7}$, constructed in \sect{MergingSubsection} and given in an ancillary file~\cite{AttachedFile},
and corresponding to 137 superficially distinct integrals, only 18 survive. These correspond to 6 superficially distinct integrals, 
five of which have ranks 2 and 4, giving the result for graph ${\bf 7}$:
\begin{align}
{\bf 7} = \null &  4 (1 - 2 \sigma^2)^2 m_1 ^4 m_2 ^4 \; 
  ( t^2 {\rm H}[1, 1, 1, 1, 1, 1, 1, 0, 0])
   - 16 \sigma^2 m_1 ^2 m_2 ^4 \;  
 (t {\rm H}[1, 1, 1, 1, 1, 1, 1, -2, 0])
\cr
 \null  &+ 16 \sigma m_1 ^3 m_2 ^3 \;  
 (t {\rm H}[1, 1, 1, 1, 1, 1, 1, -1, -1]) 
   -  16 \sigma^2 m_1 ^4 m_2 ^2 \; 
  (t {\rm H}[1, 1, 1, 1, 1, 1, 1, 0, -2])
\cr
  &+ 
  2 m_2^4 \; {\rm H}[1, 1, 1, 1, 1, 1, 1, -4, 0]
  +
  2 m_1^4 \; {\rm H}[1, 1, 1, 1, 1, 1, 1, 0, -4] \,.
  \label{G7firstpass}
\end{align}

As described in \sect{sssec:IBP}, the IBP identities express all dimensionally-regularized scalar and tensor integrals with 
the topology of the H graph and graphs related to it by propagator collapse, i.e.\ all integrals in Eq.~\eqref{allHints}, in terms 
of 51 master integrals. 
The coefficients of the  master integrals have two important features: (a) they may contain inverse powers of $t=q^2$ and (b) they may be singular as $D\rightarrow 4$. The former implies that master integrals which, on their own, do not contain classical terms may nonetheless contribute in the classical limit due to their $t$-dependent coefficient. 
The latter implies that, for a complete evaluation of the classical contribution of graph ${\bf 7}$ through ${\cal O}(\epsilon^0)$ it is necessary to evaluate 
master integrals to higher orders in the dimensional regulator. 
Examples of relations reducing integrals in Eq.~\eqref{G7firstpass}, expanded in $t$ and $\epsilon$ and illustrating their structure, are: 
\begin{align}
\label{IBPeg1}
{\rm H}[1, 1, 1, 1, 1, 1, 1, 0, -2] & = \left(\frac{m_2^2}{m_1^2} - \frac{(m_1^2 - m_2^2)}{4 m_1^4}t \right) {\rm H}[1, 1, 1, 1, 1, 1, 1, -2, 0] 
\\
& \hspace{-1truecm}
+\left(1 - \frac{m_2^2}{m_1^2}\right) \frac{1}{t} {\rm H}[1, 1, -1, 0, 1, 1, 1, 0, 0]  
- \left(1 - \frac{m_2^2}{m_1^2}\right) {\rm H}[1, 1, 0, 0, 1, 1, 1, 0, 0] 
+\dots
\cr
{\rm H}[1, 1, 1, 1, 1, 1, 1, -4, 0] & =-\left(m_1^2 - \frac{2}{3} m_1^2 \sigma^2\right) \; t {\rm H}[1, 1, 1, 1, 1, 1, 1, -2, 0]
\cr
& \hspace{-1truecm}
+\left(\frac{4}{3\epsilon}m_1^4(\sigma^2-1) + \frac{10}{3} m_1^4 - \frac{14}{9} m_1^4 \sigma^2 - \frac{16}{9} m_1^4 \sigma^4\right) 
\frac{1}{t^2} {\rm H}[1, 1, -2, 0, 1, 1, 1, 0, 0] 
 \cr
 & \hspace{-1truecm} 
 +\left( -\frac{1}{\epsilon}m_1^4(1 + 2  \sigma^2) + \frac{10}{3} m_1^4 + 8 m_1^4 \sigma^2 + \frac{8}{3} m_1^4 \sigma^4 \right) \frac{1}{t}{\rm H}[1, 1, -1, 0, 1, 1, 1, 0, 0] 
 \cr
 & \hspace{-1truecm}
 -\left(\frac{34}{9} m_1^4 \sigma^2 + \frac{8}{9} m_1^4 \sigma^4 - \frac{1}{3\epsilon}m_1^4(1 + 2  \sigma^2) \right) {\rm H}[1, 1, 0, 0, 1, 1, 1, 0, 0] 
 +\dots \,.
 \nn
\end{align}
In both these equations the ellipsis stand for terms of higher order $t$ and $\epsilon$ containing the master integrals written explicitly. They also
account for master integrals which, while having classical (and possibly superclassical) contributions to each of the 
terms in Eq.~\eqref{G7firstpass}, cancel out in the complete expression of graph ${\bf 7}$. 
The differential equation method described in \sect{sssec:diffeq}, is well-suited for
the evaluation of the remaining master integrals to any order in the expansion around four dimensions.  
The Mellin-Barnes representation method is often a very helpful alternative.
%

\begin{table}[bt]
\centering
\vskip - 1 cm 
\begin{eqnarray}
\begin{array}{c|c|c}
i & h_i & I_i \cr
\hline 
1& 4 (-3 + 4 \sigma^4) &    t^2 {\rm H}[1, 1, 1, 1, 1, 1, 1, 0, 0] \vphantom{\Big|}\cr
2& 0 &  t^{3/2} {\rm H}[1, 1, 1, 1, 1, 1, 1, 0, -1]  \vphantom{\Big|}\cr
3& 0 &  t^{3/2} {\rm H}[1, 1, 1, 1, 1, 1, 1, -1, 0]   \vphantom{\Big|}\cr
4& \frac{8}{\epsilon} - \frac{4}{3} (15 + 22 \sigma^2) &   t {\rm H}[1, 1, 1, 1, 1, 1, 1, -2, 0]  \vphantom{\Big|}\cr
5&  -(1 + 2 \sigma^2) \frac{16}{\epsilon^2} +  (17 + 94 \sigma^2) \frac{8}{3 \epsilon} + \frac{16}{9} (50 - 283 \sigma^2 - 4 \sigma^4)  &
  {\rm H}[1, 1, 0, 0, 1, 1, 1, 0, 0] \vphantom{\Big|}\cr
6&0 & t^{1/2} {\rm H}[1, 1, 0, 0, 1, 1, 1, 0, -1]  \vphantom{\Big|} \cr
7&0 & t^{1/2} {\rm H}[1, 1, 0, 0, 1, 1, 1, -1, 0]  \vphantom{\Big|} \cr
8& (1 + 2 \sigma^2) \frac{48}{\epsilon^2 } -  (1 + 2 \sigma^2) \frac{296}{\epsilon } + \frac{16}{3} (95 + 138 \sigma^2 + 4 \sigma^4)  &
   {\frac{1}{t}} {\rm H}[1, 1, -1, 0, 1, 1, 1, 0, 0] \vphantom{\Big|}\cr
9&  - (\sigma^2-1)  \frac{64}{\epsilon^2 } +  (\sigma^2-1) \frac{1024}{3 \epsilon } -  \frac{16}{9} (\sigma^2-1)(8\sigma^2+181)  &  
 {\frac{1}{t^2}} {\rm H}[1, 1, -2, 0, 1, 1, 1, 0, 0] \vphantom{\Big|}\cr  
10&  32 ( \sigma^2-1)  &    t {\rm H}[1, 1, 1, 1, 1, 1, 0, 0, 0] \vphantom{\Big|}\nn
\end{array}
\end{eqnarray}
\\ \vspace{-10pt}
\caption{\small Coefficients and master integrals obtained from reducing graph ${\bf 7}$ in \fig{MasterDiagsFigure}. 
The integrals in the third column are the master integrals defined in Eqs.~\eqref{eq:mastersH} and \eqref{eq:mastersN}. 
The definition of $I_{10}$ is provided by the third entry of the last line of this table.}
\label{coefficientsandintegralsTable}
\end{table}

Using the solution of the IBP equations and expanding the coefficients of the master integrals at small $t$ and near four dimensions, we find that graph ${\bf 7}$ with all outgoing momenta contributes in the classical limit as
\begin{equation}
{\bf 7} = m_1^4 m_2^4 \sum_{i=1}^{10} h_i I_i \,,
\end{equation}
where the coefficients $h_i$ and integrals $I_i$ are given in \tab{coefficientsandintegralsTable} and correspond to the H and ${\rm H}[1,1,0,0,1,1,1,0,0]$ master integrals defined in Eqs.~\eqref{eq:mastersH} and \eqref{eq:mastersN}.
We note here that all superclassical terms that appear in the reduction of individual integrals (cf. e.g.\ \eqref{IBPeg1}) canceled out and the leading 
small-$t$ term is the classical one.  This is, of course, as it should be because this graph does not contain any lower-loop iterations.

Graph ${\bar {\bf 7}}$ is obtained from graph ${\bf 7}$ through the relabeling $p_2\leftrightarrow p_3$. In terms of the variable $\sigma$, this transformation
amounts to the replacement
\begin{equation}
\sigma \rightarrow -\sigma - \frac{t}{2m_1m_2} \,.
\end{equation}
Because the classical contribution of the integrals as defined in Table~\ref{coefficientsandintegralsTable} is the first term in their small-$t$ expansion, 
the term linear in $t$ in the transformation above does not contribute in the classical limit. Since the coefficients $h_i$ depend quadratically on 
$\sigma$ and recalling \eqref{IbarDef} that $\bar I_i =I_i|_{p_2\leftrightarrow p_3}$, we find
\begin{equation}
{\bf 7}+ 
{\bar {\bf 7}} =  m_1^4 m_2^4 \sum_{i=1}^{10} h_i \left(I_i+\bar I_i  \right) 
\end{equation} 

Direct evaluation shows that ${\rm H}[1, 1, 1, 1, 1, 1, 0, 0, 0]$ does not, in fact, contribute in the classical limit. 
Moreover, while surviving in the classical limit, the imaginary part of $\left(I_i+\bar I_i  \right)$ with $i=1,\dots,9$ does not 
contribute to the conservative potential.
For $\operatorname{Re}\left(I_i+\bar I_i  \right)$ we use the results of the differential equation approach discussed in the 
previous section, which  we have also verified numerically.

We evaluated the $\ln(-t)$-dependent  real part of the $p_2\leftrightarrow p_3$-symmetric combination of other five integrals, 
to  ${\cal O}(\epsilon)^2$ and to 40th order in the small velocity expansion. All the singular terms cancel out, as they should. 
The resulting series resums to
\begin{equation}
{\bf 7} + 
{\bar {\bf 7}} 
= \frac{(1-2\sigma^2)^2 \, {\rm arcsinh}\sqrt{\frac{\sigma-1}{2}}}{16\pi^2 \sqrt{\sigma^2-1}} m_1^3m_2^3\, \ln {\bm q}^2 + \dots \,,
\label{relint77bar}
\end{equation}  
where we used that $\ln(-t) = \ln {\bm q}^2$ and the ellipsis stand for imaginary terms and real terms with no  $\ln {\bm q}^2$ dependence. 
It is interesting to note that, up to a factor of $m_1^4 m_2^4(1-2\sigma^2)^2$, this is just the scalar H integral, ${\rm H}[1, 1, 1, 1, 1, 1, 1, 0, 0]$; 
inspecting the coefficient of this integral in Table~\ref{coefficientsandintegralsTable} it is easy to see that this 
is a consequence of the other integrals and especially of their ${\cal O}(\epsilon)$ and 
${\cal O}(\epsilon^2)$ terms. The result \eqref{relint77bar} for the combination ${\bf 7}+\bar{\bf 7}$ agrees 
with \Eq{eq:PM_graphs} obtained by resumming the velocity expansion of these graphs.

To verify the uniqueness of the resummation we also evaluated
separately in \app{HxH} both ${\rm H}[1, 1, 1, 1, 1, 1, 1, 0, 0]$
and its image under the $p_2\leftrightarrow p_3$ remapping and without localization on the matter poles; 
the real part of their sum confirms both the resummation of the velocity
expansion and the result of the differential equation approach. The integral also has an imaginary part, which is consistent
with the possibility of a real graviton emission from graphs ${\bf 7}$ and
${\bar {\bf 7}}$. This imaginary part however does not contribute to the
conservative potential at the 3PM order.

\subsection{Mellin-Barnes Integration \label{sec:MB}}

In evaluating the integrals in the classical limit it is important to
have nontrivial cross-checks, based on completely different methods.
The Mellin-Barnes (MB) representation~\cite{Smirnov:2004ym} offers a
rather different approach compared to the nonrelativistic methods used
in \sect{sec:NRint}.  In particular, it does not rely on resumming in
velocity, and offers a useful check of that procedure.  Instead
of isolating regions of the momentum-space integration domain that
contribute to the classical limit, in this approach all regions of
loop momenta are integrated over and the extraction of classical
contributions happens at a later stage.  On the other hand, the method
can be more difficult to apply to more complicated integrals because
of issues related to the analytic continuation of the result to the physical
region~\cite{1607.07538, 1001.3243, 1609.09111}.

We use the Mellin-Barnes representation to explore the properties of the master
integrals resulting from the IBP system as well as for their
evaluation, especially for Euclidean kinematics.
In this method one first Feynman parametrizes the loop integral,
and uses the MB representation to
evaluate the Feynman parameter integrals resulting from the
standard evaluation of the loop momentum integrals.
The main identity in this method, which may be applied iteratively for 
more complicated denominator linear functions, is
\begin{equation}
\frac{1}{(a+b)^\nu}=\frac{1}{2\pi i}\int_{\beta-i\infty}^{\beta+i\infty}dw\,
\frac{\Gamma(-w)\Gamma(w+\nu)}{\Gamma(\nu)}\,
\frac{a^w}{b^{\nu+w}} \,,
\label{MBparametrization}
\end{equation}
where the integral over the Mellin-Barnes parameter $w$ is along a contour which is parallel to the imaginary axis and crosses the real axis at $-\text{Re}(\nu)< \beta < 0$. 
After applying repeatedly Eq.~\eqref{MBparametrization}, all integrals over Feynman parameters can be evaluated in terms of 
Euler gamma functions whose arguments are linear combinations of the MB parameters, the exponents of the propagators and the dimensional regulator $D=4-2\epsilon$.
A Feynman integral can have multiple MB representations; it is of course natural to choose the simplest one. Some amount of trial is necessary
to find it, especially for nonplanar integrals. 
Rather than constructing MB representations for one integral at a
time, it is convenient to derive them for classes of integrals
corresponding to a fixed graph topology but with arbitrary exponents
for each propagator, as e.g.\ in \eqn{allHints}. Apart from the
obvious possibility of reusing a once-derived good parametrization for
any choice of exponents (the limit of vanishing exponents must be
treated carefully), it provides a prescription for choosing the real
part of the integration contours such that the poles due to factors of
$\Gamma(A-w)$ are to the right of the $w$-contour while those due to
factors of $\Gamma(A+w)$ are to the left of that contour. Here $A$
stands for some (linear) combination of the other MB parameters,
dimensional regulator and propagator exponents and $\epsilon$ is fixed
in the neighborhood of some suitable negative value. If $\epsilon$ can
be chosen from the beginning in the neighborhood of the origin, then
the integral is finite.
An algorithm for the construction of MB representation of Feynman integrals is implemented in the Mathematica 
package {\tt AMBRE}~\cite{Gluza:2007rt}. See also Refs.~\cite{Smirnov:2004ym, hep-ph/9905323, hep-ph/9909506, hep-ph/0305142, hep-ph/0406053, hep-th/0505205} for further details on constructing MB representations 
and many examples. A simple example is the one-loop triangle integral. Taking all momenta to be outgoing, it is easy to find that
\begin{eqnarray}
I_{\rm T} &=&\frac{i}{(4\pi)^{D/2}} \int\frac{d^D \ell}{i\pi^{D/2}} \frac{1}{\ell^2  ( (\ell+p_1)^2-m_1^2) (\ell+q)^2} \nn
\\
&=& -\frac{i}{(4\pi)^{D/2}} \frac{1}{ (-t)^{1 + \epsilon}}\int_{{\cal C}} \frac{dz}{2\pi i}
\left(\frac{-t}{m_1^2}\right)^{-z}
\frac{  \Gamma(-\epsilon - z)^2 \Gamma(-z) \Gamma(1 + \epsilon + z) \Gamma(1 + 2 z)}{\Gamma(1 - 2 \epsilon)}
 \,,
\label{MBonelooptriangle}
\end{eqnarray}
where ${\cal C}$ is a contour parallel to the imaginary axis, which crosses the real axis at $z_0 \simeq -0.162$ and $z_0<\epsilon< 0$.

Once an MB representation of a Feynman integral is constructed, the next step is to construct its small $\epsilon$ expansion; the coefficients
of the various powers of $\epsilon$ will be integrals depending solely on external momenta so they could potentially be amenable to numerical 
evaluation (in the appropriate momentum region). Analytically, they are naturally represented as sums over the residues of the integrand, 
i.e.\ as infinite sums over various powers of ratios of external momentum invariants and masses. Residues may also exhibit logarithmic dependence on external momentum invariants.
For our purpose of extracting the classical term of an amplitude, which amounts to picking out a specific power of the $t=s_{23}=-{\bm q}^2$ 
Mandelstam invariant, this representation is particularly useful. 

Therefore, one must first analytically continue $\epsilon$ from the negative value used 
to fix the integration contours to the neighborhood of the origin, while keeping the integration contours fixed.
To do so one must carefully follow the poles of the integrand as $\epsilon$ increases; whenever a pole crosses an integration contour its residue 
must be separated and added/subtracted and subsequently treated recursively as $\epsilon$ is further increased.  
The construction of contours and the analytic continuation $\epsilon\rightarrow 0$ has been automated in the Mathematica packages {\tt MB.m}~\cite{Czakon:2005rk} and {\tt MBresolve.m}~\cite{Smirnov:2009up}.
Continuing the example of the one-loop triangle integral \eqref{MBonelooptriangle} and closing the contour either to the left or to the right,
it is then easy to see that, as $\epsilon\rightarrow 0$, no pole of the $\Gamma$ functions crosses the integration contour. Thus, we can 
simply put $\epsilon=0$ in the integrand to obtain,
\begin{eqnarray}
I_{\rm T}
&=& \frac{i}{(4\pi)^{D/2}} \frac{1}{ t}\int_{{\cal C}} \frac{dz}{2\pi i}
\left(\frac{-t}{m_1^2}\right)^{-z}
{  \Gamma( - z)^3  \Gamma(1  + z) \Gamma(1 + 2 z)} \,.
\label{epEQ0Triangle}
\end{eqnarray}
We are, of course, simply recovering the fact that the massive triangle integral we are interested is IR-finite. 

The MB representation techniques have proven to be extremely efficient, leading to the evaluation of a large variety of  massive and massless, planar 
and nonplanar Feynman integrals, at two loops and beyond, some exhibiting infrared singularities (see e.g.\ Ref.~\cite{hep-ph/9905323, hep-ph/9909506, hep-ph/0305142, hep-ph/0406053, hep-th/0505205}).
For our purpose however, it is, of course, more convenient to extract the desired terms in the small-$q$ expansion {\it before} the Mellin-Barnes integrals
fully evaluated rather than afterwards. This amounts to organizing the poles of the integrand and picking out only the ones that contribute 
to the desired order. This step typically leads to localization of some---sometimes all---of the MB integrals. The expansion in small momentum 
invariants and mass parameters has been automated in the Mathematica package {\tt MBAsymptotics.m}\footnote{ https://mbtools.hepforge.org/}. 
For the case of the one-loop triangle integral example, the small $q$ expansion can be constructed by inspection: we need to pick the poles of the 
integrand for which $z$ is negative. They are the poles of the product $ \Gamma(1  + z) \Gamma(1 + 2 z)$ in Eq.~\eqref{epEQ0Triangle},
so they occur at negative integers and half-integers, the first of which is at $z=-1/2$ and corresponds to the classical part of the one-loop 
amplitude:
\begin{eqnarray}
I_{\rm T}
&=& -\frac{i}{32 m_1 |{\bm q}|}  + {\cal O}(\ln({\bm q}^2), ({\bm q}^2)^0) \,.
\label{epEQ0Triangle2}
\end{eqnarray}
Thus, we reproduce Eq.~\eqref{onelooptriangle}.
Repeating the same steps for the box integral we find that, through ${\cal O}(\epsilon^0)$ the dependence on ${\bm q}^2$ and $p_1\cdot p_2 = m_1 m_2 \sigma$ factorizes as
\begin{equation}
I_{\rm B} = \frac{1}{\epsilon}\frac{1}{({\bm q}^2)^{1+\epsilon}}f(\sigma) \,,
\end{equation}
and therefore this integral does not have a contribution to the
classical potential, as discussed in \sect{1loopbox}.  A factorization
of a similar type also occurs for the two-loop H integral and can be
proven by similar means. 

While, in some cases, upon picking out the relevant classical terms
all MB integrals localize, in most instances some are left to be
evaluated and lead to nontrivial functions of the remaining momentum
invariants and masses. They may either be evaluated analytically or
numerically. A useful automated package that converts MB integrals into infinite sums and bypasses case-by-case analysis
is {\tt MBsums.m}~\cite{Ochman:2015fho}.
An interesting feature of these sums is that they are typically
convergent for Euclidean momenta, when all momentum invariants are
negative (for our mostly-minus metric convention). To correctly carry
out the analytic continuation to the physical region it is necessary
to track the causal $i\epsilon$ through the MB parametrization.
The numerical evaluation of MB integrals has similar properties: they
are convergent with the standard choice of contours described above
only for Euclidean momenta, as originally pointed out in
Ref.~\cite{Czakon:2005rk}. For one-dimensional MB integrals a
procedure for deforming the integration contours to obtain convergence
for physical momenta was devised in Ref.~\cite{1607.07538, 1001.3243, 1609.09111}. For
general integrals, finding contours that guarantee convergence in the
physical region remains an open problem.

We have used the MB-based integration methods to explore the structure
of the classical limit and verify the existence of classical
contributions of various one- and two-loop integrals as well as to
evaluate those whose classical contributions have none or one MB
integrals. Among them are the integrals appearing in the complete
one-loop amplitude in \sect{MassSingularitySubsection}, the
integral discussed in \sect{sec:TT} and other integrals with triangle
topology.

For example, we can construct a three-dimensional MB representation for the double triangle integral in \sect{sec:TT}, shown in \fig{fig:NRint_W}:
\begin{align}
I_{\rm TT}
 &= - \frac{1}{(4\pi)^4}\frac{1}{m_2}^2\int_{\cal C}\frac{dz_1dz_2dz_3}{(2\pi i)^3} \left(\frac{m_1^2}{-t}\right)^{1+z_1}
  \Gamma(-z_1) \Gamma(-z_3) \Gamma(1 + z_1) \Gamma(1 + z_2) 
  \cr
 & \qquad\qquad\qquad
 \times \frac{ \Gamma(-1 - z_{12}) \Gamma(z_1 - z_{23}) \Gamma(1 + z_{12} - z_3) \Gamma^2(z_3 -z_1)
          \Gamma(2 + z_{123})}{\Gamma(2 + z_{12} - z_3) \Gamma(1 - z_1 + z_3)} \,,
\end{align}   
where $z_{ij\dots} = z_i+z_j+\dots$ and the contour ${\cal C}$ crosses the real axis at  $z_1=-64/101$, $z_2=-29/51$ and
$z_3 = -22/51$. 
Upon extracting the leading term with logarithmic dependence on $t$,
two of the integrals localize. The remaining integral can be evaluated
through Cauchy's theorem and reproduces the result in
Eq.~\eqref{eq:integrated_DT}.

\begin{figure}
\begin{center}
\includegraphics[scale=.5]{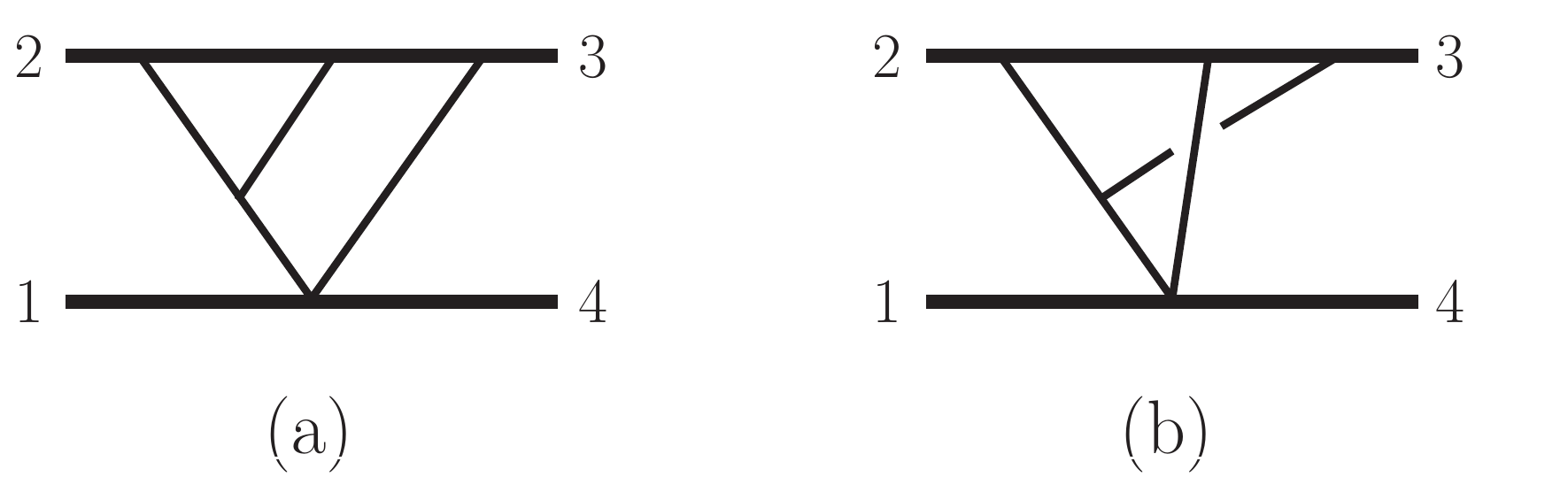}
\end{center}
\vspace*{-0.3cm}
\caption{\small Planar and nonplanar triangle integrals which appear in the cuts in \fig{TwoLoopCutsFigure}(a).
\label{VV'}}
\end{figure}

Similarly, following the steps described above we can construct MB representations which depend on three parameters for the 
integral in \fig{VV'}(a) and eight parameters for the integral in \fig{VV'}(b). Upon specification of contours however 
they become more complicated, involving several integrals of lower dimension which are needed for the contours to take the 
standard straight line form.  We do not include them explicitly here.
Despite these superficial differences, it turns out that the two integrals have identical leading $\ln(-t)$-dependent term: 
\begin{equation}
I_{\text{Fig.}\;\ref{VV'}\text{(a)}}\Big|_{\ln (-t) \; \text{term}} =I_{\text{Fig.}\;\ref{VV'}\text{(b)}}\Big|_{\ln (-t) \; \text{term}}  
= - \left(\frac{i}{(4\pi)^2}\right)^2\frac{\ln (-t)}{ 2 m_2^2  t} 
\int_{\cal C} \frac{dz}{2\pi i} \frac{\Gamma(-1 - 2 z) \Gamma^2(-z) \Gamma^3(1 + z) }{ \Gamma(-2 z) \Gamma(2 + z) } \,,
\end{equation} 
where the contour ${\cal C}$ crosses the real axis at $ z_0 = -0.897$.
Closing the contour to the right picks up poles at $z=-1/2$ and at all non-negative integers; the corresponding residues are
\begin{equation}
\text{Res}_{z=-1/2} = -\pi^2
\,, \hskip 1.5 cm 
\text{Res}_{z=n\ge 0 \cap \mathbb{Z}}= \frac{3 + 4 n}{(1+n)^2(1+2n)^2} \,,
\end{equation}
and, once resummed, they give
\begin{equation}
I_{\text{Fig.}\;\ref{VV'}\text{(a)}}\Big|_{\ln {\bm q}^2 \text{\ term}}  = -\frac{1}{768 \pi^2} \frac{\ln {\bm q}^2}{m_2^2 \, {\bm q}^2}\, .
\end{equation}

As a further example of application of the MB integration techniques we revisit briefly the integrals $ {\rm H}[1,1,0,1,1,1,1,0,0]$ and 
$ {\rm H}[1,1,0,0,1,1,1,0,0]$, defined in \eqn{allHints}. We discussed them from the point of view of the differential equation for integrals related 
to the H graph with two cut matter  propagators and restricted to the leading small-$t$ logarithmic term and found that, if the latter is chosen 
as master integral, the former can be reduced. 
It is not difficult to derive a 7-dimensional MB representation for $ {\rm H}[1,1,0,1,1,1,1,0,0] $ 
and a 5-dimensional MB representation for $ {\rm H}[1,1,0,0,1,1,1,0,0]$. 
Extracting the leading $\ln( -t)=\ln {\bm q}^2$ coefficients we find that both integrals are proportional to 
 \begin{equation}
I = -\frac{\ln(-t) }{(4\pi)^4}  \int_{\cal C} \frac{dz_1 dz_2}{(2\pi i)^2}\left(\frac{-\tau_{12}}{ m_1 m_2}\right)^{z_2}
\frac{\Gamma^3(-z_1) \Gamma(1 + z_1) \Gamma((1+z_2)/2 + z_1 ) \Gamma(-z_2) \Gamma((1 + z_2)/2) }{ 4 m_1 m_2 \; \Gamma(1 - z_1)}\,,
   \end{equation}
with the contour ${\cal C}$  crossing the real axis at  $z_1 = -0.078$ and $z_2 = -0.228$.\footnote{We note that this integral is well-defined 
for Euclidean momenta, where $\tau_{12} = 2p_1\cdot p_2 <0$. To analytically continue it to the physical region, with $\tau_{12}>0$,
it is necessary to track the causal $i\epsilon$ prescription through the MB representation.}
More specifically
 \begin{equation}
I =   {\rm H}[1,1,0,0,1,1,1,0,0]\Big|_{\ln(-t) \text{ term}}  = 
-t  {\rm H}[1,1,0,1,1,1,1,0,0]\Big|_{\ln(-t) \text{ term}}  \,,
 \end{equation}
 thus recovering the result of the differential equation approach. We note that the argument here shows that $ {\rm H}[1,1,0,1,1,1,1,0,0]$ and 
$ {\rm H}[1,1,0,0,1,1,1,0,0]$ obey this relation even without localization on matter poles.

While, as mentioned above, it is difficult to numerically evaluate the MB integrals for physical momenta, other approaches are up 
for this task.  An example is the sector decomposition strategy, as implemented in the computer codes
{\tt FIESTA}~\cite{0807.4129, 1511.03614} and {\tt SecDec}~\cite{Borowka:2012ii}.  We used the former to verify the result of the 
analytic result for the scalar H integral in \app{HxH}.

In \sect{sec:graph7} we have noticed that all integrals related to \fig{Hfig} by collapse of propagators enter the classical limit
of the amplitude only in the combination symmetrized under the remapping $p_2\leftrightarrow p_3$. As described in Ref.~\cite{StermanEikonal}, 
in the classical limit and to leading order in the small-$t$ expansion, the matter propagators organize as 
\begin{align}
\frac{1}{2l_4\cdot p_2 + i\epsilon} + \frac{1}{2l_4\cdot p_3 + i\epsilon} \simeq \frac{1}{2l_4\cdot p_2 + i\epsilon} - \frac{1}{2l_4\cdot p_2 - i\epsilon}
=-2\pi i \delta(2l_4\cdot p_2) \,,
\cr
\frac{1}{2l_1\cdot p_1 + i\epsilon} + \frac{1}{2l_1\cdot p_4 + i\epsilon} \simeq \frac{1}{2l_1\cdot p_1 + i\epsilon} - \frac{1}{2l_1\cdot p_1 - i\epsilon}
=-2\pi i \delta(2l_1\cdot p_1) \,.
\end{align}
The arguments of the two Dirac $\delta$-functions are nothing but the on-shell conditions for the two internal matter lines in the classical limit 
$|l_i^2| \sim -t \ll p_i^2$.
Thus, the symmetrized combination of integrals, $I_i+ \bar I_i$, localizes on the matter poles and this combination can be compared to the results of the differential equations and nonrelativistic integration that enforce this condition.   
We have indeed verified that the result of the differential equation approach for the leading $\ln (-t) = \ln {\bm q}^2$ term of scalar integrals  
related to \fig{Hfig} by collapse of propagators (i.e. $ {\rm H}[1,1,1,1,1,1,1,0,0]$, $ {\rm H}[1,1,0,0,1,1,1,0,0]$, $ {\rm H}[1,1,0,1,1,1,1,0,0]$, etc.) 
is reproduced by  high-precision numerical integration. Thus, the results obtained by resumming the nonrelativistic velocity expansion  
in \sect{sec:NRint} and through differential equations in \sect{sec:diffEq} and their contribution to graphs {\bf 7} and ${\bar {\bf 7}}$ in \sect{sec:graph7} are complete.

\section{Scattering Amplitude}
\label{sec:scattering}

In \sect{TwoLoopSection}, we used the double copy and generalized unitarity to construct the integrands denoted as graphs $\bm 1 - \bm 8$, which are shown in \fig{MasterDiagsFigure}.
The integration method described in \sect{sec:NRint} ultimately trivializes all one and two loop integrals by reducing them to three-dimensional bubble integrals via nonrelativistic expansion. In \sect{sec:PNexpand} below, we present results for each diagram up to 7PN.

Of course the cost of analytic tractability is that the answer is
formally computed order by order in the nonrelativistic
expansion. However, if we retain terms to sufficiently high order in
velocity, then we can resum this expansion, yielding our putative all
orders in velocity expression for the PM amplitude. The resummation
can be reliably performed because relativity imposes strong
constraints on the structure of the series in two important
ways. First, the series encountered are due to relativistic invariants
whose structure is simple enough to identify order by order in
velocity. Second, there is a bound for the highest PN order at which
new relativistic structures can appear. Beyond this order velocity
corrections are solely from structures that are already present at
lower orders. The detailed argument is presented
in \sect{sec:PMexpand}.

We employ two methods for resummation. First is resummation by
inspection. As we will see below, the series encountered are
geometric, binomial, or arcsinh, where a pattern can be clearly
identified. These series are collected
in \app{appendix:series}. Second is resummation by ansatz. A natural
ansatz can be formed from the small basis of functions encountered in
the case of scalar two-loop integrals, multiplied by arbitrary
polynomials of the external four momenta. The free coefficients are
then uniquely fixed by comparing to the first several terms in the PN
expansion. Both methods yield the same result, and we discuss various
checks below. The final result for scattering amplitudes up to 3PM are
given in \Eq{eq:PM_amp}.

\subsection{Post-Newtonian Expansion}
\label{sec:PNexpand}
We can mechanically integrate graphs $\bm 1 - \bm 8$ using the methods of \sect{sec:NRint}. This involves expanding the integrands in the nonrelativistic limit, and we do this by assigning the scalings $p_1 \cdot p_2 \sim \rho^2\,, E_i \sim \rho\,, m_i \sim \rho$, and then expanding order by order in large $\rho$. This generally leads to much simpler expressions compared to expanding in large $m_{i}$ everywhere since functions of $p_1 \cdot p_2$ and $E_i$ are kept intact to all orders in velocity. Our results up to 7PN are
\begin{align}
{\bf 1}  &= - \frac{2 m_1^6 m_2^6 (1- 2\sigma^2)^3 }{ E^2} \int \frac{d^{D-1} \bm \ell_1 }{(2\pi)^{D-1}} \frac{d^{D-1} \bm \ell_2 }{(2\pi)^{D-1}}  \frac{1}{  \bm \ell_1 ^2 (  \bm \ell_2 - \bm \ell_1)^2 (\bm \ell_2 + \bm q)^2 ( \bm \ell_1^2 + 2 \bm p \bm \ell_1 ) ( \bm \ell_2^2 + 2 \bm p \bm \ell_2 )} 
 \,,   \nn\\[15pt]
\overline{\bf 1 } &= 0 \,,  \nn\\[15pt]
{\bf 2 } &= -{  m_1^2m_2^6 (2\sigma^2 - 1)   E \bm p^2 \over 32 E_2^3 }  \int \frac{d^{D-1} \bm \ell }{(2\pi)^{D-1}}  \frac{1}{  \bm \ell^2  | \bm \ell + \bm q| ( \bm \ell^2 + 2 \bm p \bm \ell ) }   \bigg[  1 + \frac{3 \bm p^2}{2E_2^2}    + \frac{15\bm p^4}{8E_2^4}    + \frac{35\bm p^6}{16E_2^6}    +\cdots  \bigg]  \nn\\
&\quad  + {m_1^2 m_2^5 ( m_2 + 4 m_1  \pp - 5 m_2 \pp^2 - 8  m_1  \pp^3  )  \over 192 \pi^2 E_2^2 }   \bigg[  1 + \frac{\bm p^2}{E_2^2}    + \frac{\bm p^4}{E_2^4}    + \frac{\bm p^6}{E_2^6}   + \frac{\bm p^8}{E_2^8}   + \frac{\bm p^{10}}{E_2^{10}}  +\cdots  \bigg] \nn\\
&\quad - {\pp^2 m_1^4 m_2^4 \over 12 \pi^2 E_1 E_2 }    \bigg[ 1 + {\left(E_1^3+ E_2^3 \right) \over 3 E_1^2 E_2^2 E } {\bm p}^2  + \frac{\left(E_1^5+E_2^5\right)}{5 E_1^4 E_2^4 E} {\bm p}^4 + \frac{\left(E_1^7+E_2^7\right)}{7 E_1^6 E_2^6 E} {\bm p}^6 + \frac{\left(E_1^9+E_2^9\right)}{9 E_1^8 E_2^8 E} {\bm p}^8 \nn\\
&\quad + \frac{\left(E_1^{11}+E_2^{11}\right)}{11 E_1^{10} E_2^{10} E} {\bm p}^{10} + \cdots \bigg]  \nn\\
&\quad + {m_1^4 m_2^2  E \bm p^2  \over 36 \pi^2 E_1^3 }   \bigg[ 1 + \frac{6\bm p^2}{5E_1^2} + \frac{9\bm p^4}{7E_1^4}  + \frac{4\bm p^6}{3E_1^6}  + \frac{15\bm p^8}{11E_1^8} + \cdots \bigg]  \nn\\
&\quad 
+ {m_1^4 m_2^2  \bm p^2  \over 36 \pi^2 E_1^2 }   \bigg[ 1 + \frac{3\bm p^2}{5E_1^2} + \frac{3\bm p^4}{7E_1^4} + \frac{\bm p^6}{3E_1^6}  + \frac{3\bm p^8}{11E_1^8} + \cdots \bigg] \nn\\
&\quad + { m_1^4 m_2^2 \bm p^2 \over 36 \pi^2 E_1 E_2 }  \bigg[ 1 + {\left(E_1^3+ E_2^3 \right) \over 5 E_1^2 E_2^2 E } {\bm p}^2  + \frac{3\left(E_1^5+E_2^5\right)}{35 E_1^4 E_2^4 E} {\bm p}^4 + \frac{\left(E_1^7+E_2^7\right)}{21 E_1^6 E_2^6 E} {\bm p}^6 + \frac{\left(E_1^9+E_2^9\right)}{33 E_1^8 E_2^8 E} {\bm p}^8 +  \cdots \bigg] \nn\\
&\quad + {m_1^2 m_2^6  (2\pp^2 - 1) E \over 768 \pi^2 E_2^3}  \bigg[1 + {7 \bm p^2 \over 2 E_2^2} + {51 \bm p^4 \over 8 E_2^4} + {151 \bm p^6 \over 16 E_2^6}  + {1615 \bm p^8 \over 128 E_2^8}  + {4065 \bm p^{10} \over 256 E_2^{10} } + \cdots \bigg]  \,, \nn\\[15pt]
\overline{\bf 2 } &= - {\pp m_1^3 m_2^5 (2\pp^2 -  1)   \over 24 \pi^2 E_2^2  } \bigg[  1 + \frac{\bm p^2}{E_2^2}    + \frac{\bm p^4}{E_2^4}    + \frac{\bm p^6}{E_2^6}  + \frac{\bm p^8}{E_2^8}   + \frac{\bm p^{10}}{E_2^{10}} +\cdots  \bigg] \nn\\
&\quad - {\pp^2 m_1^2 m_2^6 \over 64 \pi^2 E_2^2} \bigg[ 1 - \frac{\bm p^2}{3E_2^2} - \frac{5\bm p^4}{3E_2^4} - \frac{3\bm p^6}{E_2^6}- \frac{13 \bm p^8}{3 E_2^8}   - \frac{17 \bm p^{10}}{3 E_2^{10}} + \cdots \bigg] \nn\\ 
&\quad + {\pp m_1^2 m_2^6 (2 \pp E_1 E_2 - m_1 m_2) \over 96 \pi^2 E_2^4 } \bigg[  1 + \frac{2 \bm p^2}{E_2^2}    + \frac{3 \bm p^4}{E_2^4}    + \frac{4 \bm p^6}{E_2^6} + \frac{5\bm p^8}{E_2^8}   + \frac{6\bm p^{10}}{E_2^{10}} +\cdots  \bigg] \nn\\
&\quad  - {\pp^2 m_1^4 m_2^4 \over 24\pi^2 E_1 E_2}   \bigg[ 1 + {\left(E_1^3+ E_2^3 \right) \over 3 E_1^2 E_2^2 E } {\bm p}^2  + \frac{\left(E_1^5+E_2^5\right)}{5 E_1^4 E_2^4 E} {\bm p}^4 + \frac{\left(E_1^7+E_2^7\right)}{7 E_1^6 E_2^6 E} {\bm p}^6 + \frac{\left(E_1^9+E_2^9\right)}{9 E_1^8 E_2^8 E} {\bm p}^8 \nn\\
&\quad \hskip 3 cm  + \frac{\left(E_1^{11}+E_2^{11}\right)}{11 E_1^{10} E_2^{10} E} {\bm p}^{10} + \cdots \bigg] \nn\\
&\quad + {m_1^4 m_2^2 \bm p^2 \over 72 \pi^2 E_1^2}  \bigg[  1 + \frac{3\bm p^2}{5E_1^2}    + \frac{3\bm p^4}{7E_1^4}   + \frac{\bm p^6}{3E_1^6}  + \frac{3\bm p^8}{11E_1^8}  +\cdots  \bigg]  \nn\\
&\quad
+ { E m_1^4 m_2^2 \bm p^2 \over 72 \pi^2 E_1^3 } \bigg[  1 + \frac{6\bm p^2}{5E_1^2}    + \frac{9\bm p^4}{7E_1^4}  + \frac{4\bm p^6}{3E_1^6}  + \frac{15\bm p^8}{11E_1^8}   +\cdots  \bigg]  \nn\\
&\quad + { m_1^4 m_2^2 \bm p^2 \over 72 \pi^2 E_1 E_2 }  \bigg[ 1 + {\left(E_1^3+ E_2^3 \right) \over 5 E_1^2 E_2^2 E } {\bm p}^2  + \frac{3\left(E_1^5+E_2^5\right)}{35 E_1^4 E_2^4 E} {\bm p}^4 + \frac{\left(E_1^7+E_2^7\right)}{21 E_1^6 E_2^6 E} {\bm p}^6 + \frac{\left(E_1^9+E_2^9\right)}{33 E_1^8 E_2^8 E} {\bm p}^8 + \cdots \bigg] \,, \nn\\[15pt]
{\bf 3 } &= 0 \,, \nn\\[15pt]
%
%
{\bf 4 } &= {  (2\sigma^2 - 1) (4\sigma^2 - 1) m_1^4 m_2^6 \over 8 E_2 E }  \int \frac{d^{D-1} \bm \ell }{(2\pi)^{D-1}}  \frac{1}{  \bm \ell^2  | \bm \ell + \bm q| ( \bm \ell^2 + 2 \bm p \bm \ell ) }   \bigg[  1 + \frac{ \bm p^2}{2E_2^2}    + \frac{3\bm p^4}{8E_2^4}    + \frac{5\bm p^6}{16E_2^6}    +\cdots  \bigg]  \nn\\
&\quad + {\pp m_1^3 m_2^5 (4\pp^2  - 1) \over 48 \pi^2 E_2^2 } \bigg[  1 + \frac{\bm p^2}{E_2^2}    + \frac{\bm p^4}{E_2^4}    + \frac{\bm p^6}{E_2^6}  + \frac{\bm p^8}{E_2^8}   + \frac{\bm p^{10}}{E_2^{10}} +\cdots  \bigg]  \nn\\
&\quad - { m_1^4 m_2^6 (2\pp^2 - 1 )(4\pp^2 -1 ) \over 128 \pi^2 E_2^3 E } \bigg[1 + {5 \bm p^2 \over 4 E_2^2 } + {11 \bm p^4 \over 8 E_2^4} + {93  \bm p^6 \over 64 E_2^6} + {193 \bm p^8 \over 128 E_2^8} + {793 \bm p^{10} \over 512 E_2^{10}} +\cdots \bigg]+ \{ 1 \leftrightarrow 2 \}   \,, \nn\\[15pt]
\overline{\bf 4 } &= {\pp m_1^3 m_2^5 (4\pp^2 - 1) \over 24 \pi^2 E_2^2 }  \bigg[  1 + \frac{\bm p^2}{E_2^2}    + \frac{\bm p^4}{E_2^4}    + \frac{\bm p^6}{E_2^6} + \frac{\bm p^8}{E_2^8}   + \frac{\bm p^{10}}{E_2^{10}} +\cdots  \bigg]  + \{ 1 \leftrightarrow 2 \}  \, , \nn\\[15pt]
{\bf 5 } &= -{\pp m_1^2 m_2^5 ( 12 \pp^2 m_1 + \pp m_2 - 6 m_1)  \over 96 \pi ^2 E_2^2} \bigg[  1 + \frac{\bm p^2}{E_2^2}    + \frac{\bm p^4}{E_2^4}    + \frac{\bm p^6}{E_2^6}  + \frac{\bm p^8}{E_2^8}   + \frac{\bm p^{10}}{E_2^{10}}+\cdots  \bigg]   \nn\\
&\quad - {\bm p^2 \pp m_1 m_2^5 E^2 \over 48 \pi^2  E_2^4 } \bigg[  1 + \frac{2\bm p^2}{E_2^2}    + \frac{3 \bm p^4}{E_2^4}    + \frac{4 \bm p^6}{E_2^6}  + \frac{5\bm p^8}{E_2^8}  +\cdots  \bigg]    \,, \nn\\[15pt]
{\bf 6 } &= - {\pp^2 m_1^2 m_2^6 \over 24 \pi^2 E_2^2}  \bigg[  1 + \frac{\bm p^2}{E_2^2}    + \frac{\bm p^4}{E_2^4}    + \frac{\bm p^6}{E_2^6}  + \frac{\bm p^8}{E_2^8}   + \frac{\bm p^{10}}{E_2^{10}}+\cdots  \bigg]      \,, \nn\\[15pt]
%
%
\bf 7 & + \overline{\bf 7 } = {   (2\pp^2 - 1)^2  m_1^4 m_2^4 \over  16 \pi^2 E_1 E_2}  \bigg[ 1 + {\left(E_1^3+ E_2^3 \right) \over 3 E_1^2 E_2^2 E } {\bm p}^2  + \frac{\left(E_1^5+E_2^5\right)}{5 E_1^4 E_2^4 E} {\bm p}^4 + \frac{\left(E_1^7+E_2^7\right)}{7 E_1^6 E_2^6 E} {\bm p}^6 + \frac{\left(E_1^9+E_2^9\right)}{9 E_1^8 E_2^8 E} {\bm p}^8 \nn\\
&\quad + \frac{\left(E_1^{11}+E_2^{11}\right)}{11 E_1^{10} E_2^{10} E} {\bm p}^{10} + \cdots \bigg]\,,  \nn\\[15pt]
{\bf 8 } &=  0 \, .
\label{eq:PN_graphs2}
\end{align}
Here for every equation we have omitted a factor of  $\frac12 (8 \pi G)^3 \log \bm q^2$ for the finite terms and a factor of  $(8 \pi G)^3$ for the remaining infrared divergent integrals. We have not included a factor of $1/(4E_1 E_2)$ for nonrelativistic normalization. Moreover, for diagrams $\bf 4$ and $\overline {\bf 4}$, we denote by $\{ 1 \leftrightarrow 2 \}$ contributions obtained from swapping the mass and energy subscripts in all terms. The results above are for the individual diagrams $\bm 1 - \bm 8$, and do not include contributions from topologies related by flipping the diagram. Graphs $\overline {\bf1}, \overline {\bf 2}, \overline {\bf 4}, \overline {\bf7}$ are related to graphs $\bf{1}, \bf{2}, \bf{4}, \bf{7}$ by crossing, i.e. by swapping the external momenta $p_2 \leftrightarrow p_3$ or equivalently $p_1 \leftrightarrow p_4$. We refer the reader to Table~\ref{table:assembly}.

Note that judicious further expansion of factors of $E_i$ in some places can lead to even simpler results. In particular, series resummation often yields factors of $E_i$ and $m_i$ that cancel in the final result (see \app{appendix:series}). For example, in graph $\bf 6$, expanding the factor of $m_2^2/E_2^2$ in large $m_2$ cancels the geometric series in $\bm p^2 /E_2^2$, yielding a result that is exact at 2PN. This is expected since it follows from the topology of $\bf 6$ that its nonrelativistic expansion is in $|\bm q| /m_i$, and there are no velocity corrections from integration.

We see that graphs $\bf1 \,, \overline{\bf 1}\,,  \bf3$, and $\bf 8$ are zero except for a contribution from iterations of lower order processes which will cancel in the matching with the EFT.

\subsection{Post-Minkowskian Expansion}
\label{sec:PMexpand}

The results for graphs $\bm 1 - \bm 8$ above are expressed in terms of
series that have a substantially simple structure order by order in
$\bm p^2$. This simplicity is of course due to relativistic
invariance. However, even if it is trivial by inspection to resum all
orders in velocity, it is natural to be suspicious since any finite
number of terms in a generic series has no bearing on the full
function.  Nevertheless, apart from the verification of certain graphs
by fully relativistic methods, we believe that our resummation is
trustworthy due to the following argument.

The underlying worry about resummation is that extrapolation may fail
due to the appearance of new velocity-dependent structures at some
high order. Note, however, that the velocity expansion is not an
arbitrary series. Rather, it is strongly constrained by relativistic
invariance of the underlying theory and any new velocity-dependent
structure must originate from a new fully relativistic one.

There is a simple argument from dimensional analysis that forbids new
relativistic structures from appearing at arbitrarily high order in
the PN expansion.  First, recall that the two loop scattering
amplitude has mass dimension six. Since every graph has seven
propagators and the integration measure has mass dimension eight, the
numerator of the integrand must have mass dimension twelve.  If we
assume conservatively that the entire mass dimension of the numerator
is due to $\bm p$, then the numerator scales at most as $\bm v^{12}$
which, together with the $G^3$ factor, is of 8PN order. This logic
suggests that beyond 8PN no new velocity dependent structure will
appear and all velocity corrections are from the expansion of
structures that are already present at lower orders.

The 8PN bound argued above only uses dimensional analysis and can be
refined by noting that the mass dimension of the numerator cannot be
due entirely to $\bm p$. In particular, the numerator must provide
some dimensionful factors of $\bm q$ in order to produce a classical
contribution. We thus need to consider the numerator dependence on
factors of $p \cdot \ell$ and $\ell^2$, where $p$ is an external four
momentum and $\ell$ is a loop four momentum.  Other numerator factors
which are independent of the loop variables and depend only on $p_1
\cdot p_2$ and $\bm q^2$, are effectively constant and do not affect
the integration.  Consequently, these spectator factors can be
separated from the actual integration and do not produce new factors
of $\bm p$ upon integration.  Given the $\bm q$ and $\bm v$ scaling of
graviton momenta in the potential region discussed in
\sect{QuantumVsClassicalSection}, a generic combination $(p \cdot
\ell)^a (\ell^2)^b$ scales at most as $|{\bm q}|^{a+2b} |{\bm
  v}|^{a+2b}$. In deriving this we focused on the highest power of the
velocity as this automatically covers any weaker dependence on $|{\bm
  v}|$.
The upshot is that, for the non-spectator terms in the numerator, the scaling of the numerator in $v$ is equal to the scaling of the numerator in $|{\bm q}|$, thus lowering the bound by a factor of two.

Concretely, consider a diagram with $n_G$ graviton propagators and
$n_M$ matter propagators.  Graphs ${\bf 1} - {\bf 8}$ all have exactly
$n_G+n_M = 7$ propagators.  Using this, together with the fact that
the measure scales as $|{\bm q}|^8$ and the propagator denominators
scale as $|{\bm q}|^{2 n_G + n_M}$, we learn that the numerator must
scale at most as $|{\bm q}|^{n_G -1 }$.  Any higher $\bm q$ scaling
will produce a quantum correction.  Hence the numerator scales at most
as $|{\bm v}|^{n_G-1}$. Since the maximum number for $n_G$ in our set
of diagrams is five, we obtain a global bound at 4PN as the highest
order at which new $\bm p$-dependent structures can arise.

Still the above argument can be refined further if we consider
information about individual diagrams. For example, we know from the
possible relativistic invariants of diagrams $\bf 3$ and $\bf 6$ that
they are expansions in $\bm q/m$ and not $\bm p/m \sim \bm v$. These
diagrams are therefore exact at 2PN, and in fact the series in the
result for $\bf 6$ in \Eq{eq:PN_graphs2} cancels with the factor of
$m_2^2/E_2^2$. Another example is given by diagrams such as $\bf 1$,
$\bf 2$, and $\bf 4$, which have $n_M > 2$ matter propagators. For
these diagrams the numerator terms that contribute to the static or
2PN order scale as $|\bm p|^{n_M-2}$ in order to cancel the extra
factor of $|\bm p|^{2-n_M}$ from the matter propagators. This lowers
the bound to 2PN for diagrams with four matter propagators and to 3PN
for diagrams with three matter propagators. As a final example, we
know from the mechanics of two loop integration that diagram $\bf 8$
will not develop the required $\log \bm q^2$ structure that makes a
contribution at two loop order classical, and therefore it vanishes to
all orders.

The above arguments imply that all relativistic invariants are already manifest in the 7PN results in  \Eq{eq:PN_graphs2}, and hence that resummation can be reliably performed. For most of the series encountered, there is immediately an obvious pattern that can be extended naturally to all orders, while others require some sleuthing. Details are given in \app{appendix:series}. Upon resummation, we obtain the following all orders in velocity expressions:
\begin{align}
{\bf 1}  &=  - \frac{2 m_1^6 m_2^6 (1- 2\sigma^2)^3 }{ E^2} \int \frac{d^{D-1} \bm \ell_1 }{(2\pi)^{D-1}} \frac{d^{D-1} \bm \ell_2 }{(2\pi)^{D-1}}  \frac{1}{  \bm \ell_1 ^2 (  \bm \ell_2 - \bm \ell_1)^2 (\bm \ell_2 + \bm q)^2 ( \bm \ell_1^2 + 2 \bm p \bm \ell_1 ) ( \bm \ell_2^2 + 2 \bm p \bm \ell_2 )} 
 \,,   \nn\\[3pt]
\overline{\bf 1 } &= 0 \,,  \nn\\[3pt]
{\bf 2 } &=   -{  m_1^2m_2^3 (2\sigma^2 - 1)   E \bm p^2 \over 32 }  \int \frac{d^{D-1} \bm \ell }{(2\pi)^{D-1}}  \frac{1}{  \bm \ell^2  | \bm \ell + \bm q| ( \bm \ell^2 + 2 \bm p \bm \ell ) }  - {m_1^2 m_2^3 \pp (6 m_1 \pp^2 + 3\pp m_2 -11 m_1  ) \over 192 \pi^2} \nn\\
&\quad - {m_1^4 m_2^5 (\pp^2 -1)( 2\pp^2 -1) \over 256 \pi^2 E \bm p^2}   -{ m_1^3 m_2^3 (2\pp^2 + 1) \over 12 \pi^2}  { {\rm arcsinh}  \sqrt{ \sigma - 1 \over 2}   \over \sqrt{\sigma^2 -1 } } \,,  \nn\\[3pt]
\overline{\bf 2 } &= - {m_1^2 m_2^3 \pp  (12 \pp^2 m_1   + 3\pp m_2 -10 m_1 ) \over 192 \pi^2}  -{ m_1^3 m_2^3 (2\pp^2 + 1) \over 24 \pi^2}  {  {\rm arcsinh}  \sqrt{ \sigma - 1 \over 2}   \over \sqrt{\sigma^2 -1 } }   \,, \nn\\[3pt]
{\bf 3 } &= 0 \,, \nn\\[3pt]
{\bf 4 } &=    {  (2\sigma^2 - 1) (4\sigma^2 - 1) m_1^4 m_2^4 m \over 8 E }  \int \frac{d^{D-1} \bm \ell }{(2\pi)^{D-1}}  \frac{1}{  \bm \ell^2  | \bm \ell + \bm q| ( \bm \ell^2 + 2 \bm p \bm \ell ) }  +  { \pp m_1^3 m_2^3 (4\pp ^2 -1) \over 24 \pi^2}  \nn\\
&\quad - { m_1^4 m_2^4 (2\pp^2 - 1) (4\pp^2 - 1) (E-m) \over 64 \pi^2 E \bm p^2 } \,, \nn\\[3pt]
\overline{\bf 4 } &= { \pp m_1^3 m_2^3 (4\pp^2 -1) \over 12 \pi^2} \,,  \nn\\[3pt]
{\bf 5 } &=  -\frac{2 \pp m_1^2 m_2^3 \left(14  \pp^2 m_1 + \pp m_2 - 8 m_1 \right)}{192 \pi ^2}\,, \nn\\[3pt]
{\bf 6 } &= - {\pp^2 m_1^2 m_2^4 \over 24 \pi^2}  \,,   \nn\\[3pt]
{\bf 7 } & + \overline{\bf 7  } = {m_1^3 m_2^3 (2\pp^2 -  1)^2 \over   8 \pi^2  } {  {\rm arcsinh}  \sqrt{ \sigma - 1 \over 2}   \over \sqrt{\sigma^2 -1 } } \,, \nn\\[3pt]
{\bf 8 } &=  0\, ,
\label{eq:PM_graphs}
\end{align}
again omitting the same prefactors as in \Eq{eq:PN_graphs2}.  Note
that for diagrams $\bf 4$ and $\overline {\bf 4}$ the contributions
denoted by $\{ 1 \leftrightarrow 2 \}$ in \Eq{eq:PN_graphs2} are
included in \Eq{eq:PM_graphs}.  For some diagrams such as ${\bf 7} +
\overline{\bf 7}$, we have checked the result using relativistic
integration (see \sect{sec:Rint}).

The total amplitude is then obtained by summing these contributions and including the relabeling given in Table~\ref{table:assembly}. We present here our final expression for the 3PM scattering amplitude, ${\cal M}_3$, along with expressions for the 1PM and 2PM amplitudes, ${\cal M}_1$ and ${\cal M}_2$, for completeness:
\eq{
{\cal M}_1 &=- {4\pi G \nu^2 m^2 \over \gamma^2 \xi \bm q^2} (1-2\sigma^2)  \,, \\
{\cal M}_2 &= - {3\pi^2 G^2 \nu^2 m^3 \over 2 \gamma^2 \xi |\bm q| } (1- 5\sigma^2) +{32 \pi^2 G^2 \nu^4 m^5 (1-2\sigma^2)^2 \over  \gamma^3 \xi }   \int \frac{d^{D-1} \bm \ell_1 }{(2\pi)^{D-1}}  \frac{1}{  \bm \ell^2  (\bm \ell + \bm q)^2 ( \bm \ell^2 + 2 \bm p \bm \ell ) }   \,,  \\
{\cal M}_3 &=  \frac{\pi  G^3 \nu ^2 m^4  \log \bm q^2 }{6 \gamma ^2 \xi }  \bigg[   3-6\nu  + 206 \nu  \pp -54\pp^2 + 108 \nu \pp^2  + 4 \nu \pp^3  \\[3pt]
&\quad  - {48 \nu \left(3+12 \pp^2-  4 \pp^4  \right) {\rm arcsinh} \sqrt{\pp-1 \over 2} \over \sqrt{ \pp^2 - 1}} 
- {18 \nu \gamma \left( 1 - 2 \pp^2  \right) \left( 1 - 5 \pp^2 \right)  \over  \left( 1+ \gamma \right) \left(1 +  \pp \right) }\bigg] \\[3pt]
&\quad + {8\pi^3 G^3 \nu^4 m^6  \over \gamma^4 \xi } \bigg[ {3 \gamma  \left( 1 - 2 \pp^2  \right) \left( 1 - 5 \pp^2 \right)  } \int \frac{d^{D-1} \bm \ell }{(2\pi)^{D-1}}  \frac{1}{  \bm \ell^2  | \bm \ell + \bm q| ( \bm \ell^2 + 2 \bm p \bm \ell ) }    \\
&\quad -{32 m^2 \nu^2 \left( 1 - 2 \pp^2  \right)^3}   \int \frac{d^{D-1} \bm \ell_1 }{(2\pi)^{D-1}} \frac{d^{D-1} \bm \ell_2 }{(2\pi)^{D-1}}  \frac{1}{  \bm \ell_1 ^2 (  \bm \ell_2 - \bm \ell_1)^2 (\bm \ell_2 + \bm q)^2 ( \bm \ell_1^2 + 2 \bm p \bm \ell_1 ) ( \bm \ell_2^2 + 2 \bm p \bm \ell_2 )} \bigg]  \,.
}{eq:PM_amp} 
We remind the reader that here we use center-of-mass coordinates where the incoming and outgoing particle momenta are $\pm \bm p$ and $\pm (\bm p-\bm q)$, respectively. We have included the overall normalization factor $(8\pi G)^3$ as well as the nonrelativistic normalization factor $1/4 E_1 E_2$, where $E_{1,2} = \sqrt{\bm p^2 + m_{1,2}^2}$. We define the total mass $m= m_1 +m_2$, the symmetric mass ratio $\nu = m_1 m_2 /m^2$, the total energy $E = E_1 +E_2$, the symmetric energy ratio $\xi = E_1 E_2 / E^2$, the energy-mass ratio $\gamma = E/m$, and the relativistic kinematic invariant $\pp = {p_1 \cdot p_2  \over m_1 m_2}$. 

The remaining integrals in \Eq{eq:PM_amp} are IR divergent and manifest the iterative structure through factors of the tree-level and one loop triangle scalar coefficients, given respectively by $(1-2\sigma^2)$ and $(1-5\sigma^2)$. The integrals in ${\cal M}_2$ and ${\cal M}_3$ that are proportional to $(1-2\sigma^2)^2$ and $(1-2\sigma^2)^3$ represent the double and triple iterations of tree-level exchange. The integral in ${\cal M}_3$ that is proportional to $(1-2\sigma^2)(1-5\sigma^2)$ represents the iteration of the tree-level exchange and the one loop triangle. Note that the $(1-5\sigma^2)$ factor arises from combining the IR divergent integrals in graphs $\bf 2$ and $\bf 4$. We have also computed these IR pieces by using factorization properties of the full amplitude in transverse impact parameter space. As we will see in the next section, these IR artifacts will cancel with similar contributions from the EFT. This is of course guaranteed because the two theories describe the same infrared dynamics, but nonetheless provides a nontrivial check.

As an alternative to resummation by inspection, we can build an ansatz from the small number of simple functional basis elements encountered in the case of scalar two loop integrals discussed in \sect{sec:NRint}. The basis functions correspond to the primitive topologies given by the scalar double triangle diagram, the scalar box triangle diagram, and the scalar double triangle prime diagram, which are respectively given by
\eq{
{\cal O}_1  =1\,, \quad 
{\cal O}_2  =\frac{E-m}{E \bm p^2} \,, \quad
{\cal O}_3 = {1 \over E |\bm p|} \left(  {\rm arcsinh} {|\bm p| \over m_1} +   {\rm arcsinh}  {|\bm p| \over m_2}   \right)  \,.
}{}
The ansatz is then
\eq{
{\cal M}_3^{\rm ansatz}  = \tau_1 {\cal O}_1 + \tau_2 {\cal O}_2 + \tau_3 {\cal O}_3 \,, 
}{}
where the $\tau_i$ are arbitrary polynomial functions of $m_1^2$, $m_2^2$, and $\sigma$ with correct overall mass dimension. Performing the fit to explicit results, we find that these functions are already uniquely fixed by the 6PN results. Consequently, the 7PN terms of the explicit calculation provide a nontrivial verification of the resummed amplitude in \Eq{eq:PM_amp}. The uniqueness and the simplicity of the all-orders result in \Eq{eq:PM_amp} are remarkable.

A feature of the 3PM amplitude in \eqn{eq:PM_amp} is that it contains a mass singularity due to the arcsinh factor. This factor, which is proportional to 
the sum of particle rapidities, ${\rm arctanh}\, |{{\bm p}}|/E_{1,2}$, diverges as the two masses become simultaneously small. 
Taking $m_1 \sim m_2 \rightarrow 0$ in the 3PM amplitude we find a logarithmic singularity,
\eq{
{\cal M}_3 \rightarrow - 128 \pi {G^3}  {\bm p}^4\log{\bm q}^2\, 
\log\Bigl(\frac{m_1 m_2}{4{\bm p}^2}\Bigr) + \cdots  = - 8 \pi {G^3}  s^2 \log(-t)\,
\log\Bigl(\frac{m_1 m_2}{s}\Bigr) + \cdots \, ,
}{eq:AmplitudeMassSingularity}
where we display only the singular term in the IR finite part.  At first sight this might
seem surprising because scattering amplitudes in quantum gravity do
not have such singularities~\cite{Weinberg:1965nx, 1101.1524, Akhoury:2011kq}.  However,
as we explain in \sect{MassSingularitySubsection}, this singularity arises in
the classical potential region, and an interchange of limits of small
mass with small momentum transfer prevents the cancellation of mass
singularities that would occur in the quantum theory.

\section{Effective Field Theory}
\label{sec:EFT}

The culmination of the previous sections is the 3PM scattering amplitude shown in \Eq{eq:PM_amp}.  Employing the EFT framework described in Ref.~\cite{CliffIraMikhailClassical} we can now translate the 1PM, 2PM and 3PM amplitudes into the classical Hamiltonian describing the 
conservative dynamics of a compact binary system.  
Our approach is a straightforward EFT matching calculation.   First, we compute the two particle scattering amplitudes in the EFT mediated by a generic 
classical Hamiltonian compatible with the potential in Eq.~\eqref{eq:VPR}. Next, the free coefficients of the Hamiltonian are fixed by matching the EFT 
amplitudes to the amplitudes of the full theory.   In the present section we briefly recall the details of the EFT and the matching procedure.

As we will see, the EFT framework described here offers a useful tool for checking that a given potential is consistent, or that two potentials in different gauges are physically equivalent, by computing on-shell amplitudes, which are gauge invariant and independent of field variables. In particular, in \sect{sec:checks_scattering}, we take known potentials, such as the 4PN potential from literature and the Schwarzschild potential, and verify that the resulting amplitudes agree with \Eq{eq:PM_amp} in the overlap regions. An alternative approach, based on the EOB framework, was described in Ref.~\cite{DamourTwoLoop}.

\subsection{Formalism}
The effective theory describes two nonrelativistic scalar fields $\phi_1$ and $\phi_2$, with masses $m_1$ and $m_2$, interacting through a long distance potential. The Lagrangian for this system is
\eq{
{\cal L} =& \int {d^{D-1} \bm k \over (2\pi)^{D-1}}   \, \phi_1^\dagger(-\bm k) \left(i\partial_t - \sqrt{\bm k^2 + m_1^2}\right) \phi_1(\bm k) + \int  {d^{D-1} \bm k \over (2\pi)^{D-1}}    \, \phi_2^\dagger(-\bm k) \left(i\partial_t - \sqrt{\bm k^2 + m_2^2}\right) \phi_2(\bm k) \\
&
-\int  {d^{D-1} \bm k \over (2\pi)^{D-1}}   {d^{D-1} \bm k' \over (2\pi)^{D-1}}   \, V(\bm k ,\bm k') \, \phi_1^\dagger(\bm k') \phi_1(\bm k) \phi_2^\dagger(-\bm k') \phi_2(-\bm k) \,,
}{eq:eftL}
where we work in the center of mass frame. The potential is written as a contact operator since it is generated by integrating out potential modes which are off shell. The function $V(\bm k ,\bm k')$ is parameterized by an ansatz of real rotational invariants composed of $\bm k$ and $\bm k'$. We choose a field basis where $V(\bm k ,\bm k')$ depends only on the combination $\bm k^2 +\bm k'^2 $ and the momentum transfer $|\bm k - \bm k'|$. Other choices involve combinations that vanish on shell, such as $\bm k^2 - \bm k'^2 $ and $k_0 - k_0'$, and are related to our choice by field redefinition. The explicit form of our ansatz is
\eq{
V(\bm k,\bm k') &=\sum_{n=1}^\infty { (G/2)^n (4\pi)^{( D-1)/2} \over |\bm k - \bm k'|^{D-1-n}}      \frac{\Gamma\left[ (D-1-n)/2 \right] }{\Gamma\left[ n/2 \right]} \, c_n \hspace{-0.12cm} \left( {\bm k^2 +\bm k'^2 \over 2} \right) \\[3pt]
&= \frac{4 \pi G }{|\bm k - \bm k'|^2}\, c_1\hspace{-0.12cm}\left( {\bm k^2 +\bm k'^2 \over 2} \right) + \frac{2 \pi^2 G^2 }{|\bm k - \bm k'|} \, c_2\hspace{-0.12cm}\left( {\bm k^2 +\bm k'^2 \over 2} \right) -2 \pi G^3 \log (\bm k - \bm k')^2  \, c_3\hspace{-0.12cm}\left( {\bm k^2 +\bm k'^2 \over 2} \right) + \cdots \, ,
}{eq:VPQ}
where we work in $D = 4- 2\epsilon$ dimensions, the coefficient functions $c_n$ contain all orders in the velocity expansion, and $n$ labels the order 
in the PM expansion at which a coefficient becomes relevant. In the second line, the ellipsis contains terms of higher order in the PM expansion, and we have dropped terms that are independent of the momentum transfer since these are contact interactions that do not affect long distance scattering. Note that the normalization adopted here is different from the one in Ref.~\cite{CliffIraMikhailClassical}.

For off-shell kinematics, the potential $V(\bm k ,\bm k')$ includes quantum corrections that are proportional to off-shell matter propagators through the combination $\bm k^2 +\bm k'^2$. For on-shell kinematics, the potential can be taken to be purely classical. We have checked that including quantum mechanical corrections to the on-shell potential is not necessary for determining the classical dynamics. Such quantum corrections would lead to classical terms in the amplitude that are, however, infrared divergent and would simply cancel in the matching. For the application of this EFT for determining the full quantum potential and for a discussion of matching with off-shell external kinematics see Ref.~\cite{Manohar:2000hj}.

We can Fourier transform the momentum transfer $|\bm k - \bm k'|$ to its conjugate variable $\bm r$, the distance between the two particles in position space. This gives the classical conservative potential between two point particles in \Eq{eq:VPR},
where we have set $\bm k^2 = \bm k'^2 = \bm p^2$ in $c_n({\bm k^2 +\bm k'^2 \over 2})$ for on-shell kinematics. Note that our choice of field basis in momentum space corresponds to the so-called isotropic gauge in position space in which the invariant $\bm p \cdot \bm r$ does not appear. 

We identify the interaction potential of the scalar field theory to be the conservative potential for the binary system.  Let us elaborate on the reason for this equivalence.  The interaction vertex $V$ is defined in the Lagrangian ${\cal L}$ for a second quantized field theory describing two massive scalars.  We can mechanically Legendre transform to the Hamiltonian ${\cal H} = \Pi_1 \frac{\delta {\cal L}}{\delta \phi_1}+ \Pi_2 \frac{\delta {\cal L}}{\delta \phi_2} - {\cal L}$ of the second quantized field theory, where $\Pi_i = \frac{\delta {\cal L}}{\delta \dot \phi_i}$.  Since $V$ has no dependence on $\dot \phi_i$, it does not affect the conjugate momentum in the second quantized theory.  Consequently, the interaction terms from $V$ simply flip sign when going to the Hamiltonian formalism.  We then truncate the second quantized Hamiltonian down to the two-particle subspace spanned by the massive scalars, as required by the nonrelativistic limit.  The resulting quantity is equal to the first quantized Hamiltonian for the two-particle system, whose corresponding interaction is by definition the conservative potential between classical point particles.

\subsection{Scattering Amplitude}

\begin{figure}[t]
\begin{center}
\includegraphics[scale=.7]{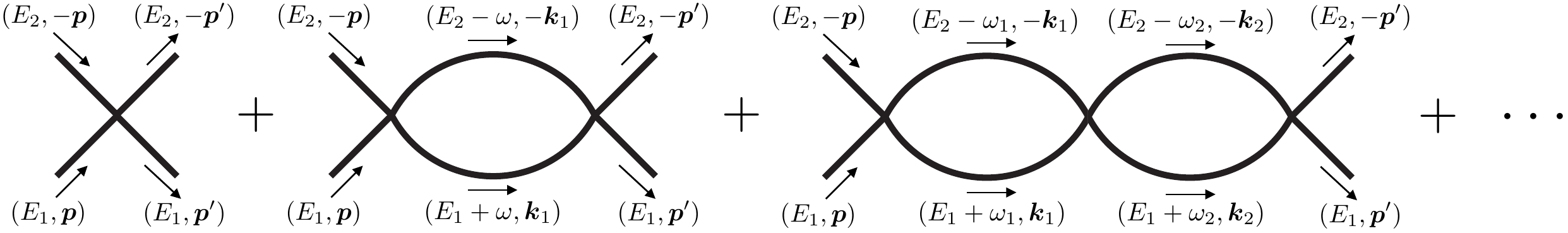}
\end{center}
\vspace*{-0.5cm}
\caption{\small The EFT scattering amplitude is given by a sum of bubble diagrams. }\label{fig:EFT_amp}
\end{figure}

The EFT amplitude is given by the sum of iterated bubbles shown in \fig{fig:EFT_amp}.  Note that particle number is conserved in such diagrams; topologies involving pair production of the massive states is kinematically forbidden in the classical nonrelativistic limit.
Since the interaction $V(\bm k ,\bm k')$ is not homogeneous in the coupling constant $G$,  the $n$PM amplitude, which scales as ${\cal O}(G^n)$, receives contributions from all diagrams with at most $n-1$ bubbles.

The Feynman rules for the EFT follow from \Eq{eq:VPQ} in the usual way. For the propagator of the scalars, it is convenient to define a ``two-body'' propagator given by the product of the two nonrelativistic propagators comprising a bubble, integrated over the energy component of the loop momentum:
\eq{
i \Delta({\bm k}) &= \int \frac{d \omega}{2\pi} \frac{i}{\omega-\sqrt{{\bm k}^2 +m_1^2}}  \frac{i}{E-\omega -\sqrt{{\bm k}^2 +m_2^2}} 
= \frac{i}{E - \sqrt{{\bm k}^2 +m_1^2}-\sqrt{{\bm k}^2 +m_2^2}}\, ,
}{}
where the second equality is obtained by closing the $\omega$ contour either upwards or downwards in the complex plane. We are able to perform this integration over the energy component of the loop momentum since the interaction vertex in \Eq{eq:VPQ} has no energy dependence. This leaves us with integration over the spatial component of the loop momenta.

Using the Feynman rules described above, we compute the $n$PM amplitude in the effective theory,
\eq{
{\cal M}^{\rm EFT}_{n} = - V(\bm p, \bm p') - \sum_{n_L=1}^{n-1} \left[   \prod_{i=1}^{n_L} \int    {d^{D-1} \bm k_i \over (2\pi)^{D-1} } \right]   V(\bm p, \bm k_1) \Delta(\bm k_1) \cdots \Delta(\bm k_{n_L})  V(\bm k_{n_L}, \bm p')\, ,
}{eq:Lbubble1}
where the second term involving the sum contributes for $n >1$. The loop momenta ${\bm k}_i$ for $i=1$ to $n_L$ are the momenta flowing through the internal matter lines as shown in \fig{fig:EFT_amp}. We can simplify this integral by expanding the integrand in the classical limit, taking the momentum transfer at each interaction point to scale as $|\bm k_{i} - \bm k_{i+1}| \sim \bm q$, while the degree to which each two-body propagator is off-shell scales as $\bm k_i^2 - \bm p^2 \sim \bm q$. Expanding the amplitude through the classical order we find 
\eq{
{\cal M}^{\rm EFT}_{n} = - V(\bm p, \bm p') - \sum_{n_L=1}^{n-1}   \left[  \prod_{i=1}^{n_L} \int  {d^{D-1} \bm k_i \over (2\pi)^{D-1} } \right]   \left[ \prod_{i=0}^{n_L} {1 \over |\bm k_i - \bm k_{i+1}|^2}  \right] \left[ \prod_{i=1}^{n_L} {1 \over \bm k_i^2 -  \bm p^2} \right] {\cal N}^{\rm EFT}_{n_L}  \, ,
}{eq:Lbubble2}
where we have introduced $\bm k_0 =\bm p$ and $\bm k_{n_L+1} =\bm p'$. The first term is the tree level contribution given by the potential evaluated on shell, and the second term captures the classical contributions from the sum of iterated bubbles. The poles $\bm k_i^2 - \bm p^2$ come from the expansion of the two-body propagators $\Delta$ in \Eq{eq:Lbubble1}, while the poles $|\bm k_i - \bm k_{i+1}|^2$ come from the expansion of the vertices $V$ in \Eq{eq:Lbubble1}. The numerator ${\cal N}^{\rm EFT}_{n_L} $ is a regular function whose momentum dependence may cancel factors in the denominator. 

We can map the integral in \Eq{eq:Lbubble2} to the general type treated in \sect{sec:NRint} by changing integration variables $\bm k_i \to \bm p + \bm \ell_i$ for $i=1$ to $n_L$. For any one of the loop momenta, the resulting form is that of \Eq{eq:integrable} with $\gamma \leq 1$ and $\alpha, \beta \leq 2$. The poles in $|\bm k_i - \bm k_{i+1}|^2$ and $\bm k_i^2 - \bm p^2$ here map to the graviton and matter propagator poles in \Eq{eq:integrable}, respectively. As described in \sect{sec:NRint}, all triangle subdiagrams are evaluated sequentially, while box subdiagrams lead to superclassical iterations, which are infrared divergent and cancel in the matching between full theory and effective theory.

Putting together all of this machinery, we obtain the EFT scattering amplitudes up to 3PM order: 
\begin{align}
{\cal M}^{\rm EFT}_{1}  & = -\frac{4 \pi G  c_1}{\bm q^2}  \,, \nn\\
{\cal M}^{\rm EFT}_{2}  &=  -\frac{2\pi^2 G^2
   c_2}{|\bm q|} + {\pi^2 G^2 \over  E \xi |\bm q|} \bigg[ (1-3 \xi ) c_1^2+4 \xi ^2 E^2 c_1 c_1' \bigg] +   \int {d^{D-1} {\bm \ell}  \over (2\pi)^{D-1} } { 32 E \xi  \pi^2 G^2 c_1^2 \over \bm \ell^2 (\bm \ell + \bm q)^2  (\bm \ell^2 + 2 \bm p \bm \ell)}  \,, 
   \nn\\
{\cal M}^{\rm EFT}_{3} 
&=   2 \pi G^3
   \log \bm q^2 c_3   - { \pi G^3 \log \bm q^2 \over  E^2 \xi }  \bigg[ (1-4 \xi ) c_1^3-8 \xi ^3 E^4 c_1  c_1' {}^2-4 \xi ^3 E^4 c_1^2 c_1'' +4 \xi ^2 E^3 c_2  c_1'  \nn\\
&\quad\hskip 3 cm  +  4\xi ^2 E^3 c_1  c_2' -2 (3-9 \xi ) \xi  E^2 c_1^2 c_1' +2E(1-3\xi)c_1  c_2   \bigg] \nn\\
&\quad  + \int {d^{D-1} {\bm \ell}  \over (2\pi)^{D-1} } {16 \pi^3 G^3 c_1 \left[ 2E \xi c_2 - (1- 3 \xi) c_1^2 - 4 \xi ^2 E^2 c_1 c_1' \right] \over \bm \ell^2 |\bm \ell + \bm q|  (\bm \ell^2 + 2 \bm p \bm \ell)} \nn\\
&\quad -    \int {d^{D-1} {\bm \ell_1}  \over (2\pi)^{D-1} } {d^{D-1} {\bm \ell_2}  \over (2\pi)^{D-1} } { 256 E^2 \xi^2 \pi^3 G^3 c_1^3 \over \bm \ell_1^2 (\bm \ell_1 + \bm \ell_2)^2 (\bm \ell_2 + \bm q)^2  (\bm \ell_1^2 + 2 \bm p \bm \ell_1) (\bm \ell_2^2 + 2 \bm p \bm \ell_2)} \,,
\label{eq:M_EFT}
\end{align}
where the dependence of the functions $c_n$ on $\bm p^2$ is kept implicit, while $c_n'$ and $c_n''$ denote first and second derivatives with respect to $\bm p^2$. The unevaluated integrals are IR divergent and manifest the iterative structure through factors of the coefficients $c_n$. The integrals in ${\cal M}_2^{\rm EFT}$ and ${\cal M}_3^{\rm EFT}$ that are proportional to $c_1^2$ and $c_1^3$ represent the double and triple iteration of the 1PM potential. The integral in ${\cal M}_3^{\rm EFT}$ that depends on the product $c_1 c_2$ represents the iteration of the 1PM potential and 2PM potential.

It is straightforward to extend these results to higher orders in the PM expansion. Note however that our construction here includes only the conservative sector of the effective theory as sufficient for extracting the classical conservative potential at 3PM order. For describing dissipative dynamics and the conservative dynamics beyond 3PM, operators encoding gravitational wave emission must be included in the effective theory. See \sect{sec:radiation} for further discussion.

\subsection{Conservative Potential from Matching}
\label{sec:matching}
By construction, the effective theory given by \Eq{eq:eftL} captures the same physics described by the full theory for two-to-two scattering of scalars interacting through exchanges of gravitons in the classical potential region. Thus, the full theory and EFT amplitudes must match order by order in the PM expansion, 
\eq{
{\cal M}_n  = {\cal M}^{\rm EFT}_n \,,
   }{eq:match}
and this allows us to successively determine the unknown coefficient functions $c_1\,, c_2 \,, c_3\,,$... that parameterize the classical potential. In particular, matching at $n$PM order determines the coefficient $c_n$ which comes from the tree-level diagram in the EFT and appears linearly in the first term of the amplitudes in Eq.~(\ref{eq:M_EFT}). The matching at $n$PM involves the lower order coefficients, $c_i$ for $i=1$ to $n-1$, through subtraction terms which include infrared divergent integrals. Performing this matching procedure, we obtain the classical conservative Hamiltonian at 3PM order
\eq{
H^{\rm 3PM}(\bm p, \bm r) &= \sqrt{\bm p^2 + m_1^2}+ \sqrt{\bm p^2 + m_2^2} 
+ V^{\rm 3PM}(\bm p, \bm r) \,,
}{eq:Hamiltonian3PM}
with potential
\eq{
 V^{\rm 3PM}(\bm p, \bm r) =  \sum_{n=1}^3   \left( \frac{G}{|\bm r|} \right)^n  c_n(\bm p^2) \,,
}{eq:Potential3PM}
where
\eq{
c_1&=  \frac{\nu ^2 m^2}{\gamma ^2 \xi } \left(1-2 \sigma ^2\right) \,, \\
c_2&=\frac{\nu ^2 m^3}{\gamma ^2 \xi } \left[ \frac{3}{4} \left(1-5 \sigma
   ^2\right)-\frac{4 \nu  \sigma  \left(1-2 \sigma
   ^2\right)}{\gamma  \xi } -
\frac{\nu ^2 (1-\xi) \left(1-2 \sigma ^2\right)^2}{2 \gamma ^3
   \xi ^2}
\right] \,, \\
c_3 &= \frac{\nu ^2 m^4}{\gamma ^2 \xi } \Biggl[   \frac{1}{12}{\left(3-6\nu  + 206 \nu  \pp -54\pp^2 + 108 \nu \pp^2  + 4 \nu \pp^3\right)}
-\frac{4 \nu  \left(3+12 \pp^2-  4 \pp^4  \right) {\rm arcsinh} \sqrt{\frac{\sigma
   -1}{2} }}{\sqrt{\sigma ^2-1}}
 \\
& \hskip 1.8 cm  \null
   -\frac{3 \nu \gamma   \left(1-2 \sigma ^2\right) \left(1-5
   \sigma ^2\right)}{2(1+\gamma)(1+\pp) }
-\frac{3 \nu  \sigma  \left(7-20 \sigma ^2\right)}{2 \gamma  \xi
   }    +\frac{2 \nu ^3 (3-4 \xi ) \sigma  \left(1-2 \sigma ^2\right)^2}{\gamma ^4 \xi ^3}          
   \\
&\hskip 1.8 cm  \null
   -\frac{\nu ^2  \left(3+8\gamma - 3\xi
   -15\pp^2 -   80 \gamma 
   \sigma ^2 +15 \xi  \sigma ^2 \right)\left(1-2 \sigma ^2\right)}{4 \gamma ^3 \xi ^2}
   +\frac{\nu ^4 (1-2 \xi) \left(1-2 \sigma ^2\right)^3}{2 \gamma
   ^6 \xi ^4} \Biggr] \, .
   }{eq:HamiltonianCoefficients} 
The variables used in Eq.~\eqref{eq:HamiltonianCoefficients} are
defined below \eqn{eq:PM_amp} and in \app{appendix:notation}.  Note that plugging in $c_1$ and $c_2$
into the IR divergent integrals in \Eq{eq:M_EFT} exactly reproduces
the IR divergent integrals in \Eq{eq:PM_amp}. This explicitly
demonstrates the cancellation of IR artifacts in the matching between
full theory and the EFT.

The Hamiltonian in \eqn{eq:Hamiltonian3PM} contains a mass
singularity, reflecting the mass singularity in the
amplitude~\eqref{eq:PM_amp}.  Taking both masses small, the
arcsinh term in $c_3$ dominates and gives
\begin{equation}
H^{\rm 3PM}(\bm p, \bm r)\rightarrow - 64\, \frac{G^3  {\bm p}^4}{|\bm r|^3}
 \ln \frac{ m_1 m_2}{4 {\bm p}^2} + \cdots
= - 4\, \frac{G^3  s^2}{|\bm r|^3}
 \ln \frac{ m_1 m_2}{s} + \cdots \, ,
\label{MassSingHamiltonian}
\end{equation}
where we display only the singular term.  As we explain
in \sect{DiscussionSection}, this singularity is consistent with the known absence of collinear 
singularities in gravitational theories \cite{Akhoury:2011kq} because the small mass and 
small momentum transfer limits do not commute.


\section{Consistency Checks}
\label{sec:checks}
Our calculation of the 3PM Hamiltonian exploits a number of novel
techniques. To validate the result, we have performed several
consistency checks against known results such as the 4PN Hamiltonian,
the Schwarzschild solution, and the 4PN and 2PM scattering
angles. These are of course not all independent but nevertheless it is
satisfying to reproduce multiple results in the literature.

As shown in \fig{fig:PN_PM}, the overlap between our 3PM Hamiltonian
and the 4PN Hamiltonian provides a nontrivial check. However,
Hamiltonians depend on the choice of coordinates and cannot be
directly compared. In \sect{sec:coordinate}, we construct a canonical
transformation that relates our 3PM Hamiltonian in
\Eq{eq:Hamiltonian3PM} and the 4PN Hamiltonian in
Ref.~\cite{JaranowskSchafer2}, thus demonstrating their equivalence in the overlap region.

Alternatively, we may check the equivalence between Hamiltonians by
comparing scattering amplitudes, which encode only physical information
and are independent of the gauge choice. In
\sect{sec:checks_scattering}, we use the EFT framework in
\sect{sec:EFT} to compute scattering amplitudes from two known
potentials, the Schwarzschild potential and the 4PN potential in
Ref.~\cite{JaranowskSchafer2}, and compare with our result in
\Eq{eq:PM_amp}. Note that the Schwarzschild solution provides a check
in the probe limit, $m_2 \ll m_1$, to all orders in velocity.

In \sect{sec:angle}, we give a general derivation of the PM scattering
angle from a generic PM Hamiltonian with potential of the form given
in \Eq{eq:VPR}. We provide results for the scattering angle as a
function of the coefficient functions $c_i$ up to 4PM order. We
evaluate this explicitly through 3PM, and compare to results in the
literature.

\subsection{Coordinate Transformation}
\label{sec:coordinate}

The first check is the equivalence of our 3PM Hamiltonian and the 4PN 
Hamiltonian in Ref.~\cite{JaranowskSchafer1, JaranowskSchafer2} in the overlap region,  i.e. up to $\mathcal{O}(G^3\, v^4)$.
We need to find the corresponding coordinate transformation on canonical variables $(\bm {r} , \bm p)$. Consider a general canonical transformation
\begin{equation}
(\bm {r} , \bm p)  \rightarrow (\bm {R},\bm {P}) = ( A \, \bm r   + B\,  \bm p  , C \, \bm p  +  D \, \bm r )\,,
\end{equation}
where $A,B,C,D$ are scalar functions in terms of $\bm {r}$, $\bm p$, masses, and $G$. 
Here we do not assume bound orbits so the velocity expansion is not correlated to the expansion in $G$.
The coordinate transformation is not arbitrary, but needs to preserve the Poisson brackets
\begin{equation}
	\lbrace R_i, P_j \rbrace = \delta_{ij}\,, \quad \textrm{and}\quad
	\lbrace R_i,R_j \rbrace = 	\lbrace P_i, P_j \rbrace =0\,,
	\label{eq:poisson}
\end{equation}
for $i,j=1,2,3$.
Given the constraints from Poisson brackets, matching the two Hamiltonians provides a highly non-trivial check for our result.

To find the allowed coordinate transformation, we use a bottom-up
approach in the spirit of Ref.~\cite{DamourPoincare}. Some simple
examples can also be found in
Ref.~\cite{BjerrumBohr:2002kt,Holstein:2008sx}.  In the PM scenario,
we assume that the results can be separately expand in $G$ and
velocity. The coordinate transformation can be parametrized as
\begin{align}
A,C &= 1 + \sum_{k,n,l} f_{A,C}(k,n,l)\,
\frac{G^k}{|{\bm r}|^k}\, \bm{p}^{2n} \left(\bm{p}\cdot \hat{\bm{r}} \right)^{2l} \,,
\label{eq:canonical_1} \\
B &= |\bm r|\,\sum_{k,n,l} f_{B}(k,n,l)\,
\frac{G^k}{|{\bm r}|^k}\, \,\bm{p}^{2n}\, 
\left(\bm{p}\cdot \hat{\bm{r}} \right)^{2l+1}\,, \label{eq:canonical_2} \\
D &= \frac{1}{|\bm r|}\, \sum_{k,n,l} f_{D}(k,n,l)\,
\frac{G^k}{|{\bm r}|^k}\,\bm{p}^{2n}\, 
\left(\bm{p}\cdot \hat{\bm{r}} \right)^{2l+1}\,,
\label{eq:canonical_3}
\end{align}
where $\hat{\bm{r}}={\bm{r}}/|\bm r|$, $f_{A,B,C,D}(k,n,l)$ are
functions of masses, and $k,n,l$ are non-negative integers.  The above
expressions are designed to have the correct classical counting in
$(\bm {R},\bm {P})$ following the discussion in
\sect{QuantumVsClassicalSection}.  They also preserve time-reversal
symmetry, under which $(\bm {r} , \bm p)\rightarrow (\bm {r} , -\bm
p)$.  The range of $k,n,l$ are bounded by perturbative structure.
First $k$ has to be positive because the two Hamiltonians have
identical kinematic energy.  To build a canonical transformation valid
up to 4PN, we only need to consider $k+n+l \le 4$ for $A,C,D$, and
$k+n+l \le 3$ for $B$.  Given the parametrization in
Eqs.~\eqref{eq:canonical_1}-\eqref{eq:canonical_3}, we can solve the constraints imposed by
the Poisson brackets in Eq.~\eqref{eq:poisson}, order by order in $G$
and velocity.\footnote{Solving the constraints order by order makes it
  technically simple, because the equations become linear in
  $f_{A,B,C,D}(k,n,l)$ at each order.}  This yields the space of
consistent coordinate transformations, which we solved up to 4PN order.

Given the canonical transformation built above, the remaining free coefficients can be adjusted to obtain a perfect
match between $H_{\rm 4PN}$ in Eq.(8.41) of
Ref.~\cite{JaranowskSchafer1, JaranowskSchafer2} and our result in
Eq.~\eqref{eq:Hamiltonian3PM},
\begin{equation}
H_{\text{4PN}} ({\bm R}, {\bm P})\Big|_{{\cal O}(G^3v^4)} = H_{\text{3PM}}({\bm r}, {\bm p})\Big|_{{\cal O}(G^3v^4)} \,,
\label{eq:H_match}
\end{equation}
valid up to $\mathcal{O}(G^3 v^4)$. 
To leading order, the functions $A,B,C,D$ in Eqs.~\eqref{eq:canonical_1} to \eqref{eq:canonical_3} are 
\begin{align}
&A  = 1 + \frac{Gm \nu}{2|\bm r|} + \cdots, \quad \! B = \frac{G (2/\nu-1)}{4m} \bm p\cdot \hat{\bm r}+ \cdots,\\
&C  = 1 - \frac{Gm \nu}{2| \bm r|} + \cdots,  \quad \! D = \frac{G m  \nu}{2 | \bm r|^2} \bm p\cdot \hat{\bm r}+ \cdots, 
\label{finaltransformation}
\end{align}
where the ellipsis stand for higher order terms available in the ancillary file
{\tt Canonical.m}~\cite{AttachedFile}.  Note that the transformation \eqref{finaltransformation}
maps the Hamiltonian in Ref.~\cite{JaranowskSchafer1, JaranowskSchafer2} to our Hamiltonian, which is the
inverse of the transformation in Ref.~\cite{3PMPRL}.  This proves that our
Hamiltonian is physically equivalent to that of Ref.~\cite{JaranowskSchafer1, JaranowskSchafer2} in the
region where both are valid.

As an additional but redundant check we have also verified that our
3PM potential produces the correct expressions for the binding energy
of a circular orbit for the relevant overlapping 2PN contributions.

\subsection{Comparison of Scattering Amplitudes}
\label{sec:checks_scattering}

In the previous section we employed a canonical transformation to establish the equivalence of Hamiltonians in different gauges. In this section, we alternatively check the equivalence by comparing scattering amplitudes computed from known expressions for the potential in the literature. To compute the scattering amplitude from a given potential we employ the EFT framework described in \sect{sec:EFT}.

The Hamiltonian for a point particle in a Schwarzschild background~\cite{WS93,Jaranowski:1997ky} is
\eq{
H^{\rm Sch} = m \nu \left(  \left(1-\frac{G m}{2 |\bm r| } \right) \left(1+\frac{ G m}{ 2 |\bm r|} \right)^{-1} \sqrt{1+ \left(1+\frac{Gm}{2 |\bm r|} \right)^{-4} {\bm p^2\over m^2 \nu^2}}-1   \right),
}{eq:H_Schwarzchild}
where $m=m_1+m_2$ and $\nu = m_1 m_2/m^2$. Taking the probe limit, $m_2 \ll m_1$, and Fourier transform of the potential term in \Eq{eq:H_Schwarzchild} yields the potential in \Eq{eq:VPQ} with the coefficients
\eq{
c_1(\bm p^2) &= -\frac{m_1}{E_2} (2\bm p^2 + m_2^2) \,,  \\
c_2(\bm p^2) &= \frac{m_1^2}{4E_2^3} (9\bm p^4 +13 \bm p^2 m_2^2 + 2 m_2^4) \,, \\
c_3(\bm p^2) &= -\frac{m_1^3}{4E_2^5}(8\bm p^6 +20 \bm p^4 m_2^2 + 15 \bm p^2 m_2^4+ m_2^6) \,. 
}{eq:c_Schwarzchild}
The coefficients $c_i$ are proportional to $m_1^i$ so that each term is proportional to the $i$th power of the Schwarzschild radius of the heavy mass.  Since $H^{\rm Sch}$ does not depend on $\bm p \cdot {\bm r}$, it is in the same isotropic gauge as our result and a direct comparison can be made.
The expressions in \Eq{eq:c_Schwarzchild} agree with those in \Eq{eq:HamiltonianCoefficients} upon taking the probe limit.

It of course follows that the amplitudes also agree. To use the classical potential described by the coefficients in \Eq{eq:c_Schwarzchild} as a Feynman vertex we simply make the replacement 
\eq{
\bm p^2 \to {\bm k^2 +\bm k'^2 \over 2} \,, \qquad |\bm q| \to |\bm k - \bm k'| \,,
}{eq:repvertex1}
where $\bm k$ and $\bm k'$ are off shell. The rest of the computation follows the procedure described in \sect{sec:EFT}. Here we simply plug the coefficients in \Eq{eq:c_Schwarzchild} into the EFT amplitudes in \Eq{eq:M_EFT}, yielding
\eq{
{\cal M}^{\rm Sch}_1 &= {4\pi G m_1(2 E_2^2 -m_2^2)  \over \bm q^2 E_2}  \,, \\
{\cal M}^{\rm Sch}_2 &=  {3\pi^2 G^2 m_1^2(5E_2^2 - m_2^2)  \over 2 | \bm q| E_2} \,,  \\
{\cal M}^{\rm Sch}_3 &= -   {\pi G^3  \log \bm q^2\, m_1^3 (18E_2^2 - m_2^2 )  \over 2 E_2} \, ,
}{}
which agree with the results in \Eq{eq:PM_amp} upon taking the probe limit.

Now we discuss the computation of the scattering amplitude from the 4PN Hamiltonian in Ref.~\cite{JaranowskSchafer2}, which depends on $\bm p \cdot \hat{\bm r}$ and is thus in a gauge different from ours. In momentum space this 4PN classical potential takes the form 
\eq{
V(\bm p, \bm q) = {G \over \bm q^2} \, b_1\hspace{-0.12cm} \left( \bm p^2 , (\bm p \cdot \hat{\bm q})^2 \right) + {G^2 \over |\bm q|} \, b_2\hspace{-0.12cm} \left( \bm p^2 , (\bm p \cdot \hat{\bm q})^2 \right) + {G^3 \log {\bm q}^2} \, b_3\hspace{-0.12cm} \left( \bm p^2 , (\bm p \cdot \hat{\bm q})^2 \right) + \cdots\,,
}{eq:VB}
where $\hat{\bm q} = \bm q / | \bm q |$ and  the ellipsis denotes higher order terms. The coefficients $b_i$ truncate to 4PN and scale as $\sim \bm q^0$. The dependence on $(\bm p \cdot \hat{\bm q})^2$ arises from the Fourier transform of the $\bm p \cdot \hat{\bm r}$ terms.

To define a Feynman vertex from the 4PN classical potential in \Eq{eq:VB} we make the replacements in \Eq{eq:repvertex1} together with
\eq{
\bm p \cdot \hat{\bm q}  \to  \frac12 {\bm k^2 - \bm k'^2 \over |\bm k - \bm k'|} \,, 
}{eq:repvertex2}
where $\bm k$ and $\bm k'$ are off shell. On shell we have $\bm k^2 = \bm k'^2$ and the right-hand side vanishes, while $\bm p \cdot \bm q  = \bm q^2/2$ and the left hand side is dropped as a quantum contribution.

We then follow the general procedure in \sect{sec:EFT} for computing the EFT bubble diagrams, except in this case the expansion of the integrand that puts it in the form of  \Eq{eq:integrable} is more subtle due to the $ (\bm k_i^2 - \bm k_{i+1}^2)/|\bm k_i - \bm k_{i+1}|$ terms from \Eq{eq:repvertex2}. Unlike the $\bm k_i^2 + \bm k_{i+1}^2$ terms, these are not expanded in the classical limit since $ (\bm k_i^2 - \bm k_{i+1}^2)/|\bm k_i - \bm k_{i+1}| \sim \bm q^0$. Nonetheless we can proceed by noting that a term of the form $ (\bm k_i^2 - \bm k_{i+1}^2)^n/|\bm k_i - \bm k_{i+1}|^n$ yields upon integration contributions that scale as $\bm p^n + \bm p^{n-2} \bm q^2 + \cdots + \bm q^n$. This counting, together with the fact that the two loop amplitude has superclassical terms that are enhanced by $\bm q^{-2}$ relative to the classical scaling, implies that terms with $n=6$ are necessary for the comparison between the 4PN and 3PM amplitudes. Indeed, our final scattering amplitude computed from the 4PN Hamiltonian in Ref.~\cite{JaranowskSchafer2} agrees with the 3PM amplitude in \Eq{eq:PM_amp} where these regions overlap.


\subsection{Scattering Angle}
\label{sec:angle}

\begin{figure}[tb]
\begin{center}
\includegraphics[scale=.6]{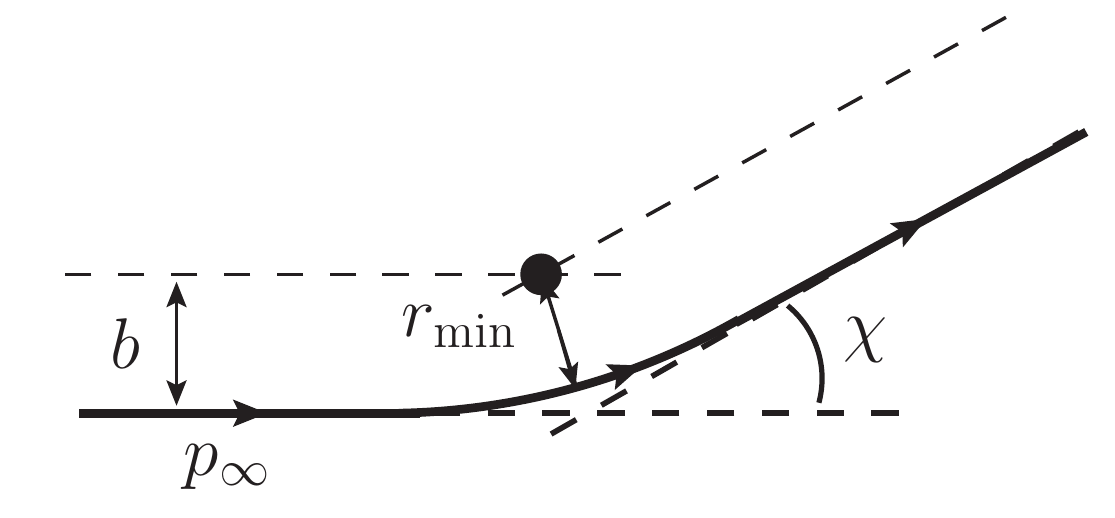}
\end{center}
\vskip -.5 cm
\caption{\small The scattering angle in center of mass coordinates. 
	The black circle denotes the center of mass.
	The thick solid line is a schematic representation of the trajectory, with $\chi$ being the angle of deflection in the final state.
  }
\label{AngleFigure}
\end{figure}

Armed with the two-body Hamiltonian, obtaining the scattering angle for two black holes (ignoring radiation effects) is
straightforward.
As explained in Refs.~\cite{Damour:2016gwp, Antonelli:2019ytb} the scattering angle is a useful stepping stone to obtain an effective one-body Hamiltonian~\cite{gr-qc/9811091, gr-qc/0001013}.

Before specializing to the case of our 3PM Hamiltonian, consider first the general problem of an arbitrary central-field Hamiltonian, $H({\bm r}^2, {\bm p}^2)$,
describing the interaction of two particles in the center of mass frame. We shall assume that, as in our case, ${\bm r}$ and 
${\bm p}$ are canonically-conjugate to each other.
The goal is to find (an integral representation for) the scattering angle
in terms of the total energy of the system and the angular momentum or, equivalently, the impact parameter. 

As in the classical case of Newtonian scattering, it is convenient to use polar coordinates. As for that case, 
the central nature of the interaction implies that the scattering process occurs in a plane as shown in \Fig{AngleFigure}, so it suffices to specify 
only the two planar polar coordinates, ${\bm r}= r {\bm e_r}$ and the corresponding momenta, 
${\bm p} = p_r {\bm e_r}+p_\theta {\bm e_\theta}$. Here ${\bm e_r}$ and ${\bm e_\theta}$ are unit vectors in the radial and angular directions, respectively.
In these coordinates, Hamilton's equations are
\begin{align}
\dot r {\bm e_r} + r\dot\theta {\bm e_\theta} &= 2 (p_r {\bm e_r} + p_\theta {\bm e_\theta} ) F({\bm r}^2, {\bm p}^2) \,,
 \qquad F({\bm r}^2, {\bm p}^2)  = 
\frac{\partial H({\bm r}^2, {\bm p}^2)}{ \partial {\bm p}^2} \,,
\nn \\
 (\dot p_r -p_\theta \dot\theta) {\bm e_r} + (\dot p_\theta +p_r \dot \theta) {\bm e_\theta} &= -2 r {\bm e_r} \,K({\bm r}^2, {\bm p}^2)   \,, \hskip 2.1 cm  \;
  K({\bm r}^2, {\bm p}^2) = \frac{\partial H({\bm r}^2, {\bm p}^2)}{ \partial {\bm r}^2} \,,  \hskip .3 cm 
\end{align}
and the two conservation laws are
\begin{equation}
J = |{\bm r}\times {\bm p}|= |{\bm b}\times {\bm p}_\infty| = r p_\theta = b p_\infty\,,
\hskip 1.5 cm 
E =H(r^2, p_r^2 + p_\theta^2)
\,.
\label{conservations}
\end{equation}
where $b\equiv |\bm b|$ is the impact parameter.
We denoted the norm of the three-momentum at infinity by $p_\infty$; it is related to the total energy in the usual way
for a scattering process,
\begin{eqnarray}
E = \sqrt{p_\infty^2+m_1^2}+  \sqrt{p_\infty^2+m_2^2}  \,,
\label{Epinf}
\end{eqnarray}
The two components of the first of Hamilton's equations  and the $e_\theta$ component of the second suffice to determine the trajectory
in terms of the radial momentum $p_r$,
\be
\frac{dr}{d\theta} = \frac{r^2}{J} p_r \,,
\label{trajectory}
\ee
which in turn is determined given by conservation of energy \eqref{conservations},
\begin{equation}
E =  H(r^2, p_r^2 + J^2/r^2) \,.
\label{encons}
\end{equation}
We denote this solution by $p_r(r)$. The scattering angle is then obtained by integrating the trajectory equation \eqref{trajectory}:
\be
\chi = -\pi + 2 J \int_{r_\text{min}}^{\infty} \frac{dr}{r^2 \sqrt{p_r(r)^2}} \,.
\label{angle}
\ee
We denoted the minimum distance between the two particles by $r_\text{min}$. At this point the radial momentum $p_r$ changes 
sign, and therefore must vanish:
 \be
0 = p_r(r_\text{min}) \,.
\ee
We shall use this relation to determine $r_\text{min}$.\footnote{An alternative is to extract it from energy conservation, $E = H(r_\text{min}^2, J^2/r_\text{min}^2) $.}

\subsubsection{Scattering Angle in the Post-Minkowskian Expansion}

As discussed before, the conservative Hamiltonian for a system of two spinless compact bodies in the post-Minkowskian 
expansion, neglecting radiation effects, has 
the form \eqref{eq:VPR}
\begin{align}
\label{Hamiltonian}
H({\bm r}^2, {\bm p}^2) =& \sqrt{{\bm p}^2+m_1^2}+  \sqrt{{\bm p}^2+m_2^2} 
\\
&+ c_1({\bm p}^2) \frac{G }{|\bm r|}+c_2({\bm p}^2) \left(\frac{G }{|\bm r|}\right)^2
+c_3({\bm p}^2) \left(\frac{G }{|\bm r|}\right)^3+c_4({\bm p}^2) \left(\frac{G }{|\bm r|}\right)^4+\dots \,.
\nonumber
\end{align}
The dependence on the radial coordinate is sufficiently simple to allow us to find the scattering angle for 
arbitrary coefficients $c_i({\bm p}^2)$, whose index reflects the PM order at which they appear. Since this structure of the Hamiltonian 
relies only on the spinless nature of the particles  and Lorentz invariance of their interactions, such a general 
expression can also be used to explore extensions of General Relativity by further fields and/or further interactions.

The radial momentum $p_r$ as a function of the radial coordinate $r\equiv |\bm r|$, obtained by solving perturbatively \eqref{encons}, is
\be
p^2_r(r) = \frac{p_\infty^2 r^2 - J^2}{r^2}+ P_1 \frac{G }{r}+P_2 \left(\frac{G }{r}\right)^2
+P_3 \left(\frac{G }{r}\right)^3 + P_4 \left(\frac{G }{r}\right)^4+{\cal O}(G^5)\, .
\label{prr}
\ee
The coefficients $P_i$ will be listed shortly. The vanishing of $p_r(r)$  determines the minimum distance $r_\text{min}$:
\begin{align}
r_\text{min}^2=\null & 
b^2 
- b^3 P_1\frac{G }{J^2} 
+ \frac{b^4}{2} (P_1^2 - 2 p_\infty^2 P_2) \frac{G ^2}{J^4}
\cr
 &- \frac{b^5}{8} (P_1^3 - 4 p_\infty^2 P_1 P_2 + 8 p_\infty^4 P_3) \frac{G ^3}{J^6}
 - b^6 p_\infty^6 P_4\frac{G ^4}{J^8}
 + {\cal O}(G^5) \,.
 \label{rmin}
\end{align}

Using Eqs.~\eqref{prr} and \eqref{rmin} in Eq.~\eqref{angle}, it is not
difficult to find the scattering angle through fourth order
in the PM expansion\footnote{The expansion on the integrand in \Eq{angle} must be carried out carefully to avoid spurious singularity.}
, in terms of the expansion coefficients in Eq.~\eqref{prr}
\begin{align}
\chi = \sum_{i \ge 1} \chi^{i\text{PM}}= & \frac{ P_1}{p_\infty} \left(\frac{G }{J} \right)
                + \frac{\pi}{2}  P_2  \left(\frac{G }{J} \right)^2
                - \frac{  P_1^3 - 12 p_\infty^2 P_1 P_2 - 24 p_\infty^4 P_3}{12p_\infty^3}  \left(\frac{G }{J} \right)^3 
                \nn \\
              & \null 
                + \frac{3\pi}{8}   (P_2^2 + 2 P_1 P_3 +2p_\infty^2 P_4)  \left(\frac{G }{J} \right)^4
                +{\cal O}\left((G/J)^5\right)
               \, ,
\label{theangle}
\end{align}
where the $\chi^{i\text{PM}}$ is the angle at the order $(G/J)^i$.
The dependence on the momentum at infinity may be traded for dependence on the total energy by inverting Eq.~\eqref{Epinf}:
\begin{equation}
p_\infty^2 = \frac{1}{4E^2}(E^2 - (m_1 - m_2)^2) (E^2 - (m_1 + m_2)^2) \,.
\end{equation}

Lastly, the coefficients $P_i$ of the $G $ expansion of the radial momentum are:
\def\cbar{{\bar c}}
\begin{align}
P_1 = \null &  -2 E \xi  \cbar{}_1 \,,
\label{P1}
\\[2pt]
P_2 =\null & -2 E \xi  \cbar{}_2 +(1-3 \xi ) \cbar{}_1^2+4 E^2 \xi ^2 \cbar{}_1 \cbar'{}_1 \,,
 \label{P2}
     \\[2pt]
P_3 = \null & -2 E \xi \cbar{}_3 +2 (1-3 \xi )\cbar{}_1 \cbar{}_2
   -4E^3 \xi ^3  \cbar{}_1 \left(2  \cbar'{}_1^2+ \cbar{}_1 \cbar''{}_1\right)
   +4 E^2 \xi ^2 \left(\cbar{}_2
   \cbar'{}_1+\cbar{}_1 \cbar'{}_2\right) 
   \nonumber\\[0pt]
  & -6 E (1-3 \xi ) \xi 
   \cbar{}_1^2 \cbar'{}_1
   +\frac{(1 - 4 \xi) \cbar{}_1^3}{E} \,,
      \label{P3}
      \\[2pt]
P_4 = \null & -2 E \xi  \cbar{}_4
  +\frac{8}{3} E^4 \xi ^4 \cbar{}_1 \left(6 \cbar'{}_1^3+9 \cbar{}_1
   \cbar''{}_1 \cbar'{}_1+\cbar{}_1^2 \cbar'''{}_1\right) +\frac{5 (1 - 4 \xi) \cbar{}_1^4}{4 E^2}+\frac{3 (1-4 \xi ) \cbar{}_1^2
   \cbar{}_2}{E}  \hskip 2 cm 
   \nonumber\\[2pt]
&   -4 E^3 \xi ^3 \left(2 \cbar{}_2 \left(\cbar'{}_1^2+\cbar{}_1 \cbar''{}_1\right)+\cbar{}_1 \left(4 \cbar'{}_1 \cbar'{}_2+\cbar{}_1 \cbar''{}_2\right)\right)
+(1-3 \xi )\cbar{}_2^2 
\nonumber\\[2pt]
&  +4 E^2 \xi ^2 \left(2 (1-3 \xi )
   \cbar''{}_1 \cbar{}_1^3+6 (1-3 \xi )\cbar'{}_1^2
  \cbar{}_1^2+\cbar'{}_3 \cbar{}_1+\cbar{}_3
   \cbar'{}_1+\cbar{}_2 \cbar'{}_2\right)
  \nonumber\\[2pt]
&  -6 E (1-3 \xi ) \xi 
   \cbar{}_1 \left(2 \cbar{}_2\cbar'{}_1+\cbar{}_1
   \cbar'{}_2\right)
+2 \cbar{}_1 \left((1-5 \xi )^2 \cbar'{}_1
   \cbar{}_1^2+(1-3 \xi ) \cbar{}_3\right)  \,,  
   \label{P4}
 \end{align}  
where $\cbar_i \equiv c_i(p_\infty^2)$,  the primes denote derivatives with respect to the argument and, as before, $E=E_1+E_2$
and $\xi = E_1E_2/E^2$.  The $c_i$ are the coefficients in the potential in \eqn{eq:Hamiltonian3PM} and are given in \eqn{eq:HamiltonianCoefficients}.
The expressions relating $P_{i\ge 5}$ to the coefficients of the Hamiltonian are lengthier, but may be derived 
without difficulty by inverting Eq.~\eqref{conservations}, iteratively in Newton's constant.
        
We note that all coefficients $P_i$ have the structure 
\be
P_i = -2E\xi  \cbar_i + \dots \,,
\ee
where the ellipsis only depend on Hamiltonian coefficients $c_j$ with $j<i$ and their derivatives; this may be easily understood 
from the special structure of the Hamiltonian \eqref{Hamiltonian}.
We also note that there exists a close relation between the coefficients $P_i$ and the EFT amplitudes at the same order in $G$. 
Indeed, comparing eqs.~\eqref{P1}, \eqref{P2} and \eqref{P3} with \eqref{eq:M_EFT}  it is easy to see that $P_i$ is proportional to the
IR-finite part of the EFT amplitudes. We expect that this feature will continue to higher PM orders.  

\subsubsection{Angle Through 3PM and Comparison with Known Results}

Using \eqref{P1}, \eqref{P2}, \eqref{P3} and  \eqref{theangle} as well as the Hamiltonian \eqref{eq:Hamiltonian3PM}-\eqref{eq:HamiltonianCoefficients}, 
it is straightforward to obtain that, through 3PM order, the conservative part of the scattering angle is~\cite{3PMPRL, Antonelli:2019ytb}
\begin{align}
\chi^\text{1PM}&\null =
 \frac{2(2\sigma^2-1)}{\sqrt{\sigma^2-1}}\frac{ \nu m^2 G }{J} \,,
\nn \\
\chi^\text{2PM} &\null =
\frac{3\pi}{4}  \frac{5\sigma^2-1}{\sqrt{1+2\nu (\sigma-1)}}\frac{\nu^2 m^4 G^2}{J^2} \,,
\nn \\
\chi^\text{3PM}& \null =
-\biggl[
 \frac{1}{12} \left(  \frac{2(2\sigma^2-1)}{\sqrt{\sigma^2-1}}\right)^3 
 - \frac{2}{\pi} \left(  \frac{2(2\sigma^2-1)}{\sqrt{\sigma^2-1}}\right) 
   \biggl( \frac{3\pi}{4} \frac{5\sigma^2-1}{\sqrt{1+2\nu (\sigma-1)}}\biggr)
 \nn \\
& +4 \, \frac{\sqrt{ \sigma^2-1}}{2 (\sigma - 1) \nu + 1}
\biggl(
-\frac{3 \nu \left(1-5 \sigma^2\right) \left(1-2 \sigma^2\right) (2 \nu (\sigma-1)+1)}{2 (\sigma+1) \left(2 \nu (\sigma-1)+\sqrt{2 \nu
   (\sigma-1)+1}+1\right)}
   \cr
&\qquad \qquad   -\frac{4 \nu \left(-4 \sigma^4+12 \sigma^2+3\right)} {\sqrt{\sigma^2-1}} \,
{\rm arcsinh}\!\left(\frac{\sqrt{\sigma-1}}{\sqrt{2}}\right)
   \cr
&\qquad \qquad  
   +\frac{1}{12} \left(4 \nu \sigma^3+108 \nu \sigma^2+206 \nu \sigma-54 \sigma^2-6 \nu +3\right)
\biggl)
 \biggr] 
\frac{ \nu^3 m^6 G ^3}{J^3} \,. \hskip 2 cm 
\end{align}
Using the observation that $P_i$ are proportional to the IR-finite parts ${\cal M}'_i$ of the EFT amplitudes evaluated at $p=p_\infty$,
 the angle can be rewritten very compactly as~\cite{3PMPRL}
\begin{align}
2\pi \chi =  {d_1 \over  J } + {d_2  \over J^2  }  + \frac{1}{J^3}
 \biggl( \vphantom{\frac{1}{\pi^2}} {  -4  d_3     }   
 + {  d_1 d_2 \over \pi^2    }  
-{ d_1^3 \over 48 \pi^2  }  \biggr) \,,
\label{angle_ito_amplitudes}
\end{align}
where  $d_i$ are defined in terms of  ${\cal M}'_i$ as
\begin{equation}
d_1 = m \gamma \xi {\bm q}^2 {\cal M}_{1}'/|{\bm p}| \,,
\hskip 1 cm 
d_2 =m \gamma \xi |\bm q| {\cal M}_{2}' \,,
\hskip 1 cm 
d_3 =  m \gamma \xi |{\bm p}| {\cal M}_{3}' / \log {\bm q}^2 \, ,
\end{equation}
and $p\rightarrow p_\infty$ is implicitly understood.

The 2PM angle has been obtained through a variety of methods in
Refs.~\cite{Damour:2016gwp, DamourTwoLoop,
  BjerrumClassical,Cristofoli:2019neg}, and our result is in
agreement.  Moreover, the 4PN terms in the 3PM angle reproduce the
corresponding terms in \cite{Bini:2017wfr}. This is, of course, to be
expected since we have already shown that the 4PN part of our
Hamiltonian reproduces (up to a canonical transformation) the ${\cal
  O}(G ^3)$ part of the known 4PN Hamiltonian.

One may parametrize the higher-PM angle in several different ways; given the expected relation between the $P_i$ coefficients in Eq.~\eqref{prr} and the 
classical limit of the IR-finite parts of the $i$-loop amplitude, it seems convenient to parameterize them recursively in terms of the 
lower-PM angle and $P_i$. For example, the 4PM angle without radiation effect can be written as: 
\begin{align}
\chi^\text{4PM} & \null =  \frac{3 \pi p_\infty^2 G ^4}{4 J^4} P_4 + \frac{3 \pi}{8} \chi^\text{1PM} \chi^\text{3PM}
+\frac{3}{2\pi} (\chi^\text{2PM} )^2 -\frac{3}{4} (\chi^\text{1PM})^2\chi^\text{2PM} +\frac{\pi}{32}(\chi^\text{1PM})^4  \,.
\end{align}
It would be interesting to understand the relation between the method
used here to compute the scattering angle and the one based on the
eikonal limit of scattering amplitudes~\cite{StermanEikonal}. In particular, it
would be interesting to see how terms proportional to the IR-finite part of scattering amplitudes
\eqref{angle_ito_amplitudes} arise from the recent amplitude-based methods~\cite{OConnellObservables,BjerrumClassical,Cristofoli:2019neg}.
 
\section{Discussion}
\label{DiscussionSection}

In this section we discuss various features and subtleties in our
results.  In particular, we describe the mass singularity that appears
in our 3PM results, as well as subtleties associated with infrared
singularities.  In general relativity infrared singularities are quite
tame compared to gauge theory, because they are a simple exponential
of one-loop divergence~\cite{Weinberg:1965nx, 1101.1524, Akhoury:2011kq}.  We therefore do
not anticipate serious difficulties at higher loops when applying
four-dimensional methods to construct integrands relevant for extracting
classical physics.  In this section we analyze the one- and two-loop
situations, to show that straightforward application of
four-dimensional helicity methods indeed gives the
same final result as a more careful treatment, where all steps are
computed in $D=4-2\eps$ dimensions, as required when using conventional
dimensional regularization.

We also comment on radiative contributions to
conservative dynamics, which we show are not relevant at 3PM order,
but should become important at the next order in the post-Minkowskian
approximation.

\subsection{Mass Singularities and Collinear Structure}
\label{MassSingularitySubsection}

A feature of the 3PM amplitude \eqref{eq:PM_amp},
Hamiltonian \eqref{eq:Hamiltonian3PM} and scattering
angle \eqref{angle_ito_amplitudes} is that as both masses are taken to
vanish a logarithmic mass singularity develops, as
displayed in \eqn{eq:AmplitudeMassSingularity}.
At first sight this might seem surprising because quantum
amplitudes do not have mass or, equivalently, collinear
singularities~\cite{Weinberg:1965nx, 1101.1524, Akhoury:2011kq}.  Since our 
results originate from a quantum amplitude, one might be
surprised by its appearance in our final results.

In this section we trace the origin of the mass singularity as due to
an inability to interchange the classical and massless limits
originating from our hierarchy of scales \eqref{eq:classical_limit},
that takes the momentum transfer $|\bm q| \ll m_i$, while assuming
that both masses $m_i$ have fixed finite values.

As explained in \sect{QuantumVsClassicalSection}, we defined the PM
potential such that, at each order in Newton's constant, its velocity
expansion matches all corresponding terms in the PN expansion.
Indeed, as discussed in \sect{sec:checks}, the known 4PN
Hamiltonian~\cite{JaranowskSchafer1, JaranowskSchafer2} confirms the
first three terms of the small-velocity expansion of the arcsinh term
which becomes logarithmically-divergent in the massless limit.  One
may attempt to define a Hamiltonian for which the limit is smooth by
e.g.  resumming the various logarithmic terms to all orders in $G$
expansion, but, given the above, it is unclear how such a result would
match known PN potentials, without introducing extraneous
contributions.  Of course, a function that is smooth in the massless
limit and contains all information necessary for constructing the
classical potential is the full quantum amplitude.  However, this
function is likely highly nontrivial to construct and manipulate.

We verified that it is not possible to remove this mass singularity of
the classical potential by a canonical transformation, which is consistent with
it appearing also in the amplitude and in the scattering angle.
This implies that the massless limit is generally discontinuous---and
in fact may be ill-defined if taken after arriving at the classical
potential.  On the other hand, if the massless limit, is taken prior
to taking the classical limit we do expect to smoothly match onto
other purely massless computations, such as the scattering angle 
in the high-energy limit obtained via eikonal methods in Ref.~\cite{ACV}.

From an intuitive perspective, it is perhaps not so surprising that
the massless limit is subtle.  After all, we usually think of a
potential as mediating an instantaneous force between two bodies.  We
naturally expand the potential about the static limit in which the
relative velocity is zero.  However, if interacting particles are
moving at the speed of light, this notion of instantaneous potential
becomes unclear.

Our non-relativistic integration methods explicitly assume, following
the construction of the classical limit
in \sect{QuantumVsClassicalSection}, that the momentum transfer is
much smaller than the mass in order to define potential-mode
gravitons.  Although, not relying on a velocity expansion, when using
relativistic methods for evaluating the integrals in \sect{sec:Rint}
we make a similar assumption, by first carrying out the asymptotic
expansion at small-momentum transfer.
Once the integration is carried out, one naturally expects to encounter
difficulties when attempting to interchange the order of limits and compare
the result with one obtained in a massless theory.
Similar interchange of limits issues appear, for example, in factorization limits
of infrared divergent theories~\cite{Bern:1995ix}.
Indeed, the massless limit of our results at two loops, included in
Eq.~\eqref{eq:AmplitudeMassSingularity} does not match those of
Ref.~\cite{ACV}, obtained in the high energy limit.

In order to sharpen our understanding of the mass singularity
appearing in our results, we analyze its appearance
in the case in the first subleading quantum terms in the small-$q$
expansion of the one-loop amplitude, as noted in
Refs.~\cite{PierreLogM, Paolo2PM}. 
We will do so as follows:
\begin{itemize}

\item We explicitly confirm the well-known cancellation of mass
     singularities in the full one-loop quantum amplitude.

\item We explicitly show that the small-$q$ limit does not commute with 
the massless limit for the first quantum correction to the amplitude.

\end{itemize}
Our conclusion will be that the 3PM classical potential indeed has a
discontinuity in the $m \rightarrow 0$ limit.

We note that while the mass singularity is not of phenomenological
interest in the inspiral region because the logarithm is of order
unity, it is of some theoretical interest to understand its origin and,
even better, to attempt to exploit it.  Typically, mass
singularities in gauge theories can be resummed to all orders in the
coupling.  This could conceivably form the basis of new approximation schemes.  In
any case, in this section, we explain the appearance of a mass
singularity within our framework.

\subsubsection{Absence of Mass Singularity for \texorpdfstring{$\bm q^2 \gg m^2$}{$\mathbf q^2 \gg m^2$}}
\label{12.1.1}
A full analysis of the mass singularity at two-loop case is
nontrivial, in part, because of the difficulty of carrying out the
loop integration without making explicit use of the classical limit.
We therefore instead carry out a detailed one-loop analysis.  This
offers important insight, because it also exhibits a similar mass
singularity, albeit starting in the first subleading quantum
contribution~\cite{PierreLogM, Paolo2PM}.  The complete one-loop quantum amplitude is
straightforward to evaluate, allowing us to
track both the cancellation of the mass singularities in the full
quantum theory~\cite{Akhoury:2011kq} and to track how these
cancellations break down when expanding around the classical limit.
The result is that the expansion around the classical limit does not
commute with the small-mass limit.

We first show that, for fixed kinematical invariants $s$ and $t$, the
one-loop quantum amplitude does not contain a $\ln( m_1 m_2)$
singularity for small $m_1$ and $m_2$, as expected.  While modern
methods can help~\cite{BCJ, GeneralizedDoubleCopy}, we used Feynman
diagrams since this calculation is simple enough, followed by standard
loop integration methods~\cite{Chetyrkin:1981qh, Laporta:2001dd}.
In carrying out this computation, we ignore bubble-on-external-leg
contributions, since these do not carry the mass singularities: such
contributions are suppressed by factors of mass-squared, as follows
from simple dimensional analysis and the fact that the only kinematic
invariant they can depend on is $p_i^2 = m_i^2$.  Because the
calculation of the integrand and its integration is straightforward,
we do not include the details here, but merely quote the result
focusing on collinear and mass singularities.

For one-loop calculations, it is possible to organize the amplitude as
linear combinations of scalar ``master'' integrals.  Our results is
given by\footnote{As mentioned above, this is not the complete
one-loop amplitude because one needs to add the
bubble-on-external-line contributions and matter self-energy graphs,
which we ignore here.}
\begin{equation}
\mathcal M {}^{(1)}= \sum_{i=1}^{17} a_i \, I^{(1)}_i \,,
\label{eq:amp}
\end{equation}
where $I^{(1)}_i$ is the master integral of the one-loop graph $i$ in
Table~\ref{table:oneloop_topology}. 
The master integral is defined as
\begin{equation}
I^{(1)}_i=\int d^D \ell \, \frac{ e^{\gamma_E \epsilon}}{i \pi^{D/2}}\, \frac{1}{\prod_{j\in i} D_j} \,,
\label{eq:integral_normalization}
\end{equation}
where $\prod_{j} D_j$ is the product of all inverse propagators
corresponding to the edges of the graph $i$ in
Table~\ref{table:oneloop_topology}, and $\gamma_E$ is the Euler
constant.  The coefficients $a_i$ in \eqn{eq:amp} depend on external
masses and Mandelstam invariants.
A property that will be useful shortly is that all coefficients are
regular in the small $m_{1}$ and $m_{2}$ limit.
%
%

\begin{table}[t]
	\centering 
	\begin{tabular*}{0.9\textwidth}{l @{\extracolsep{\fill}} l}
		\hline\hline
		\hspace{3.82cm}
		\includegraphics[scale=0.37,trim={0 21.2cm 5cm 2cm},clip]{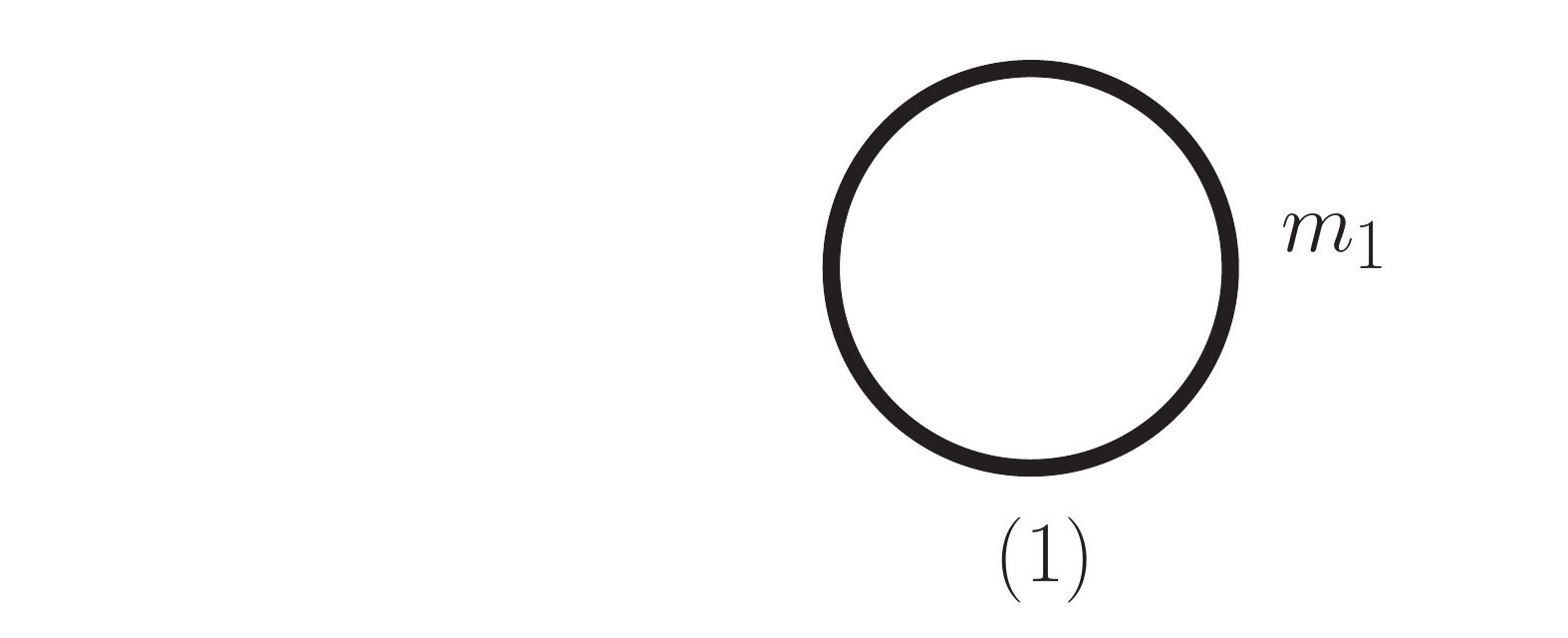} 
		\\
		\hline
		\hspace{0.65cm}
		\includegraphics[scale=0.37,trim={0 9.2cm 0cm 1cm},clip]{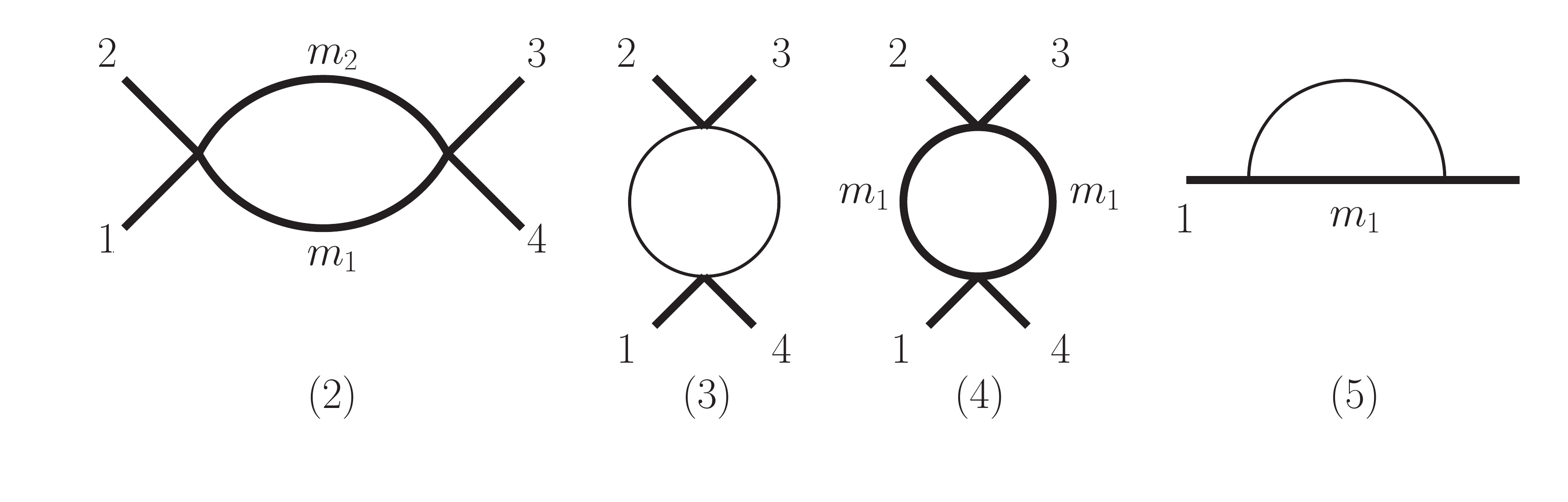} 
		\vspace{-0.5cm}
		\\
		\hline
		\hspace{2.50cm}
		\includegraphics[scale=0.37,trim={2.5cm 11cm 3cm 1.5cm},clip]{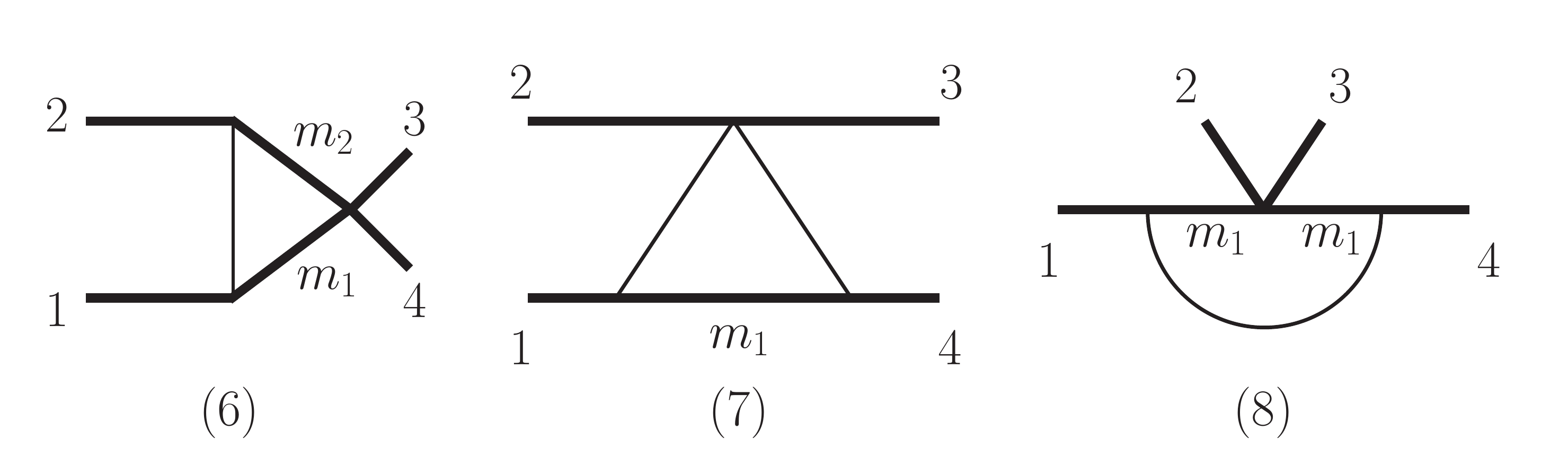}
		\\
		\hline
		\hspace{3.78cm}
		\includegraphics[scale=0.37,trim={0 20.8cm 0cm 1cm},clip]{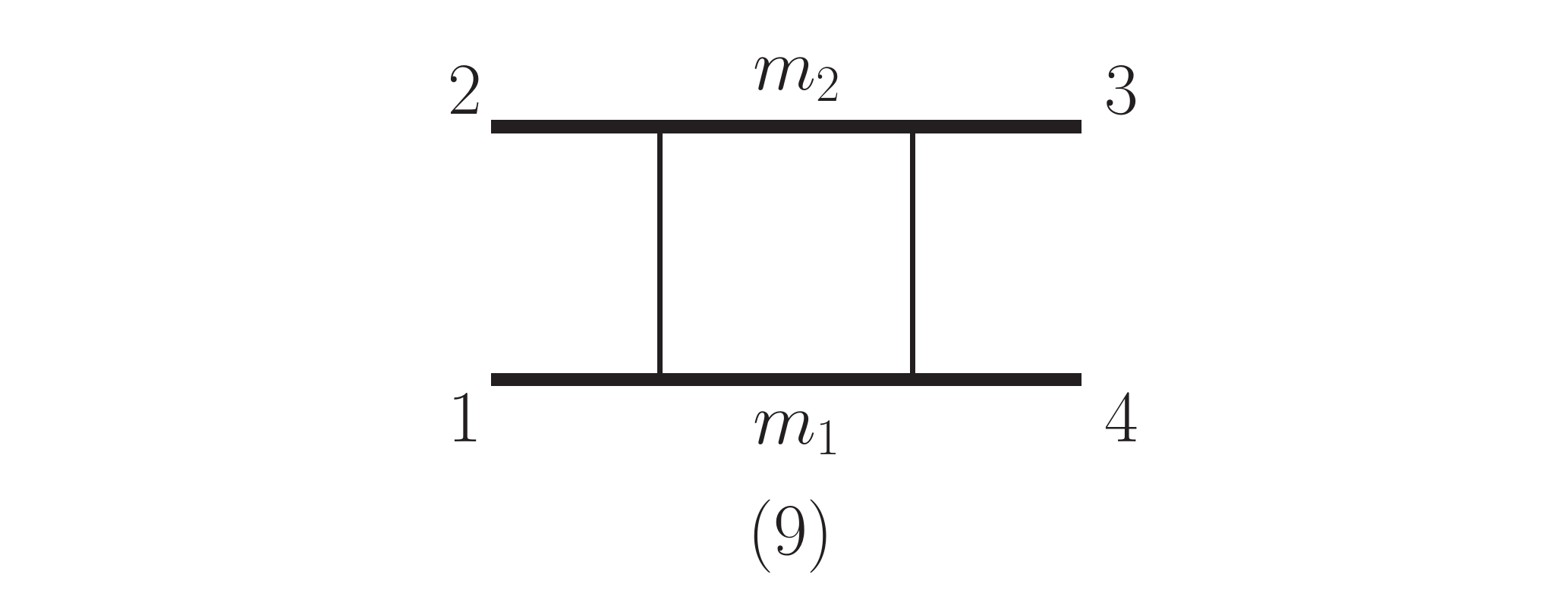} 
		\\
		\hline \hline
	\end{tabular*}
	\caption{\small Master integrals $I^{(1)}_1$ to $I^{(1)}_9$
	for the one-loop scattering amplitude in the full quantum
	theory.  The thin lines are massless and thick lines are
	massive. The internal masses are specified explicitly.  The
	other master integrals are obtained from the displayed one via
	relabeling.  The $u$-channel integrals, $I^{(1)}_{10,11,12}$,
	are given by $I^{(1)}_{2,6,9}$ with $2\leftrightarrow 3$.  The
	master integrals $I^{(1)}_{13,14,15,16,17}$ are obtained from
	$I^{(1)}_{1,4,5,7,8}$ by the map $(1,4)\leftrightarrow (2,3)$
	as well as exchanging internal masses $m_1\leftrightarrow m_2$.
	} \label{table:oneloop_topology}
\end{table}

All the analytic expressions of the scalar tadpole, bubble, triangle,
and box integrals can be found in Ref.~\cite{Ellis:2007qk}.  We derive
their small-mass logarithmic singularities using the Mellin-Barnes
method briefly reviewed in \sect{sec:MB} and implemented in the
computer codes {\tt MB.m} and {\tt MBasymptotics.m}; the resulting
expressions are, of course, consistent with the integrals listed in
Ref.~\cite{Ellis:2007qk}.
Integrals $I^{(1)}_1,I^{(1)}_2,I^{(1)}_3,I^{(1)}_4,I^{(1)}_{10},
I^{(1)}_{13}$ and $I^{(1)}_{14}$ are smooth when $m_i\rightarrow 0$.
$I^{(1)}_1$ and $I^{(1)}_{13}$ in fact vanish in this limit, being
proportional to $m_i^2$. The other five bubble integrals are also
regular in the massless limit, because in this limit the total
momentum flowing through them is fixed and nonzero.
An important property of all 17 integrals is that they do not exhibit
any power-like singularities in the small-mass limit with fixed momentum 
transfer.

The singular terms as $m_i\rightarrow 0$ for the remaining ten
integrals are collected in the third column of
Table~\ref{table:log_dept_oneloop_ints_and_coefficients}.  Since none
of them exhibits a power-like mass singularity to study the $\ln
m_{i}$ behavior of the amplitude it is sufficient to evaluate the
integral coefficients with vanishing masses. These coefficients are
collected in the second column of
Table~\ref{table:log_dept_oneloop_ints_and_coefficients}.

\begin{table}[t]
	\centering 
	\begin{tabular}{c|c|c}
	$i$ & $a_i$ & $I^{(1)}_i$ \\[2pt]
	\hline 
$5$ & $0$ & $\displaystyle{  \frac 1 \epsilon + 2 - 2 \ln m_1} $
$\displaystyle{\vphantom{\Big|_{A_A}^{A^A}}}$
\\
$6$ & $ \frac{\pi^2 s^3} {8}$ & $ \displaystyle{\frac {1}{\epsilon \,s} \,\left(\ln m_1+\ln m_2\right) - \frac{1}{s} \left((\ln m_1)^2+(\ln m_2)^2 \right) + \dotsc}$ 
$\displaystyle{\vphantom{\Big|_{A_A}^{A^A}}}$
\\
$7$ & 
$\frac{1}{16} \pi^2 t (s^2 - su + u^2 )$ 
& $\displaystyle{ \frac{2}{t} \left( (\ln m_1)^2 - \ln (-t) \ln m_1\right) + \dotsc }$
$\displaystyle{\vphantom{\Big|_{A_A}^{A^A}}}$
\\
$8$ & $-\frac{1}{16} \pi^2 st u$ & $\displaystyle{ \frac{2}{t}\left( \frac {1} {\epsilon } \ln m_1-  (\ln m_1)^2 \right)+ \dotsc }$ 
$\displaystyle{\vphantom{\Big|_{A_A}^{A^A}}}$
\\
$9$ & $\frac 1 {16} \pi^2 s^4$ & $\displaystyle{   \frac{2}{s t}\left(\frac{1}{\epsilon} -  \ln(-t))\right)(\ln m_1 +\ln m_2)+\dotsc}$
$\displaystyle{\vphantom{\Big|_{A_A}^{A^A}}}$
\\
$11$ & $\frac 1 8 \pi^2 u^3$ & $\displaystyle{ I^{(1)}_6\Big|_{(2\leftrightarrow 3)}}$
$\displaystyle{\vphantom{\Big|_{A_A}^{A^A}}}$
\\
$12$ & $ \frac 1 {16} \pi^2 u^4$ & $\displaystyle{ I^{(1)}_9\Big|_{(2\leftrightarrow 3)}}$
$\displaystyle{\vphantom{\Big|_{A_A}^{A^A}}}$
\\
$15$ & $0$ & $\displaystyle{ I^{(1)}_5\Big|_{((1,4)\leftrightarrow (2,3); m_1\leftrightarrow m_2)}}$
$\displaystyle{\vphantom{\Big|_{A_A}^{A^A}}}$
\\
$16$ & 
$\frac{1}{16} \pi^2 t (s^2 - su + u^2 )$ 
& $\displaystyle{I^{(1)}_7\Big|_{((1,4)\leftrightarrow (2,3); m_1\leftrightarrow m_2)}}$
$\displaystyle{\vphantom{\Big|_{A_A}^{A^A}}}$
\\
$17$ & $-\frac{1}{16} \pi^2 st u$ & $\displaystyle{I^{(1)}_8\Big|_{((1,4)\leftrightarrow (2,3); m_1\leftrightarrow m_2)}}$
$\displaystyle{\vphantom{\Big|_{A_A}^{A^A}}}$
	\end{tabular}
	\caption{\small Master integrals in the physical region and
          their coefficients in the small-mass limit. Integrals not
          included explicitly do not have any $\ln m_{i}$
          singularities.  Ellipsis contain terms which are regular in
          the small-mass limit but may contain divergence in the
          dimensional regulator $\epsilon$.  }
	\label{table:log_dept_oneloop_ints_and_coefficients}
\end{table}

It is not difficult to see that by multiplying the second and third
column and adding the resulting ten terms all the simple and double
mass logarithms cancel out, as required by the general argument of
Ref.~\cite{Akhoury:2011kq}. For example, the complete soft and
collinearly-divergent terms in the sum of integrals $I^{(1)}_6$ and
$I^{(1)}_{11}$, proportional to $(s^2+u^2)\ln(m_1 m_2)/\epsilon$,
cancel against similar terms in the sum of integrals $I^{(1)}_9$ and
$I^{(1)}_{12}$. The remainder, proportional to $s u\ln(m_1
m_2)/\epsilon$, cancels against analogous terms in the sum of
$I^{(1)}_8$ and $I^{(1)}_{17}$. Relations between Mandelstam
invariants are, of course, essential. A second cancellation that is
easy to see by inspection is that of the double logarithms $\ln(-t)\ln(
m_1 m_2)$, which are given only by $I^{(1)}_7+I^{(1)}_{16}, I^{(1)}_9$
and $I^{(1)}_{12}$. The first combination of integrals gives the
double logarithms under discussion while the sum of the second
integrals, each of which has the double logarithms, produces the
necessary overall coefficient.

All integrals in
Table~\ref{table:log_dept_oneloop_ints_and_coefficients} are essential
for the complete cancellation of collinear singularities.  However,
not all of them survive in the small-$q$ limit. This can be seen by
examining the $t$-dependence of the combination of the integrals and
their coefficients which needs to yield $t^{-1/2}$ or $\ln(-t)$
dependence to contribute to the classical 2PM potential or the first
quantum correction to it, respectively.  For example, the combination
$a_6 I^{(1)}_6$ does not appear in the construction of 2PM potential,
either at the classical and the first quantum order, because it only
depends on the $s$ Mandelstam invariant.


\subsubsection{Appearance of Mass Singularity for \texorpdfstring{$\bm q^2\ll m^2$}{$\mathbf q^2\ll m^2$}}

Now that we understand in detail how the mass singularity cancels in
the complete one-loop quantum amplitude, we can explore the
breakdown of this cancellation when expanding around the classical limit.
As noted in Refs.~\cite{PierreLogM,Paolo2PM}, while there is no mass
singularity in the 2PM amplitude used to derive the classical potential,
there is one in the first subleading quantum  correction to it.
This provides a much simpler case to study than at two loops, which
nonetheless has similar features.  We will first identify the
mass singularity and recall how it cancels in the quantum theory,
and then proceed to show that it no longer does if the amplitude is first
series-expanded in small $|\bm q|$ to extract the classical potential.

In Ref.~\cite{Paolo2PM}, the first quantum correction to the classical
potential is shown to have a $\log(m_1 m_2 /s)$ singularity.  It
originates from the one-loop scalar box integral $I^{(1)}_9$ in
Table~\ref{table:oneloop_topology}, which has no classical
limit. Interestingly, the same type of function appears at two loops.
The complete expression of the box
integral with two internal massive lines is~\cite{Beenakker:1988jr,
Smirnov:2004ym, Ellis:2007qk}
\begin{align}
I^{(1)}_9 = \frac{2}{t\sqrt{(s-\Delta m^2)(s-M^2)}}
\left(\frac{1}{\epsilon}-\log\left(\frac{-t}{\mu^2}\right) \right)
\Biggl(\log\Biggl(\frac{1-\sqrt{\frac{s-M^2}{s-\Delta m^2}}}{1+\sqrt{\frac{s-M^2}{s-\Delta m^2}}}\,\Biggr) +i\pi \Biggr)\,,
\end{align}
where $\Delta m^2 = (m_1-m_2)^2$ and $M^2= (m_1+m_2)^2$. 
The dependence on $t = -|\bm q|^2$ and on mass are factorized, so the
small-$t$ expansion and the massless limits commute trivially.
However, this integral exhibits a logarithmic singularity in the
small-mass limit
\begin{equation}
	I^{(1)}_9 = \frac{2}{t\, s}
	\left(\frac{1}{\epsilon}-\ln(-t) \right)
	\ln\left(\frac{m_1 m_2}{s}\right) + \dotsc \, ,
	\label{eq:box_collinear}
\end{equation}
where the ellipsis stand for a power series in $m_i m_j/s$.
Combining this with the cross-box integral, $I^{(1)}_{12}$, obtained
by the replacement $s\rightarrow u = -s-t + M^2+\Delta m^2$, and using
the integral coefficients $a_9$ and $a_{12}$ in
Table~\ref{table:log_dept_oneloop_ints_and_coefficients} exposes the
complete mass singularity of the first subleading quantum correction
to the amplitude:
\begin{equation}
 a_9 I^{(1)}_9 + a_{12}I^{(1)}_{12} =  
 - \frac{\pi^2}{8}(s^2-su+u^2) \left(\frac{1}{\epsilon}-\log(-t) \right)  \ln\left(\frac{m_1 m_2}{s}\right) + \dotsc \, ,
 \label{boxlogm}
\end{equation}
where the ellipsis contain terms that are regular in the small-$m_i$
limit as well as higher orders in a small-$t$ expansion.
This reproduces the mass singularity as found in
Ref.~\cite{Paolo2PM} in the leading quantum correction to the 2PM
amplitude.

As discussed above in \sect{12.1.1}, in the small-mass limit taken at
fixed $t$ and $s$, the overlap of soft and collinear singularities in
the contribution of $I^{(1)}_{9}$ and $I^{(1)}_{12}$, $\ln(m_1
m_2)/\epsilon$, cancels against the sum of $I^{(1)}_6, I^{(1)}_{11},
I^{(1)}_8$ and $I^{(1)}_{17}$, while the double logarithm,
$\ln(-t)\ln(m_1 m_2)$, cancels against the sum of $I^{(1)}_7$ and
$I^{(1)}_{16}$.  Thus, to understand the fate of collinear
singularities in an expansion around the classical limit it suffices
to expand these integrals first at small $t$ and then at small $m_i$.
Using {\tt MB.m} and {\tt MBasymptotics.m} one finds that $I^{(1)}_6$
and $I^{(1)}_{11}$ are given by the same expressions in
Table~\ref{table:log_dept_oneloop_ints_and_coefficients}. The other
four integrals are different. $I^{(1)}_7$ and $I^{(1)}_8$ are given by
\begin{eqnarray}
\bigl( I^{(1)}_7 \bigr|_{t\to 0} \bigr) \bigr|_{m_i \to 0} &=&-\frac{\pi^2}{2m_1\sqrt{-t}} -\frac{1}{2m_1^2}\left( \ln (-t) - 2\ln m_1 - 2\right)+\dotsc
\cr
\bigl( I^{(1)}_8 \bigr|_{t\to 0} \bigr) \bigr|_{m_i \to 0}  &=& \frac{1}{2m^2_1} \left( \frac{1}{\epsilon} - 2\ln m_1 \right)+\dotsc \, ,
\label{expansions_classical}
\end{eqnarray}
and $I^{(1)}_{16}$ and $I^{(1)}_{17}$ are given by the appropriate
relabelings.  The ellipsis stand for terms which are regular in the limit $t\rightarrow 0, |t|\ll m^2_i$ and $s$ fixed; on dimensional grounds all 
these terms are increasingly more singular as $m_i\rightarrow 0$.
Comparing with the relevant entries in
Table~\ref{table:log_dept_oneloop_ints_and_coefficients} we see that
$I^{(1)}_7, I^{(1)}_8, I^{(1)}_{16}$ and $I^{(1)}_{17}$ have different
expansions around the small-mass limit and around the classical
limit. In particular, in the later expansion they contain
stronger-than-logarithmic singularities in the mass while missing the
soft-collinear overlap, $\ln m_i/\epsilon$, and double-logarithms
$\ln(-t)\ln m_i$. It is therefore clear that the cancellation of
collinear singularities in the small-mass expansion is broken in the
expansion around the classical limit. Thus, the small-$m$ and
small-$t$ expansions do not commute.  Because the $1/t$ appearing in
the small-mass limit is replaced by a $1/m_1^2$ in the small
$t$ limit, the contributions from $I^{(1)}_7$ and $I^{(1)}_8$, and their
coefficients in the expansion around the classical result are subsubleading in the 
expansion around the classical limit and thus unable to cancel the mass singularity \eqref{boxlogm}
appearing in the first subleading quantum correction.

The expansions \eqref{expansions_classical} may be confirmed by starting from the known analytic 
expressions for $I^{(1)}_7$ and $I^{(1)}_8$ in the physical $t<0$
region; for example, the latter integral is given by~\cite{Ellis:2007qk}:
\begin{align}
I^{(1)}_8=\frac{x_t}{m_1^2 \,(1-x_t^2)}\left[
\log x_t \left(-\frac{1}{\epsilon}-\frac{1}{2} \log x_t +2 \log(1-x_t^2) +\log\frac{m_1^2}{\mu^2}
\right) -\frac{\pi^2}{6} +\textrm{Li}_2(x_t^2)+2 \textrm{Li}_2(1-x_t)
\right]	,
\end{align}
where $\mu$ is the dimensional regularization scale and
\begin{equation}
x_t = \frac{-t}{4m_1^2} \left(\sqrt{1+\frac{4m_1^2}{-t}}-1 \right)^2  .
\end{equation}

While we have not carried out the corresponding calculation at two
loops, the one-loop calculation described in detail here illustrates that,
generically, the small-$m$ and small-$t$ expansions do {\it not} commute and
moreover that the latter expansion contains remnants of
collinearly-singular terms that cancel in the former, in agreement
with the general arguments of Ref.~\cite{Akhoury:2011kq}. 
Based on our one-loop analysis, we therefore conclude that mass singularities 
in the 3PM classical potential are not surprising.


\subsection{Four- vs.~$D$-Dimensional Integrands}
\label{sec:evanescence}

An important issue which can help streamline future calculations is
whether we can compute the integrand for the classical potential using
only four-dimensional methods or whether it is necessary to be more careful
and use $D$-dimensional methods when dimensionally 
regularizing infrared singularities.

When using four-dimensional helicity methods to construct the
integrand the following choices are implicitly made:
\begin{enumerate}

\item Terms containing Gram determinants or terms that vanish in four but not in $D$ dimensions are 
not included.  Equivalently, terms containing loop momenta that are outside the four-dimensional subspace 
are dropped.

\item The state-counting parameter is implicitly taken to be $D_s = 4$.

\end{enumerate}
This is to be contrasted to conventional dimensional regularization
which keeps all such Gram determinants, and takes $D_s = 4 -
2 \eps$.  Here we show that through two loops these differences in prescription have no
effect on the final classical potential.  Some points which appear to be generic underlie this conclusion:
\begin{itemize}

\item Terms with $(-2 \eps)$-dimensional components of loop momentum are suppressed in $\eps$ and cannot contribute unless they interfere with $1/\eps$ singularities.  
However, after extracting an explicit factor of $\epsilon$, it turns out that remaining integrals are equivalent to 
integrals in higher dimensions that cannot be infrared singular. If a contribution remains 
it must come from an ultraviolet divergence whose origin is quantum mechanical.

\item The dependence on the state-counting $D_s$ parameter may strike
infrared singularities and thus modify the integrand by finite terms.  However, even if some terms were
to remain, the $D_s$ prescription needs to be applied consistently and
uniformly to different orders in the PM expansion, including in the
EFT.  The net effect is that the iteration terms in the EFT should
automatically subtract any prescription differences.

\item The $(-2\eps)$-dimensional terms, whether from $D_s$ or $(-2\eps)$-dimensional components of loop momenta,
are of the wrong form to generate classical contributions.  In particular, at 3PM order they
do not generate the required $\ln {\bm q}^2$ terms.

\end{itemize}
While a detailed proof beyond 3PM order is beyond the scope of this
paper, the fact that the underlying ideas appear to be generic
suggests that helicity methods will also be sufficient at higher orders.

\subsubsection{Integrand Comparison at One Loop}

Before turning to the 3PM case, it is useful to understand how four-dimensional helicity methods 
for constructing the integrand lead to the correct 2PM potential.  As emphasized earlier,  at first sight one might worry 
that, in the presence  of ${\cal O}(1/\epsilon)$ infrared singularities, terms of ${\cal O}(\epsilon)$ originating from 
differences between four- and $D$-dimensional methods may yield  errors in finite terms.
We will show that this does not happen.

As already noted in \sect{OneLoopDCutsSection}, there are only four
independent momenta in the one-loop integrand: three external and one
loop momentum. Thus, it is not possible to form a Gram determinant
that vanishes in four dimensions but not in general
dimensions. Consequently, four-dimensional helicity methods must fully
reconstruct the $D$-dimensional integrand, up to the choice of the
$D_s$ parameter, which four-dimensional helicity methods implicitly
fix to $D_s = 4$.  This reduces the question of whether the classical
potential obtained through helicity methods drops any important terms
to that of whether we can use $D_s = 4$ in the simplified physical
state projector in \eqn{SimplifiedGravitonProjector} without affecting
the potential.

As we discussed in \sect{sec:EFT}, the triangle integrals are infrared-finite and the only infrared divergence comes 
from box integrals. Thus, only their contribution is sensitive to the value of $D_s$.
The part of the box-integral coefficient proportional to $(D_s-4)$ can be obtained from \eqn{OneloopDCutA} by cutting the
$i/\t_{15}$ propagator, so that  all four box propagators are cut. Combining this with the $D_s = 4$ result \eqref{D4BoxContribution}, 
gives the $D_s$-dependent coefficient of the box integral
\begin{align}
C_{\rm box} = &
((s-m_1^2 + m_2^2)^2 - 2 m_1^2 m_2^2)^2 +
\frac{4 m_1^2 m_2^2(D_s -4)}{(D_s-2)}
\Big( (s-m_1^2-m_2^2)^2 - m_1^2 m_2^2 \frac{D_s}{D_s-2}   \Bigr)\nn \\
 = &
\left( (s-m_1^2 -m_2^2)^2 - \frac{4}{D_s-2} m_1^2 m_2^2 \right)^2 .
\end{align}
Its essential property is that it is the square of the numerator of the tree amplitude in $D_s$ dimensions,
\begin{equation}
M_4(1^s, 2^s, 3^s, 4^s) = \frac{i}{t} \Bigl( (s-m_1^2 -m_2^2)^2 - \frac{4}{D_s-2} m_1^2 m_2^2\Bigr)\,.
\end{equation}
That is, the only modification that survives as $\epsilon\rightarrow
0$ is that the coefficient of the four-dimensional box
integral \eqref{D4BoxContribution} is the $D_s$-dimensional tree
amplitude instead of the four-dimensional one.
Because the box contains the only infrared singularity of the one-loop
amplitude which, by construction, is the same as that of the EFT, it
follows that the complete $D_s$ dependence of the one-loop amplitude
is reproduced by the iterated $D_s$-dependent tree amplitude of the
EFT.
The net effect is that, when the iterated tree including $D_s$
dependence is subtracted to determine the 2PM potential, the $D_s$
dependence is subtracted as well.
We therefore find that the integrand constructed in four dimensions
and the $D$-dimensional one give the same 2PM classical potential,
without even needing to look further at detailed form of the
integration.


Although, as we argued above, four-dimensional methods fully
reconstruct the complete one-loop integrand as a function of momenta
without dropping any kinematic terms containing Gram determinants, by
two loops this is no longer true.  Moreover, this is no longer true
even at one loop if the external states carry spin or if one is
interested in higher-point amplitudes such as those containing
outgoing gravitons.  It is therefore instructive to understand
the effect of such terms on the classical potential, if they were
present at one loop.  This also serves as a preview of the two-loop
analysis, which is similar.
The types of terms we are interested in are proportional to Gram
determinants formed from five or more vectors so that they vanish in
four dimensions. Equivalently, they can also be characterized as being
proportional to components of loop momenta that lie outside of four
dimensions.  These two view points are equivalent because any Gram
determinant that vanishes in four dimensions must be proportional to
the extra-dimensional components of loop momenta.

To analyze such numerator factors at one loop order,  we separate the loop momentum, following Ref.~\cite{BernMorgan}, into the 
four-dimensional component, $\bar \ell$, and the $(-2\eps)$ component, $\lambda$, as 
\begin{equation}
\ell =  \bar \ell + \lambda \, .
\end{equation}
Taking $\eps <0$ we have that 
\begin{equation}
\ell^2 =  \bar \ell^2 - \lambda^2\,,
\end{equation}
where $\bar \ell \cdot \lambda = 0$ because $\bar\ell$ and $\lambda$ lie in orthogonal subspaces.
In general, when constructing integrands using four-dimensional
helicity methods, we may drop terms\footnote{We also drop the
$\lambda^2$ in propagators, but in practice these are trivial to
restore since the denominators must be those of $D$-dimensional Feynman
propagators.} proportional to $\lambda^2$.

The question we would like to address is whether these dropped 
${\cal O}(\lambda^2)$ kinematic terms might lead to incorrect results for the
classical potential.
To this end, we use the fact that any one-loop integral in $D=4-2\eps$
dimensions with numerator proportional to $\lambda^{2r}$ can be
expressed in terms of an integral in shifted dimension, $D\rightarrow
D+2r$, whose numerator no longer has the factor depending on the
$(-2\epsilon)$ components of the loop momentum (see Eq.~(A.15) of
Ref.~\cite{BernMorgan}):
\begin{align}
\int d^{4-2\epsilon}\ell \, \lambda^{2r} f(\ell^\mu, p_i^\mu, \lambda^2) 
 & = 
- \frac{1}{\pi} \eps (1-\eps) \ldots (r-1-\eps) \int d^{4 + 2r  -2\epsilon}\ell \; f(\ell^\mu, p_i^\mu, \lambda^2) \,.
  \label{eq:oneLoopDimShiftFactor}
\end{align}
From this relation it is clear that the effect of adding a single
factor of $\lambda^2$ in the numerator is equivalent to shifting the
integration dimension from $D= 4-2\eps$ to $D= 6-2\eps$ dimensions,
and multiplying the resulting integral by a compensating
$\epsilon$-dependent factor.

Consider such a term in an amplitude. Since infrared divergences are
absent in dimensions $D>4$, the only possible source of $1/\epsilon$
poles to compensate the overall $\epsilon$ factor in
Eq.~\eqref{eq:oneLoopDimShiftFactor} is an UV divergence.  General
renormalization theory requires that, because at one loop there are no
subdivergences, all UV divergences are local.  Consequently, so are
the finite terms generated by multiplication by $\epsilon$,
cf. Eq.~\eqref{eq:oneLoopDimShiftFactor}.
Since only terms proportional to $(-t)^{-1/2}$ contribute to the
classical potential, we conclude that box integrals with factors of
$\lambda^{2r>0}$ in the numerator will affect the amplitude only at
some subleading order in the expansion around the classical limit.
As we already noted, while such terms do not appear in the elastic
scattering of two spinless particles at one-loop, they do occur when
spin is added to the problem or when the scattering is allowed to be
inelastic.

\subsubsection{ Integrand Comparison at Two Loops}

Consider now the possible dependence of the two-loop amplitude on the treatment of $D$ and $D_s$. 
The simplest case is that of graph $\bm 8$ in \fig{MasterDiagsFigure}, which factorizes into two one-loop integrals,
and is therefore already covered by the one-loop argument above.

The genuine two-loop four-point diagrams are of course more intricate. Here the diagram integrands differ not only because 
of dependence on the $D_s$ parameter, but also because of the 
appearance of the Gram determinant $G_5$, defined in \eqn{GramDete}, that vanishes in~$D=4$. 
As at one loop, we begin by discussing the possible kinematic contributions, and then proceed to the $D_s$ effects.

At first sight, the appearance of $G_5$ seems to be a rather serious complication. Fortunately,
its  effect at two loops is essentially identical to that of $\lambda^2$ at one loop.
This can be understood using the Baikov representation of Feynman integrals~\cite{hep-ph/9603267, hep-ph/9611449, 1104.3993, 1701.07356}. 
For our purpose we do not need the full power of this formalism, and instead follow a simpler 
formulation given in e.g.\ Ref.~\cite{Larsen:2015ped}.
For any two-loop integral with four external legs, the external
momenta span a three-dimensional subspace, due to momentum
conservation. We decompose each of the two loop momenta, $\ell_i^\mu
, \, i=1,2$, into a sum of their projection onto these three
dimensions (not to be confused with the three spatial dimensions used
in the NR integration), denoted by ${\bar l}_i^\mu$, and the
projection into the orthogonal complement subspace of
dimension\footnote{At one loop our $\lambda^\mu$ lives in a
$(-2\epsilon)$ subspace, but this is only a minor difference and is
due to our desire to match formulas known in the literature.} 
 $(1 - 2\epsilon)$, denoted by $\lambda_i^\mu$.
Introducing the notation
\begin{equation}
\lambda_i \cdot \lambda_j = \lambda_{ij} \, ,
\end{equation}
the two-loop integration measure can be rewritten as
\begin{align}
\int d^{4 - 2\epsilon} \ell_1 \, d^{4 - 2\epsilon} \ell_2 &=
\frac{2^{-1 - 2\epsilon } \pi^{2 - \epsilon}} { \pi^4 \Gamma(-2\epsilon)}
\int_0^\infty d\lambda_{11} \int_0^\infty d\lambda_{22} \int_{-\sqrt{\lambda_{11} \lambda_{22}}}^{\sqrt{\lambda_{11} \lambda_{22}}} d\lambda_{12} \nonumber\\
& \quad \times
(\lambda_{11} \lambda_{22} - \lambda_{12}^2)^{\frac{-2 -2\epsilon }{2}} \int d^3 \bar \ell_1 d^3 \bar \ell_2 \, ,
\end{align}
which holds for any integrand.

Note that $(\lambda_{11} \lambda_{22} - \lambda_{12}^2)={\rm Gram}(\lambda_1,\lambda_2)$ vanishes
if the original integration dimension is exactly 4, because in that case $\lambda_{1}^\mu$ and $\lambda_{2}^\mu$ are 
both one-dimensional.
In fact, $(\lambda_{11} \lambda_{22} - \lambda_{12}^2)$ can be shown to be
proportional to $G_5$; the constant of proportionality depends on
Mandelstam invariants and masses of external legs.  So, up to overall
factors, one power of $G_5$ in the numerator has the effect of
increasing the integration dimension by 2, while also introducing an
extra factor of $\epsilon$, similar to the one-loop situation in
Eq.~\eqref{eq:oneLoopDimShiftFactor}.  As at one-loop, when extracting
the classical potential we localize the energy integrals to matter
poles, leaving behind a spatial integration in one lower dimension for
each loop.
Taking advantage of the lack of one-loop sub-divergences of UV
origin in the NR integration with gravitons in the potential region
and near odd spatial dimensions, a finite $\mathcal O(\epsilon^0)$
contribution can only arise from an overall two-loop UV divergence.
General renormalization theory implies that, if they exist at all,
these potential UV divergences are local. We therefore can obtain at
most four-scalar contact-term discrepancies between integrands
constructed in $D=4$ and in $D=4-2\eps$ dimensions, from graphs $\bm
1$, $\bm 2$, $\bm 4$, $\bm 5$, and $\bm 7$ in \fig{MasterDiagsFigure}.
Graphs $\bm 3$ and $\bm 6$ contain a two-loop triangle graph attached
by a propagator to a three-point vertex.  A similar argument as above
implies that the triangle integrals can at most give a local two-loop
divergence (in shifted dimension).  Classical (small-$q$)
power counting further implies that the coefficient of this divergence
is at least of the order of ${\cal O}(|{\bm q}|)$.  The additional
tree-level vertex and propagator will dress it by a factor of the form
$P_2(p_i)/{\bm q}^2$, where $P_2$ is some quadratic polynomial
resulting from the tree-level scalar field stress tensor. Thus, as for
the other graphs, Gram determinant terms in graphs $\bm 3$ and $\bm 6$
cannot have the requisite $\ln {\bm q}^2$ dependence to contribute to
the 3PM conservative classical potential while also scaling at most as
${\cal O}(q^0)$.

Unlike the one-loop amplitude, several of the two-loop graphs exhibit
$D_s$ dependence and, moreover, for some of them it does not enter in
a factorized fashion.  A detailed analysis is therefore needed to
track, graph by graph, the fate of the ${\cal O}(\epsilon)$ terms.
Nevertheless, infrared singularities of amplitudes have a universal
structure to all loop orders based on the one loop
singularities~\cite{Weinberg:1965nx, 1101.1524, Akhoury:2011kq} and are identical in the
complete theory and the EFT.
Because of this we expect that, as at one loop, the two- and
higher-loop amplitude terms that depend on the prescription used for
the state-counting parameter $D_s$ are automatically compensated by
analogous terms in the EFT amplitudes and therefore do not contribute
to the conservative potential.

The independence of the final results on the treatment of
$(-2\eps)$-dimensional components of loop momenta and $D_s$ can also
be understood through an analysis of the mechanics of the appearance
of $\epsilon/\epsilon$-type contributions in the nonrelativistic
integration method and matching with the EFT described
in \sects{sec:NRint}{sec:EFT}.
At two loops, the energy master integrals are
finite. The two $(D-1)$-momentum integrals are then done in succession
and, as before, the first one is finite. The second one is done in
$D-1=3-2 \epsilon$ dimensions and has both ultraviolet and infrared
artifacts.  The latter correspond to iterations and are canceled at
the integrand level in the matching of the full theory and the
effective theory.
The only $1/\epsilon$ factor which may give nontrivial contributions
is the ultraviolet divergence from the second space-like momentum
integral. To understand whether it can contribute nontrivially it is
useful to recall that the $\log {\bm q}^2$ signaling a contribution to
the classical potential arises from $|{\bm q}|^{\epsilon}/\epsilon$ in a
regularized integral.  However, terms of ${\cal O}(\epsilon)$ from
$D_s$ or elsewhere can produce finite terms only when multiplying the
$1/\epsilon$; such terms will therefore not have the necessary
$\log \bm q^2$ factor and will not affect to the classical potential.

As a direct check, following the methods described
in \sects{OneLoopDCutsSection}{TwoLoopSection}, we have explicitly
derived the 3PM classical potential starting from a loop integrand
constructed in $D=D_s= 4 - 2\eps$ dimensions and compared it to the
potential derived from an integrand obtained using four-dimensional
helicity methods; the resulting potentials are identical,
confirming our discussion above.  The fact that, through two loops, we
can construct the necessary integrand using efficient helicity methods makes it
promising that the same will be true at higher loop orders.

\subsection{Radiative Contributions to the Conservative Potential}
\label{sec:radiation}

The method of nonrelativistic integration described
in \sect{sec:NRint} incorporates the effects of potential-mode
gravitons only.  As discussed in \sect{sec:graviton_kinematics},
potential modes are never on shell, and thus can only make a
real contribution to the scattering and hence to the conservative
potential.  However, we are still left with the converse question: can
radiation modes also contribute to the conservative dynamics?  As we
explain, the answer is negative at our 3PM order of interest.  However, radiation-mode 
contributions to the conservative potential 
should become nontrivial at 4PM order.

To understand why, let us recapitulate some known facts about the
dynamics of a binary inspiral in the language of EFT.  In the standard
picture, the gravitational modes are split between near-zone
(potential mode) and far-zone (radiation mode) degrees of freedom.
Potential gravitons have spatial momentum of order $1/r$, which is the
minimum distance between the constituents, while radiation gravitons
have spatial momentum of order $v/r$.  Integrating out the near-zone
gravitons yields an EFT for the two-body system describing a pair of
compact objects interacting via a conservative gravitational
potential.  Afterwards, we integrate out the far-zone gravitons to
obtain a one-body EFT describing the entire binary.

Let us consider the effects of radiation gravitons order by order.
The leading effect of radiation is the Burke-Thorne radiation-reaction
force~\cite{Burke:1970dnm, Thorne:1969rba, Burke:1970wx}, which is generated by the one-loop self energy
diagram in the one-body EFT.  This process corresponds to the real
emission of a graviton from the binary system.  The effect is purely
dissipative, in the sense that the induced response is purely odd
under time reversal.  Consequently, Burke-Thorne radiation-reaction
does not contribute to the conservative dynamics.

The next-to-leading order effect arises from a two-loop self-energy
diagram in the one-body EFT.  This process is the so-called tail
effect~\cite{TailEffect1, TailEffect2, TailEffect3, TailEffect4, TailEffect5, TailEffect6}, corresponding to graviton radiation emitted
from the binary which scatters off the monopole gravitational
potential of the system before falling back in.  The tail effect has
both time-even and time-odd contributions.  The time-even piece
produces the leading contribution to the conservative potential from
radiation modes.  This term scales as $\sim G^2 \dddot{I}^2$ where
$\dddot{I}$ is the triple time derivative of the center of mass
quadrupole moment, which at leading order is $I \sim (r^i r^j - 1/3 r^2
\delta^{ij})$.  The time derivatives necessarily produce at least one
acceleration and one velocity, so $\dddot{I} \sim Gv$, so the tail
effect occurs at order $G^4 v^2$
\cite{JaranowskSchafer1,arXiv:1703.06433, arXiv:1903.05118}.  Even including arbitrarily
higher order PN corrections to $I$ we reach the same conclusion.
Simply counting powers in $G$, we see that the tail effect induces
conservative radiation reaction force starting at 4PM order.

The tail effect implies a breakdown of the naive split between near-
and far-zone dynamics.  In the original works on the PN
potential, the tail effect appears as a
non-local in time ``hereditary'' effect~\cite{TailEffect1, TailEffect2, TailEffect3, TailEffect4, TailEffect5, TailEffect6, Foffa:2019eeb}. 
  From the
point of view of EFT~\cite{arXiv:1703.06433, arXiv:1903.05118} the tail effect appears as an
 infrared divergence in the PN expansion. This divergence is nonlocal in $r$ and
thus cannot be absorbed by any counterterm.  Given the understanding of this subtlety in the PN expansion we do not expect any difficulties dealing with it at 4PM order.

\section{Conclusion}\label{sec:conclusion}

This paper details a general framework for deriving the conservative
dynamics of a compact binary system using modern methods from
scattering amplitudes and effective field theory. This framework has
been used in our previous work to obtain the 3PM conservative two-body
Hamiltonian~\cite{3PMPRL}, which we described in detail in the present
paper.  This result is of interest~\cite{Antonelli:2019ytb} to see
whether velocity corrections lead to improvements when building accurate waveform
models for LIGO/Virgo data analysis.

Our approach exploits the double-copy
construction~\cite{KLT,BCJ,BCJLoop}, which expresses gravitational
scattering amplitudes in terms of simpler gauge-theory
ones. Generalized unitarity~\cite{hep-ph/9403226, hep-ph/9409265, hep-ph/9708239, hep-th/0412103, 0705.1864} then
efficiently builds loop integrands for multi-loop scattering processes
that encode the classical dynamics. The input gauge-theory tree
amplitudes are remarkably compact when using four-dimensional helicity
states~\cite{Berends:1981rb, Berends:1981uq, Xu:1986xb,KhozeMassive}.  Using a battery of
standard and nonstandard integration methods, including relativistic
and nonrelativistic approaches, we then integrate these expressions
and obtain the 3PM amplitude for classical scattering mediated by
potential gravitons.  The nonrelativistic approach of
Ref.~\cite{CliffIraMikhailClassical}, in particular, efficiently
targets the classical contributions and displays excellent scaling to
higher loop orders.  By matching the amplitude to one computed in a
low-energy effective field theory~\cite{RothsteinClassical,
  CliffIraMikhailClassical}, we then extract the 3PM conservative
potential.

We described various cross-checks of our result for the 3PM potential
against existing literature, including showing that our expressions are
equivalent to known PN results where these expansions overlap.  Our
result has already been analyzed in an initial
study~\cite{Antonelli:2019ytb} for its potential to improve
gravitational wave template models for LIGO/Virgo, when combining it
with state of the art PN
calculations~\cite{JaranowskSchafer1,JaranowskSchafer2,arXiv:1703.06433, arXiv:1903.05118},
effective one-body models~\cite{gr-qc/9811091, gr-qc/0001013} and numerical
relativity~\cite{gr-qc/0507014, gr-qc/0511048, gr-qc/0511103}.  While preliminary, this study shows interesting
promise.

While the 3PM potential we have computed is now state of the art in PM
order, it is natural to think about pushing forward to 4PM.  We expect
that our approach to integrand construction via double copy and
unitarity should scale well to quite high loop orders, as highlighted
by their application to supergravity theories up to five-loop
order~\cite{0905.2326, SimplifyingBCJ, 1309.2498, 1409.3089, 1708.06807, 1804.09311}.  Moreover, our method of nonrelativistic
integration should also scale well to higher
loops~\cite{CliffIraMikhailClassical}.  In fact, since it mirrors the
integration approach of NRGR---which has been amenable to integration
up to the five-loop 5PN \cite{1902.10571, 1902.11180}---we believe the same will be
true in our case.

As we proceed to higher orders in the PM expansion we do expect to
encounter new issues and subtleties.  One expected subtlety that
should arise at 4PM is the appearance of conservative
radiation-reaction, whereby radiation graviton modes back react on the
conservative dynamics~\cite{arXiv:1703.06433, arXiv:1903.05118}.  We expect this will appear
through nontrivial infrared divergence structure in the 4PM amplitude
and in the effective field theory mapping; this deserves a dedicated
study as part of any 4PM computation.

There are a number of other issues that warrant further attention.
Through 3PM we showed that, in line with expectations, we can compute
the integrand using four-dimensional techniques and then afterwards
apply dimensional regularization to deal with infrared singularities,
without missing any pieces.  While it seems plausible that this should
be generally true, can we find an all-orders proof?  If true then we
could use four-dimensional helicity amplitudes to construct loop
integrands, without needing to deal with $D$-dimensional integrand
constructions.

Prior to carrying out the integration we merged the unitarity cuts
into a single integrand. When extracting the classical potential the
regions of integration that contribute effectively have matter
cuts reimposed. This suggests that there should, in fact,
be a method that allows us to extract the classical potential at high
loop orders directly from the unitarity cuts without merging
them into a single integrand.

While our nonrelativistic integration methods are efficient and more
importantly have good scaling properties, it would be helpful to also have
a fully-relativistic integration method.  For example, one would like
a relativistic method that does not involve using integration by
parts~\cite{Chetyrkin:1981qh, Laporta:2001dd}, which tends to lead to large systems
with poor scaling properties. The Mellin-Barnes representation could be
such an approach, but this has difficulties stemming from nontrivial analytic
continuations into the physical region~\cite{1607.07538, 1001.3243, 1609.09111}.  On
the other hand, we know that after integration final results are
remarkably simple, strongly suggesting that, in fact, much more
efficient relativistic techniques can be developed to perform the loop 
integration in the classical limit.

Furthermore, the present work is purely on a single aspect of the
binary inspiral which is the conservative dynamics.  However, the
ideas we have described here are extendable and continuously linked
to a large array of connected phenomena that are crucial for
extracting useful physics from the LIGO/Virgo experiment.  For
example, here we have only considered nonspinning objects described by
scalars, but the initial intrinsic angular momentum of the objects can
be important for the gravitational-wave signal.  Another effect that
we have neglected is related to the finite size of the constituents of
the binary system, which is particularly critical for understanding
neutron-star mergers and the extraction of the nuclear equation of
state.  Conveniently, spin and finite-size effects can be both
systematically incorporated into any relativistic calculation in the
usual way, as highlighted by recent work in this direction
\cite{1709.00590, 1709.06016, 1805.10809, 1812.06895, 1812.00956, 1812.08752, 1906.09260, 1906.10071}.  The other obvious future direction is to include
dissipation effects.  On-shell radiation has been studied in terms of
concrete double-copy procedure~\cite{1603.05737, 1611.03493, 1705.09263, 1711.09493, 1712.08684, 1712.09250, 1803.02405, 1806.07388, 1903.12419} and also soft
theorems~\cite{1801.07719, 1804.09193, 1806.01872, 1808.03288, 1812.08137, 1906.08288, 1907.10021}, albeit the incorporation of effective
field theory has not yet been explored.

A surprising feature of our results for the 3PM amplitude, potential
and scattering angle is that they contain a term which is singular in
the small-mass limit, introducing a discontinuity with massless
results.  At first sight this might seem to violate the absence of
such mass singularities in the full quantum amplitude, but as we
explained in some detail in \sect{DiscussionSection}, they can appear
in the classical limit due to an interchange of limits issue.  A
rather intriguing prospect would be to understand the appearance of
the mass singularity well enough to predict its coefficient to any
order in the PM expansion.

Another aspect worthy of future study pertains to possible resummation
in Newton's constant $G$.  This is a natural question, given
that we have analytic expressions now for the 1PM, 2PM, and 3PM
potential, amplitude and scattering angle.  A closely related
question is whether the mass logarithms discussed in
\Sec{MassSingularitySubsection} can also be resummed in some way.  Given the
appearance of new structures at 3PM order it may be that a proper
attempt at resummation of the expansion in $G$ would require having at least the 4PM
result at hand.  In any event, we look forward to gaining a deeper
understanding of gravitational perturbation theory towards the goal of
improving predictions of gravitational radiation from compact
astrophysical sources.  It may turn out that the most useful and novel
aspect of the present work will be its potential for
unraveling the systematic structure of the PM expansion of general
relativity.

\section*{Acknowledgments}

We thank Alessandra Buonanno, Simon Caron-Huot, Thibault Damour, Paolo Di Vecchia,
Michael Enciso, David Kosower, Andr\'es Luna, Aneesh Manohar, Smadar
Naoz, Julio Parra-Martinez, Rafael Porto, Jan Steinhoff, George
Sterman, Pierre Vanhove, Gabriele Veneziano, Justin Vines, Mark Wise, and Zahra Zahraee for many
helpful discussions and encouragement.  In addition, we especially
thank Ira Rothstein for his many insightful comments throughout this
project.  Z.B. is supported by the U.S. Department of Energy (DOE)
under Award Number DE-SC0009937.  C.C.~is supported by the DOE under
grant no.~DE-SC0011632.  R.R.~is supported by the U.S. Department of
Energy (DOE) under grant no.~DE-SC0013699.  C.H.S.~is supported by the
Mani L. Bhaumik Institute for Theoretical Physics. M.P.S.~is supported
by the DOE under grant no.~DE-SC0011632 and the McCone Fellowship at
the Walter Burke Institute.  M.Z.~is supported by the Swiss National
Science Foundation under contract SNF200021 179016 and the European
Commission through the ERC grant pertQCD.

\newpage
\appendix
\section{Notations and Conventions}
\label{appendix:notation}

\vspace{-0.2cm}
We collect some notation here for easy reference.\\[3pt]
\renewcommand{\arraystretch}{1.04}
\begin{tabular}{ | l | l| }
  \hline		
  Signature &$(+,-,-,-)$ $\vphantom{\displaystyle{\big|}}$\\ 
    Spacetime dimension & $D = 4 -2\epsilon$  \\	
 Spacetime dimension for state counting & $D_s$  \\	
    Newton's constant & $G$  \\
  Color-ordered gauge-theory amplitude & $A$  \\
  Color-dressed gauge-theory amplitude & $\colorA{A}$  \\
  Gravitational amplitude italic & $M$  \\ 
  $n$PM amplitude & ${\cal M}_n $ \\
    $n$PM EFT amplitude & ${\cal M}_n^{\rm EFT} $ \\
Eight diagrams in Fig.~\ref{MasterDiagsFigure} & $\bf 1, \bf 2, ...$ \\
Diagrams related by crossing & $\overline{\bf 1}, \overline{\bf 2}, ...$ \\
    Relative position & $\bm r$ \\
    Relative velocity & $\bm v$ \\
    Impact parameter & $\bm b$ \\
    Large angular momentum & $J$ \\
  Mass of particle 1 & $m_1$  \\	
    Mass of particle 2 & $m_2$  \\	
  Incoming four-momentum of particle 1 & $(E_1, {\bm p})=-p^{\mu}_1$  \\	
  Incoming four-momentum of particle 2 & $(E_2, -{\bm p})=-p^{\mu}_2$  \\	
  Outgoing four-momentum of particle 1 & $(E_1, {\bm p}')=p^{\mu}_4$  \\	
  Outgoing four-momentum of particle 2 & $(E_2, -{\bm p}')=p^{\mu}_3$  \\		
Four-momentum transfer &$q^\mu = (0, \bm q) =(0, \bm p - \bm p')=(p_2+p_3)^{\mu}$ \\
Mandelstam variables & $s=(p_1+p_2)^2,\, u=(p_1+p_3)^2, \,t=(p_2+p_3)^2=-{\bm q}^2$ \\
  Total energy & $E = E_1+E_2$  \\
  Symmetric energy ratio &  $ \xi = E_1 E_2 / E^2$  \\
  Total mass & $m = m_1 + m_2$  \\ 
  Symmetric mass ratio & $\nu = m_1 m_2 /m^2$ \\
  Energy mass ratio & $\gamma = E/m$ \\
  Invariant product & $\sigma = p_1 \cdot p_2/m_1 m_2$ \\ 
  Full relativistic integrand & ${\cal I}$  \\	
    Spatial integrand & $\widetilde  {\cal I}$  \\	
   Matter field loop four-momentum & $k^\mu =(\varepsilon, \bm k)$ \\
   Graviton loop four-momentum & $\ell^\mu =(\omega, \bm \ell)$ \\
   Number of loops & $n_L$  \\
   Number of matter propagators & $n_M$  \\
   Number of graviton propagators & $n_G$  \\
      Integrand numerator & ${\cal N}$  \\
            Effective numerator & $\widetilde {\cal N}$  \\
            Matter pole & $\omega_P$  \\
            Antimatter pole & $\omega_A$  \\		
  Scattering angle & $\chi$ $\vphantom{\displaystyle{\big|}}$ \\
      \hline  
\end{tabular}

\section{Tree Amplitudes for Unitarity Cuts}
\label{TreeAmplitudesAppendix} 

The starting point for our construction of the 3PM two-body
Hamiltonian is gauge-theory amplitudes for a scalar coupled to
gluons.
In this appendix we give the four- and five-point gauge-theory tree amplitudes 
needed as input into the cut construction of gravity amplitudes through the KLT and double-copy relations.  
We do so for both the four-dimensional and for the dimensionally-regularized $D$-dimensional theories.

\subsection{Four-Dimensional Tree Amplitudes}

While we will, at times, use generalized unitarity cuts containing
three-point tree amplitudes, they have the disadvantage that they
contain a reference vector needed to define the helicity states.  As
noted in \sect{IntegrandsSection}, it is more convenient to instead
start with the rather clean forms of the four- and five-point trees
and generate cuts containing three-point tree amplitudes by taking
residues on the appropriate cut propagators. By doing so we bypass the
need of explicit three-point tree amplitudes.

The independent two-gluon two-scalar tree amplitudes needed in the KLT relations \eqref{KLT} are:
\begin{align}
A^\tree_4(1^s,2^+,3^+,4^s) & = i \frac{ m^2 \spb{2}.{3} }
  { \spa{2}.{3} \t_{12} }\, , \nonumber \\
A^\tree_4(1^s,2^+,3^-,4^s) & =
    i \frac{\sand{3}.{1}.{2}^2}{s_{23} \t_{12}}\,,
\label{SGGS}
\end{align}
where $\t_{12} = 2 p_1\cdot p_2 = s_{12}^2- m^2$, $s_{ij} = (p_i+p_j)^2$ and legs $1$ and
$2$ are massive scalar legs with mass $m$.  The `$\pm$' superscripts
refer to the helicities of the two gluons legs and the `$s$' superscript labels a scalar leg.  
The four-gluon amplitudes are~\cite{Parke:1986gb, Mangano:1987xk,hep-th/0509223, hep-ph/9601359}:
\begin{align}
A^\tree_4(1^-, 2^-, 3^+, 4^+) & =  i\, \frac{\spa{1}.{2}^4}{\spa{1}.{2}\spa{2}.{3} \spa{3}.{4} \spa{4}.{1}}  \,, \nonumber\\
A^\tree_4(1^-, 2^+, 3^-, 4^+) & =  i\, \frac{\spa{1}.{3}^4}{\spa{1}.{2}\spa{2}.{3} \spa{3}.{4} \spa{4}.{1}}  \,.
\label{GGGG}
\end{align}

The relevant tree-level helicity amplitudes at
five-points are $A(s,+,+,+,s)$, $A(s,+,+,-,s)$, and $A(s,+,-,+,s)$. While the
first one is already rendered in a compact form by Ref.~\cite{KhozeMassive}, the
other two contain undesirable spurious singularities that complicate loop
integration. Compact forms for all required five-point tree amplitudes are:
\begin{align}
A^\tree(1^s,2^+,3^+,4^+,5^s) &= -i\,\frac{m^2\, [4| 5 1 |2]}{\t_{12}\, \spa2.3\,\spa3.4\, \t_{45}} \,,\nn\\
A^\tree(1^s,2^+,3^+,4^-,5^s) &= i\,\frac{\t_{45}[34]\,\sandmm 4.{5 1}.4 \,\sand4.1.2
        +m^2\, \spb2.3 \spa3.4\left(\spb4.3 \sandmm4.{5 1}.4 - s_{23} \sand4.5.3\right)}
          {\t_{12}\, \spa2.3\, s_{34}\, \t_{45}\, s_{51}} \,, \nn\\
A^\tree(1^s,2^+,3^-,4^+,5^s) &= i\,
\frac{1}{\t_{12}\, s_{23}\, s_{34}\, \t_{45} \, s_{51}}
\biggl[ -[23][34]  \,\sand3.1.2 \, \sand 3.5.4 \,\sandmm 3.{5 1}.3
\label{SGGGS} 
  \\
& \null \hskip 4 cm 
     + m^2\, \spa2.3 \spa3.4 \spb2.4^2  
  \left(\spb3.4 \sand 3.1.2 - \spb2.4 \t_{45}\right) \biggr] \,,\nn
\end{align}
where $\t_{12}= 2p_1\cdot p_2=s_{12}-m^2$ and $\t_{45}= 2p_4\cdot
p_5=s_{45}-m^2$ are the inverse propagators and $p_{ij}= p_i+p_j$.
We use these expressions in \sects{OneLoopSection}{TwoLoopSection} to obtain compact forms for 
the four-dimensional generalized cuts of one- and two-loop four-scalar amplitudes. 
Here $m$ is the mass of the scalar legs 1 and 5. (The second amplitude in \eqn{SGGGS} corrects 
a relative sign compared to  Ref.~\cite{KhozeMassive}).

To use the KLT relations, we also need amplitudes where the massive scalars are in
different positions in the color ordering.  They can be obtained simply through the $U(1)$ decoupling identities,
\begin{equation}
A^\tree(1^s, 2, 3, 5^s, 4) = - A^\tree(1^s, 4, 2, 3, 5^s) - A^\tree(1^s, 2, 4, 3, 5^s) -
A^\tree(1^s, 2, 3, 4, 5^s) \, .
\end{equation}
which hold for any helicity configuration.  We will also use reflection symmetry
\begin{equation}
A^\tree(1^s, 2, 5^s, 4, 3) = - A^\tree(5^s, 2, 1^s, 3, 4)  \,,
\end{equation}
and cyclic symmetry
\begin{equation}
A^\tree(1^s, 2, 3, 5^s, 4) =  A^\tree(5^s, 4, 1^s, 2, 3)  \,.
\end{equation}

\subsection{$D$-Dimensional Tree Amplitudes}

\begin{figure}
\begin{center}
\includegraphics[scale=.5]{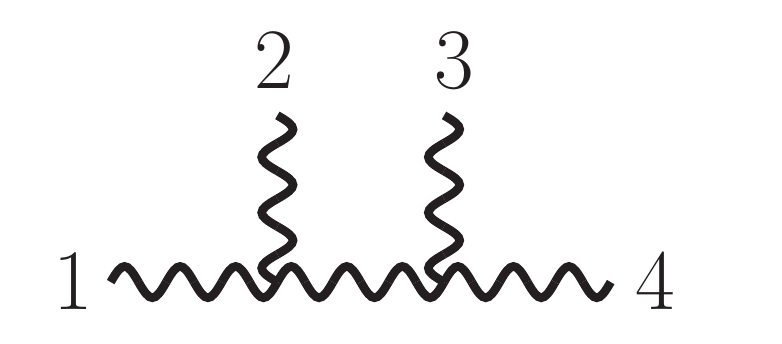}
\end{center}
\vskip -.5cm
\caption{\small The four-point cubic diagram whose numerator \eqref{s_num_nice} determines all other diagram numerators.}
\label{DDM4PtFigure}
\end{figure}

We will also need $D$-dimensional versions of tree amplitude in our
explicit checks that the four-dimensional trees are sufficient to capture the full classical conservative potential.
We express the $D$-dimensional amplitudes in terms of diagrams with only
cubic vertices.  Using BCJ duality, a simple way to specify all the
numerators is in terms of the so-called DDM basis~\cite{DelDuca:1999rs}, where only a single numerator is sufficient to generate the full amplitude.  For example, \Eq{ImprovedNumerators} yields the two scalar two gluon amplitude.
The BCJ numerator for the four-gluon amplitude corresponding to the diagram in \fig{DDM4PtFigure} is
\begin{align}
 n_s &= n(1,2,3,4)\nn \\
 & = \frac{i}{2}\, \Big\lbrace
(\pol_1 \cdot \pol_4) \left(
2(p_1 \cdot \pol_2)(p_{12} \cdot \pol_3) + 2 (p_{34} \cdot \pol_2)(p_4 \cdot \pol_3)
-s (\pol_2 \cdot \pol_3)
\right) \nonumber \\
&\quad
+ ( 4(\pol_3 \cdot \pol_4)(p_2 \cdot \pol_1)(p_{3} \cdot \pol_2)-4(\pol_3 \cdot \pol_4)
(p_3 \cdot \pol_1)(p_{12} \cdot \pol_2)-4(\pol_2 \cdot \pol_4)(p_2 \cdot \pol_1)(p_{12} \cdot \pol_3)) \nonumber \\
&\quad
+(4(\pol_2 \cdot \pol_1)(p_3 \cdot \pol_4)(p_{2} \cdot \pol_3)-4(\pol_2 \cdot \pol_1)
(p_2 \cdot \pol_4)(p_{34} \cdot \pol_3)-4(\pol_3 \cdot \pol_1)(p_3 \cdot \pol_4)(p_{34} \cdot \pol_2)) \nonumber \\
&\quad
+2s\, (\pol_1 \cdot \pol_3)(\pol_2 \cdot \pol_4)
+2u\, (\pol_1 \cdot \pol_2)(\pol_3 \cdot \pol_4)
-4(\pol_2 \cdot \pol_3)(p_2 \cdot \pol_1)(p_{3} \cdot \pol_4) \Big\rbrace \,.
\label{s_num_nice}
\end{align}
Note that under dimensional reduction in Eq.~\eqref{eq:reduction_1} and
Eq.~\eqref{eq:reduction_2}, this numerator reduces to $s$-channel
numerator in Eq.~\eqref{ImprovedNumerators}.  The $u$-channel
numerator is given by swapping $2\leftrightarrow3$,
\begin{equation}
n_u =  n(1,3,2,4) \,,
\end{equation}
and the $t$-channel is given by Jacobi identity \eqref{BCJDuality}
\begin{equation}
n_t =  n(1,2,3,4) - n(1,3,2,4) \,.
\end{equation}
In terms of these numerators, the four-point tree-level gauge-theory amplitude is
\begin{equation}
{\cal A}^\tree_4(1,2,3,4) = \frac{ n_s c_s}{s}+ \frac{n_t c_t}{t} 
+ \frac{n_u c_u}{u}\,,
\label{GaugeFourAmplitude}
\end{equation}
while the gravity amplitude is 
\begin{equation}
M^\tree_4(1,2,3,4) = i \Bigl(\frac{n_s^2}{s}+ \frac{n_t^2}{t} + \frac{n_u^2}{u}\Bigr)\,.
\label{GravitFourAmplitude}
\end{equation}

\begin{figure} \begin{center}
\includegraphics[scale=.5]{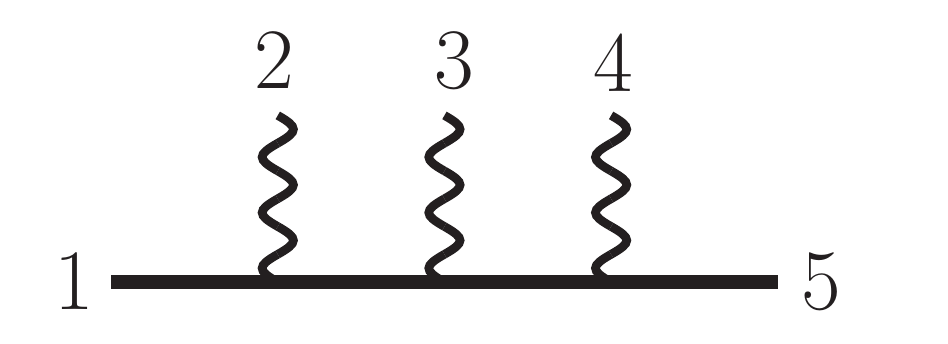}
\end{center}
\vskip -.5cm
\caption{\small The five-point diagram whose kinematic numerator determines all other diagram numerators.}
\label{DDM5ptFigure}
\end{figure}

We also need the $D$-dimensional two-scalar three-gluon tree amplitude.
Using again the DDM basis and BCJ duality, we need only specify the numerator corresponding to 
the diagram in \fig{DDM5ptFigure},
\begin{align}
n(1^s,2,3,4,5^s) = \frac{i}{\sqrt{8}} \,\Big\lbrace &
-4(p_1 \cdot \pol_2) (p_{12} \cdot \pol_3) (p_{123} \cdot \pol_4)
-2(\pol_2 \cdot \pol_3)(p_{12} \cdot p_3)\left((p_1-p_5+2p_2) \cdot \pol_4\right) \nonumber \\
&
+4(p_{345} \cdot \pol_2)(p_{45} \cdot \pol_3)(p_{5} \cdot \pol_4)
+2(\pol_3 \cdot \pol_4)(p_{12} \cdot p_3)\left((p_1-p_5-2p_4) \cdot \pol_2\right) \nonumber \\
&
+4(\pol_2 \cdot \pol_4)\left((p_{2} \cdot \pol_3) (p_{4} \cdot p_5)-(p_{4} \cdot \pol_3) (p_{1} \cdot p_2) \right) \nonumber \\
&
+(\pol_2 \cdot \pol_4)((p_{1}-p_5) \cdot \pol_3)(p^2_{12}+p^2_{45}-2m^2)
\Big\rbrace,
\label{eq:5pt_num_nice}
\end{align}
where $p_{ijk}=p_i+p_j+p_k$.
The propagators and color factors can be read from the diagrams. The
diagrams where the scalars are held fixed and the gluon legs are
permuted are obtained simply by relabeling the gluon legs $2,3,4$
in \eqn{eq:5pt_num_nice}.  The remaining 9 diagram, for a total of 15,
are obtained by using the BCJ identities for the kinematic numerator factors $n$.  As usual,
the full amplitudes are assembled using
Eqs.\eqref{CubicRepresentation} or \eqref{DoubleCopy}.

The $D$-dimensional tree-level amplitudes with two scalar lines
presented here have the useful property that the on-shell Ward
identities,
\begin{equation}
 \mathcal{A}|_{\pol_i \rightarrow p_i} \rightarrow 0 \,,
\label{GeneralizedWardIdentity}
\end{equation}
hold for the gluon or graviton $i=2,3,\dotsc n-1$ without using
transverse property on other legs.  Whenever the Ward identity holds
without using the transverse properties on any other legs, we will
call this a ``generalized Ward identity''.
An amplitude with this property is obtained by appropriately adding
terms that vanish due to on-shell conditions to a generic form that
amplitude. The existence of such amplitudes has been exploited in
the calculations of Refs.~\cite{3PMPRL,Paolo2PM} to help simplify
$D$-dimensional constructions of the integrand.  We use this property
in the calculations in \sects{OneLoopDCutsSection}{TwoLoopSection} to
simplify physical-state projectors that appear in $D$-dimensional unitarity cuts.

\section{Series Resummation}
\label{appendix:series}
We collect here the series encountered in \Eq{eq:PN_graphs2} and their resummation. We have the simple geometric series and its square:
\eq{
&1 + \frac{\bm p^2}{E_2^2}    + \frac{\bm p^4}{E_2^4}    + \frac{\bm p^6}{E_2^6}  +\cdots   \equiv  
{E_2^2 \over m_2^2}  \,, \\[3pt]
  &1 + \frac{2\bm p^2}{E_2^2}    + \frac{3 \bm p^4}{E_2^4}    + \frac{4 \bm p^6}{E_2^6}  +\cdots  
  \equiv  
 {E_2^4 \over m_2^4} \,,  \\[3pt]
    &1 - \frac{\bm p^2}{3E_2^2} - \frac{5\bm p^4}{3E_2^4} - \frac{3\bm p^6}{E_2^6}+ \cdots  \equiv 
   {7 E_2^2 \over 3m_2^2}  - {4E_2^4 \over 3 m_2^4} \,, 
   }{}
   where the third series is a linear combination of the first two. For the resummation of the IR pieces we encountered the series
      \eq{
   &1 + \frac{\bm p^2}{2E_2^2}    + \frac{3\bm p^4}{8E_2^4}    + \frac{5\bm p^6}{16E_2^6}+ \cdots   \equiv  
{E_2 \over m_2}  \,, \\
   &1 + \frac{3 \bm p^2}{2E_2^2}    + \frac{15\bm p^4}{8E_2^4}    + \frac{35\bm p^6}{16E_2^6}   + \cdots   \equiv  
{E_2^3 \over m_2^3}  \,,
}{}
where the first one is the square root of the geometric series above
and is also encountered in the box-triangle example in
\sect{sec:BTdiagram}.
   
We have series that involve the expansion of square roots and hence binomial coefficients, which we denote here as ${\rm B}[m,n]$:
   \eq{
     &1 + {5 \bm p^2 \over 4 E_2^2 } + {11 \bm p^4 \over 8 E_2^4} + {93  \bm p^6 \over 64 E_2^6} + \cdots  \equiv 2\sum_{n=0}^\infty \left( 1 + (-1)^n{\rm B}[ -1/2, n+1]   \right) { \bm p^{2n} \over E_2^{2n} }  = {2E_2^3 (E_2 - m_2) \over m_2^2 \bm p^2}\, , \\[3pt]
 &1 + {7 \bm p^2 \over 2 E_2^2} + {51 \bm p^4 \over 8 E_2^4} + {151 \bm p^6 \over 16 E_2^6} + \cdots \equiv  \sum_{n=0}^\infty  \left(  4(n+1)  -3 (-1)^n {\rm B}[ -3/2,n] \right){\bm p^{2n} \over E_2^{2n} }   ={ E_2^3 (4E_2 - 3m_2) \over m_2^4}  \, .
   }{}
The first series above appears in the example of the two-loop scalar box-triangle integral.

Finally we have series that resum to ${\rm arcsinh} {|\bm p| \over m_{i}}$, which is the rapidity of particle $i=1$ or 2: 
\eq{
   &1 + {\left(E_1^3+ E_2^3 \right) \over 3 E_1^2 E_2^2 E } {\bm p}^2  + \frac{\left(E_1^5+E_2^5\right)}{5 E_1^4 E_2^4 E} {\bm p}^4 + \frac{\left(E_1^7+E_2^7\right)}{7 E_1^6 E_2^6 E} {\bm p}^6 + \cdots  \equiv {1 \over E} \sum_{n=0}^\infty {(E_1^{2n+1} + E_2^{2n+1} )  \bm p^{2n}  \over (2n+1)  E_1^{2n} E_2^{2n}}  \\
   & \hspace{9.75cm} = {E_1E_2  \left[   {\rm arcsinh}  {|\bm p| \over m_1}  +  {\rm arcsinh}{|\bm p| \over m_2}  \right]  \over  E |\bm p| } \, , \\[3pt]
  &1 + {\left(E_1^3+ E_2^3 \right) \over 5 E_1^2 E_2^2 E } {\bm p}^2  + \frac{3\left(E_1^5+E_2^5\right)}{35 E_1^4 E_2^4 E} {\bm p}^4 + \cdots \equiv  {3 \over E} \sum_{n=0}^\infty {(E_1^{2n+1} + E_2^{2n+1} )  \bm p^{2n}  \over (2n+1) (2n+3)  E_1^{2n} E_2^{2n}}  \\
  & \hspace{7.15cm} =  {3E_1 E_2 \over 2 \bm p^2}\left[ 1 +  {  m_1^2 \, {\rm arcsinh} {|\bm p| \over m_1} +  m_2^2 \, {\rm arcsinh}  {|\bm p| \over m_2}   \over   E |\bm p|  } \right] \,, \\[3pt]
 &1 + \frac{3\bm p^2}{5E_1^2}    + \frac{3\bm p^4}{7E_1^4}     +\cdots  \equiv 3\sum_{n=0}^\infty {\bm p^{2n} \over (2n+3) E_1^{2n} }  =  -{3 E_1^2 \over \bm p^2}  + {3 E_1^3 {\rm arcsinh} {|\bm p| \over m_1}  \over |\bm p|^3 } \,, \\[3pt]
    &1 + \frac{6\bm p^2}{5E_1^2}    + \frac{9\bm p^4}{7E_1^4}     +\cdots \equiv3\sum_{n=0}^\infty {(n+1) \bm p^{2n} \over (2n+3) E_1^{2n} }  = {3 E_1^4 \over 2m_1^2 \bm p^2}  - {3 E_1^3 {\rm arcsinh} {|\bm p| \over m_1}  \over 2 |\bm p|^3 } \,.
}{}
The first series above appears in the example of the two loop scalar flipped double triangle and scalar H integrals.

Note that a few of the series shown in this appendix are related to each other by the operator ${d \over dx}  x ( \cdot )$, where $x$ is the small parameter. Moreover, notice that the resummation often yields factors of $E_i$ that cancel with those multiplying the series in \Eq{eq:PN_graphs2}. In particular, combining those factors into the series and then expanding in large $m_i$ can lead to even simpler series.

\section{Evaluation of H Diagram and Imaginary Part}
\label{HxH}

We extract the small-$t$ limit of the H and ${\overline {\rm H}}$ integral
which has been calculated for the fully quantum case in
Ref.~\cite{Bianchi:2016yiq} for the case of equal masses (motivated by
massive-quark scattering). We start from results in the Euclidean
region in the aforementioned reference, written in terms of harmonic
polylogarithms, and with different notations for the kinematic
variables and perform the necessary analytic continuation. The
material here differs from \sect{sec:diffEq} in two respects: (1) here
we do not give results for tensor integrals; (2) here we present the
full $\ln(-t)$ terms of the H integral and ${\overline {\rm H}}$
integral \emph{separately}, without adding them together to localize
on matter poles, and without dropping imaginary contributions.
This gives us an important confirmation on our techniques, because
the classical limit is taken only at the end, after all loop integrations
are carried out in a fully relativistic fashion.

The $\ln(-t)$ term of the scalar H integral in the small-$t$ limit,
starting from Eq.~(4.1) of Ref.~\cite{Bianchi:2016yiq}, is
\begin{equation}
I_{\rm H} \big|_{\ln(-t)} = - \frac{1}{192 \pi^4} \frac{1} {m^2  t^2} \frac{1}{\sqrt{\sigma^2 -1}}
 \operatorname{arcsinh} \sqrt{\frac{\sigma -1}{2}}
\left[ 4 \operatorname{arcsinh}^2\sqrt{\frac{\sigma -1}{2}}
 - 6 i \pi \operatorname{arcsinh}\sqrt{\frac{\sigma -1}{2}} - 2\pi^2 \right] ,
\label{eq:resultHcont}
\end{equation}
where we have divided by
$(-256\pi^4)$ to account for the difference in normalization
compared to the one used here and $\sigma$ is defined  in \eqn{eq:sigma_def}, taking $m_1 = m_2 = m$.
The result for the
scalar ${\overline {\rm H}}$ integral is
\begin{equation}
I_{\overline {\rm H}} \big|_{\ln(-t)} = \frac{1}{192 \pi^4} \frac {1} {m^2 t^2}\frac{1}{\sqrt{\sigma^2 -1}}
 \operatorname{arcsinh} \sqrt{\frac{\sigma -1}{2}} 
\left[ 4 \operatorname{arcsinh}^2 \sqrt{\frac{\sigma -1}{2}}  + \pi^2 \right] .
\label{eq:resultXH}
\end{equation}
Adding Eq.~\eqref{eq:resultHcont} and Eq.~\eqref{eq:resultXH} for the
H and ${\overline {\rm H}}$ diagram, the $\log(-t)$ coefficient in the sum is
\begin{equation}
\left( I_{\rm H} + I_{\overline {\rm H}} \right) \big|_{\log (-t)} = \frac{1}{64 \pi^3}
\frac{1}{m^2 t^2} \frac{1}{\sqrt{\sigma^2 -1}} \operatorname{arcsinh} \sqrt{\frac{\sigma -1}{2}} 
\left[ \pi + 2i \operatorname{arcsinh} \sqrt{\frac{\sigma -1}{2}} \right] .
\label{eq:HplusXH}
\end{equation}
The real part is in perfect agreement with our previous results
in \eqns{eq:Hresummed}{eq:resultI1} from expansions around the
potential region, after setting $m_1 = m_2 = m$ (and accounting for
the $1/t^2$ extracted from the earlier expressions).
Eq.~\eqref{eq:HplusXH} also has a nonzero imaginary part as expected,
corresponding to the possibility that the internal graviton propagator
not attached to the matter lines goes on shell.  This term, however,
does not contribute to the conservative classical potential.  The
agreement of the real part of the $\log(-t)$ coefficient with our
result is a direct confirmation of the appearance of a mass
singularity in the classical part of this integral, starting with a
full quantum evaluation.



\end{document}